\documentclass[longauth]{aa}  

\usepackage{graphicx}
\usepackage{xcolor}
\usepackage{float}
\usepackage{chngcntr}
\usepackage{textcomp}

\usepackage[colorlinks]{hyperref}  
\hypersetup{citecolor=blue} 

\usepackage{txfonts}
\begin{document}

   \title{Investigating 39 Galactic Wolf-Rayet stars with VLTI/GRAVITY\thanks{Based on observations collected at the European Southern Observatory under ESO program IDs 111.24JN (PI: H. Sana) and 109.23CN (PI: I. Waisberg).}}

   \subtitle{Uncovering A Long Period Binary Desert}

    \author{K. Deshmukh \inst{\ref{inst:kul},\ref{inst:lgi}} \and 
          H. Sana \inst{\ref{inst:kul},\ref{inst:lgi}}  \and 
          A. Mérand \inst{\ref{inst:eso}} \and
          E. Bordier \inst{\ref{inst:koln}} \and 
          N. Langer \inst{\ref{inst:aifa},\ref{inst:mpifr}}\and
          J. Bodensteiner \inst{\ref{inst:eso}} \and 
          K. Dsilva \inst{\ref{inst:ulb}} \and 
          A. J. Frost \inst{\ref{inst:esoc}} \and \\
          E. Gosset \inst{\ref{inst:liege}} \and 
          J. -B. Le Bouquin \inst{\ref{inst:gren}} \and 
          R. R. Lefever \inst{\ref{inst:zah}} \and
          L. Mahy \inst{\ref{inst:rob}} \and 
          L. R. Patrick \inst{\ref{inst:madrid}} \and
          M. Reggiani \inst{\ref{inst:kul}} \and \\
          A. A. C. Sander \inst{\ref{inst:zah}} \and 
          T. Shenar \inst{\ref{inst:tau}} \and 
          F. Tramper \inst{\ref{inst:madrid}} \and 
          J. I. Villaseñor \inst{\ref{inst:mpia}} \and
          I. Waisberg \inst{\ref{inst:ind},\ref{inst:wis}}
          }

   \institute{{Institute of Astronomy, KU Leuven, Celestijnlaan 200D, 3001 Leuven, Belgium \label{inst:kul}}\\ 
              \email{kunalprashant.deshmukh@kuleuven.be}
        \and 
             {Leuven Gravity Institute, KU Leuven, Celestijnenlaan 200D, box 2415
3001 Leuven, Belgium \label{inst:lgi}}
        \and 
             {European Southern Observatory, Karl-Schwarzschild-Straße 2, 85748 Garching, Germany \label{inst:eso}}
        \and 
             {I. Physikalisches Institut, Universität zu Köln, Zülpicher Str. 77, 50937, Köln, Germany \label{inst:koln}}
        \and 
            {Argelander Institut für Astronomie, Auf dem Hügel 71, DE-53121 Bonn, Germany \label{inst:aifa}}
        \and
            {Max-Planck-Institut für Radioastronomie, Auf dem Hügel 69, DE-53121 Bonn, Germany \label{inst:mpifr}}
        \and 
            {Institut d’Astronomie et d’Astrophysique, Université Libre de Bruxelles CP 226, Boulevard du Triomphe, 1050 Brussels, Belgium \label{inst:ulb}}
        \and 
            {European Southern Observatory, Santiago, Chile \label{inst:esoc}}
        \and 
            {Groupe d'Astrophysique des Hautes Energies, STAR, Université de Liège, Quartier Agora (B5c, Institut d'Astrophysique et de Géophysique), Allée du 6 Août 19c, B-4000 Sart Tilman, Liège, Belgium \label{inst:liege}}
        \and 
            {Univ. Grenoble Alpes, CNRS, IPAG, 38000 Grenoble, France \label{inst:gren}}
        \and 
            {Zentrum für Astronomie der Universität Heidelberg, Astronomisches Rechen-Institut, Mönchhofstr. 12-14, 69120 Heidelberg, Germany \label{inst:zah}}
        \and 
            {Royal Observatory of Belgium, B-1180 Brussels, Belgium \label{inst:rob}}
        \and 
            {Departamento de Astrofísica, Centro de Astrobiología, (CSIC-INTA), Ctra. Torrejón a Ajalvir, km 4, 28850 Torrejón de Ardoz, Madrid, Spain \label{inst:madrid}}
        \and 
            {The School of Physics and Astronomy, Tel Aviv University, Tel Aviv 6997801, Israel \label{inst:tau}}
        \and 
            {Max-Planck-Institut für Astronomie, Königstuhl 17, D-69117 Heidelberg, Germany \label{inst:mpia}}
        \and
            {Independent Researcher \label{inst:ind}}
        \and
            {Department of Particle Physics and Astrophysics, Weizmann Institute of Science, Rehovot 76100, Israel \label{inst:wis}}
             }

   \date{Received; Accepted}

 
  \abstract
   {Wolf-Rayet stars (WRs) represent \text{one of} the final evolutionary stages of massive stars and are thought to be immediate progenitors of stellar-mass black holes. Their multiplicity characteristics form an important anchor point in single and binary population models for predicting gravitational-wave progenitors. Recent spectroscopic campaigns have suggested incompatible multiplicity fractions and period distributions for N- and C-rich Galactic WRs (WNs and WCs) at short as well as long orbital periods, in contradiction with evolutionary model predictions.}
   {In this work, we employed long-baseline infrared interferometry to investigate the multiplicity of WRs at long periods and explore the nature of their companions. We present a magnitude-limited ($K<9$; $V<14$) survey of 39 Galactic WRs, including 11 WN, 15 WC and 13 H-rich WN (WNh) stars.}
   {We used the $K$-band instrument GRAVITY at the Very Large Telescope Interferometer (VLTI) in Chile. The sensitivity of GRAVITY at spatial scales of $\sim$1--200 milliarcseconds and flux contrast of $1\%$ allowed an exploration of periods in the range $10^{2}-10^{5}$ d and companions down to $\sim$5$\,M_\odot$. We carried out a companion search for all our targets, with the aim of either finding wide companions or calculating detection limits. We also explored the rich GRAVITY dataset beyond a multiplicity search to look for other interesting properties of the WR sample.}
   {We detected wide companions with VLTI/GRAVITY for only four stars in our sample: WR~48, WR~89, WR~93 and WR~115. Combining our results with spectroscopic studies, we arrived at observed multiplicity fractions of $f^{\rm WN}_{\rm obs} = 0.55\pm0.15$, $f^{\rm WC}_{\rm obs} = 0.40\pm0.13$ and $f^{\rm WNh}_{\rm obs} = 0.23\pm0.12$. The multiplicity fractions and period distributions of WNs and WCs are consistent in our sample. For single WRs, we placed upper limits on the mass of potential companions down to $\sim$5$\,M_\odot$ for WNs and WCs, and $\sim$7$\,M_\odot$ for WNh stars. In addition, we also found other features in the GRAVITY dataset such as (i) a diffuse extended component contributing significantly to the $K$-band flux in over half the WR sample; (ii) five known spectroscopic binaries resolved in differential phase data, essentially constituting an alternate detection method for close binaries; and (iii) spatially resolved winds in four stars: WR~16, WR~31a, WR~78 and WR~110.}
   {Our survey reveals a lack of intermediate (few 100s~d) and long- (few years to decades) period WR systems. The 200-d peak in the period distributions of WR+OB and BH+OB binaries predicted by Case B mass-transfer binary evolution models is not seen in our data. The rich companionship of their O-type progenitors in this separation range suggests that the WR progenitor stars expand and interact with their companions, most likely through unstable mass-transfer, resulting in either a short-period system or a merger.}


   \titlerunning{Interferometry of Wolf-Rayet Stars}
   
   \maketitle
%

\section{Introduction}

Massive stars with initial masses greater than $8\, M_\odot$ live eventful lives that often conclude with powerful explosions and leave behind a compact object such as a neutron star or stellar-mass black hole \citep{2002Woosley}. They enrich their surroundings with gas, dust, nucleosynthesis products and ionizing radiation, strongly influencing the local and galactic environment through their lives and deaths \citep{1997Haiman,2013Nomoto,2014Hopkins}. Most massive stars reside in binary/multiple star systems and undergo interaction with a companion in the form of mass transfer or merger \citep{2012Sana, 2014Sana, 2017Moe}. The multiplicity of massive stars therefore plays an important role in their evolution \citep{2012Langer,2023Marchant}. Understanding the multiplicity properties of massive stars at different evolutionary stages can provide strong constraints on the nature of past interactions and predictions of their future evolution.

Wolf-Rayet stars (WRs) represent one of the final stages of stars with initial masses more than $\sim$$20\,M_\odot$, directly preceding black hole formation \citep{1976Conti,2021Higgins,2022Pauli,2023Aguilera}. Observationally, WRs are characterized by strong, broad emission lines in their spectra originating from strong optically thick winds. A large fraction of WRs are hydrogen-free, and are termed classical WRs or cWRs. They are classified as nitrogen-rich (WN), carbon-rich (WC) or, very rarely, oxygen-rich (WO) based on the dominating spectral lines, in particular the emission line strengths and their ratios \citep{1968aSmith,2007Crowther}. WN stars are often subdivided into early- and late-type WNs, namely WNE (earlier than WN6) and WNL (WN6 and later), respectively. Another sub-type of Wolf-Rayet stars are the hydrogen-rich WN stars. A suffix of 'h' was added to the spectral classification to denote the presence of hydrogen in WR spectra \citep{1996Smith}.  They tend to be similar to late-type WN stars, and show a continuity of spectral properties to very massive O stars. Unlike cWRs, a large fraction of WNh stars are likely very massive main-sequence stars that have strong winds that give them a WR-like spectroscopic appearance \citep{2007Crowther}. In this work, we refer to and treat WNh stars separately from cWRs.

Classical WRs are evolved, post main-sequence objects (typically core He-burning) that have been stripped of their hydrogen envelope. It has long been thought that this stripping happens either through stellar winds \citep{1976Conti,1987Abbott,2014Smith} or via interaction with a close companion \citep{1967Paczynski,1998Vanbeveren}. In the single-star channel, also referred to as the modified Conti scenario, the formation and evolution of WRs is thought to follow the path O $\rightarrow$ RSG/LBV $\rightarrow$ WN  $\rightarrow$ WC ($\rightarrow$ WO in rare cases), where RSG and LBV stand for red supergiants and luminous blue variables. Despite large uncertainties, single-star evolution models generally cannot reproduce the luminosity distribution of observed WR populations, suggesting that the binary channel plays a significant role in producing WRs \citep{2018Schootemeijer,2019Shenar,2019Hamann}. The contribution of each of the two scenarios to the Galactic WR population remains to be determined \citep{2014Neugent,2019Shenar,2020Shenar,2024Schootemeijer}. Understanding the multiplicity of massive stars in the Wolf-Rayet phase can help anchor the role of binary evolution in the formation of WRs.

Over the years, many studies have addressed the multiplicity of WRs in the Galaxy \citep{1980VanbeverenConti,2001vanderHucht}. The latter studied a sample of 227 WRs and arrived at an observed binary fraction of 0.4 based on photometry, spectroscopy, visual binaries and dilution of emission lines due to a companion. The heterogeneity of the instruments and methods used was a limiting factor for appropriate bias correction and subsequent derivation of the intrinsic binary/companion fraction. Presently, the Galactic WR Catalog v1.30\footnote{\url{ http://pacrowther.staff.shef.ac.uk/WRcat/}} \citep[originally Appendix 1 in][]{2015RossloweCrowther} has 679 stars. Taking a more systematic approach than previous studies, \citet{kd1,kd2,kd3} considered a homogeneous, magnitude-limited sample of 39 Galactic WRs (27 WNs and 12 WCs, where WNs also include H-rich WNs) to establish their intrinsic binary fraction. In particular, they used spectroscopic data to determine radial velocity variation in WRs and employed well understood bias correction to derive their intrinsic binary fraction.

WN binaries were found to have a period distribution dominated at short periods (P < 100 d), similar to that of the OB stars \citep{kd2,kd3}. This is not compatible with the intermediate RSG/LBV phase that can have radii large enough to initiate interactions \citep{2022Mahy}, while common envelope transition towards short-period WNs can be excluded from theoretical considerations. \citet{kd1} found an observed binary fraction of $0.58\pm0.14$ for WCs, while \citet{kd3} found $0.41\pm0.09$ for WNs (1$\sigma$ errors from binomial statistics). Based on a Bayesian inference analysis of the additional radial velocity (RV) noise, they predicted the period distributions for WN and WC stars to differ significantly, with WNs skewed towards shorter periods and WCs skewed towards longer periods \citep{kd3}. They also predicted a higher intrinsic binary fraction for WCs ($0.96^{+0.04}_{-0.22}$) as compared to WNs ($0.52^{+0.14}_{-0.12}$), although the difference between the two could be originating from small number statistics. These results are based on modeling radial velocity variations seen in the WR spectral lines, which are heavily affected by wind variability. This confines the ability of spectroscopy to only detect binaries with short periods, high mass ratios and high inclinations. The binary detection probability drops at large orbital periods (P > 1 year) and the Bayesian framework developed for the bias correction at long periods is subject to assumptions on the shape of the period and mass-ratio distributions \citep{kd2,kd3}. With optical long-baseline interferometry, it is possible to detect binaries with longer periods that are usually beyond the reach of spectroscopy. Interferometry has been previously used to study hot dust in WRs \citep{2010Rajagopal}, wind collisions in WR+O star binaries \citep{2017Lamberts} and determining orbits of such binaries \citep{2011Monnier,2021Richardson,2024Richardson}. Interferometry can also serve as a powerful and less time-consuming complementary method to search for new binary companions at long periods \citep[e.g.][]{2010Rajagopal,2016Richardson}, constraining the multiplicity better and resolving the potential period distribution discrepancy between WNs and WCs.

The evolutionary connection between WNs and WCs is thought to be following the path WNL $\rightarrow$ WNE $\rightarrow$  WC. This evolution is a direct consequence of the strong winds of WRs that continuously strip the outer layers at high mass-loss rates, revealing the products of hydrogen burning (WNs) and helium burning (WCs) \citep{1943Gamow, 1991Lamers}. The orbital period of WR binaries changes during this phase due to a combination of mass loss and angular momentum loss via winds, the former increasing the period by a factor of at most $\sim$2 \citep{kd3}. Such an increase is not sufficient to explain the predicted period distribution discrepancy of WNs and WCs, indicating that WC binaries should preferably evolve from WN binaries with longer orbital periods, and the short-period WN binaries avoid evolving into WC binaries altogether. While the latter is a problem that is tackled best by spectroscopy, the binarity at longer periods can be better studied with interferometry.

With an aim to investigate the multiplicity of WRs at long periods, we turned to the Very Large Telescope Interferometer (VLTI) operated by the European Southern Observatory (ESO). Equipped with several infrared instruments, the VLTI is capable of resolving objects with angular separations in the range of $\sim$1-200 milliarcseconds (mas). For a typical $\sim$2~kpc distance for Galactic WRs, this corresponds to orbital periods in the range $10^2 - 10^5$ days, extending the period parameter space well beyond that of spectroscopic measurements ($< 10^{3}$ days). A census of WR multiplicity at long periods is an important step towards understanding not only the origin of such systems, but also their evolution into potential compact object binaries. 

In this paper, we present our results on the multiplicity of 39 Galactic WRs based on interferometric observations with the VLTI, and discuss their implications on the formation and evolution of WRs. The paper is organized as follows: Section \ref{sec:obs} describes the observations and data reduction of our WR sample, followed by Section \ref{sec:data} detailing our data analysis approach. In Section \ref{sec:results}, we describe the results of our analysis, including detected binaries as well as other features found in the dataset. Section \ref{sec:fbin} focuses on the multiplicity of our sample as well as the limitations of our observations. Finally, we present the possible interpretations of our results and the conclusion in Section \ref{sec:evol}.

\section{Sample, Observations and Data Reduction}
\label{sec:obs}

We chose VLTI/GRAVITY + Auxiliary Telescopes (ATs) as our instrument to investigate the multiplicity of WRs, owing to the unique spectro-interferometric capabilities of the instrument in the $K$-band (1.95 -- 2.45 {\textmu}m). We selected all WRs with $K$ < 9 visible during the observing run, and excluded targets that were previously proposed/observed with GRAVITY. An exception to this was WR~48, which has been observed with GRAVITY before and is a candidate triple system. This added up to a total of 49 WRs - 23 WCs and 26 WNs+WNhs. 

A total of 39 WRs out of the requested 49 were observed. Observations for the 10 remaining targets were hampered owing to technical issues or poor observing conditions. Additionally, of the WRs excluded from our proposal, archival GRAVITY data for 4 WRs were also included in our sample for this work (1 WC and 3 WN+WNh stars). All data were obtained in the medium ($R\sim500$) or high ($R\sim4000$) resolution mode of GRAVITY, depending on the $K$-band magnitudes of the WRs (high resolution for brighter targets).

We performed the data reduction and visibility calibration using the standard GRAVITY Pipeline v1.6.0 \citep{2014Lapeyrere}. The calibrator stars selected for program 111.24JN were FIII stars within a $2^{\circ}$ radius and $\Delta K < 2$, ideally brighter than the WR. For the targets from the archival program 109.23CN, the calibrators chosen were early-type A dwarf stars. A total of 4 targets in our sample were not reduced successfully because of bad data quality. To summarise, a total of 14 targets were excluded from our analysis due to issues with observations/data reduction, comprising of 9 WC stars and 5 WN+WNh stars. Nevertheless, our sample still covers all spectral types of WRs uniformly, as listed in Table\,\ref{tab:summ}. 

Our final reduced sample was hence comprised of 39 WRs. This also included 4 stars analysed by \citet{kd1,kd2,kd3} as part of their study. A full list of targets along with their corresponding calibration stars, times of observation and spectral resolution used are summarised in the Appendix (Table \ref{tab:allobs}). The data quality varied significantly across the sample. The last column of Table \ref{tab:summ} lists a qualitative data evaluation for every target as good, average or poor based on visual inspection and signal-to-noise ratio (S/N) considerations. Although not quantified, this classification is meant to convey the overall quality of observations.

\section{Data Analysis}
\label{sec:data}

The GRAVITY instrument provides a $K$-band spectrum of the target as well as interferometric observables such as visibility amplitude (|V|), closure phases (T3PHI) and differential phases (DPHI). We used all the available data types to qualitatively assess every system. A brief description of this qualitative approach is discussed in Appendix \ref{sec:qual}. This was followed by geometric modeling to look for any resolved companions or other features. PMOIRED\footnote{\url{https://github.com/amerand/PMOIRED}} \citep{2022SPIE12183E..1NM} was our primary analysis tool for modeling the spectro-interferometric data. PMOIRED enables parametric modeling where one can add individual components such as circular and elliptical uniform disks, Gaussian point sources, and many more complex structures, and fit such models to the data given some prior parameter values or grids. We also used the telluric correction feature in the module (based on a grid of atmospheric spectra generated from molecfit, \citealp{2015A&A...576A..77S}) to obtain the continuum normalized flux (NFLUX) for every target. 

\begin{figure}
    \includegraphics[width=\columnwidth]{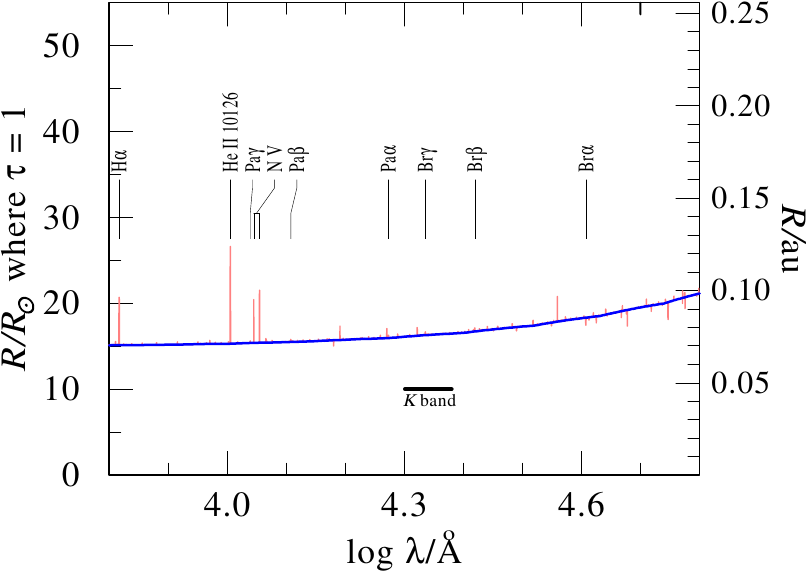}
    \caption{Radius in solar radii where the optical depth is equal unity based on a PoWR atmosphere model for WR\,78. The blue curve only takes the continuum opacity into account while the light red curve also accounts for the line opacities.}
    \label{fig:wr78rtau1}
\end{figure}

WRs typically have strong optically thick winds that can extend up to several stellar radii. In the $K$-band, the Potsdam Wolf-Rayet Models (PoWR) predict the extent to be few tens of solar radii. An example for WR\,78 is given in Fig.\,\ref{fig:wr78rtau1}, illustrating that in the $K$-band the continuum opacity already starts to extend to slightly larger radii than at optical wavelengths. Lines can be optically thick to much larger radii, but as evident from Fig.\,\ref{fig:wr78rtau1} this effect is not prominent in the $K$-band. Considering that WRs in our sample are typically $\sim$2 kpc away, this extent corresponds to an angular size much less than 1 mas, and thus mostly unresolvable for the VLTI. Consequently, our initial expectation was for all WRs to be approximately point sources. Nevertheless, we used a Gaussian component to model the WR and left the full width at half maximum (fwhm) as a free parameter in case the WR winds were more extended than expected. To search for potential resolved companions to WRs, we assumed main-sequence stars which can be treated as point sources.

Some WRs are also known dust producers, which is apparent through the additional flux contribution of dust seen in infrared photometry. The spatial scales for such dusty regions around WRs can reach several thousands of au, making them resolvable with the VLTI. In our sample of 39 WRs, only one star is a known dust producer (WR~113; WC8d+O8-9IV). Nevertheless, such emission in the context of GRAVITY data can be factored in as a fully resolved flux spanning the whole field of view. Such an emission component was included agnostically for all targets and ignored in case it did not contribute significantly.

The $K$-band spectra for WRs are characterised by emission lines from certain elements corresponding to their spectral type, which are thoroughly discussed in literature \citep{1997Figer, 2006Crowther, 2018Rosslowe, 2018Clark}. For WN stars the Br$\gamma$ line (often blended with \ion{He}{i} and \ion{He}{ii}) is almost always present in emission, along with a component of the \ion{He}{i} + \ion{N}{iii}\,2.11 {\textmu}m and the \ion{He}{ii}\,2.189 {\textmu}m features. WC stars, on the other hand, typically show the \ion{C}{iv}\,2.07-2.08 {\textmu}m and \ion{C}{iii}\,2.114 {\textmu}m (blended with \ion{He}{i}\,2.113 {\textmu}m) features. Some WCs also show a \ion{C}{iv} emission line around 2.43 {\textmu}m \citep{2018Clark}. Additionally, WRs show a further variety of spectral lines varying from star to star depending on the composition and velocities of their winds, sometimes even showing different spectral features for the same spectral type \citep{2021Clark}. Nevertheless, we mainly focus on the multiplicity aspect of WRs in our sample in this work, and only comment on the spectral lines wherever necessary.

As described in Appendix \ref{sec:qual} and shown in Fig.\,\ref{fig:int-basics}, the presence of spectral lines may or may not manifest itself in |V| data depending on what components the system is made up of. Moreover, fitting |V| and T3PHI data alone is typically sufficient to model the presence (or absence) of a resolved companion. We therefore used |V|-T3PHI data for companion search assuming flat spectra, and only invoked NFLUX and DPHI for targets showing spectral features in any of |V|, T3PHI or DPHI data. 

We converged on the final model for every WR using reduced chi-squared ($\chi^{2}_{\rm red}$) minimisation in PMOIRED. We examined the data visually first to look for any obvious signatures of a companion. This was followed by finding best-fit models without and with a companion to look for any significant improvement in the $\chi^{2}_{\rm red}$ value of the latter. When including NFLUX and DPHI data, we only modeled the most prominent spectral features that were sufficient to explain the data. A side-effect of this was much higher values for $\chi^{2}_{\rm red}$ due to unmodeled features, particularly for targets observed in the high resolution mode. This primarily concerns the binary detections and is further described in Appendix\,\ref{sec:grid}. Furthermore, the residuals from our model fits were not always flat at zero, often showing some baseline-dependent signals. This could arise due to the complex spatial properties of the target beyond our modeling capability, or due to imperfect visibility calibration. We only modeled companion(s) and fully resolved flux components; exploring more detailed structure in GRAVITY snapshot data is beyond the scope of this work. 

The errors obtained by PMOIRED do not account for systematics and are overly optimistic. Consequently, once we converged on the final model, we performed bootstrapping on the model to get realistic error-bars on our best-fit parameters \citep{2019Lachaume}. As an additional note, all error-bars noted in this paper, obtained from model fitting or otherwise, are 1$\sigma$ errors.

\section{Results from Interferometry}
\label{sec:results}

In this section, we discuss the results of our interferometric data analysis along with more details about modeling for the different categories of WRs found in our GRAVITY sample.

\subsection{Unresolved Stars}
\label{aps}

\begin{figure}
    \includegraphics[width=\columnwidth]{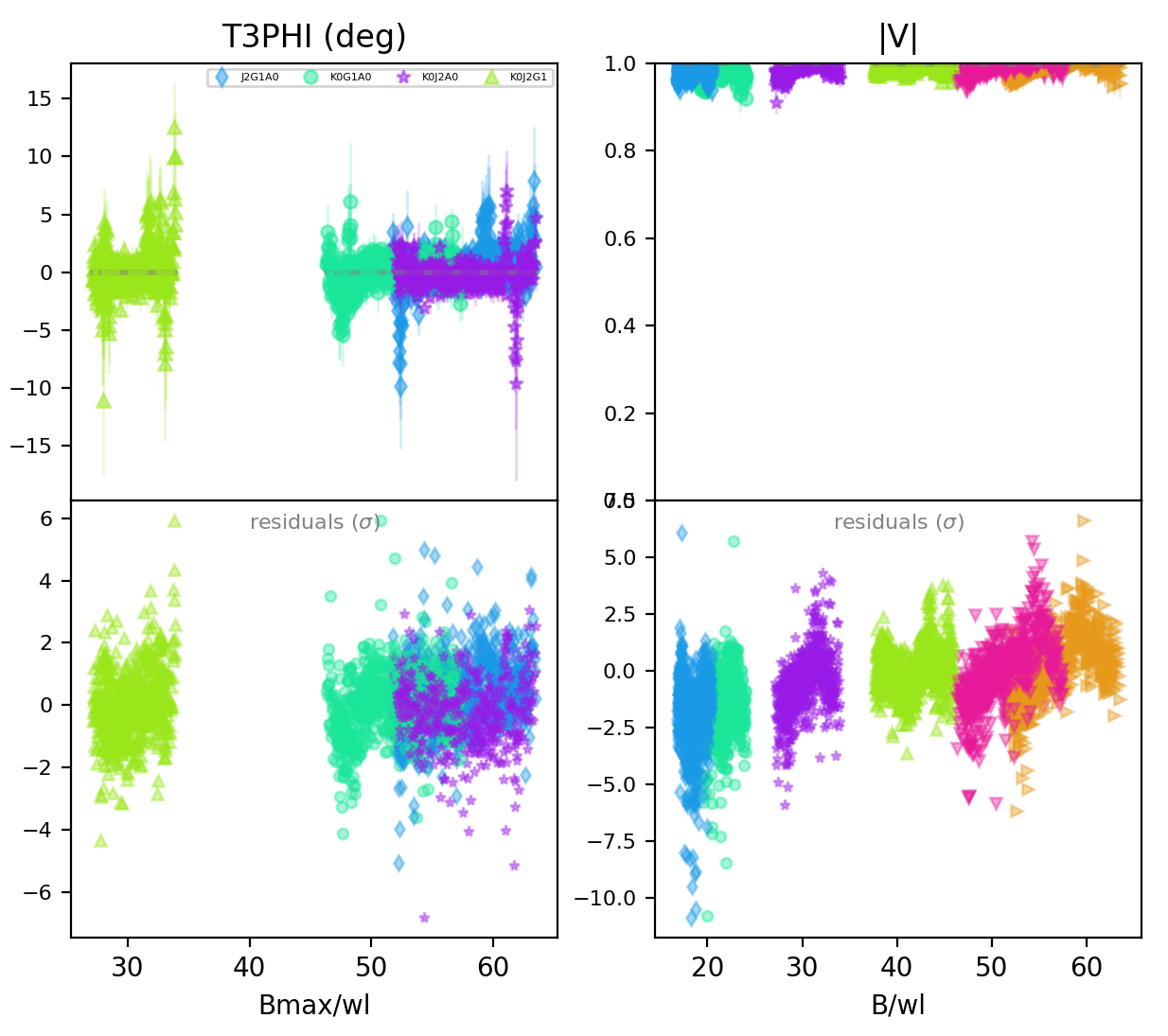}
    \caption{|V|-T3PHI data for the WN7h + O9III-V spectroscopic binary system WR~22. The different colors in T3PHI represent four combinations of 3 ATs each, while those in |V| represent the six baselines of 2 ATs each. The top panels show the best fit model (gray) along with the data, while the bottom panels show the residuals. The flat-at-zero T3PHI and flat-at-unity |V| data are typical of a single unresolved point source.}
    \label{fig:wr22}
\end{figure}

Single stars are expected to be mostly unresolved point sources in GRAVITY data due to their small angular sizes. Similarly, stars in binaries with angular separations smaller than the VLTI resolution limit ($\sim$1 mas) can also resemble single stars in interferometric data. We combined these two categories into one, termed as ``unresolved stars'' since they were indistinguishable and hence treated identically in our analysis. Here, unresolved refers to the lack of a resolved companion in |V|-T3PHI. In terms of interferometric observables, these stars were expected to show mostly flat, close to unity visibility amplitudes and zero closure and differential phases. A total of 29 WRs in our sample fall under this broad category. A good example of this is WR~22 (Fig.\,\ref{fig:wr22}), which is a WN7h + O9III-V spectroscopic binary (period $\sim$80 days) showing point-source-like properties. The best-fit model was a point source component alone.

\begin{figure}
  \includegraphics[width=\columnwidth]{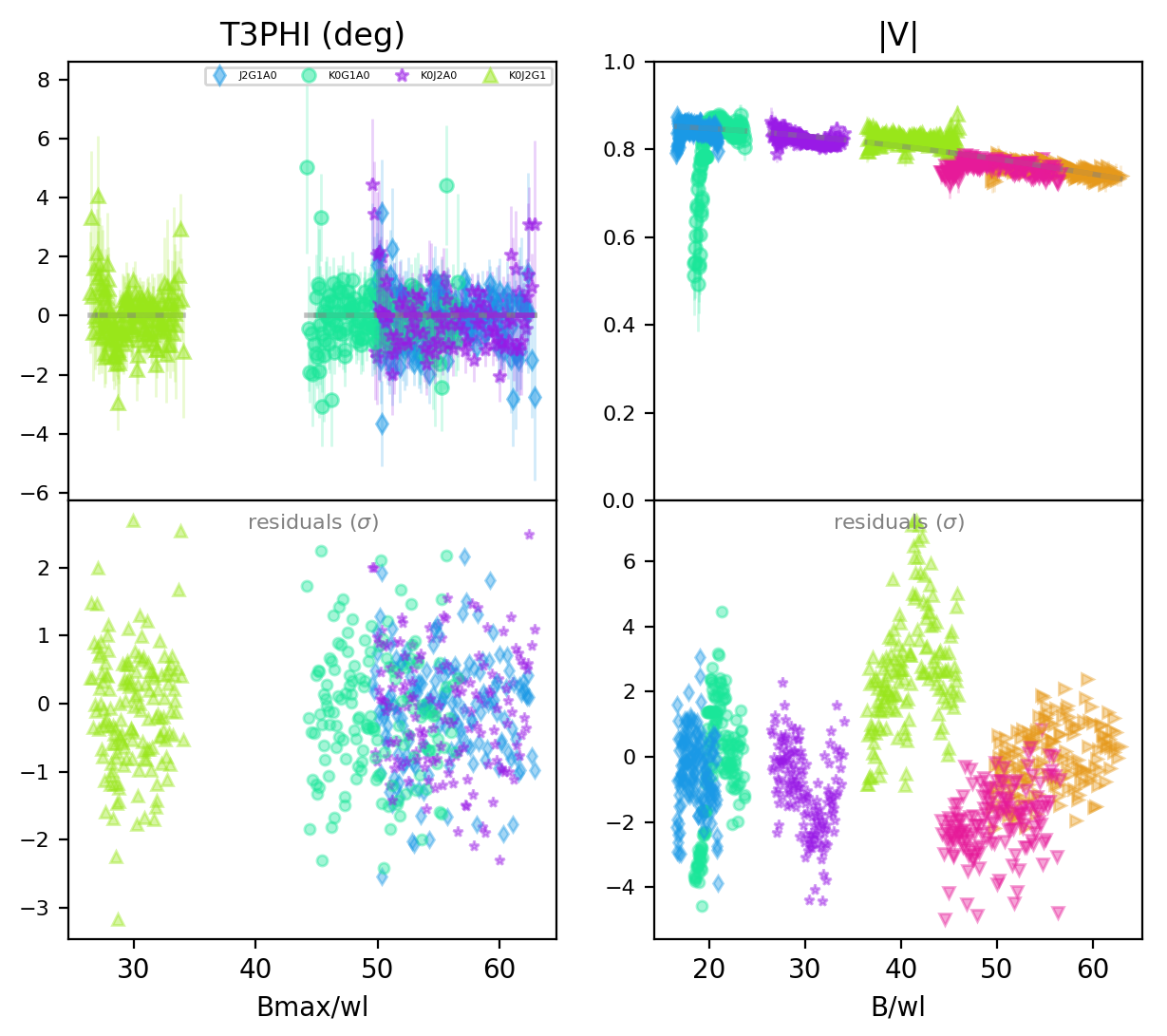}
  \caption{|V|-T3PHI data for the WN4 star WR~18 (similar to Fig.\,\ref{fig:wr22}). The flat-at-zero T3PHI and flat but vertically offset |V| data are best explained by a combination of an unresolved point source and a fully resolved component.}
  \label{fig:wr18_1}
\end{figure}

We expected all unresolved stars to show point-source-like features  just as WR~22. Yet several of them show a significant downward shift in |V| at all baselines, indicative of a second, fully resolved component. WR~18, a single WN4 star, is a good representative for such cases (Fig.\,\ref{fig:wr18_1}). It was modeled using a point source and a fully resolved component uniformly spanning the whole field of view. Such a component could not be constrained in size and its exact spatial extent remains uncertain.

\subsection{Resolved Wide Binaries in Visibility Amplitude and Closure Phase}
\label{sec:widecomp}

In our GRAVITY data, the presence of a wide companion with an angular separation of about 1--200 mas was likely to produce a strong signal in |V| and T3PHI, down to a contrast of $\Delta K\sim5$. In particular, a non-zero T3PHI implies asymmetry in the target and motivates the search for a companion. For the central WR, we used a model similar to the unresolved stars described in Sec. \ref{aps}. To search for a companion, we added a second point source with the $\Delta{\rm East}$ or $\Delta{\rm E}$ and $\Delta{\rm North}$ or $\Delta{\rm N}$ coordinates (with respect to the primary) searched over a grid and flux ratio set as a free parameter \citep{2015Gallenne}. The search was done in two steps - an initial larger and coarser grid to get an approximate solution, followed by a smaller finer grid to constrain the companion well. Details of the grid search are summarised in Appendix \ref{sec:grid}.

Only four targets were found to have potential wide companions. It is important to note that our observations represent a single epoch for each target, and follow-up observations would be necessary to confirm whether a companion is indeed gravitationally bound to the central WR. That being said, the probability of chance alignment of a random star in a 200 mas radius around the WR is very low. For a quantitative estimate, we followed the method described for Equation 8 from \citet{2014Sana} to estimate the probability of spurious association for companions within 200 mas and brighter than $\Delta K < 5$, finding a $\lesssim$0.1\% probability of chance alignment. 

Below we describe the four systems in more detail:

\paragraph{WR~89:}
\label{sec:wr89}

It is classified as a WN8h+abs single star \citep{1996Smith}, where abs stands for absorption lines of unknown origin, though they could in principle also be intrinsic to the WR. It was identified as a thermal radio source by \citet{2009Montes} and a strong X-ray source by \citet{2013Naze}. The latter supported the possibility of WR~89 being a colliding wind binary based on its properties, but could not confirm it using archival optical spectra. This hinted at a potential long-period binary (years to decades) and necessitated long-term spectroscopic monitoring to confirm binarity.

\begin{figure}
  \includegraphics[width=\columnwidth]{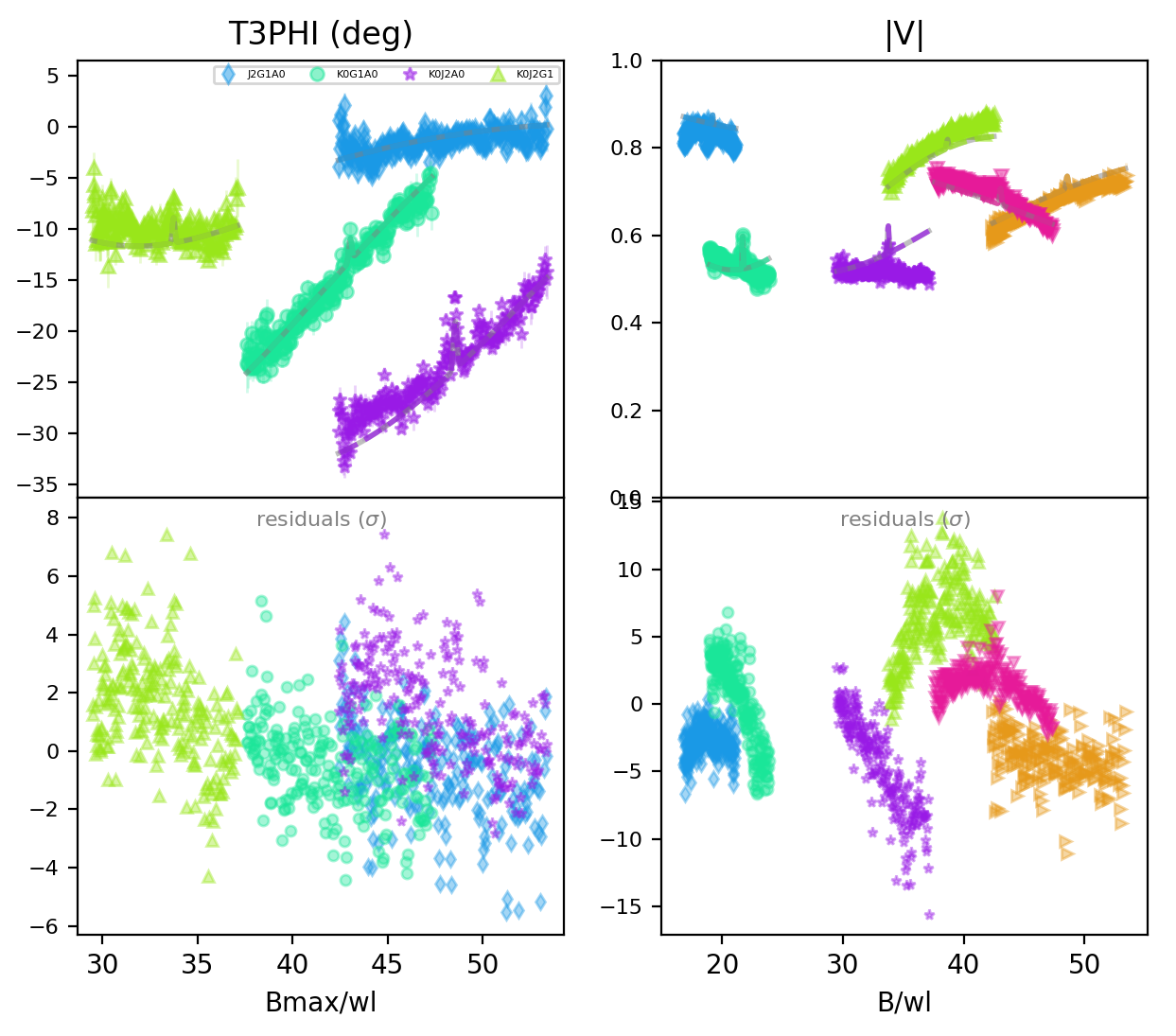}
  \caption{|V|-T3PHI data for the WN8h+abs star WR~89 (similar to Fig.\,\ref{fig:wr22}). The T3PHI and |V| data deviate drastically from symmetric/point-source-like behaviour, and are best modeled by a wide binary system with two distinct point sources.}
  \label{fig-wr89_1}
\end{figure}

In GRAVITY data, WR~89 shows clear deviation from point-source-like behaviour (see Fig.\,\ref{fig-wr89_1}). We performed a grid search for a potential binary companion and found a global minimum for $\chi^{2}_{\rm red}$ at $\Delta{\rm E}$ = $0.02\pm0.08$ mas and $\Delta{\rm N}$ = $-5.18\pm0.10$ mas relative to the central primary. The flux of the companion is $0.24\pm0.02$ times that of the WR star. Based on a \textit{Gaia} DR3 parallax of $0.27\pm0.02$ mas, the projected separation of the companion is about $19\pm1$ au.

\paragraph{WR~115:}

It is classified as a WN6-w single star \citep{2019Hamann}, where the -w stands for weak lines. It also shows spectral line variability potentially caused by corotating interaction regions \citep{2009StLouis}. A speckle-imaging companion search by \citet{2022Shara} identified a candidate early B dwarf companion with a projected separation of 0.20 arcseconds (200 mas) and position angle of 320 degrees East of North. Whether or not this supposed companion is bound to the central WR could only be confirmed with follow-up astrometric observations.

In GRAVITY data, we see clear signs of a companion in |V|, T3PHI as well as DPHI. We performed a grid search for a potential binary companion and found a global minimum for $\chi^{2}_{\rm red}$ at $\Delta{\rm E}$ = $-124.80\pm0.04$ mas and $\Delta{\rm N}$ = $154.93\pm0.02$ mas relative to the central primary. In terms of projected separation and position angle, this translates to $\sim$199 mas and 321 degrees East of North respectively. This is largely consistent with \citet{2022Shara}, although follow-up observations in the future would be necessary to confirm the orbit. The flux of the companion is $0.05\pm0.01$ times that of the WR star. The \textit{Gaia} parallax for WR~115 is $0.58\pm0.46$ mas, and cannot be used to constrain the physical projected separation due to its large error-bar.

\paragraph{WR~48:}

Also known as $\theta$ Mus A, WR~48 was initially classified as a spectroscopic binary with a 19-day orbital period \citep{1977Moffat,2002Hill}. This was based on a composite spectrum (WR+O at least) that showed features of a WC6 star moving at the orbital period in short-period binary, and features of an O star that were virtually stationary, possibly unrelated to the binary. \citet{1999Hartkopf} performed a speckle interferometric survey of WRs, in which they detected an O supergiant wide companion to WR~48, although with some uncertainty over their relative positions.

In GRAVITY data, we detected this potential outer companion at $\Delta{\rm E}$ = $1.59\pm0.03$ mas and $\Delta{\rm N}$ = $-12.56\pm0.06$ mas relative to the central binary. It has a flux $1.27\pm0.10$ times that of the central binary. Based on the \textit{Gaia} parallax of $0.46\pm0.07$ mas, the projected separation is $28\pm4$ au. One more epoch of speckle interferometry data \citep{2012Hartkopf} and two more epochs of archival GRAVITY data are available for $\theta$ Mus A. A more comprehensive analysis of the object using all available data will be covered in Gosset et al. (in prep.).

\paragraph{WR~93:}

It was classified as a WC7 + O7-9 spectroscopic binary by \citet{1988Hucht} based on its spectra, although no orbital period has been determined to date. It belongs to one of the embedded clusters in the \ion{H}{ii} complex NGC 9357 \citep{2014Lima}.

We detected a strong binary signal in GRAVITY data for WR~93. We performed the routine grid search for a companion, finding the best fit for $\Delta{\rm E}$ = $5.60\pm0.03$ mas and $\Delta{\rm N}$ = $13.19\pm0.03$ mas. The flux of the companion is $0.20\pm0.01$ times that of the primary, and is likely the O7-9 companion known from spectroscopy. Based on the \textit{Gaia} parallax of $0.49\pm0.03$ mas for WR~93, the projected separation is about $29\pm2$ au.

\vspace{5mm}
\noindent More details about the companion grid search along with plots for these WRs are presented in Appendix \ref{sec:grid}.

\begin{figure*}
\centering
  \includegraphics[trim={0.5cm 1.5cm 0.5cm 1.5cm}, clip, height=90mm]{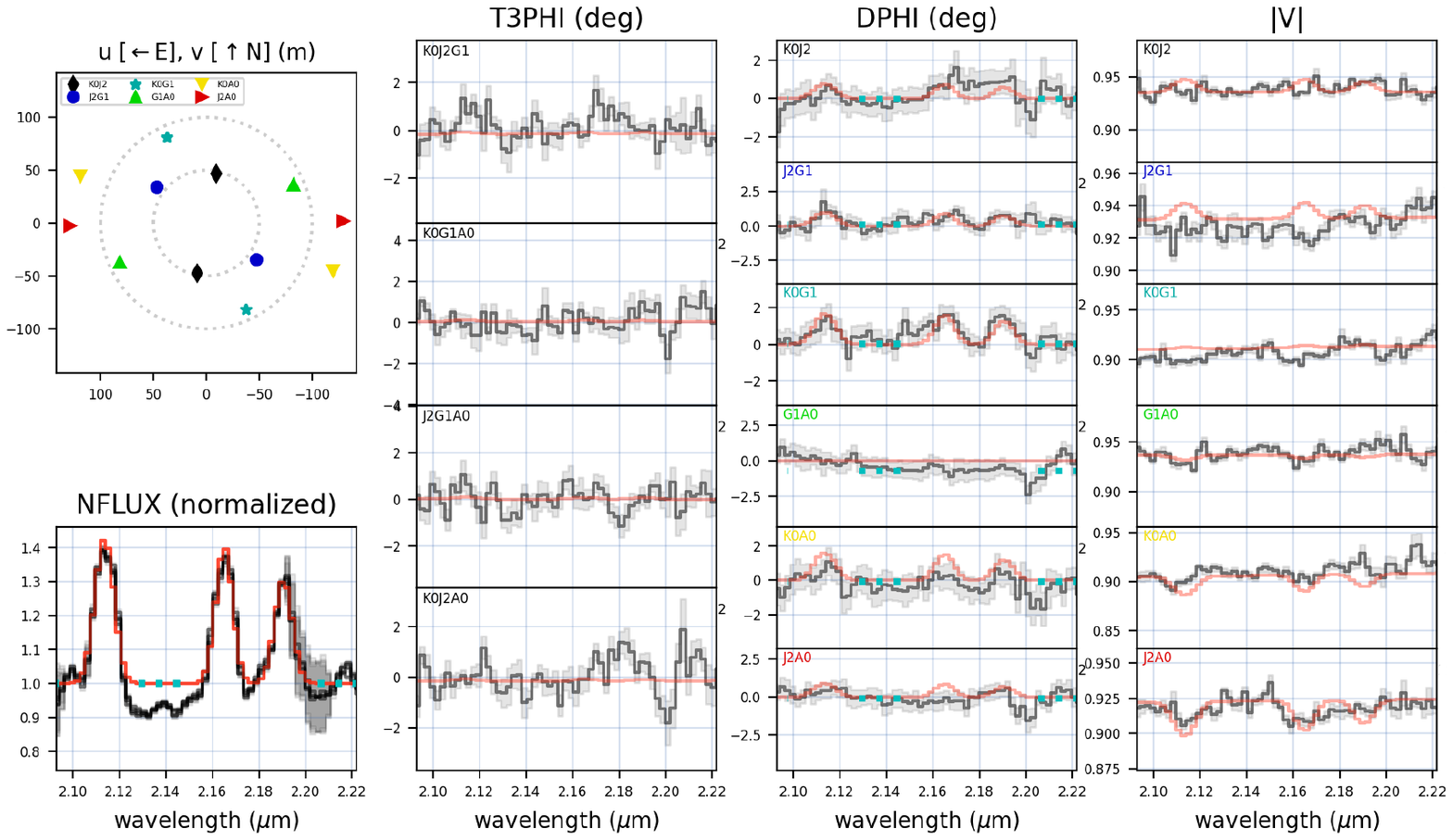}
  \caption{GRAVITY data for WR~98, a WN8/WC7 + O8-9 spectroscopic binary, zoomed in at 2.10 to 2.22 {\textmu}m along with the best-fit model shown in red. The top left panel shows the u-v coverage for the observation. The remaining panels, from left to right, show NFLUX, T3PHI, DPHI and |V| respectively, color-coded according to baselines. The data are zoomed in to a wavelength range of around 2.10-2.22 {\textmu}m, with three prominent emission lines present at 2.11, 2.16 and 2.19 {\textmu}m. The DPHI data (along with some of |V| data) show slight but clear features corresponding to the spectral lines. The best-fit model for such a system is a very close binary consisting of the line-emitting WR and a flat-spectrum companion, along with a fully resolved component. }
  \label{fig:wr98_1}
\end{figure*}

\begin{figure*}
\centering
  \includegraphics[trim={0.5cm 1.5cm 0.5cm 1.5cm}, clip, height=90mm]{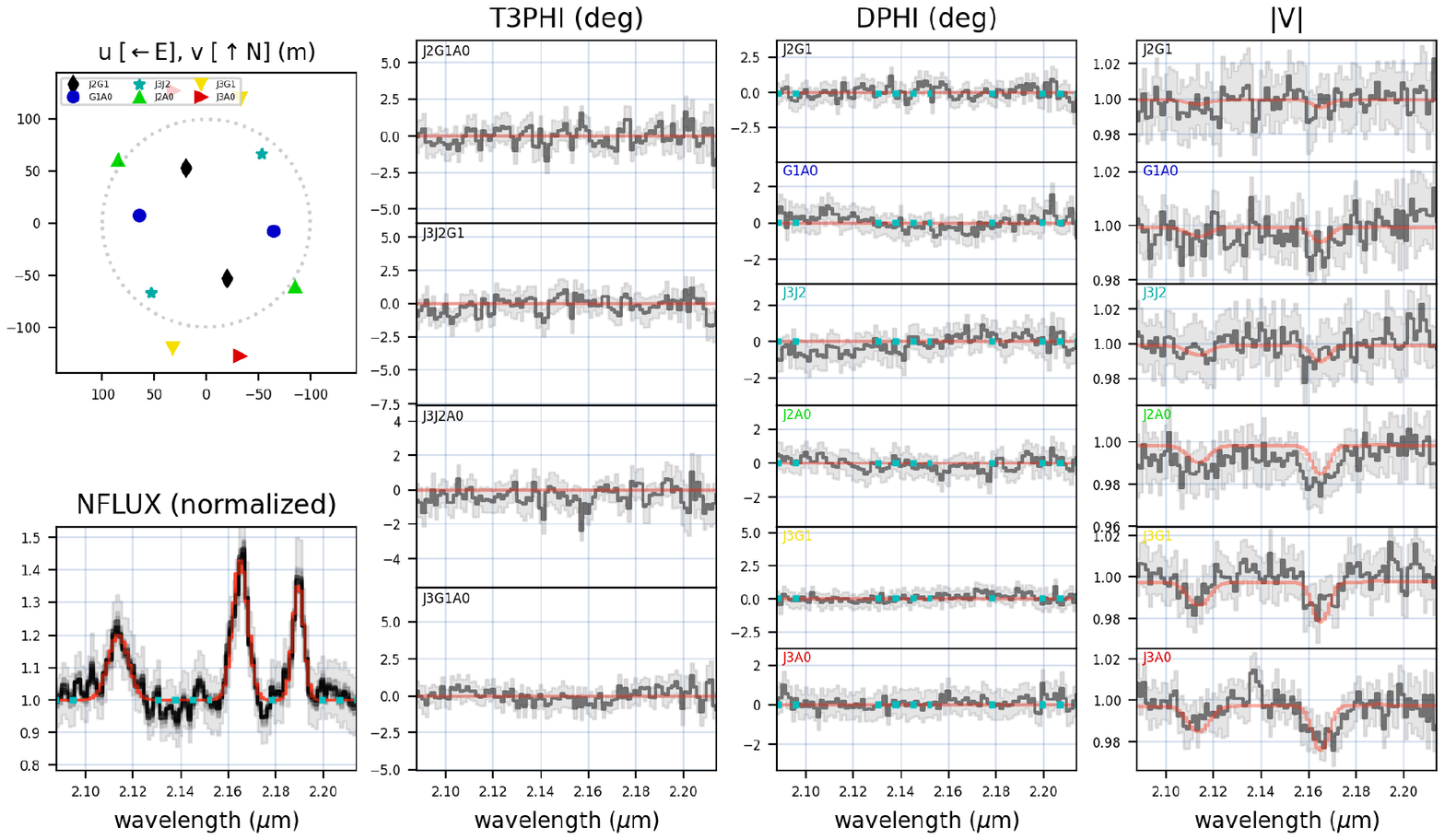}
  \caption{GRAVITY data for WR~78, a single WN7h star, along with the best-fit model shown in red. The top left panel shows the u-v coverage for the observation. The remaining panels, from left to right, show NFLUX, T3PHI, DPHI and |V| respectively, color-coded according to baselines. The data are zoomed in to a wavelength range of around 2.09-2.21 {\textmu}m, with three strong emission lines present around 2.11, 2.16 and 2.19 {\textmu}m. The |V| data show clear features corresponding to the first two spectral lines, decreasing in prominence from longer to shorter baselines. The best-fit model for such a system is an unresolved point source emitting the 2.19 {\textmu}m spectral line and a slightly resolved Gaussian (fwhm$\approx$$0.51\pm0.02$ mas) emitting the 2.11 and 2.16 {\textmu}m spectral lines.}
  \label{wr78_1}
\end{figure*}

\begin{figure}
  \includegraphics[width=\columnwidth]{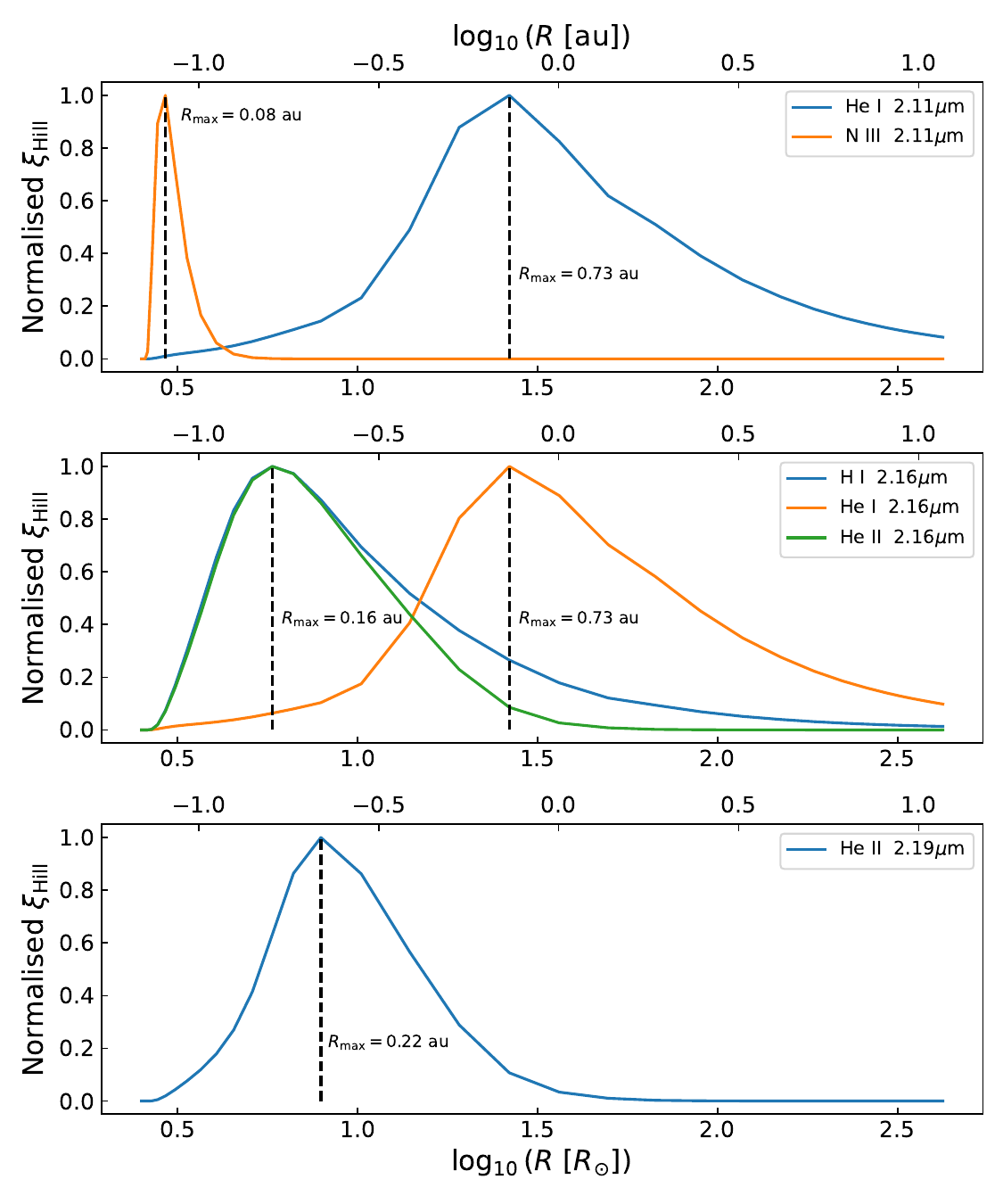}
  \caption{Line formation strengths $\xi$ \citep{1987Hillier} at 2.11 (top), 2.16 (middle) and 2.19 $\mu$m per radius for the PoWR atmosphere model used in Fig. \ref{fig:wr78rtau1}. The $\xi$ values are normalized for better line-to-line comparison. The radii where most line formation occurs $R_\mathrm{max}$ are highlighted by the dashed black lines.}
  \label{fig:line-formation}
\end{figure}

\subsection{Resolved Spectroscopic Binaries in Differential Phase}
\label{sec:DPHI}

A total of 12 stars in our sample have been previously classified as confirmed spectroscopic binaries with OB companions and measured periods ranging from 6-80 days. For a typical parallax of 2 kpc, this translates to separations <0.5 mas which is too small to be resolved with the VLTI when using |V| and T3PHI. However, since WRs have emission lines in the $K$-band and OB stars do not, we could potentially get a non-zero signal in the differential phase data (DPHI) at the emission line wavelengths. This is because DPHI is sensitive to spatial asymmetries in the wavelength regime - in this case, owing to the presence of emission lines in WRs and lack thereof in OB stars. This has been shown in the fitting of other binary systems. For example, in \cite{2022Frost} the properties of a decretion disk around a Be star were determined from fits to DPHI, despite its size being below the VLTI resolution limit and there being no drop in the |V| to fit. 

This method of companion detection was applied to the whole sample regardless of their current classification. For the purpose of modeling in this case, the NFLUX and DPHI data were also included and models were updated to include spectral lines. A minimum binary separation of $\sim$$(\lambda/B)\times e_{\rm DPHI}/360^{\circ}$ radian is necessary for a detection in DPHI, where $\lambda$ is the wavelength of observation, $B$ is the baseline and $e_{\rm DPHI}$ is the error level in DPHI data. For our GRAVITY observations assuming $e_{\rm DPHI}\sim$1$^{\circ}$, the minimum separation comes out to be $\gtrsim$0.01 mas. To model such a companion, a second point source with a flat spectrum was added with a grid search region spanning $2\times2$ mas centred on the primary star. Only five stars out of the sample of 39 were found to have a DPHI signal corresponding to any spectral line(s). WR~98, a WN8/WC7 + O8-9 spectroscopic binary, is an example of such a binary detection as shown in Figure \ref{fig:wr98_1}. Three broad emission lines at 2.11 {\textmu}m (\ion{He}{i} + \ion{N}{iii} blend), 2.16 {\textmu}m (Br$\gamma$ + \ion{He}{i} + \ion{He}{ii} blend) and 2.19 {\textmu}m (\ion{He}{ii}), typical for WN stars as discussed in Section \ref{sec:data}, can be seen in NFLUX as well as DPHI. The features in DPHI indicate an asymmetry in all spectral lines, hinting towards a close binary as discussed above. 

The remaining four stars were WR~9, WR~42, WR~47 and WR~79, which are discussed in more detail in Appendix\,\ref{sec:all}. All five binaries detected using this method are known spectroscopic binaries. The remaining seven binaries were not detected likely due to their smaller projected separation at the time of observation, higher flux contrast, and/or poor S/N of GRAVITY data. Nevertheless, we demonstrate the potential of this method for short-period binary detection in WRs, in addition to the typical |V|-T3PHI detections for wide binaries.

\subsection{Resolved Line-emitting Regions}
\label{sec:LER}

WRs are characterized by strong optically thick winds that dominate their spectra, resulting in broad emission lines. The different emission lines are produced by a multitude of processes, including resonance scattering, collisional excitation, (dielectronic) recombination, and continuum fluorescence, with complex interdependencies \citep[see, e.g.,][for a more detailed discussion and examples]{2015Hillier}. The dominating process for the origin of a specific line further depends on the detailed, non-local thermodynamic equilibrium (non-LTE) population numbers and can change for different conditions and thus different WR sub-types. In the dense winds of WR stars, ionization changes in the outer wind are common, giving rise to the appearance of emission lines of many different ionization stages in the spectrum. For the purpose of this study, the recombination of helium from the fully ionized \ion{He}{iii} to \ion{He}{ii} is particularly important as it is responsible for the prominent appearance of \ion{He}{i} lines in the $K$-band.

The spectro-interferometric capabilities of VLTI/GRAVITY offer a unique insight into extended line-emitting regions. For a target with multiple components contributing to the flux, spectroscopy alone can be limited in distinguishing the flux coming from different components. Combining with interferometry, however, it is possible to do so. Taking a WR star for example, if the total flux is a sum of the continuum originating from the (unresolved) photosphere and emission lines originating from the extended (resolved) winds, the |V| data will reveal their presence as two different components. Such spectral disentangling can provide a physical measurement of the extent of line-emitting regions, that has only been estimated from radiation hydrodynamical models before.

Figure \ref{wr78_1} shows the GRAVITY data for the WN7h star WR~78. The spectrum shows WN-like broad emission lines at 2.11 {\textmu}m (\ion{He}{i} + \ion{N}{iii} blend), 2.16 {\textmu}m (Br$\gamma$ + \ion{He}{i} + \ion{He}{ii} blend) and 2.19 {\textmu}m (\ion{He}{ii}) {\textmu}m. The T3PHI and DPHI data are flat at zero, implying no point asymmetry in the object. The overall level of |V| is essentially unity, characteristic of an unresolved point source. However, two spectral features can be seen corresponding to the lines at 2.11 {\textmu}m and 2.16 {\textmu}m, while such a feature is absent for 2.19 {\textmu}m. The likely explanation for such |V| data is the presence of two components in the system - (i) an unresolved point source responsible for the continuum and the line 2.19 {\textmu}m (uniform disk with fixed radius = 0.001 mas); (ii) a slightly resolved component only emitting the lines 2.11 and 2.16 {\textmu}m (gaussian with fwhm = free parameter). On fitting such a model to the data, we obtain an fwhm of $0.51\pm0.02$ mas. WR~78 has a parallax of $0.56\pm0.03$ mas in \textit{Gaia} DR3, based on which, the fwhm or ``size'' of the resolved component is $0.89\pm0.05$ au or $191\pm11\,R_\odot$. 

Using a PoWR atmosphere model of WR~78, the line formation strengths per radius $\xi$ \citep[defined in][]{1987Hillier} are shown in Figure\,\ref{fig:line-formation} for the observed lines in Figure\,\ref{wr78_1}. The values are normalized, as the absolute values of $\xi$ carry little physical meaning, but the figure illustrates where the line formation is highest in the stellar wind. It is evident that the peak of the formation for the \ion{He}{i} lines occurs just a bit outside of the region where He recombines from \ion{He}{iii} to \ion{He}{ii} as discussed above. Notably, the main formation further happens well within 1 au for all the considered lines, including those we resolve in our data. The reason for this is likely twofold: firstly, imperfections in the model for WR~78 might slightly underestimate the necessary stellar radius by incorrectly estimating the optically thick wind onset region and the velocity stratification \citep[see also][]{2023Lefever}. This is not unlikely as our model is not a dynamically-consistent one \citep{2020Sander,2023Sander} and could affect the radial scale of the line-formation calculations. Secondly, even if the peak of line formation is unresolved, the emission strength can still be strong enough farther out in the wind to be resolved with GRAVITY, \text{which seems to be the case for \ion{He}{i} lines}.

In our sample of 39 stars, a total of four WRs (all of them late WN/WNh stars) were found to have resolved line emitting regions. It is important to note here that our observations were taken in the snapshot mode of GRAVITY and higher S/N observations with more uv-coverage can provide an even more detailed look into the line-forming regions. Such observations could be undertaken and discussed in a future study.

The wide variety of interferometric data for WRs in our sample could further motivate separate studies for each category described above. A target-by-target summary for all stars in our sample is provided in Appendix\,\ref{sec:all}. WR~113 being the only dust producing WR in our sample shows a distinct signature, also discussed in detail in Appendix\,\ref{sec:all}. In the following sections, we focus on the multiplicity analysis and subsequent interpretation for our WR sample.

\section{Multiplicity Fraction}
\label{sec:fbin}

\subsection{Observed Multiplicity Fraction}

The multiplicity of WRs as noted in literature is largely based on spectroscopic and photometric observations. These methods are less likely to detect companions to WRs that are lower in mass, inclination and/or in wider orbits due to their smaller effect on the Doppler motion of the WR. With our GRAVITY observations, we have taken a complementary approach of detecting the wider companions with interferometry. By doing so, we opened up a new parameter space covering two orders of magnitude in projected separations (three orders in orbital period) down to $K$-band flux contrast as low as $\sim$1\%.

Table \ref{tab:fbin} summarises the multiplicity status for our sample classified according to their spectral classes - WNh (H-rich WN), WNL (late-type WN), WNE (early-type WN) and WC. Here, WNL refers to WRs with spectral types WN6 or later, while WNE refers to earlier spectral types.

For every spectral class, we have reported the number of stars that were - a) classified as single; b) confirmed double-lined spectroscopic binaries (SB2) but the companion was unresolved in GRAVITY data; c) confirmed SB2s with the secondary resolved in GRAVITY DPHI data; d) found to have a wide companion resolved in GRAVITY |V|-T3PHI data. For the special case of WR~48 which is likely a triple system, it is both a confirmed short-period binary and has a wide companion, thus being included twice. The observed multiplicity fraction, which is calculated as the total number of WRs that have companions divided by the sample size, is listed in the last column, along with error-bars derived using binomial statistics.

The observed multiplicity fraction for WRs altogether was found to be $0.38\pm0.08$. Based on spectral classes, the observed fraction for WNL ($0.50\pm0.20$), WNE ($0.60\pm0.22$) and WC ($0.40\pm0.13$) were found to be consistent with each other within the error-bars. WNh stars, however, turned out to have a \text{relatively} lower multiplicity fraction ($0.23\pm0.12$) than their classical counterparts. In the next subsection, we discuss the important physical limitations of our interferometric survey before proceeding to Section \ref{sec:evol}, where we interpret the multiplicity fractions from Table \ref{tab:fbin} in an evolutionary context.

   \begin{table*}[ht]
      \caption[]{Combined multiplicity properties from spectroscopy (literature) and interferometry (this work) for our sample of 39 WRs.}
         \label{tab:fbin}
         \renewcommand{\arraystretch}{1.15}
\begin{tabular*}{\textwidth}{@{\extracolsep{\fill}} cccccccc }

\hline
\begin{tabular}{@{}c@{}} WR \\ Class  \end{tabular} 
& \begin{tabular}{@{}c@{}} Total \\ Stars \end{tabular}
& \begin{tabular}{@{}c@{}} Single \\ Stars \end{tabular}
& \begin{tabular}{@{}c@{}} Unresolved \\ Spectroscopic \\ Companions \end{tabular}
& \begin{tabular}{@{}c@{}} Resolved \\ Spectroscopic \\ Companions \end{tabular}
& \begin{tabular}{@{}c@{}} Resolved \\ Wide \\ Companions \end{tabular} 
& \begin{tabular}{@{}c@{}} Total WRs \\ With \\ Companions \end{tabular} 
& \begin{tabular}{@{}c@{}} Observed \\ Multiplicity \\ Fraction \end{tabular} \\

\hline

\textbf{WNh} & 13 & 10 & 2 & 0 & 1 & 3 & \textbf{0.23$\pm$0.12} \\

\hline

\textbf{WN} & 11 & 5 & 3 & 2 & 1 & 6 & \textbf{0.55$\pm$0.15} \\

\ \ \ \ \ \ \ \ \ \ WNL & 6 & 3 & 0 & 2 & 1 & 3 & 0.50$\pm$0.20 \\

\ \ \ \ \ \ \ \ \ \ WNE & 5 & 2 & 3 & 0 & 0 & 3 & 0.60$\pm$0.22 \\

\hline

\textbf{WC} & 15 & 9 & 1 & 4 & 2$^\ast$ & 6 & \textbf{0.40$\pm$0.13} \\

\hline
\textbf{cWR} & 26 & 14 & 4 & 6 & 3 & 12 & \textbf{0.46$\pm$0.10} \\
\hline

\textbf{All WR} & 39 & 24 & 6 & 6 & 4 & 15 & \textbf{0.38$\pm$0.08} \\

\hline
\end{tabular*}

\medskip
\textbf{Notes:} Rows are divided into WNh, WN and WC, with WN further divided into WNL and WNE. The last column lists the observed multiplicity fraction along with errors calculated using binomial statistics (*WR~48 being a triple candidate is counted twice).

   \end{table*}

\subsection{Determining Physical Parameters for Detected Companions and Detection Limits}

\begin{figure*}[ht]
  \includegraphics[width=\textwidth]{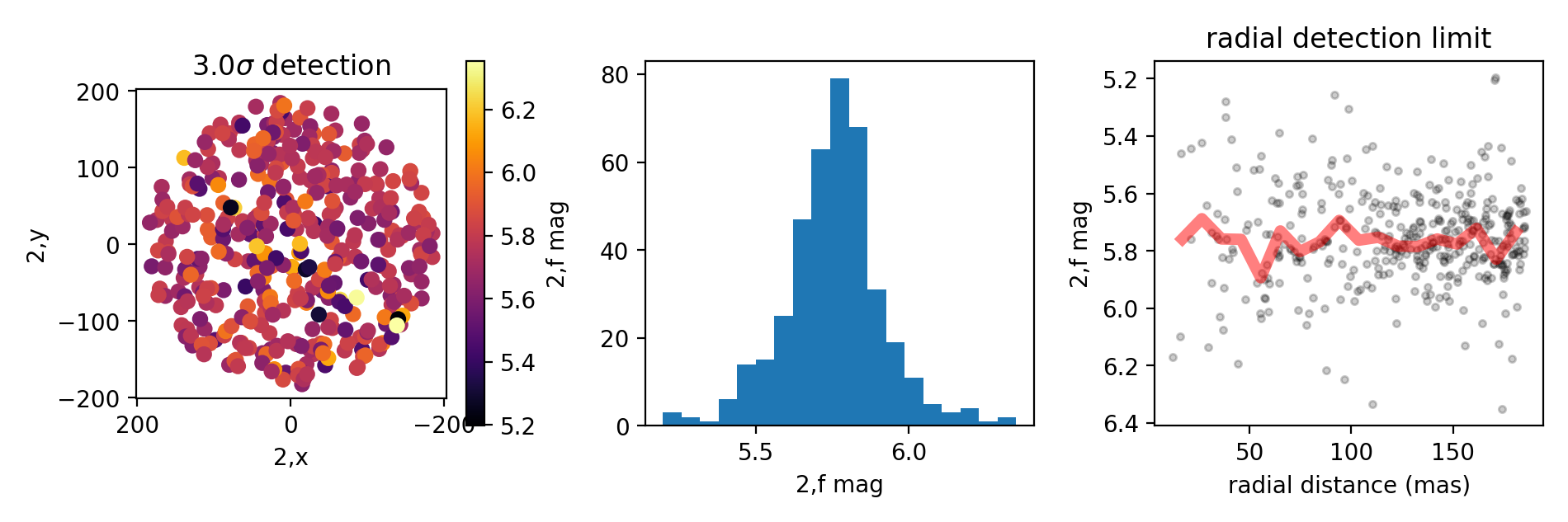}
  \caption{The companion detection limit plots for WR~18. The first plot shows a randomly created dense grid about 200 mas in radius around the central primary star that was used to inject a second point source and calculate the flux ratio needed for a $3\sigma$ detection. The scale for flux ratio is given in magnitude with reference to the primary. The second plot shows a histogram of limiting magnitudes over all the grid points. Lastly, the third plot shows detection limit as a function of radial distance, along with a rolling average in red.}
  \label{fig:wr18_2}
\end{figure*}

Our magnitude-limited GRAVITY observations of 39 WRs form a homogeneous sample {in terms of the parameter space explored}, making it possible to uniformly assess the overall properties and limitations of our detection methods. A large fraction of WRs in our sample was unresolved in GRAVITY data, including some known close binaries as well as stars classified as single. For all such targets, we calculated a $3\sigma$ detection limit for a wide companion using PMOIRED. This was done by injecting a point source randomly within 200 mas around the central primary and estimating the flux required for a $3\sigma$ detection \citep[as described in][]{2011Absil}. A graphical example of this for WR~18 is shown in Figure \ref{fig:wr18_2}. The detection limit was expressed as magnitude contrast with respect to the central WR. It tends to vary across different positions with respect to the central star, resulting in an approximate Gaussian distribution as seen in the second plot of Figure \ref{fig:wr18_2}. As a conservative estimate for the limiting magnitude contrast, we picked the 99 percentile value at the lower end of the distribution. For example, in case WR~18, the 99 percentile value is 5.057 magnitude.

This contrast in $K$-band magnitudes or corresponding flux ratios could be interpreted physically, based on certain assumptions and provided we have a good understanding of the fundamental stellar parameters of the stars. This could be done for both detection limits as well as detected wide companions to some extent, as described in the following text.

\citet{2019Hamann} and \citet{2019Sander} have reported fundamental stellar parameters for a large number of Galactic WN (including WNh) and WC stars respectively using updated \textit{Gaia} DR2 distances. This includes 24 WRs from our sample, most of which are classified as single stars. For WR binaries in which the companion contributes significantly to the spectrum, making it composite, the determination of stellar parameters was not possible. We used all information available from the two studies for our WR sample, with the primary goal of obtaining corresponding absolute magnitudes for detected companions or detection limits. A brief description is given below.

\citet{2019Hamann} have listed the following stellar parameters for WN and WNh stars - the narrow-band color excess ($E_{b-v}$), relative visibility ($R_{V}$) associated with the applied reddening law, and absolute $v$-band magnitude ($M_{v}$). Firstly, we converted the narrow-band color excess to broadband color excess using the equation from \citet{1982Turner}: 

\begin{equation}
    E_{B-V} = 1.21\times E_{b-v}.
\end{equation}

We used this to calculate the $V$-band extinction and subsequently the $K$-band extinction following the equations from \citet{1999Fitzpatrick}:

\begin{equation}
    A_{V} = R_{V}\times E_{B-V},
\end{equation}
\begin{equation}
    A_{K} = 0.12\times A_{V}.
\end{equation}

Finally, we employed the apparent $K$-band magnitudes listed on the Galactic Wolf Rayet Catalog v1.30 and \textit{Gaia} DR3 distances to compute absolute $K$-band magnitudes using the equation:

\begin{equation}
    M_{K} = m_{K} - A_{K} - 5\times \text{log}_{10}\left(\frac{d}{10}\right),
\end{equation}
where $d$ is the distance to the star expressed in parsec. We performed a similar exercise based on \citet{2019Sander} for the WC stars in our sample. In this case, however, the $R_{V}$ parameter was not reported. The uncertainty in $R_{V}$ only propagates into $A_{K}$ with a factor of 0.12. Consequently, we assumed the standard value of $R_{V}$ = 3.1 for all WRs as a reasonable approximation, and determined their absolute $K$-band magnitudes.

\subsubsection{Detected Companions}
\label{sec:detectedcomp}

As described in Section \ref{sec:widecomp}, we detected wide companions in four WRs in our sample - WR~48, 89, 93 and 115. WR~89 and 115 have been classified and analysed by \citet{2019Hamann} as single stars, owing to the lack of any companion contribution to their spectra. Using the $K$-band flux ratio we obtained from interferometry along with the absolute $K$-band magnitudes of the WRs, we calculated the absolute $K$-band magnitudes of their wide companions. Subsequently, assuming the companion is a main-sequence OB star, we used synthetic photometry from \citet{2006Martins,2013Pecaut} to estimate the spectral types of the companions. Following is a short summary:

\begin{itemize}
    \setlength\itemsep{1em}
    \item \textbf{WR~89}: The absolute $K$ magnitude for the WR was found to be --6.607$\pm$0.179, while that for its companion was $-5.057\pm0.179$. Based on the latter, we estimate the companion to be approximately an O5III giant.
    \item \textbf{WR~115}: The absolute $K$ magnitude for the WR was found to be --4.595$\pm$1.736, while that for the wide companion was --1.342$\pm$1.736. The large error-bars in this case are primarily due to the uncertain \textit{Gaia} parallax of WR~115 (0.58$\pm$0.46). Consequently, it is not possible to constrain the spectral type of the companion. Nevertheless, based on the reported parallax, we estimated the companion to be a B2V dwarf.
    \item \textbf{WR~48, WR~93}: Both WRs have a composite spectrum and were excluded from the analysis of \citet{2019Sander}. As a result, the above mentioned method could not be applied to these stars to determine the nature of their companions.
\end{itemize}

It is important to note that our estimates are based on the assumption that the companion is on the main sequence. Since we are basing the spectral type solely on one data point (absolute $K$ magnitude), we also explored the possibility of a late-type pre-main-sequence companion that is equally bright. Based on the pre-main-sequence tracks computed by \citet{2000Siess}, F/G/K-type stars can have absolute $K$ magnitudes similar to those calculated above for our WR companions, provided they are very young (age < 1 Myr). Such stars are unlikely to reside in binaries with WRs, which are likely a few Myr old. We therefore exclude the possibility of pre-main-sequence companions, although future observations in a different wavelength band could potentially help confirm the nature of the companions.

\subsubsection{Detection Limits}
\label{sec:detlim}

For WRs with no resolved companions, we computed the upper limits for companion detection in GRAVITY data. The limits were determined in terms of magnitude contrast with respect to the unresolved primary. We converted this contrast to a limiting flux ratio in the $K$-band. With the knowledge of absolute $K$ magnitudes of the WRs and limiting flux ratios, we derived the corresponding absolute $K$ magnitudes for the companion. Essentially, any companion brighter than this limiting magnitude would have yielded at least a 3$\sigma$ detection. Therefore, we use this limiting absolute magnitude to infer upper limits on the mass of the companion.

\begin{figure}
  \includegraphics[width=\columnwidth]{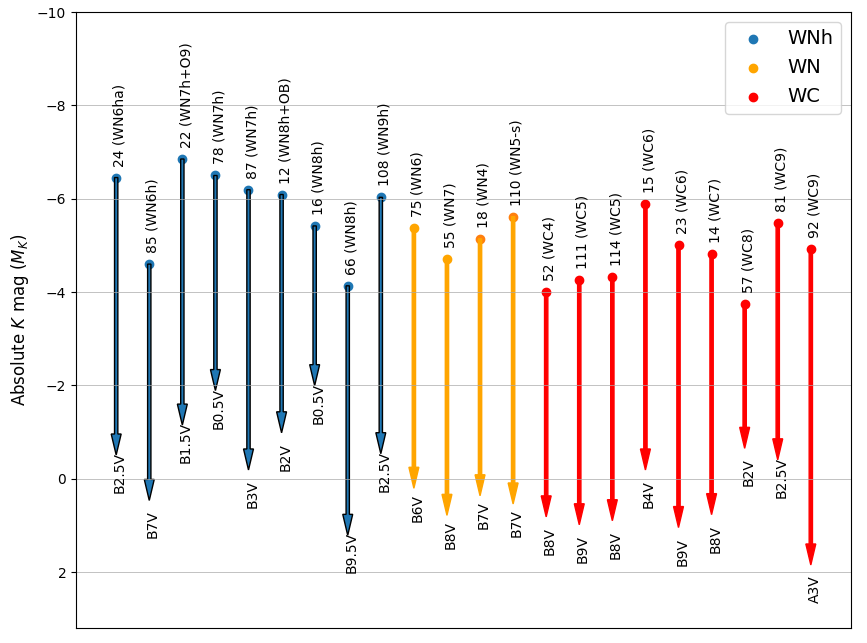}
  \caption{Absolute $K$-band magnitudes for WRs determined using stellar parameters (top points), limiting magnitude contrast from GRAVITY data (arrows), and the absolute $K$-band magnitudes for limiting companions along with corresponding spectral types (bottom). WNh stars (blue) have, on average, higher luminosities and consequently earlier limiting companion spectral types. WN (orange) and WC (red) stars show lower luminosities and later limiting companion spectral types.}
  \label{fig:det_lim}
\end{figure}

Taking a conservative approach, we assumed the companion to be a main-sequence dwarf. We again employed synthetic photometry provided by \citet{2013Pecaut} to estimate the spectral type of the limiting companion. Figure \ref{fig:det_lim} illustrates the absolute $K$ magnitudes of the WRs, limiting magnitude contrasts and the absolute $K$ magnitudes of the companions along with their spectral types. There is a scatter in the limiting magnitudes owing to the scatter in WR magnitudes as well as limiting contrasts, the latter being dependent on data quality. Nevertheless, the sample as a whole has a median detection limit around spectral type B4-5V, or about $5\,M_\odot$. The WNh stars have, on average, higher limits than cWRs due to their higher luminosity, and a median detection limit around spectral type B2V or mass of about $7\,M_\odot$.

\section{Evolutionary Implications}
\label{sec:evol}

The multiplicity fraction of \text{WRs}, along with the period distribution of binaries is an informative indicator of their evolutionary history. In this work, we extended the discovery parameter space far beyond spectroscopic studies and were able to observationally constrain the multiplicity of WRs up to orbital periods of $\sim$$10^5$ days and down to companion masses of $\sim$$5\,M_\odot$. Many WRs in literature have been classified as candidate binaries based on the presence of absorption features or the dilution of emission lines. We only classified a WR as a binary if it either has a known orbital period or it is detected in our interferometric survey. Based on this definition, Table \ref{tab:fbin} summarises the observed multiplicity fraction of WRs across different spectral classes. Our results therefore represent lower limits on WR multiplicity in the Galaxy based on conservative binary detection criteria. Here, we discuss the physical interpretation of our results, and  corresponding evolutionary implications.

\subsection{The Curious Case of WNh Stars}

The multiplicity fraction of the 13 WNh stars in our sample is particularly low ($0.23\pm0.12$) compared to cWR stars ($0.46\pm0.10$). As mentioned in the introduction, a significant fraction of WNh stars are thought to be very massive main-sequence stars  displaying the WR phenomenon due to their high luminosities (log$(L/L_\odot) \gtrsim 6$) and subsequent strong winds. In such a scenario, the formation of WNh stars does not require binary interaction, allowing for a lower multiplicity fraction compared to cWRs. For example, WR~12 and WR~22 are known spectroscopic binaries with high luminosities \citep{2019Hamann}. The newly detected wide binary WR~89 shares similar properties, consisting of a high luminosity WNh star and an O star companion unlikely to have interacted before. 

However, for the WNh stars that are truly single, the need of a high luminosity is even more pronounced. For the stars classified as single in our sample that were also reported in \citet{2019Hamann}, we compared the WNh stars to their hydrogen-poor counterparts (WNs). While the WNh stars are on average more luminous, there is a significant overlap between the two populations. In particular, not all WNh stars classified as single are significantly more luminous than WN stars. \citet{1994Langer}  discussed the possibility of WNh stars, H-rich or H-poor, being pre- or post-LBV evolutionary phase of very massive stars respectively. There are also observed connections with late-type WNh stars being seen after LBV variability stopped, e.g.\ Romano's star in M33 \citep{2020Smit} or HD5980 in the Small Magellanic Cloud \citep{2010Koenigsberger}. Alternative formation channels could also be contributing to the WNh population, e.g. \citet{2024Li} discuss the formation of WNh stars via binary mergers, which is indeed a form of binary interaction but results in single WNh stars. While the signatures of massive binary mergers have been reported before \citep{2023Shenar,2024Frost}, analogs for WNh stars are not known, potentially due to their violent winds.

WNh stars are most likely progenitors of cWR through the wind stripping channel. As such, it is surprising that their multiplicity is lower than that of the WNs. The higher the luminosity of the WR star, the higher is the companion detection limit. As discussed in Sect.\,\ref{sec:detlim}, companions later than spectral type B2~V are unlikely to be detected for WNh stars in our survey. Such companions can also avoid spectroscopic detection due to their extreme mass ratios and lower luminosity. A combination of  these factors might possibly explain the lower multiplicity fraction of WNh stars but would require quantitative modeling for \text{confirmation}.

\subsection{Comparing WN and WC multiplicity}

The spectroscopic survey of \citet{kd1,kd2,kd3} focused on a sample of 39 Galactic WRs to investigate multiplicity properties and evolutionary link between WN and WC stars. Since, spectroscopy is most efficient at short periods, the short-period WR binaries were most easily detected. They found a lack of short-period ($\lesssim$100 days) WC binaries compared to their WN counterparts. In our sample however, the number of short-period WNs and WCs is comparable (see literature review in Appendix\,\ref{sec:all}). Consequently, the inconsistency between WNs and WCs at short periods might be of statistical origin.

As discussed in the introduction, \citet{kd3} also used Bayesian inference to model the excess radial velocity (RV) noise and to predict binary properties of WRs well beyond the spectroscopic regime. They inferred the multiplicity fraction of WCs to be much higher than WNs, especially at longer periods ($\gtrsim$100 days), which formed the rationale of our interferometric detection campaign. However, as seen in Table~\ref{tab:fbin}, we find no such overabundance of wide binaries for WCs compared to WNs. It is therefore reasonable to conclude that the evolutionary link of WNs evolving into WCs does not face any tension based on our results.

Furthermore, the lowest luminosity WR stars spend little to no time in the WC phase \citep{2007Crowther,2022Aguilera}, suggesting a slightly higher occurrence of WN binaries. The values in Table\,\ref{tab:fbin} might reflect this expectation, although the overlapping error-bars do not establish it conclusively.

Finally, 14 stars were excluded from our analysis due to poor observations/data reduction, 9 of which are WCs (including 6 WCd stars) while only 3 are WNs. The exclusion of these stars from our sample  potentially introduces a bias in the multiplicity statistics, particularly due to the link between dust-production and binarity \citep{2020Lau}. While the different spectral types are represented  in our sample, relative sample sizes of different sub-categories of WRs might not be representative of the total WR population.

\subsection{Comparing cWR Binaries with O star Binaries}

With GRAVITY, we explored the angular separation range from $\sim$1 to 200~mas, corresponding projected separations from  $\sim$1 to 300~au. The binaries in this range were detected based on just one epoch, and do not have a known orbit yet. \citet{2014Sana} presented a similar high angular resolution multiplicity survey of O-type stars using the VLTI/PIONIER and NACO/SAM instruments. They detected binaries with angular separations of $\sim$$1 - 250$ mas and magnitude contrasts $\Delta H \lesssim 4-5$. Tramper et al. (in prep.) further analysed the O star sample to derive the projected separations (au) and mass ratios of the detected binaries. They also report the spectral types and corresponding masses for the O stars as well as their companions.

Massive O-type binaries are the likely progenitors of cWR binaries and a comparison between the two populations can be insightful to investigate their evolutionary connection. Only the massive O stars are thought to evolve into cWRs, with the single-star channel suggesting a mass threshold of $\sim$$25\,M_\odot$ in the Galaxy, although it is highly uncertain with estimates ranging from $18-33\,M_\odot$ \citep{2008Eldridge,2016Eldridge,2015Georgy,2018Limongi,2003Dray}. On the other hand, the binary channel allows for lower mass progenitors, down to $\sim$$20\,M_{\odot}$ \citep{2020Shenar}. To keep a conservative approach, we set a common threshold at 25 $M_\odot$ for any O star to be considered a potential cWR progenitor regardless of the channel.

To fix the parameter space for comparison of cWR and O star binaries, we converted the angular separation (mas) to physical projected separation (au) for WR binaries using their \textit{Gaia} DR3 parallax measurements. Since our study probes the $\sim$$1 - 300$ au projected separation range, we only selected the O star binaries detected in this range. Furthermore, all companions to O stars detected in this separation range have derived spectral types earlier than B3V (Tramper et al., in prep.), quite similar to the detection limits for the cWRs in our sample as shown in Figure \ref{fig:det_lim}. The possibility of pre-main-sequence companions can be ruled out on similar grounds as in Sect.\,\ref{sec:detectedcomp}. We note here that WNh stars are excluded from the comparison since they do not resemble H-depleted WN or WC stars in an evolutionary context. 

\begin{figure}[H]
  \includegraphics[width=\columnwidth]{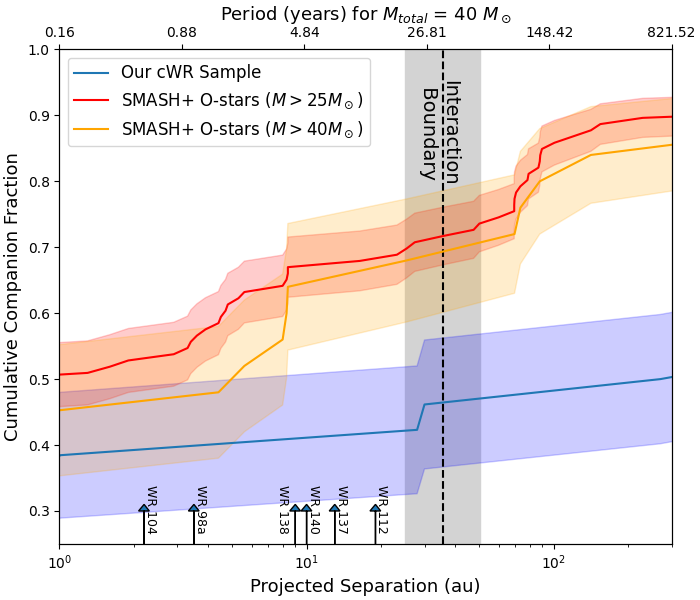}
  \caption{Cumulative companion fraction and corresponding binomial error regions plotted against the projected binary separation for cWR stars in our sample (blue) and O star primaries more massive than $25\,M_\odot$ (red) and $40\,M_\odot$ (orange) from the SMASH+ survey (Tramper et al. in prep.). The bottom horizontal-axis range is bound by the limits of the GRAVITY detection regime. The top horizontal-axis shows the orbital period for corresponding binary separations assuming a total system mass of $40\,M_\odot$. Furthermore, the black dotted line represents an interaction boundary for the relatively extreme case of $M_1$ = $40\,M_\odot$, q = 1 and e = 0.3. The gray shaded region spans the eccentricity range from e = 0 to e = 0.5. Also shown are a few WRs binaries from literature approximately placed on the plot for illustration.}
  \label{fig:smash_comp}
\end{figure}

Figure \ref{fig:smash_comp} shows the cumulative companion fraction (total number of companions/sample size) plotted against the projected separation for our cWR sample and the massive O star sample. Also included is an additional line for a higher mass cut of 40 $M_\odot$ for the O star sample. While the bottom horizontal-axis shows the projected separation - a direct observable from interferometry; the top horizontal-axis provides an approximate period scale, assuming a total system mass of 40 $M_\odot$. The figure also shows an interaction boundary, which was calculated as the Roche radius of the primary for an extreme case of a 40 $M_\odot$ primary, mass ratio of 1 and eccentricity of 0.3. We equated the maximum radial extent of a 40 $M_\odot$ star \citep{2023Marchant} to the Roche radius and found the corresponding binary separation. This was also done for eccentricities of 0 and 0.5 to show an approximate spread in the extent of this boundary. Finally, a few cWR binaries not in our sample are indicated with arrows in the plot at approximated separations for illustration. These include WR~98a \citep{2016Hendrix}, WR~104 \citep{2023Soulain}, WR~112 \citep{2020aLau}, WR~137, WR~138 \citep{2016Richardson} and WR~140 \citep{2022Lau}, all of which have known orbital periods.

Since the bottom horizontal-axis of the plot is based on the GRAVITY regime, all unresolved binaries (known from spectroscopic studies) lie below the projected separation of 1 au. Up to this separation, the companion fraction for cWRs is $\sim$38\% while that of O stars is $\sim$50\%. As such, this does not necessarily cause any evolutionary tension between the two populations, considering that a significant fraction of close O star binaries might end up merging - thus ``destroying'' the binary. However, as we uncover the new parameter space of separations ranging from $1-300$ au, the multiplicity fractions of the two populations seem to diverge even more. Up to 300 au, the companion fraction is $\sim$90\% for O stars and only $\sim$50\% for WRs.  This divergence can be discussed in two parts separated by the binary  interaction limit defined by the maximum extent of an evolved massive star, thus about 25-50~au ($\approx 5-10\times 10^3$~R$_{\odot}$).

Up to the interaction boundary, the multiplicity fraction of O stars steadily increases while there are virtually no detected cWR binaries in this range. In other words, most O star binaries in this regime seem to not have WR binary successors. Such dearth of WRs in wide binaries has also been reported in the Small \citep{2024Schootemeijer} and  Large Magellanic Cloud (LMC) \citep{2001Bartzakos,2019Shenar}, although only from spectroscopy. For our Galactic sample, a possible explanation for this can be that massive O stars might tend to expand considerably in their later evolutionary stages, initiating binary interaction. In case of stable mass transfer, the growing cumulative O star multiplicity fraction with period should qualitatively be preserved to the cWR stage  since the orbital period is modified gradually. However,  a drastic change such as that  seen in Figure\,\ref{fig:smash_comp} could reflect unstable mass transfer, with the interaction resulting in either a short period WR+OB binary or in a merger that may or may not evolve into the WR stage, depending on the mass of the merger product and the efficiency of the wind-stripping mechanism. In either case, the result of such interactions could be a lack of long-period cWR binaries. As mentioned earlier and indicated in Fig.\,\ref{fig:smash_comp}, some cWR binaries are found in literature with orbital separations in the range of 2 to 20~au. However, some of them have non-zero eccentricities, raising doubts about their post-interaction status.

Beyond the interaction boundary, we again see a steady increase in the multiplicity fraction of O stars (for mass thresholds of 25 as well as 40 $M_\odot$) while there is virtually no increase for cWRs. If such O stars indeed do not interact and form cWRs by themselves via the single-star channel, their companions should be preserved, likely still on the main sequence. Additionally, their orbital separations should only change by a factor of two even in the extreme case of the system losing half of its total mass via primary wind mass-loss. The lack of cWR binaries in this regime is therefore concerning for the single-star channel. Based on our results, the mass threshold to produce cWR stars via the single-star channel might be even higher than 40 $M_\odot$ \citep[see e.g.\ figure~10 in][]{2012Langer}. Alternatively, these wide, less gravitationally bound companions might be lost due to dynamical interactions with their environment. Detailed simulations, accounting for the local birth density and dynamical history of each system would be needed to investigate the likelihood of such a scenario and whether it can explain the lack of very wide cWR binaries compared to their O star counterparts.

\subsection{Implications for Black Hole Binaries}

The growing catalog of merging black hole (BH) binaries \citep{2019Abbott,2021Abbott,2023Abbott} has garnered significant interest in their possible formation channels. Massive binary evolution is considered a strong potential contributor to such BH binaries. \citet{2020Langer} investigated the properties of OB star-black hole (OB+BH) systems at LMC metallicity using detailed binary evolution models as these represent a crucial intermediate stage in the formation of BH+BH binaries.

The evolutionary stage immediately preceding an OB+BH system is thought to be an OB+cWR binary, where the cWR undergoes a supernova/collapse to form the BH \citep[see e.g.\ figure~1 in][]{2020Langer}. For example, \citet{2020Langer} assume the BH mass to be equal to the He-core mass of the cWR, and no momentum kick during its formation. As a result, the OB+BH systems are virtually OB+cWR systems at the time of collapse. Under these assumptions, our cWR binary sample offers a glimpse into the initial orbital properties of the OB+BH phase and allows for a direct comparison between observations and theoretical computations.

According to \citet{2020Langer}, the period distribution of OB+BH binary models at LMC metallicity is expected to peak around 200 days \citep[figure~6 in][]{2020Langer}, and is dominated by systems that have undergone Case B mass transfer. \citet{2022Pauli} conducted a similar population study for WRs in the LMC, finding peaks around 10 and 30 days in the period distributions of WN and WC binaries respectively. However, they also found a significant population of WR binary models at periods up to few 100 days \citep[figures 6 and 7 in][]{2022Pauli}. While these simulations have been computed with LMC metallicity, \citet{2022Janssens} showed that one expects only a small difference of these predicted period distributions between Milky Way and LMC (see their figure B.1).

In literature, the lack of observed wide WR binaries predicted by such simulations has been so far attributed to the limitations of spectroscopic campaigns at longer periods. With GRAVITY being sensitive to periods in the range of months to decades, the present observational campaign provides a systematical exploration of the period range of OB+BH progenitors predicted from simulations. However, our survey failed to reveal any OB+cWR binaries in the proximity of the 200-day period peak predicted by evolutionary models. Similarly, no OB+cWR binaries were found in a projected separation range of $\sim$$1-20$ au, corresponding to a period range of few tens of days to a few thousand days. Of course, neutron star or black hole companions to the cWR stars can not be excluded based on our survey. While a few cWR binaries have been found in this range by other studies (see Figure \ref{fig:smash_comp}), there is no sign of a period peak around 200 days. 

The OB+BH binaries predicted in this period range were all products of Case B mass transfer. The lack of OB+WR binaries that we report might thus warrant a revision of this mass transfer phase. In particular, the absence of OB+cWR binaries suggests that most of the long-period interacting O star binaries undergo unstable mass transfer, potentially leading to mergers \citep{2014Justham,2024Menon}.

\subsection{Single cWRs and others caveats}

The high percentage ($\sim$54\%) of single cWRs is another significant result of our study. Even after expanding the discovery parameter space by two orders of magnitude in separation and flux contrast thanks to optical long-baseline interferometry, we could not find any new binaries aside from WR~89 (which was strongly suspected to be a binary from X-ray observations).  It thus seems that there is a significant lack of WR binaries in general, and of long-period ones in particular (see again Fig.~\ref{fig:smash_comp}).

How to produce so many single cWR stars is not immediately evident. As there are few single O stars, many of the single cWRs could have a binary evolution history. However, binary mergers are not necessarily helpful in all cases, since they increase the H-rich envelope mass in the initial primary stars. Initial secondary stars could evolve into cWRs as single stars after their companions have been ejected from the orbit following a supernova. Such systems, however, will likely evolve through a previous O+cWR phase. Tailored population synthesis calculations are required to investigate whether a consistent solution is possible. 

One possibility is that our sample is affected by a significant selection bias. For example, the VLTI-ATs required optically bright sources for the STRAP injections ($V<14$) which introduces a bias against significantly reddened objects and indeed some known WC dust emitters (WR~98a, WR~112) were not included in this sample because of this limitation. While it would be good to complement the current survey with a dedicated VLTI+UTs survey that can observe reddened targets, it remains to be seen whether this possible bias alone is sufficient to explain the lack of long-period cWR binaries. Indeed, the numbers of dusty WR that we missed seems insufficient to reconcile observations and theory. In addition, their immediate WN progenitors, should not suffer from such biases and do not reveal further binaries in that period range.

\section{Conclusion}

We conducted an interferometric survey of 39 Galactic WRs with VLTI/GRAVITY to investigate their multiplicity properties. We detected wide companions for only four WRs in our sample, namely WR~48, WR~89, WR~93 and WR~115. In addition, we found four WRs with spatially resolved line-emitting regions and five WRs with resolved companions in differential phase data. We also detected a fully resolved component contributing significantly to the $K$-band flux for most WRs in our sample.

Combining with spectroscopic studies, we determined the observed multiplicity fraction for our sample to be $0.38\pm0.08$; with $0.23\pm0.12$ for WNh, $0.55\pm0.15$ for WNs and $0.40\pm0.13$ for WCs. Here, the classification of binaries was based on a known orbital period and/or a wide companion detected in our study. Companion detection limits were also calculated for the GRAVITY data and converted from flux contrasts to absolute magnitudes assuming main-sequence dwarf companions. Subsequently, we found the median detection limit for cWRs to be around $5\,M_\odot$, and that for WNh stars around $7\,M_\odot$. 

The evolutionary implications of our results span the different spectral classes of WRs, and their connection to the predecessors and descendants of WR binaries (OB binaries and OB+BH binaries, respectively). The low multiplicity fraction of WNh stars could be potentially attributed to their higher luminosities, enabling them to form via the single-star channel, or making it more difficult to detect companions as compared to cWRs. WNh stars could also be products of binary mergers that inevitably appear as single stars now.

Within the Northern cWR population, \citet{kd3} reported potential tension between the period distributions of WNs and WCs based on a modeling of the RV excess noise. If confirmed, cWR stars could not be explained by the conventional WN$\rightarrow$WC evolutionary link between them. However, we found the WN and WC multiplicity properties to be consistent within error-bars, and suggest that the tension reported in \citet{kd3} is either due to limited sample sizes or an additional RV variability signal not well modeled in \citet{kd3}. 

We also compared our cWR population with the massive O star population from the SMASH+ survey \cite{2014Sana}, and found a stark contrast in their multiplicity properties at long periods ($10^{2}-10^{5}$ days). There is a significant lack of cWR wide binaries as compared to massive O stars, indicating that binary interactions might be ``destroying'' wide O star binaries, instead of evolving them into wide cWR binaries. On the other hand, at very long periods beyond the regime of binary interaction, many of the wide O star binaries were expected to evolve into wide cWR binaries based on single star physics. The lack of such cWR binaries might raise doubts on the effectiveness of the Conti scenario, although not conclusively yet.

As for their eventual fate, many OB+cWR binaries are expected to evolve into OB+BH binaries via supernova/collapse of the cWR star. Binary evolution models for OB+BH stars in the LMC predict a peak in their period distribution around 200 days. Assuming minimal mass-loss and no momentum kick during BH formation, this period distribution can then also represent the parent period distribution, i.e.\ for the OB+cWR stars. With GRAVITY, we found no OB+cWR binaries in/around this period range, or even up to periods of a few years. We inferred that long-period interacting O star binaries might not evolve into cWR+OB stars, and instead might merge through unstable mass transfer, possibly impeding the common-envelope channel towards double-BH merger at solar metallicity. Finally, we also discussed the high observed fraction of single cWRs stemming from observational biases or their true single nature.

\begin{acknowledgements}
The research leading to these results has received funding from the European Research Council (ERC) under the European Union's Horizon 2020 and Horizon Europe research and innovation programme (grant agreement numbers 772225: MULTIPLES; and 101054731: Stellar-BHs-SDSS-V).
A.A.C.S., and R.R.L. are supported by the German \emph{Deut\-sche For\-schungs\-ge\-mein\-schaft, DFG\/} in the form of an Emmy Noether Research Group -- Project-ID 445674056 (SA4064/1-1, PI Sander).
L.R.P. acknowledges support by grants PID2019-105552RB-C41 and PID2022-137779OB-C41 funded by MCIN/AEI/10.13039/501100011033 by ``ERDF A way of making Europe''. F.T. gratefully acknowledges support by grant PID2022-137779OB-C41, funded by the Spanish Ministry of Science, Innovation and Universities/State Agency of Research MICIU/AEI/10.13039/501100011033.

\end{acknowledgements}

\bibliographystyle{aa} 
\bibliography{aanda} 

\begin{thebibliography}{130}
\expandafter\ifx\csname natexlab\endcsname\relax\def\natexlab#1{#1}\fi

\bibitem[{{Abbott} {et~al.}(2019){Abbott}, {Abbott}, {Abbott}, {Abraham}, {Acernese}, {Ackley}, {Adams}, {Adhikari}, {Adya}, {Affeldt}, {Agathos}, {Agatsuma}, {Aggarwal}, {Aguiar}, {Aiello}, {Ain}, {Ajith}, {Allen}, {Allocca}, {Aloy}, {Altin}, {Amato}, {Ananyeva}, {Anderson}, {Anderson}, {Angelova}, {Antier}, {Appert}, {Arai}, {Araya}, {Areeda}, {Ar{\`e}ne}, {Arnaud}, {Arun}, {Ascenzi}, {Ashton}, {Aston}, {Astone}, {Aubin}, {Aufmuth}, {AultONeal}, {Austin}, {Avendano}, {Avila-Alvarez}, {Babak}, {Bacon}, {Badaracco}, {Bader}, {Bae}, {Baker}, {Baldaccini}, {Ballardin}, {Ballmer}, {Banagiri}, {Barayoga}, {Barclay}, {Barish}, {Barker}, {Barkett}, {Barnum}, {Barone}, {Barr}, {Barsotti}, {Barsuglia}, {Barta}, {Bartlett}, {Bartos}, {Bassiri}, {Basti}, {Bawaj}, {Bayley}, {Bazzan}, {B{\'e}csy}, {Bejger}, {Belahcene}, {Bell}, {Beniwal}, {Berger}, {Bergmann}, {Bernuzzi}, {Bero}, {Berry}, {Bersanetti}, {Bertolini}, {Betzwieser}, {Bhandare}, {Bidler}, {Bilenko}, {Bilgili}, {Billingsley}, {Birch}, {Birney}, {Birnholtz},
  {Biscans}, {Biscoveanu}, {Bisht}, {Bitossi}, {Bizouard}, {Blackburn}, {Blackman}, {Blair}, {Blair}, {Blair}, {Bloemen}, {Bode}, {Boer}, {Boetzel}, {Bogaert}, {Bondu}, {Bonilla}, {Bonnand}, {Booker}, {Boom}, {Booth}, {Bork}, {Boschi}, {Bose}, {Bossie}, {Bossilkov}, {Bosveld}, {Bouffanais}, {Bozzi}, {Bradaschia}, {Brady}, {Bramley}, {Branchesi}, {Brau}, {Briant}, {Briggs}, {Brighenti}, {Brillet}, {Brinkmann}, {Brisson}, {Brockill}, {Brooks}, {Brown}, {Brunett}, {Buikema}, {Bulik}, {Bulten}, {Buonanno}, {Buskulic}, {Bustamante Rosell}, {Buy}, {Byer}, {Cabero}, {Cadonati}, {Cagnoli}, {Cahillane}, {Calder{\'o}n Bustillo}, {Callister}, {Calloni}, {Camp}, {Campbell}, {Canepa}, {Cannon}, {Cao}, {Cao}, {Capocasa}, {Carbognani}, {Caride}, {Carney}, {Carullo}, {Casanueva Diaz}, {Casentini}, {Caudill}, {Cavagli{\`a}}, {Cavalier}, {Cavalieri}, {Cella}, {Cerd{\'a}-Dur{\'a}n}, {Cerretani}, {Cesarini}, {Chaibi}, {Chakravarti}, {Chamberlin}, {Chan}, {Chao}, {Charlton}, {Chase}, {Chassande-Mottin}, {Chatterjee},
  {Chaturvedi}, {Chatziioannou}, {Cheeseboro}, {Chen}, {Chen}, {Chen}, {Cheng}, {Cheong}, {Chia}, {Chincarini}, {Chiummo}, {Cho}, {Cho}, {Cho}, {Christensen}, {Chu}, {Chua}, {Chung}, {Chung}, {Ciani}, {Ciobanu}, {Ciolfi}, {Cipriano}, {Cirone}, {Clara}, {Clark}, {Clearwater}, {Cleva}, {Cocchieri}, {Coccia}, {Cohadon}, {Cohen}, {Colgan}, {Colleoni}, {Collette}, {Collins}, {Cominsky}, {Constancio}, {Conti}, {Cooper}, {Corban}, {Corbitt}, {Cordero-Carri{\'o}n}, {Corley}, {Cornish}, {Corsi}, {Cortese}, {Costa}, {Cotesta}, {Coughlin}, {Coughlin}, {Coulon}, {Countryman}, {Couvares}, {Covas}, {Cowan}, {Coward}, {Cowart}, {Coyne}, {Coyne}, {Creighton}, {Creighton}, {Cripe}, {Croquette}, {Crowder}, {Cullen}, {Cumming}, {Cunningham}, {Cuoco}, {Canton}, {D{\'a}lya}, {Danilishin}, {D'Antonio}, {Danzmann}, {Dasgupta}, {Da Silva Costa}, {Datrier}, {Dattilo}, {Dave}, {Davier}, {Davis}, {Daw}, {DeBra}, {Deenadayalan}, {Degallaix}, {De Laurentis}, {Del{\'e}glise}, {Del Pozzo}, {DeMarchi}, {Demos}, {Dent}, {De Pietri}, {Derby},
  {De Rosa}, {De Rossi}, {DeSalvo}, {de Varona}, {Dhurandhar}, {D{\'\i}az}, {Dietrich}, {Di Fiore}, {Di Giovanni}, {Di Girolamo}, {Di Lieto}, {Ding}, {Di Pace}, {Di Palma}, {Di Renzo}, {Dmitriev}, {Doctor}, {Donovan}, {Dooley}, {Doravari}, {Dorrington}, {Downes}, {Drago}, {Driggers}, {Du}, {Ducoin}, {Dupej}, {Dwyer}, {Easter}, {Edo}, {Edwards}, {Effler}, {Ehrens}, {Eichholz}, {Eikenberry}, {Eisenmann}, {Eisenstein}, {Essick}, {Estelles}, {Estevez}, {Etienne}, {Etzel}, {Evans}, {Evans}, {Fafone}, {Fair}, {Fairhurst}, {Fan}, {Farinon}, {Farr}, {Farr}, {Fauchon-Jones}, {Favata}, {Fays}, {Fazio}, {Fee}, {Feicht}, {Fejer}, {Feng}, {Fernandez-Galiana}, {Ferrante}, {Ferreira}, {Ferreira}, {Ferrini}, {Fidecaro}, {Fiori}, {Fiorucci}, {Fishbach}, {Fisher}, {Fishner}, {Fitz-Axen}, {Flaminio}, {Fletcher}, {Flynn}, {Fong}, {Font}, {Forsyth}, {Fournier}, {Frasca}, {Frasconi}, {Frei}, {Freise}, {Frey}, {Frey}, {Fritschel}, {Frolov}, {Fulda}, {Fyffe}, {Gabbard}, {Gadre}, {Gaebel}, {Gair}, {Gammaitoni}, {Ganija}, {Gaonkar},
  {Garcia}, {Garc{\'\i}a-Quir{\'o}s}, {Garufi}, {Gateley}, {Gaudio}, {Gaur}, {Gayathri}, {Gemme}, {Genin}, {Gennai}, {George}, {George}, {Gergely}, {Germain}, {Ghonge}, {Ghosh}, {Ghosh}, {Ghosh}, {Giacomazzo}, {Giaime}, {Giardina}, {Giazotto}, {Gill}, {Giordano}, {Glover}, {Godwin}, {Goetz}, {Goetz}, {Goncharov}, {Gonz{\'a}lez}, {Gonzalez Castro}, {Gopakumar}, {Gorodetsky}, {Gossan}, {Gosselin}, {Gouaty}, {Grado}, {Graef}, {Granata}, {Grant}, {Gras}, {Grassia}, {Gray}, {Gray}, {Greco}, {Green}, {Green}, {Gretarsson}, {Groot}, {Grote}, {Grunewald}, {Gruning}, {Guidi}, {Gulati}, {Guo}, {Gupta}, {Gupta}, {Gustafson}, {Gustafson}, {Haegel}, {Halim}, {Hall}, {Hall}, {Hamilton}, {Hammond}, {Haney}, {Hanke}, {Hanks}, {Hanna}, {Hannam}, {Hannuksela}, {Hanson}, {Hardwick}, {Haris}, {Harms}, {Harry}, {Harry}, {Haster}, {Haughian}, {Hayes}, {Healy}, {Heidmann}, {Heintze}, {Heitmann}, {Hello}, {Hemming}, {Hendry}, {Heng}, {Hennig}, {Heptonstall}, {Hernandez Vivanco}, {Heurs}, {Hild}, {Hinderer}, {Hoak}, {Hochheim},
  {Hofman}, {Holgado}, {Holland}, {Holt}, {Holz}, {Hopkins}, {Horst}, {Hough}, {Howell}, {Hoy}, {Hreibi}, {Huang}, {Huerta}, {Huet}, {Hughey}, {Hulko}, {Husa}, {Huttner}, {Huynh-Dinh}, {Idzkowski}, {Iess}, {Ingram}, {Inta}, {Intini}, {Irwin}, {Isa}, {Isac}, {Isi}, {Iyer}, {Izumi}, {Jacqmin}, {Jadhav}, {Jani}, {Janthalur}, {Jaranowski}, {Jenkins}, {Jiang}, {Johnson}, {Johnson-McDaniel}, {Jones}, {Jones}, {Jones}, {Jonker}, {Ju}, {Junker}, {Kalaghatgi}, {Kalogera}, {Kamai}, {Kandhasamy}, {Kang}, {Kanner}, {Kapadia}, {Karki}, {Karvinen}, {Kashyap}, {Kasprzack}, {Katsanevas}, {Katsavounidis}, {Katzman}, {Kaufer}, {Kawabe}, {Keerthana}, {K{\'e}f{\'e}lian}, {Keitel}, {Kennedy}, {Key}, {Khalili}, {Khan}, {Khan}, {Khan}, {Khan}, {Khazanov}, {Khursheed}, {Kijbunchoo}, {Kim}, {Kim}, {Kim}, {Kim}, {Kim}, {Kim}, {Kimball}, {King}, {King}, {Kinley-Hanlon}, {Kirchhoff}, {Kissel}, {Kleybolte}, {Klika}, {Klimenko}, {Knowles}, {Koch}, {Koehlenbeck}, {Koekoek}, {Koley}, {Kondrashov}, {Kontos}, {Koper}, {Korobko}, {Korth},
  {Kowalska}, {Kozak}, {Kringel}, {Krishnendu}, {Kr{\'o}lak}, {Kuehn}, {Kumar}, {Kumar}, {Kumar}, {Kumar}, {Kuo}, {Kutynia}, {Kwang}, {Lackey}, {Lai}, {Lam}, {Landry}, {Lane}, {Lang}, {Lange}, {Lantz}, {Lanza}, {Lartaux-Vollard}, {Lasky}, {Laxen}, {Lazzarini}, {Lazzaro}, {Leaci}, {Leavey}, {Lecoeuche}, {Lee}, {Lee}, {Lee}, {Lee}, {Lee}, {Lee}, {Lehmann}, {Lenon}, {Leroy}, {Letendre}, {Levin}, {Li}, {Li}, {Li}, {Li}, {Lin}, {Linde}, {Linker}, {Littenberg}, {Liu}, {Liu}, {Lo}, {Lockerbie}, {London}, {Longo}, {Lorenzini}, {Loriette}, {Lormand}, {Losurdo}, {Lough}, {Lousto}, {Lovelace}, {Lower}, {L{\"u}ck}, {Lumaca}, {Lundgren}, {Lynch}, {Ma}, {Macas}, {Macfoy}, {MacInnis}, {Macleod}, {Macquet}, {Maga{\~n}a-Sandoval}, {Maga{\~n}a Zertuche}, {Magee}, {Majorana}, {Maksimovic}, {Malik}, {Man}, {Mandic}, {Mangano}, {Mansell}, {Manske}, {Mantovani}, {Marchesoni}, {Marion}, {M{\'a}rka}, {M{\'a}rka}, {Markakis}, {Markosyan}, {Markowitz}, {Maros}, {Marquina}, {Marsat}, {Martelli}, {Martin}, {Martin}, {Martynov}, {Mason},
  {Massera}, {Masserot}, {Massinger}, {Masso-Reid}, {Mastrogiovanni}, {Matas}, {Matichard}, {Matone}, {Mavalvala}, {Mazumder}, {McCann}, {McCarthy}, {McClelland}, {McCormick}, {McCuller}, {McGuire}, {McIver}, {McManus}, {McRae}, {McWilliams}, {Meacher}, {Meadors}, {Mehmet}, {Mehta}, {Meidam}, {Melatos}, {Mendell}, {Mercer}, {Mereni}, {Merilh}, {Merzougui}, {Meshkov}, {Messenger}, {Messick}, {Metzdorff}, {Meyers}, {Miao}, {Michel}, {Middleton}, {Mikhailov}, {Milano}, {Miller}, {Miller}, {Millhouse}, {Mills}, {Milovich-Goff}, {Minazzoli}, {Minenkov}, {Mishkin}, {Mishra}, {Mistry}, {Mitra}, {Mitrofanov}, {Mitselmakher}, {Mittleman}, {Mo}, {Moffa}, {Mogushi}, {Mohapatra}, {Montani}, {Moore}, {Moraru}, {Moreno}, {Morisaki}, {Mours}, {Mow-Lowry}, {Mukherjee}, {Mukherjee}, {Mukherjee}, {Mukund}, {Mullavey}, {Munch}, {Mu{\~n}iz}, {Muratore}, {Murray}, {Nagar}, {Nardecchia}, {Naticchioni}, {Nayak}, {Neilson}, {Nelemans}, {Nelson}, {Nery}, {Neunzert}, {Ng}, {Ng}, {Nguyen}, {Nichols}, {Nielsen}, {Nissanke}, {Nitz},
  {Nocera}, {North}, {Nuttall}, {Obergaulinger}, {Oberling}, {O'Brien}, {O'Dea}, {Ogin}, {Oh}, {Oh}, {Ohme}, {Ohta}, {Okada}, {Oliver}, {Oppermann}, {Oram}, {O'Reilly}, {Ormiston}, {Ortega}, {O'Shaughnessy}, {Ossokine}, {Ottaway}, {Overmier}, {Owen}, {Pace}, {Pagano}, {Page}, {Pai}, {Pai}, {Palamos}, {Palashov}, {Palomba}, {Pal-Singh}, {Pan}, {Pang}, {Pang}, {Pankow}, {Pannarale}, {Pant}, {Paoletti}, {Paoli}, {Papa}, {Parida}, {Parker}, {Pascucci}, {Pasqualetti}, {Passaquieti}, {Passuello}, {Patil}, {Patricelli}, {Pearlstone}, {Pedersen}, {Pedraza}, {Pedurand}, {Pele}, {Penn}, {Perego}, {Perez}, {Perreca}, {Pfeiffer}, {Phelps}, {Phukon}, {Piccinni}, {Pichot}, {Piergiovanni}, {Pillant}, {Pinard}, {Pirello}, {Pitkin}, {Poggiani}, {Pong}, {Ponrathnam}, {Popolizio}, {Porter}, {Powell}, {Prajapati}, {Prasad}, {Prasai}, {Prasanna}, {Pratten}, {Prestegard}, {Privitera}, {Prodi}, {Prokhorov}, {Puncken}, {Punturo}, {Puppo}, {P{\"u}rrer}, {Qi}, {Quetschke}, {Quinonez}, {Quintero}, {Quitzow-James}, {Raab}, {Radkins},
  {Radulescu}, {Raffai}, {Raja}, {Rajan}, {Rajbhandari}, {Rakhmanov}, {Ramirez}, {Ramos-Buades}, {Rana}, {Rao}, {Rapagnani}, {Raymond}, {Razzano}, {Read}, {Regimbau}, {Rei}, {Reid}, {Reitze}, {Ren}, {Ricci}, {Richardson}, {Richardson}, {Ricker}, {Riemenschneider}, {Riles}, {Rizzo}, {Robertson}, {Robie}, {Robinet}, {Rocchi}, {Rolland}, {Rollins}, {Roma}, {Romanelli}, {Romano}, {Romel}, {Romie}, {Rose}, {Rosi{\'n}ska}, {Rosofsky}, {Ross}, {Rowan}, {R{\"u}diger}, {Ruggi}, {Rutins}, {Ryan}, {Sachdev}, {Sadecki}, {Sakellariadou}, {Salafia}, {Salconi}, {Saleem}, {Salemi}, {Samajdar}, {Sammut}, {Sanchez}, {Sanchez}, {Sanchis-Gual}, {Sandberg}, {Sanders}, {Santiago}, {Sarin}, {Sassolas}, {Sathyaprakash}, {Saulson}, {Sauter}, {Savage}, {Schale}, {Scheel}, {Scheuer}, {Schmidt}, {Schnabel}, {Schofield}, {Sch{\"o}nbeck}, {Schreiber}, {Schulte}, {Schutz}, {Schwalbe}, {Scott}, {Scott}, {Seidel}, {Sellers}, {Sengupta}, {Sennett}, {Sentenac}, {Sequino}, {Sergeev}, {Setyawati}, {Shaddock}, {Shaffer}, {Shahriar}, {Shaner},
  {Shao}, {Sharma}, {Shawhan}, {Shen}, {Shink}, {Shoemaker}, {Shoemaker}, {ShyamSundar}, {Siellez}, {Sieniawska}, {Sigg}, {Silva}, {Singer}, {Singh}, {Singhal}, {Sintes}, {Sitmukhambetov}, {Skliris}, {Slagmolen}, {Slaven-Blair}, {Smith}, {Smith}, {Somala}, {Son}, {Sorazu}, {Sorrentino}, {Souradeep}, {Sowell}, {Spencer}, {Srivastava}, {Srivastava}, {Staats}, {Stachie}, {Standke}, {Steer}, {Steinke}, {Steinlechner}, {Steinlechner}, {Steinmeyer}, {Stevenson}, {Stocks}, {Stone}, {Stops}, {Strain}, {Stratta}, {Strigin}, {Strunk}, {Sturani}, {Stuver}, {Sudhir}, {Summerscales}, {Sun}, {Sunil}, {Suresh}, {Sutton}, {Swinkels}, {Szczepa{\'n}czyk}, {Tacca}, {Tait}, {Talbot}, {Talukder}, {Tanner}, {T{\'a}pai}, {Taracchini}, {Tasson}, {Taylor}, {Thies}, {Thomas}, {Thomas}, {Thondapu}, {Thorne}, {Thrane}, {Tiwari}, {Tiwari}, {Tiwari}, {Toland}, {Tonelli}, {Tornasi}, {Torres-Forn{\'e}}, {Torrie}, {T{\"o}yr{\"a}}, {Travasso}, {Traylor}, {Tringali}, {Trovato}, {Trozzo}, {Trudeau}, {Tsang}, {Tse}, {Tso}, {Tsukada}, {Tsuna},
  {Tuyenbayev}, {Ueno}, {Ugolini}, {Unnikrishnan}, {Urban}, {Usman}, {Vahlbruch}, {Vajente}, {Valdes}, {van Bakel}, {van Beuzekom}, {van den Brand}, {Van Den Broeck}, {Vander-Hyde}, {van Heijningen}, {van der Schaaf}, {van Veggel}, {Vardaro}, {Varma}, {Vass}, {Vas{\'u}th}, {Vecchio}, {Vedovato}, {Veitch}, {Veitch}, {Venkateswara}, {Venugopalan}, {Verkindt}, {Vetrano}, {Vicer{\'e}}, {Viets}, {Vine}, {Vinet}, {Vitale}, {Vo}, {Vocca}, {Vorvick}, {Vyatchanin}, {Wade}, {Wade}, {Wade}, {Walet}, {Walker}, {Wallace}, {Walsh}, {Wang}, {Wang}, {Wang}, {Wang}, {Wang}, {Ward}, {Warden}, {Warner}, {Was}, {Watchi}, {Weaver}, {Wei}, {Weinert}, {Weinstein}, {Weiss}, {Wellmann}, {Wen}, {Wessel}, {We{\ss}els}, {Westhouse}, {Wette}, {Whelan}, {White}, {Whiting}, {Whittle}, {Wilken}, {Williams}, {Williamson}, {Willis}, {Willke}, {Wimmer}, {Winkler}, {Wipf}, {Wittel}, {Woan}, {Woehler}, {Wofford}, {Worden}, {Wright}, {Wu}, {Wysocki}, {Xiao}, {Yamamoto}, {Yancey}, {Yang}, {Yap}, {Yazback}, {Yeeles}, {Yu}, {Yu}, {Yuen}, {Yvert},
  {Zadro{\.Z}ny}, {Zanolin}, {Zappa}, {Zelenova}, {Zendri}, {Zevin}, {Zhang}, {Zhang}, {Zhang}, {Zhao}, {Zhou}, {Zhou}, {Zhu}, {Zimmerman}, {Zlochower}, {Zucker}, {Zweizig}, {LIGO Scientific Collaboration}, \& {Virgo Collaboration}}]{2019Abbott}
{Abbott}, B.~P., {Abbott}, R., {Abbott}, T.~D., {et~al.} 2019, Physical Review X, 9, 031040

\bibitem[{{Abbott} \& {Conti}(1987)}]{1987Abbott}
{Abbott}, D.~C. \& {Conti}, P.~S. 1987, \araa, 25, 113

\bibitem[{{Abbott} {et~al.}(2021){Abbott}, {Abbott}, {Abraham}, {Acernese}, {Ackley}, {Adams}, {Adams}, {Adhikari}, {Adya}, {Affeldt}, {Agathos}, {Agatsuma}, {Aggarwal}, {Aguiar}, {Aiello}, {Ain}, {Ajith}, {Akcay}, {Allen}, {Allocca}, {Altin}, {Amato}, {Anand}, {Ananyeva}, {Anderson}, {Anderson}, {Angelova}, {Ansoldi}, {Antelis}, {Antier}, {Appert}, {Arai}, {Araya}, {Areeda}, {Ar{\`e}ne}, {Arnaud}, {Aronson}, {Arun}, {Asali}, {Ascenzi}, {Ashton}, {Aston}, {Astone}, {Aubin}, {Aufmuth}, {AultONeal}, {Austin}, {Avendano}, {Babak}, {Badaracco}, {Bader}, {Bae}, {Baer}, {Bagnasco}, {Baird}, {Ball}, {Ballardin}, {Ballmer}, {Bals}, {Balsamo}, {Baltus}, {Banagiri}, {Bankar}, {Bankar}, {Barayoga}, {Barbieri}, {Barish}, {Barker}, {Barneo}, {Barnum}, {Barone}, {Barr}, {Barsotti}, {Barsuglia}, {Barta}, {Bartlett}, {Bartos}, {Bassiri}, {Basti}, {Bawaj}, {Bayley}, {Bazzan}, {Becher}, {B{\'e}csy}, {Bedakihale}, {Bejger}, {Belahcene}, {Beniwal}, {Benjamin}, {Bennett}, {Bentley}, {Bergamin}, {Berger}, {Bergmann}, {Bernuzzi},
  {Berry}, {Bersanetti}, {Bertolini}, {Betzwieser}, {Bhandare}, {Bhandari}, {Bhattacharjee}, {Bidler}, {Bilenko}, {Billingsley}, {Birney}, {Birnholtz}, {Biscans}, {Bischi}, {Biscoveanu}, {Bisht}, {Bitossi}, {Bizouard}, {Blackburn}, {Blackman}, {Blair}, {Blair}, {Blair}, {Blanch}, {Bobba}, {Bode}, {Boer}, {Boetzel}, {Bogaert}, {Boldrini}, {Bondu}, {Bonilla}, {Bonnand}, {Booker}, {Boom}, {Bork}, {Boschi}, {Bose}, {Bossilkov}, {Boudart}, {Bouffanais}, {Bozzi}, {Bradaschia}, {Brady}, {Bramley}, {Branchesi}, {Brau}, {Breschi}, {Briant}, {Briggs}, {Brighenti}, {Brillet}, {Brinkmann}, {Brockill}, {Brooks}, {Brooks}, {Brown}, {Brunett}, {Bruno}, {Bruntz}, {Buikema}, {Bulik}, {Bulten}, {Buonanno}, {Buscicchio}, {Buskulic}, {Byer}, {Cabero}, {Cadonati}, {Caesar}, {Cagnoli}, {Cahillane}, {Calder{\'o}n Bustillo}, {Callaghan}, {Callister}, {Calloni}, {Camp}, {Canepa}, {Cannon}, {Cao}, {Cao}, {Carapella}, {Carbognani}, {Carney}, {Carpinelli}, {Carullo}, {Carver}, {Casanueva Diaz}, {Casentini}, {Caudill}, {Cavagli{\`a}},
  {Cavalier}, {Cavalieri}, {Cella}, {Cerd{\'a}-Dur{\'a}n}, {Cesarini}, {Chaibi}, {Chakravarti}, {Chan}, {Chan}, {Chandra}, {Chanial}, {Chao}, {Charlton}, {Chase}, {Chassande-Mottin}, {Chatterjee}, {Chattopadhyay}, {Chaturvedi}, {Chatziioannou}, {Chen}, {Chen}, {Chen}, {Chen}, {Cheng}, {Cheong}, {Chia}, {Chiadini}, {Chierici}, {Chincarini}, {Chiummo}, {Cho}, {Cho}, {Cho}, {Choate}, {Christensen}, {Chu}, {Chua}, {Chung}, {Chung}, {Ciani}, {Ciecielag}, {Cie{\'s}lar}, {Cifaldi}, {Ciobanu}, {Ciolfi}, {Cipriano}, {Cirone}, {Clara}, {Clark}, {Clark}, {Clarke}, {Clearwater}, {Clesse}, {Cleva}, {Coccia}, {Cohadon}, {Cohen}, {Colleoni}, {Collette}, {Collins}, {Colpi}, {Constancio}, {Conti}, {Cooper}, {Corban}, {Corbitt}, {Cordero-Carri{\'o}n}, {Corezzi}, {Corley}, {Cornish}, {Corre}, {Corsi}, {Cortese}, {Costa}, {Cotesta}, {Coughlin}, {Coughlin}, {Coulon}, {Countryman}, {Cousins}, {Couvares}, {Covas}, {Coward}, {Cowart}, {Coyne}, {Coyne}, {Creighton}, {Creighton}, {Croquette}, {Crowder}, {Cudell}, {Cullen}, {Cumming},
  {Cummings}, {Cunningham}, {Cuoco}, {Cury{\l}o}, {Canton}, {D{\'a}lya}, {Dana}, {DaneshgaranBajastani}, {D'Angelo}, {Danila}, {Danilishin}, {D'Antonio}, {Danzmann}, {Darsow-Fromm}, {Dasgupta}, {Datrier}, {Dattilo}, {Dave}, {Davier}, {Davies}, {Davis}, {Daw}, {Dean}, {DeBra}, {Deenadayalan}, {Degallaix}, {De Laurentis}, {Del{\'e}glise}, {Del Favero}, {De Lillo}, {De Lillo}, {Del Pozzo}, {DeMarchi}, {De Matteis}, {D'Emilio}, {Demos}, {Denker}, {Dent}, {Depasse}, {De Pietri}, {De Rosa}, {De Rossi}, {DeSalvo}, {de Varona}, {Dhurandhar}, {D{\'\i}az}, {Diaz-Ortiz}, {Didio}, {Dietrich}, {Di Fiore}, {DiFronzo}, {Di Giorgio}, {Di Giovanni}, {Di Giovanni}, {Di Girolamo}, {Di Lieto}, {Ding}, {Di Pace}, {Di Palma}, {Di Renzo}, {Divakarla}, {Dmitriev}, {Doctor}, {D'Onofrio}, {Donovan}, {Dooley}, {Doravari}, {Dorrington}, {Downes}, {Drago}, {Driggers}, {Du}, {Ducoin}, {Dupej}, {Durante}, {D'Urso}, {Duverne}, {Dwyer}, {Easter}, {Eddolls}, {Edelman}, {Edo}, {Edy}, {Effler}, {Eichholz}, {Eikenberry}, {Eisenmann},
  {Eisenstein}, {Ejlli}, {Errico}, {Essick}, {Estell{\'e}s}, {Estevez}, {Etienne}, {Etzel}, {Evans}, {Evans}, {Ewing}, {Fafone}, {Fair}, {Fairhurst}, {Fan}, {Farah}, {Farinon}, {Farr}, {Farr}, {Fauchon-Jones}, {Favata}, {Fays}, {Fazio}, {Feicht}, {Fejer}, {Feng}, {Fenyvesi}, {Ferguson}, {Fernandez-Galiana}, {Ferrante}, {Ferreira}, {Fidecaro}, {Figura}, {Fiori}, {Fiorucci}, {Fishbach}, {Fisher}, {Fishner}, {Fittipaldi}, {Fitz-Axen}, {Fiumara}, {Flaminio}, {Floden}, {Flynn}, {Fong}, {Font}, {Forsyth}, {Fournier}, {Frasca}, {Frasconi}, {Frei}, {Freise}, {Frey}, {Frey}, {Fritschel}, {Frolov}, {Fronz{\'e}}, {Fulda}, {Fyffe}, {Gabbard}, {Gadre}, {Gaebel}, {Gair}, {Gais}, {Galaudage}, {Gamba}, {Ganapathy}, {Ganguly}, {Gaonkar}, {Garaventa}, {Garc{\'\i}a-Quir{\'o}s}, {Garufi}, {Gateley}, {Gaudio}, {Gayathri}, {Gemme}, {Gennai}, {George}, {George}, {George}, {Gergely}, {Ghonge}, {Ghosh}, {Ghosh}, {Ghosh}, {Giacomazzo}, {Giacoppo}, {Giaime}, {Giardina}, {Gibson}, {Gier}, {Gill}, {Giri}, {Glanzer}, {Gleckl}, {Godwin},
  {Goetz}, {Goetz}, {Gohlke}, {Goncharov}, {Gonz{\'a}lez}, {Gopakumar}, {Gossan}, {Gosselin}, {Gouaty}, {Grace}, {Grado}, {Granata}, {Granata}, {Grant}, {Gras}, {Grassia}, {Gray}, {Gray}, {Greco}, {Green}, {Green}, {Gretarsson}, {Griggs}, {Grignani}, {Grimaldi}, {Grimes}, {Grimm}, {Grote}, {Grunewald}, {Gruning}, {Guerrero}, {Guidi}, {Guimaraes}, {Guix{\'e}}, {Gulati}, {Guo}, {Gupta}, {Gupta}, {Gupta}, {Gustafson}, {Gustafson}, {Guzman}, {Haegel}, {Halim}, {Hall}, {Hamilton}, {Hammond}, {Haney}, {Hanke}, {Hanks}, {Hanna}, {Hannam}, {Hannuksela}, {Hannuksela}, {Hansen}, {Hansen}, {Hanson}, {Harder}, {Hardwick}, {Haris}, {Harms}, {Harry}, {Harry}, {Hartwig}, {Hasskew}, {Haster}, {Haughian}, {Hayes}, {Healy}, {Heidmann}, {Heintze}, {Heinze}, {Heinzel}, {Heitmann}, {Hellman}, {Hello}, {Helmling-Cornell}, {Hemming}, {Hendry}, {Heng}, {Hennes}, {Hennig}, {Hennig}, {Hernandez Vivanco}, {Heurs}, {Hild}, {Hill}, {Hines}, {Hochheim}, {Hofgard}, {Hofman}, {Hohmann}, {Holgado}, {Holland}, {Hollows}, {Holmes}, {Holt},
  {Holz}, {Hopkins}, {Horst}, {Hough}, {Howell}, {Hoy}, {Hoyland}, {Huang}, {H{\"u}bner}, {Huddart}, {Huerta}, {Hughey}, {Hui}, {Husa}, {Huttner}, {Hutzler}, {Huxford}, {Huynh-Dinh}, {Idzkowski}, {Iess}, {Imperato}, {Inchauspe}, {Ingram}, {Intini}, {Isi}, {Iyer}, {JaberianHamedan}, {Jacqmin}, {Jadhav}, {Jadhav}, {James}, {Jani}, {Janssens}, {Janthalur}, {Jaranowski}, {Jariwala}, {Jaume}, {Jenkins}, {Jeunon}, {Jiang}, {Johns}, {Johnson-McDaniel}, {Jones}, {Jones}, {Jones}, {Jones}, {Jones}, {Jonker}, {Ju}, {Junker}, {Kalaghatgi}, {Kalogera}, {Kamai}, {Kandhasamy}, {Kang}, {Kanner}, {Kapadia}, {Kapasi}, {Karathanasis}, {Karki}, {Kashyap}, {Kasprzack}, {Kastaun}, {Katsanevas}, {Katsavounidis}, {Katzman}, {Kawabe}, {K{\'e}f{\'e}lian}, {Keitel}, {Key}, {Khadka}, {Khalili}, {Khan}, {Khan}, {Khazanov}, {Khetan}, {Khursheed}, {Kijbunchoo}, {Kim}, {Kim}, {Kim}, {Kim}, {Kim}, {Kim}, {Kimball}, {King}, {Kinley-Hanlon}, {Kirchhoff}, {Kissel}, {Kleybolte}, {Klimenko}, {Knowles}, {Knyazev}, {Koch}, {Koehlenbeck},
  {Koekoek}, {Koley}, {Kolstein}, {Komori}, {Kondrashov}, {Kontos}, {Koper}, {Korobko}, {Korth}, {Kovalam}, {Kozak}, {Kr{\"a}mer}, {Kringel}, {Krishnendu}, {Kr{\'o}lak}, {Kuehn}, {Kumar}, {Kumar}, {Kumar}, {Kumar}, {Kuns}, {Kwang}, {Lackey}, {Laghi}, {Lalande}, {Lam}, {Lamberts}, {Landry}, {Lane}, {Lang}, {Lange}, {Lantz}, {Lanza}, {La Rosa}, {Lartaux-Vollard}, {Lasky}, {Laxen}, {Lazzarini}, {Lazzaro}, {Leaci}, {Leavey}, {Lecoeuche}, {Lee}, {Lee}, {Lee}, {Lee}, {Lehmann}, {Leon}, {Leroy}, {Letendre}, {Levin}, {Li}, {Li}, {Li}, {Li}, {Li}, {Linde}, {Linker}, {Linley}, {Littenberg}, {Liu}, {Liu}, {Llorens-Monteagudo}, {Lo}, {Lockwood}, {London}, {Longo}, {Lorenzini}, {Loriette}, {Lormand}, {Losurdo}, {Lough}, {Lousto}, {Lovelace}, {L{\"u}ck}, {Lumaca}, {Lundgren}, {Ma}, {Macas}, {MacInnis}, {Macleod}, {MacMillan}, {Macquet}, {Maga{\~n}a Hernandez}, {Maga{\~n}a-Sandoval}, {Magazz{\`u}}, {Magee}, {Majorana}, {Maksimovic}, {Maliakal}, {Malik}, {Man}, {Mandic}, {Mangano}, {Mansell}, {Manske}, {Mantovani},
  {Mapelli}, {Marchesoni}, {Marion}, {M{\'a}rka}, {M{\'a}rka}, {Markakis}, {Markosyan}, {Markowitz}, {Maros}, {Marquina}, {Marsat}, {Martelli}, {Martin}, {Martin}, {Martinez}, {Martinez}, {Martynov}, {Masalehdan}, {Mason}, {Massera}, {Masserot}, {Massinger}, {Masso-Reid}, {Mastrogiovanni}, {Matas}, {Mateu-Lucena}, {Matichard}, {Matiushechkina}, {Mavalvala}, {Maynard}, {McCann}, {McCarthy}, {McClelland}, {McCormick}, {McCuller}, {McGuire}, {McIsaac}, {McIver}, {McManus}, {McRae}, {McWilliams}, {Meacher}, {Meadors}, {Mehmet}, {Mehta}, {Melatos}, {Melchor}, {Mendell}, {Menendez-Vazquez}, {Mercer}, {Mereni}, {Merfeld}, {Merilh}, {Merritt}, {Merzougui}, {Meshkov}, {Messenger}, {Messick}, {Metzdorff}, {Meyers}, {Meylahn}, {Mhaske}, {Miani}, {Miao}, {Michaloliakos}, {Michel}, {Middleton}, {Milano}, {Miller}, {Millhouse}, {Mills}, {Milotti}, {Milovich-Goff}, {Minazzoli}, {Minenkov}, {Mir}, {Mishkin}, {Mishra}, {Mistry}, {Mitra}, {Mitrofanov}, {Mitselmakher}, {Mittleman}, {Mo}, {Mogushi}, {Mohapatra}, {Mohite},
  {Molina}, {Molina-Ruiz}, {Mondin}, {Montani}, {Moore}, {Moraru}, {Morawski}, {Moreno}, {Morisaki}, {Mours}, {Mow-Lowry}, {Mozzon}, {Muciaccia}, {Mukherjee}, {Mukherjee}, {Mukherjee}, {Mukherjee}, {Mukund}, {Mullavey}, {Munch}, {Mu{\~n}iz}, {Murray}, {Nadji}, {Nagar}, {Nardecchia}, {Naticchioni}, {Nayak}, {Neil}, {Neilson}, {Nelemans}, {Nelson}, {Nery}, {Neunzert}, {Nitz}, {Ng}, {Ng}, {Nguyen}, {Nguyen}, {Nguyen}, {Nichols}, {Nissanke}, {Nocera}, {Noh}, {North}, {Nothard}, {Nuttall}, {Oberling}, {O'Brien}, {O'Dell}, {Oganesyan}, {Ogin}, {Oh}, {Oh}, {Ohme}, {Ohta}, {Okada}, {Olivetto}, {Oppermann}, {Oram}, {O'Reilly}, {Ormiston}, {Ortega}, {O'Shaughnessy}, {Ossokine}, {Osthelder}, {Ottaway}, {Overmier}, {Owen}, {Pace}, {Pagano}, {Page}, {Pagliaroli}, {Pai}, {Pai}, {Palamos}, {Palashov}, {Palomba}, {Pan}, {Panda}, {Pang}, {Pankow}, {Pannarale}, {Pant}, {Paoletti}, {Paoli}, {Paolone}, {Parker}, {Pascucci}, {Pasqualetti}, {Passaquieti}, {Passuello}, {Patel}, {Patricelli}, {Payne}, {Pechsiri}, {Pedraza},
  {Pegoraro}, {Pele}, {Penn}, {Perego}, {Perez}, {P{\'e}rigois}, {Perreca}, {Perri{\`e}s}, {Petermann}, {Petterson}, {Pfeiffer}, {Pham}, {Phukon}, {Piccinni}, {Pichot}, {Piendibene}, {Piergiovanni}, {Pierini}, {Pierro}, {Pillant}, {Pilo}, {Pinard}, {Pinto}, {Piotrzkowski}, {Pirello}, {Pitkin}, {Placidi}, {Plastino}, {Pluchar}, {Poggiani}, {Polini}, {Pong}, {Ponrathnam}, {Popolizio}, {Porter}, {Poverman}, {Powell}, {Pracchia}, {Prajapati}, {Prasai}, {Prasanna}, {Pratten}, {Prestegard}, {Principe}, {Prodi}, {Prokhorov}, {Prosposito}, {Prudenzi}, {Puecher}, {Punturo}, {Puosi}, {Puppo}, {P{\"u}rrer}, {Qi}, {Quetschke}, {Quinonez}, {Quitzow-James}, {Raab}, {Raaijmakers}, {Radkins}, {Radulesco}, {Raffai}, {Rafferty}, {Rail}, {Raja}, {Rajan}, {Rajbhandari}, {Rakhmanov}, {Ramirez}, {Ramirez}, {Ramos-Buades}, {Rana}, {Rao}, {Rapagnani}, {Rapol}, {Ratto}, {Raymond}, {Razzano}, {Read}, {Regimbau}, {Rei}, {Reid}, {Reitze}, {Rettegno}, {Ricci}, {Richardson}, {Richardson}, {Richardson}, {Ricker}, {Riemenschneider},
  {Riles}, {Rizzo}, {Robertson}, {Robinet}, {Rocchi}, {Rocha}, {Rodriguez}, {Rodriguez-Soto}, {Rolland}, {Rollins}, {Roma}, {Romanelli}, {Romano}, {Romel}, {Romero}, {Romero-Shaw}, {Romie}, {Ronchini}, {Rose}, {Rose}, {Rose}, {Rosell}, {Rosi{\'n}ska}, {Rosofsky}, {Ross}, {Rowan}, {Rowlinson}, {Roy}, {Roy}, {Ruggi}, {Ryan}, {Sachdev}, {Sadecki}, {Sadiq}, {Sakellariadou}, {Salafia}, {Salconi}, {Saleem}, {Samajdar}, {Sanchez}, {Sanchez}, {Sanchez}, {Sanchis-Gual}, {Sanders}, {Sandles}, {Santiago}, {Santos}, {Saravanan}, {Sarin}, {Sassolas}, {Sathyaprakash}, {Sauter}, {Savage}, {Savant}, {Sawant}, {Sayah}, {Schaetzl}, {Schale}, {Scheel}, {Scheuer}, {Schindler-Tyka}, {Schmidt}, {Schnabel}, {Schofield}, {Sch{\"o}nbeck}, {Schreiber}, {Schulte}, {Schutz}, {Schwarm}, {Schwartz}, {Scott}, {Scott}, {Seglar-Arroyo}, {Seidel}, {Sellers}, {Sengupta}, {Sennett}, {Sentenac}, {Sequino}, {Sergeev}, {Setyawati}, {Shaffer}, {Shahriar}, {Sharifi}, {Sharma}, {Sharma}, {Shawhan}, {Shen}, {Shikauchi}, {Shink}, {Shoemaker},
  {Shoemaker}, {Shukla}, {ShyamSundar}, {Sieniawska}, {Sigg}, {Singer}, {Singh}, {Singh}, {Singha}, {Singhal}, {Sintes}, {Sipala}, {Skliris}, {Slagmolen}, {Slaven-Blair}, {Smetana}, {Smith}, {Smith}, {Somala}, {Son}, {Soni}, {Soni}, {Sorazu}, {Sordini}, {Sorrentino}, {Sorrentino}, {Soulard}, {Souradeep}, {Sowell}, {Spencer}, {Spera}, {Srivastava}, {Srivastava}, {Staats}, {Stachie}, {Steer}, {Steinhoff}, {Steinke}, {Steinlechner}, {Steinlechner}, {Steinmeyer}, {Stevenson}, {Stolle-McAllister}, {Stops}, {Stover}, {Strain}, {Stratta}, {Strunk}, {Sturani}, {Stuver}, {S{\"u}dbeck}, {Sudhagar}, {Sudhir}, {Suh}, {Summerscales}, {Sun}, {Sun}, {Sunil}, {Sur}, {Suresh}, {Sutton}, {Swinkels}, {Szczepa{\'n}czyk}, {Tacca}, {Tait}, {Talbot}, {Tanasijczuk}, {Tanner}, {Tao}, {Tapia}, {Tapia San Martin}, {Tasson}, {Taylor}, {Tenorio}, {Terkowski}, {Thirugnanasambandam}, {Thomas}, {Thomas}, {Thomas}, {Thompson}, {Thondapu}, {Thorne}, {Thrane}, {Tiwari}, {Tiwari}, {Tiwari}, {Toland}, {Tolley}, {Tonelli}, {Tornasi},
  {Torres-Forn{\'e}}, {Torrie}, {e Melo}, {T{\"o}yr{\"a}}, {Tran}, {Trapananti}, {Travasso}, {Traylor}, {Tringali}, {Tripathee}, {Trovato}, {Trudeau}, {Tsai}, {Tsang}, {Tse}, {Tso}, {Tsukada}, {Tsuna}, {Tsutsui}, {Turconi}, {Ubhi}, {Udall}, {Ueno}, {Ugolini}, {Unnikrishnan}, {Urban}, {Usman}, {Utina}, {Vahlbruch}, {Vajente}, {Vajpeyi}, {Valdes}, {Valentini}, {Valsan}, {van Bakel}, {van Beuzekom}, {van den Brand}, {Van Den Broeck}, {Vander-Hyde}, {van der Schaaf}, {van Heijningen}, {Vardaro}, {Vargas}, {Varma}, {Vass}, {Vas{\'u}th}, {Vecchio}, {Vedovato}, {Veitch}, {Veitch}, {Venkateswara}, {Venneberg}, {Venugopalan}, {Verkindt}, {Verma}, {Veske}, {Vetrano}, {Vicer{\'e}}, {Viets}, {Vijaykumar}, {Villa-Ortega}, {Vinet}, {Vitale}, {Vo}, {Vocca}, {Vorvick}, {Vyatchanin}, {Wade}, {Wade}, {Wade}, {Walet}, {Walker}, {Wallace}, {Wallace}, {Walsh}, {Wang}, {Wang}, {Wang}, {Wang}, {Ward}, {Warner}, {Was}, {Washington}, {Watchi}, {Weaver}, {Wei}, {Weinert}, {Weinstein}, {Weiss}, {Wellmann}, {Wen}, {We{\ss}els},
  {Westhouse}, {Wette}, {Whelan}, {White}, {White}, {Whiting}, {Whittle}, {Wilken}, {Williams}, {Williams}, {Williamson}, {Willis}, {Willke}, {Wilson}, {Wimmer}, {Winkler}, {Wipf}, {Woan}, {Woehler}, {Wofford}, {Wong}, {Wrangel}, {Wright}, {Wu}, {Wysocki}, {Xiao}, {Yamamoto}, {Yang}, {Yang}, {Yang}, {Yap}, {Yeeles}, {Yoon}, {Yu}, {Yu}, {Yuen}, {Zadro{\.Z}ny}, {Zanolin}, {Zelenova}, {Zendri}, {Zevin}, {Zhang}, {Zhang}, {Zhang}, {Zhang}, {Zhao}, {Zhao}, {Zheng}, {Zhou}, {Zhou}, {Zhu}, {Zimmerman}, {Zlochower}, {Zucker}, {Zweizig}, {LIGO Scientific Collaboration}, \& {Virgo Collaboration}}]{2021Abbott}
{Abbott}, R., {Abbott}, T.~D., {Abraham}, S., {et~al.} 2021, Physical Review X, 11, 021053

\bibitem[{{Abbott} {et~al.}(2023){Abbott}, {Abbott}, {Acernese}, {Ackley}, {Adams}, {Adhikari}, {Adhikari}, {Adya}, {Affeldt}, {Agarwal}, {Agathos}, {Agatsuma}, {Aggarwal}, {Aguiar}, {Aiello}, {Ain}, {Ajith}, {Akcay}, {Akutsu}, {Albanesi}, {Allocca}, {Altin}, {Amato}, {Anand}, {Anand}, {Ananyeva}, {Anderson}, {Anderson}, {Ando}, {Andrade}, {Andres}, {Andri{\'c}}, {Angelova}, {Ansoldi}, {Antelis}, {Antier}, {Appert}, {Arai}, {Arai}, {Arai}, {Araki}, {Araya}, {Araya}, {Areeda}, {Ar{\`e}ne}, {Aritomi}, {Arnaud}, {Arogeti}, {Aronson}, {Arun}, {Asada}, {Asali}, {Ashton}, {Aso}, {Assiduo}, {Aston}, {Astone}, {Aubin}, {Austin}, {Babak}, {Badaracco}, {Bader}, {Badger}, {Bae}, {Bae}, {Baer}, {Bagnasco}, {Bai}, {Baiotti}, {Baird}, {Bajpai}, {Ball}, {Ballardin}, {Ballmer}, {Balsamo}, {Baltus}, {Banagiri}, {Bankar}, {Barayoga}, {Barbieri}, {Barish}, {Barker}, {Barneo}, {Barone}, {Barr}, {Barsotti}, {Barsuglia}, {Barta}, {Bartlett}, {Barton}, {Bartos}, {Bassiri}, {Basti}, {Bawaj}, {Bayley}, {Baylor}, {Bazzan},
  {B{\'e}csy}, {Bedakihale}, {Bejger}, {Belahcene}, {Benedetto}, {Beniwal}, {Bennett}, {Bentley}, {Benyaala}, {Bergamin}, {Berger}, {Bernuzzi}, {Berry}, {Bersanetti}, {Bertolini}, {Betzwieser}, {Beveridge}, {Bhandare}, {Bhardwaj}, {Bhattacharjee}, {Bhaumik}, {Bilenko}, {Billingsley}, {Bini}, {Birney}, {Birnholtz}, {Biscans}, {Bischi}, {Biscoveanu}, {Bisht}, {Biswas}, {Bitossi}, {Bizouard}, {Blackburn}, {Blair}, {Blair}, {Blair}, {Bobba}, {Bode}, {Boer}, {Bogaert}, {Boldrini}, {Bonavena}, {Bondu}, {Bonilla}, {Bonnand}, {Booker}, {Boom}, {Bork}, {Boschi}, {Bose}, {Bose}, {Bossilkov}, {Boudart}, {Bouffanais}, {Bozzi}, {Bradaschia}, {Brady}, {Bramley}, {Branch}, {Branchesi}, {Brandt}, {Brau}, {Breschi}, {Briant}, {Briggs}, {Brillet}, {Brinkmann}, {Brockill}, {Brooks}, {Brooks}, {Brown}, {Brunett}, {Bruno}, {Bruntz}, {Bryant}, {Bulik}, {Bulten}, {Buonanno}, {Buscicchio}, {Buskulic}, {Buy}, {Byer}, {Davies}, {Cadonati}, {Cagnoli}, {Cahillane}, {Bustillo}, {Callaghan}, {Callister}, {Calloni}, {Cameron}, {Camp},
  {Canepa}, {Canevarolo}, {Cannavacciuolo}, {Cannon}, {Cao}, {Cao}, {Capocasa}, {Capote}, {Carapella}, {Carbognani}, {Carlin}, {Carney}, {Carpinelli}, {Carrillo}, {Carullo}, {Carver}, {Diaz}, {Casentini}, {Castaldi}, {Caudill}, {Cavagli{\`a}}, {Cavalier}, {Cavalieri}, {Ceasar}, {Cella}, {Cerd{\'a}-Dur{\'a}n}, {Cesarini}, {Chaibi}, {Chakravarti}, {Subrahmanya}, {Champion}, {Chan}, {Chan}, {Chan}, {Chan}, {Chan}, {Chandra}, {Chanial}, {Chao}, {Chapman-Bird}, {Charlton}, {Chase}, {Chassande-Mottin}, {Chatterjee}, {Chatterjee}, {Chatterjee}, {Chaturvedi}, {Chaty}, {Chatziioannou}, {Chen}, {Chen}, {Chen}, {Chen}, {Chen}, {Chen}, {Chen}, {Chen}, {Cheng}, {Cheong}, {Cheung}, {Chia}, {Chiadini}, {Chiang}, {Chiarini}, {Chierici}, {Chincarini}, {Chiofalo}, {Chiummo}, {Cho}, {Cho}, {Choudhary}, {Choudhary}, {Christensen}, {Chu}, {Chu}, {Chu}, {Chua}, {Chung}, {Ciani}, {Ciecielag}, {Cie{\'s}lar}, {Cifaldi}, {Ciobanu}, {Ciolfi}, {Cipriano}, {Cirone}, {Clara}, {Clark}, {Clark}, {Clarke}, {Clearwater}, {Clesse}, {Cleva},
  {Coccia}, {Codazzo}, {Cohadon}, {Cohen}, {Cohen}, {Colleoni}, {Collette}, {Colombo}, {Colpi}, {Compton}, {Constancio}, {Conti}, {Cooper}, {Corban}, {Corbitt}, {Cordero-Carri{\'o}n}, {Corezzi}, {Corley}, {Cornish}, {Corre}, {Corsi}, {Cortese}, {Costa}, {Cotesta}, {Coughlin}, {Coulon}, {Countryman}, {Cousins}, {Couvares}, {Coward}, {Cowart}, {Coyne}, {Coyne}, {Creighton}, {Creighton}, {Criswell}, {Croquette}, {Crowder}, {Cudell}, {Cullen}, {Cumming}, {Cummings}, {Cunningham}, {Cuoco}, {Cury{\l}o}, {Dabadie}, {Canton}, {Dall'Osso}, {D{\'a}lya}, {Dana}, {Daneshgaranbajastani}, {D'Angelo}, {Danila}, {Danilishin}, {D'Antonio}, {Danzmann}, {Darsow-Fromm}, {Dasgupta}, {Datrier}, {Dattilo}, {Dave}, {Davier}, {Davis}, {Davis}, {Daw}, {de Alarc{\'o}n}, {Dean}, {Debra}, {Deenadayalan}, {Degallaix}, {de Laurentis}, {Del{\'e}glise}, {Del Favero}, {de Lillo}, {de Lillo}, {Del Pozzo}, {Demarchi}, {de Matteis}, {D'Emilio}, {Demos}, {Dent}, {Depasse}, {de Pietri}, {De Rosa}, {de Rossi}, {Desalvo}, {de Simone}, {Dhurandhar},
  {D{\'\i}az}, {Diaz-Ortiz}, {Didio}, {Dietrich}, {di Fiore}, {di Fronzo}, {di Giorgio}, {di Giovanni}, {di Giovanni}, {di Girolamo}, {di Lieto}, {Ding}, {di Pace}, {di Palma}, {di Renzo}, {Divakarla}, {Dmitriev}, {Doctor}, {D'Onofrio}, {Donovan}, {Dooley}, {Doravari}, {Dorrington}, {Drago}, {Driggers}, {Drori}, {Ducoin}, {Dupej}, {Durante}, {D'Urso}, {Duverne}, {Dwyer}, {Eassa}, {Easter}, {Ebersold}, {Eckhardt}, {Eddolls}, {Edelman}, {Edo}, {Edy}, {Effler}, {Eguchi}, {Eichholz}, {Eikenberry}, {Eisenmann}, {Eisenstein}, {Ejlli}, {Engelby}, {Enomoto}, {Errico}, {Essick}, {Estell{\'e}s}, {Estevez}, {Etienne}, {Etzel}, {Evans}, {Evans}, {Ewing}, {Fafone}, {Fair}, {Fairhurst}, {Farah}, {Farinon}, {Farr}, {Farr}, {Farrow}, {Fauchon-Jones}, {Favaro}, {Favata}, {Fays}, {Fazio}, {Feicht}, {Fejer}, {Fenyvesi}, {Ferguson}, {Fernandez-Galiana}, {Ferrante}, {Ferreira}, {Fidecaro}, {Figura}, {Fiori}, {Fishbach}, {Fisher}, {Fittipaldi}, {Fiumara}, {Flaminio}, {Floden}, {Fong}, {Font}, {Fornal}, {Forsyth}, {Franke},
  {Frasca}, {Frasconi}, {Frederick}, {Freed}, {Frei}, {Freise}, {Frey}, {Fritschel}, {Frolov}, {Fronz{\'e}}, {Fujii}, {Fujikawa}, {Fukunaga}, {Fukushima}, {Fulda}, {Fyffe}, {Gabbard}, {Gabella}, {Gadre}, {Gair}, {Gais}, {Galaudage}, {Gamba}, {Ganapathy}, {Ganguly}, {Gao}, {Gaonkar}, {Garaventa}, {Garc{\'\i}a}, {Garc{\'\i}a-N{\'u}{\~n}ez}, {Garc{\'\i}a-Quir{\'o}s}, {Garufi}, {Gateley}, {Gaudio}, {Gayathri}, {Ge}, {Gemme}, {Gennai}, {George}, {George}, {Gerberding}, {Gergely}, {Gewecke}, {Ghonge}, {Ghosh}, {Ghosh}, {Ghosh}, {Ghosh}, {Giacomazzo}, {Giacoppo}, {Giaime}, {Giardina}, {Gibson}, {Gier}, {Giesler}, {Giri}, {Gissi}, {Glanzer}, {Gleckl}, {Godwin}, {Goetz}, {Goetz}, {Gohlke}, {Golomb}, {Goncharov}, {Gonz{\'a}lez}, {Gopakumar}, {Gosselin}, {Gouaty}, {Gould}, {Grace}, {Grado}, {Granata}, {Granata}, {Grant}, {Gras}, {Grassia}, {Gray}, {Gray}, {Greco}, {Green}, {Green}, {Gretarsson}, {Gretarsson}, {Griffith}, {Griffiths}, {Griggs}, {Grignani}, {Grimaldi}, {Grimm}, {Grote}, {Grunewald}, {Gruning}, {Guerra},
  {Guidi}, {Guimaraes}, {Guix{\'e}}, {Gulati}, {Guo}, {Guo}, {Gupta}, {Gupta}, {Gupta}, {Gustafson}, {Gustafson}, {Guzman}, {Ha}, {Haegel}, {Hagiwara}, {Haino}, {Halim}, {Hall}, {Hamilton}, {Hammond}, {Han}, {Haney}, {Hanks}, {Hanna}, {Hannam}, {Hannuksela}, {Hansen}, {Hansen}, {Hanson}, {Harder}, {Hardwick}, {Haris}, {Harms}, {Harry}, {Harry}, {Hartwig}, {Hasegawa}, {Haskell}, {Hasskew}, {Haster}, {Hattori}, {Haughian}, {Hayakawa}, {Hayama}, {Hayes}, {Healy}, {Heidmann}, {Heidt}, {Heintze}, {Heinze}, {Heinzel}, {Heitmann}, {Hellman}, {Hello}, {Helmling-Cornell}, {Hemming}, {Hendry}, {Heng}, {Hennes}, {Hennig}, {Hennig}, {Hernandez}, {Hernandez Vivanco}, {Heurs}, {Hild}, {Hill}, {Himemoto}, {Hines}, {Hiranuma}, {Hirata}, {Hirose}, {Hochheim}, {Hofman}, {Hohmann}, {Holcomb}, {Holland}, {Holley-Bockelmann}, {Hollows}, {Holmes}, {Holt}, {Holz}, {Hong}, {Hopkins}, {Hough}, {Hourihane}, {Howell}, {Hoy}, {Hoyland}, {Hreibi}, {Hsieh}, {Hsu}, {Huang}, {Huang}, {Huang}, {Huang}, {Huang}, {Huang}, {H{\"u}bner},
  {Huddart}, {Hughey}, {Hui}, {Hui}, {Husa}, {Huttner}, {Huxford}, {Huynh-Dinh}, {Ide}, {Idzkowski}, {Iess}, {Ikenoue}, {Imam}, {Inayoshi}, {Ingram}, {Inoue}, {Ioka}, {Isi}, {Isleif}, {Ito}, {Itoh}, {Iyer}, {Izumi}, {Jaberianhamedan}, {Jacqmin}, {Jadhav}, {Jadhav}, {James}, {Jan}, {Jani}, {Janquart}, {Janssens}, {Janthalur}, {Jaranowski}, {Jariwala}, {Jaume}, {Jenkins}, {Jenner}, {Jeon}, {Jeunon}, {Jia}, {Jin}, {Johns}, {Johnson-McDaniel}, {Jones}, {Jones}, {Jones}, {Jones}, {Jones}, {Jonker}, {Ju}, {Jung}, {Jung}, {Junker}, {Juste}, {Kaihotsu}, {Kajita}, {Kakizaki}, {Kalaghatgi}, {Kalogera}, {Kamai}, {Kamiizumi}, {Kanda}, {Kandhasamy}, {Kang}, {Kanner}, {Kao}, {Kapadia}, {Kapasi}, {Karat}, {Karathanasis}, {Karki}, {Kashyap}, {Kasprzack}, {Kastaun}, {Katsanevas}, {Katsavounidis}, {Katzman}, {Kaur}, {Kawabe}, {Kawaguchi}, {Kawai}, {Kawasaki}, {K{\'e}f{\'e}lian}, {Keitel}, {Key}, {Khadka}, {Khalili}, {Khan}, {Khazanov}, {Khetan}, {Khursheed}, {Kijbunchoo}, {Kim}, {Kim}, {Kim}, {Kim}, {Kim}, {Kim}, {Kimball},
  {Kimura}, {Kinley-Hanlon}, {Kirchhoff}, {Kissel}, {Kita}, {Kitazawa}, {Kleybolte}, {Klimenko}, {Knee}, {Knowles}, {Knyazev}, {Koch}, {Koekoek}, {Kojima}, {Kokeyama}, {Koley}, {Kolitsidou}, {Kolstein}, {Komori}, {Kondrashov}, {Kong}, {Kontos}, {Koper}, {Korobko}, {Kotake}, {Kovalam}, {Kozak}, {Kozakai}, {Kozu}, {Kringel}, {Krishnendu}, {Kr{\'o}lak}, {Kuehn}, {Kuei}, {Kuijer}, {Kulkarni}, {Kumar}, {Kumar}, {Kumar}, {Kumar}, {Kume}, {Kuns}, {Kuo}, {Kuo}, {Kuromiya}, {Kuroyanagi}, {Kusayanagi}, {Kuwahara}, {Kwak}, {Lagabbe}, {Laghi}, {Lalande}, {Lam}, {Lamberts}, {Landry}, {Lane}, {Lang}, {Lange}, {Lantz}, {La Rosa}, {Lartaux-Vollard}, {Lasky}, {Laxen}, {Lazzarini}, {Lazzaro}, {Leaci}, {Leavey}, {Lecoeuche}, {Lee}, {Lee}, {Lee}, {Lee}, {Lee}, {Lee}, {Lehmann}, {Lema{\^\i}tre}, {Leonardi}, {Leroy}, {Letendre}, {Levesque}, {Levin}, {Leviton}, {Leyde}, {Li}, {Li}, {Li}, {Li}, {Li}, {Li}, {Lin}, {Lin}, {Lin}, {Lin}, {Lin}, {Linde}, {Linker}, {Linley}, {Littenberg}, {Liu}, {Liu}, {Liu}, {Liu}, {Llamas},
  {Llorens-Monteagudo}, {Lo}, {Lockwood}, {Loh}, {London}, {Longo}, {Lopez}, {Portilla}, {Lorenzini}, {Loriette}, {Lormand}, {Losurdo}, {Lott}, {Lough}, {Lousto}, {Lovelace}, {Lucaccioni}, {L{\"u}ck}, {Lumaca}, {Lundgren}, {Luo}, {Lynam}, {Macas}, {Macinnis}, {MacLeod}, {MacMillan}, {Macquet}, {Hernandez}, {Magazz{\`u}}, {Magee}, {Maggiore}, {Magnozzi}, {Mahesh}, {Majorana}, {Makarem}, {Maksimovic}, {Maliakal}, {Malik}, {Man}, {Mandic}, {Mangano}, {Mango}, {Mansell}, {Manske}, {Mantovani}, {Mapelli}, {Marchesoni}, {Marchio}, {Marion}, {Mark}, {M{\'a}rka}, {M{\'a}rka}, {Markakis}, {Markosyan}, {Markowitz}, {Maros}, {Marquina}, {Marsat}, {Martelli}, {Martin}, {Martin}, {Martinez}, {Martinez}, {Martinez}, {Martinovic}, {Martynov}, {Marx}, {Masalehdan}, {Mason}, {Massera}, {Masserot}, {Massinger}, {Masso-Reid}, {Mastrogiovanni}, {Matas}, {Mateu-Lucena}, {Matichard}, {Matiushechkina}, {Mavalvala}, {McCann}, {McCarthy}, {McClelland}, {McClincy}, {McCormick}, {McCuller}, {McGhee}, {McGuire}, {McIsaac}, {McIver},
  {McRae}, {McWilliams}, {Meacher}, {Mehmet}, {Mehta}, {Meijer}, {Melatos}, {Melchor}, {Mendell}, {Menendez-Vazquez}, {Menoni}, {Mercer}, {Mereni}, {Merfeld}, {Merilh}, {Merritt}, {Merzougui}, {Meshkov}, {Messenger}, {Messick}, {Meyers}, {Meylahn}, {Mhaske}, {Miani}, {Miao}, {Michaloliakos}, {Michel}, {Michimura}, {Middleton}, {Milano}, {Miller}, {Miller}, {Miller}, {Millhouse}, {Mills}, {Milotti}, {Minazzoli}, {Minenkov}, {Mio}, {Mir}, {Miravet-Ten{\'e}s}, {Mishra}, {Mishra}, {Mistry}, {Mitra}, {Mitrofanov}, {Mitselmakher}, {Mittleman}, {Miyakawa}, {Miyamoto}, {Miyazaki}, {Miyo}, {Miyoki}, {Mo}, {Modafferi}, {Moguel}, {Mogushi}, {Mohapatra}, {Mohite}, {Molina}, {Molina-Ruiz}, {Mondin}, {Montani}, {Moore}, {Moraru}, {Morawski}, {More}, {Moreno}, {Moreno}, {Mori}, {Morisaki}, {Moriwaki}, {Morr{\'a}s}, {Mours}, {Mow-Lowry}, {Mozzon}, {Muciaccia}, {Mukherjee}, {Mukherjee}, {Mukherjee}, {Mukherjee}, {Mukherjee}, {Mukund}, {Mullavey}, {Munch}, {Mu{\~n}iz}, {Murray}, {Musenich}, {Muusse}, {Nadji}, {Nagano},
  {Nagano}, {Nagar}, {Nakamura}, {Nakano}, {Nakano}, {Nakashima}, {Nakayama}, {Napolano}, {Nardecchia}, {Narikawa}, {Naticchioni}, {Nayak}, {Nayak}, {Negishi}, {Neil}, {Neilson}, {Nelemans}, {Nelson}, {Nery}, {Neubauer}, {Neunzert}, {Ng}, {Ng}, {Nguyen}, {Nguyen}, {Nguyen}, {Quynh}, {Ni}, {Nichols}, {Nishizawa}, {Nissanke}, {Nitoglia}, {Nocera}, {Norman}, {North}, {Nozaki}, {Siles}, {Nuttall}, {Oberling}, {O'Brien}, {Obuchi}, {O'Dell}, {Oelker}, {Ogaki}, {Oganesyan}, {Oh}, {Oh}, {Oh}, {Ohashi}, {Ohishi}, {Ohkawa}, {Ohme}, {Ohta}, {Okada}, {Okutani}, {Okutomi}, {Olivetto}, {Oohara}, {Ooi}, {Oram}, {O'Reilly}, {Ormiston}, {Ormsby}, {Ortega}, {O'Shaughnessy}, {O'Shea}, {Oshino}, {Ossokine}, {Osthelder}, {Otabe}, {Ottaway}, {Overmier}, {Pace}, {Pagano}, {Page}, {Pagliaroli}, {Pai}, {Pai}, {Palamos}, {Palashov}, {Palomba}, {Pan}, {Pan}, {Panda}, {Pang}, {Pang}, {Pankow}, {Pannarale}, {Pant}, {Panther}, {Paoletti}, {Paoli}, {Paolone}, {Parisi}, {Park}, {Park}, {Parker}, {Pascucci}, {Pasqualetti}, {Passaquieti},
  {Passuello}, {Patel}, {Pathak}, {Patricelli}, {Patron}, {Paul}, {Payne}, {Pedraza}, {Pegoraro}, {Pele}, {Arellano}, {Penn}, {Perego}, {Pereira}, {Pereira}, {Perez}, {P{\'e}rigois}, {Perkins}, {Perreca}, {Perri{\`e}s}, {Petermann}, {Petterson}, {Pfeiffer}, {Pham}, {Phukon}, {Piccinni}, {Pichot}, {Piendibene}, {Piergiovanni}, {Pierini}, {Pierro}, {Pillant}, {Pillas}, {Pilo}, {Pinard}, {Pinto}, {Pinto}, {Piotrzkowski}, {Piotrzkowski}, {Pirello}, {Pitkin}, {Placidi}, {Planas}, {Plastino}, {Pluchar}, {Poggiani}, {Polini}, {Pong}, {Ponrathnam}, {Popolizio}, {Porter}, {Poulton}, {Powell}, {Pracchia}, {Pradier}, {Prajapati}, {Prasai}, {Prasanna}, {Pratten}, {Principe}, {Prodi}, {Prokhorov}, {Prosposito}, {Prudenzi}, {Puecher}, {Punturo}, {Puosi}, {Puppo}, {P{\"u}rrer}, {Qi}, {Quetschke}, {Quitzow-James}, {Qutob}, {Raab}, {Raaijmakers}, {Radkins}, {Radulesco}, {Raffai}, {Rail}, {Raja}, {Rajan}, {Ramirez}, {Ramirez}, {Ramos-Buades}, {Rana}, {Rapagnani}, {Rapol}, {Ray}, {Raymond}, {Raza}, {Razzano}, {Read}, {Rees},
  {Regimbau}, {Rei}, {Reid}, {Reid}, {Reitze}, {Relton}, {Renzini}, {Rettegno}, {Reza}, {Rezac}, {Ricci}, {Richards}, {Richardson}, {Richardson}, {Riemenschneider}, {Riles}, {Rinaldi}, {Rink}, {Rizzo}, {Robertson}, {Robie}, {Robinet}, {Rocchi}, {Rodriguez}, {Rolland}, {Rollins}, {Romanelli}, {Romano}, {Romel}, {Romero-Rodr{\'\i}guez}, {Romero-Shaw}, {Romie}, {Ronchini}, {Rosa}, {Rose}, {Rosi{\'n}ska}, {Ross}, {Rowan}, {Rowlinson}, {Roy}, {Roy}, {Roy}, {Rozza}, {Ruggi}, {Ruiz-Rocha}, {Ryan}, {Sachdev}, {Sadecki}, {Sadiq}, {Sago}, {Saito}, {Saito}, {Sakai}, {Sakai}, {Sakellariadou}, {Sakuno}, {Salafia}, {Salconi}, {Saleem}, {Salemi}, {Samajdar}, {Sanchez}, {Sanchez}, {Sanchez}, {Sanchis-Gual}, {Sanders}, {Sanuy}, {Saravanan}, {Sarin}, {Sassolas}, {Satari}, {Sathyaprakash}, {Sato}, {Sato}, {Sauter}, {Savage}, {Sawada}, {Sawant}, {Sawant}, {Sayah}, {Schaetzl}, {Scheel}, {Scheuer}, {Schiworski}, {Schmidt}, {Schmidt}, {Schnabel}, {Schneewind}, {Schofield}, {Sch{\"o}nbeck}, {Schulte}, {Schutz}, {Schwartz}, {Scott},
  {Scott}, {Seglar-Arroyo}, {Sekiguchi}, {Sekiguchi}, {Sellers}, {Sengupta}, {Sentenac}, {Seo}, {Sequino}, {Sergeev}, {Setyawati}, {Shaffer}, {Shahriar}, {Shams}, {Shao}, {Sharma}, {Sharma}, {Shawhan}, {Shcheblanov}, {Shibagaki}, {Shikauchi}, {Shimizu}, {Shimoda}, {Shimode}, {Shinkai}, {Shishido}, {Shoda}, {Shoemaker}, {Shoemaker}, {Shyamsundar}, {Sieniawska}, {Sigg}, {Singer}, {Singh}, {Singh}, {Singha}, {Sintes}, {Sipala}, {Skliris}, {Slagmolen}, {Slaven-Blair}, {Smetana}, {Smith}, {Smith}, {Soldateschi}, {Somala}, {Somiya}, {Son}, {Soni}, {Soni}, {Sordini}, {Sorrentino}, {Sorrentino}, {Sotani}, {Soulard}, {Souradeep}, {Sowell}, {Spagnuolo}, {Spencer}, {Spera}, {Srinivasan}, {Srivastava}, {Srivastava}, {Staats}, {Stachie}, {Steer}, {Steinhoff}, {Steinlechner}, {Steinlechner}, {Stevenson}, {Stops}, {Stover}, {Strain}, {Strang}, {Stratta}, {Strunk}, {Sturani}, {Stuver}, {Sudhagar}, {Sudhir}, {Sugimoto}, {Suh}, {Sullivan}, {Sullivan}, {Summerscales}, {Sun}, {Sun}, {Sunil}, {Sur}, {Suresh}, {Sutton}, {Suzuki},
  {Suzuki}, {Swinkels}, {Szczepa{\'n}czyk}, {Szewczyk}, {Tacca}, {Tagoshi}, {Tait}, {Takahashi}, {Takahashi}, {Takamori}, {Takano}, {Takeda}, {Takeda}, {Talbot}, {Talbot}, {Tanaka}, {Tanaka}, {Tanaka}, {Tanaka}, {Tanaka}, {Tanasijczuk}, {Tanioka}, {Tanner}, {Tao}, {Tao}, {Mart{\'\i}n}, {Taranto}, {Tasson}, {Telada}, {Tenorio}, {Terhune}, {Terkowski}, {Thirugnanasambandam}, {Thomas}, {Thomas}, {Thomas}, {Thompson}, {Thondapu}, {Thorne}, {Thrane}, {Tiwari}, {Tiwari}, {Tiwari}, {Toivonen}, {Toland}, {Tolley}, {Tomaru}, {Tomigami}, {Tomura}, {Tonelli}, {Torres-Forn{\'e}}, {Torrie}, {E Melo}, {T{\"o}yr{\"a}}, {Trapananti}, {Travasso}, {Traylor}, {Trevor}, {Tringali}, {Tripathee}, {Troiano}, {Trovato}, {Trozzo}, {Trudeau}, {Tsai}, {Tsai}, {Tsang}, {Tsang}, {Tsao}, {Tse}, {Tso}, {Tsubono}, {Tsuchida}, {Tsukada}, {Tsuna}, {Tsutsui}, {Tsuzuki}, {Turbang}, {Turconi}, {Tuyenbayev}, {Ubhi}, {Uchikata}, {Uchiyama}, {Udall}, {Ueda}, {Uehara}, {Ueno}, {Ueshima}, {Unnikrishnan}, {Uraguchi}, {Urban}, {Ushiba}, {Utina},
  {Vahlbruch}, {Vajente}, {Vajpeyi}, {Valdes}, {Valentini}, {Valsan}, {van Bakel}, {van Beuzekom}, {van den Brand}, {van den Broeck}, {Vander-Hyde}, {van der Schaaf}, {van Heijningen}, {Vanosky}, {van Putten}, {van Remortel}, {Vardaro}, {Vargas}, {Varma}, {Vas{\'u}th}, {Vecchio}, {Vedovato}, {Veitch}, {Veitch}, {Venneberg}, {Venugopalan}, {Verkindt}, {Verma}, {Verma}, {Veske}, {Vetrano}, {Vicer{\'e}}, {Vidyant}, {Viets}, {Vijaykumar}, {Villa-Ortega}, {Vinet}, {Virtuoso}, {Vitale}, {Vo}, {Vocca}, {von Reis}, {von Wrangel}, {Vorvick}, {Vyatchanin}, {Wade}, {Wade}, {Wagner}, {Walet}, {Walker}, {Wallace}, {Wallace}, {Walsh}, {Wang}, {Wang}, {Wang}, {Ward}, {Warner}, {Was}, {Washimi}, {Washington}, {Watchi}, {Weaver}, {Webster}, {Weinert}, {Weinstein}, {Weiss}, {Weller}, {Weller}, {Wellmann}, {Wen}, {We{\ss}els}, {Wette}, {Whelan}, {White}, {Whiting}, {Whittle}, {Wilken}, {Williams}, {Williams}, {Williams}, {Williamson}, {Willis}, {Willke}, {Wilson}, {Winkler}, {Wipf}, {Wlodarczyk}, {Woan}, {Woehler}, {Wofford},
  {Wong}, {Wu}, {Wu}, {Wu}, {Wu}, {Wysocki}, {Xiao}, {Xu}, {Yamada}, {Yamamoto}, {Yamamoto}, {Yamamoto}, {Yamamoto}, {Yamashita}, {Yamazaki}, {Yang}, {Yang}, {Yang}, {Yang}, {Yang}, {Yap}, {Yeeles}, {Yelikar}, {Ying}, {Yokogawa}, {Yokoyama}, {Yokozawa}, {Yoo}, {Yoshioka}, {Yu}, {Yu}, {Yuzurihara}, {Zadro{\.z}ny}, {Zanolin}, {Zeidler}, {Zelenova}, {Zendri}, {Zevin}, {Zhan}, {Zhang}, {Zhang}, {Zhang}, {Zhang}, {Zhang}, {Zhao}, {Zhao}, {Zhao}, {Zhao}, {Zheng}, {Zhou}, {Zhou}, {Zhu}, {Zhu}, {Zimmerman}, {Zlochower}, {Zucker}, {Zweizig}, {Ligo Scientific Collaboration}, \& {Kagra Collaboration}}]{2023Abbott}
{Abbott}, R., {Abbott}, T.~D., {Acernese}, F., {et~al.} 2023, Physical Review X, 13, 041039

\bibitem[{{Absil} {et~al.}(2011){Absil}, {Le Bouquin}, {Berger}, {Lagrange}, {Chauvin}, {Lazareff}, {Zins}, {Haguenauer}, {Jocou}, {Kern}, {Millan-Gabet}, {Rochat}, \& {Traub}}]{2011Absil}
{Absil}, O., {Le Bouquin}, J.~B., {Berger}, J.~P., {et~al.} 2011, \aap, 535, A68

\bibitem[{{Aguilera-Dena} {et~al.}(2022){Aguilera-Dena}, {Langer}, {Antoniadis}, {Pauli}, {Dessart}, {Vigna-G{\'o}mez}, {Gr{\"a}fener}, \& {Yoon}}]{2022Aguilera}
{Aguilera-Dena}, D.~R., {Langer}, N., {Antoniadis}, J., {et~al.} 2022, \aap, 661, A60

\bibitem[{{Aguilera-Dena} {et~al.}(2023){Aguilera-Dena}, {M{\"u}ller}, {Antoniadis}, {Langer}, {Dessart}, {Vigna-G{\'o}mez}, \& {Yoon}}]{2023Aguilera}
{Aguilera-Dena}, D.~R., {M{\"u}ller}, B., {Antoniadis}, J., {et~al.} 2023, \aap, 671, A134

\bibitem[{{Antokhin} {et~al.}(1995){Antokhin}, {Bertrand}, {Lamontagne}, \& {Moffat}}]{1995Antokhin}
{Antokhin}, I.~I., {Bertrand}, J.~F., {Lamontagne}, R., \& {Moffat}, A.~F.~J. 1995, in IAU Symposium, Vol. 163, Wolf-Rayet Stars: Binaries; Colliding Winds; Evolution, ed. K.~A. {van der Hucht} \& P.~M. {Williams}, 62

\bibitem[{{Bartzakos} {et~al.}(2001){Bartzakos}, {Moffat}, \& {Niemela}}]{2001Bartzakos}
{Bartzakos}, P., {Moffat}, A.~F.~J., \& {Niemela}, V.~S. 2001, \mnras, 324, 18

\bibitem[{{Chen{\'e}} {et~al.}(2011){Chen{\'e}}, {Moffat}, {Cameron}, {Fahed}, {Gamen}, {Lef{\`e}vre}, {Rowe}, {St-louis}, {Muntean}, {De La Chevroti{\`e}re}, {Guenther}, {Kuschnig}, {Matthews}, {Rucinski}, {Sasselov}, \& {Weiss}}]{2011Chene}
{Chen{\'e}}, A.~N., {Moffat}, A.~F.~J., {Cameron}, C., {et~al.} 2011, \apj, 735, 34

\bibitem[{{Chen{\'e}} \& {St-Louis}(2011)}]{2011CheneStLouis}
{Chen{\'e}}, A.~N. \& {St-Louis}, N. 2011, \apj, 736, 140

\bibitem[{{Clark} {et~al.}(2018){Clark}, {Lohr}, {Patrick}, {Najarro}, {Dong}, \& {Figer}}]{2018Clark}
{Clark}, J.~S., {Lohr}, M.~E., {Patrick}, L.~R., {et~al.} 2018, \aap, 618, A2

\bibitem[{{Clark} {et~al.}(2021){Clark}, {Patrick}, {Najarro}, {Evans}, \& {Lohr}}]{2021Clark}
{Clark}, J.~S., {Patrick}, L.~R., {Najarro}, F., {Evans}, C.~J., \& {Lohr}, M. 2021, \aap, 649, A43

\bibitem[{{Cohen} {et~al.}(1975){Cohen}, {Barlow}, \& {Kuhi}}]{1975Cohen}
{Cohen}, M., {Barlow}, M.~J., \& {Kuhi}, L.~V. 1975, \aap, 40, 291

\bibitem[{{Conti}(1976)}]{1976Conti}
{Conti}, P. 1976, in Proc. 20th Colloq. Int. Astrophys., Univ. Liege, 193

\bibitem[{{Crowther}(2007)}]{2007Crowther}
{Crowther}, P.~A. 2007, \araa, 45, 177

\bibitem[{{Crowther} \& {Bohannan}(1997)}]{1997CrowtherBohannanOf}
{Crowther}, P.~A. \& {Bohannan}, B. 1997, \aap, 317, 532

\bibitem[{{Crowther} {et~al.}(2006){Crowther}, {Hadfield}, {Clark}, {Negueruela}, \& {Vacca}}]{2006Crowther}
{Crowther}, P.~A., {Hadfield}, L.~J., {Clark}, J.~S., {Negueruela}, I., \& {Vacca}, W.~D. 2006, \mnras, 372, 1407

\bibitem[{{Crowther} {et~al.}(1995){Crowther}, {Smith}, \& {Willis}}]{1995crowther}
{Crowther}, P.~A., {Smith}, L.~J., \& {Willis}, A.~J. 1995, in Wolf-Rayet Stars: Binaries; Colliding Winds; Evolution, ed. K.~A. {van der Hucht} \& P.~M. {Williams}, Vol. 163, 152

\bibitem[{{Davis} {et~al.}(1981){Davis}, {Moffat}, \& {Niemela}}]{1981Davis}
{Davis}, A.~B., {Moffat}, A.~F.~J., \& {Niemela}, V.~S. 1981, \apj, 244, 528

\bibitem[{{Dray} \& {Tout}(2003)}]{2003Dray}
{Dray}, L.~M. \& {Tout}, C.~A. 2003, \mnras, 341, 299

\bibitem[{{Drissen} {et~al.}(1992){Drissen}, {Robert}, \& {Moffat}}]{1992Drissen}
{Drissen}, L., {Robert}, C., \& {Moffat}, A. F.~J. 1992, \apj, 386, 288

\bibitem[{{Dsilva} {et~al.}(2020){Dsilva}, {Shenar}, {Sana}, \& {Marchant}}]{kd1}
{Dsilva}, K., {Shenar}, T., {Sana}, H., \& {Marchant}, P. 2020, \aap, 641, A26

\bibitem[{{Dsilva} {et~al.}(2022){Dsilva}, {Shenar}, {Sana}, \& {Marchant}}]{kd2}
{Dsilva}, K., {Shenar}, T., {Sana}, H., \& {Marchant}, P. 2022, \aap, 664, A93

\bibitem[{{Dsilva} {et~al.}(2023){Dsilva}, {Shenar}, {Sana}, \& {Marchant}}]{kd3}
{Dsilva}, K., {Shenar}, T., {Sana}, H., \& {Marchant}, P. 2023, \aap, 674, A88

\bibitem[{{Eldridge} {et~al.}(2008){Eldridge}, {Izzard}, \& {Tout}}]{2008Eldridge}
{Eldridge}, J.~J., {Izzard}, R.~G., \& {Tout}, C.~A. 2008, \mnras, 384, 1109

\bibitem[{{Eldridge} \& {Stanway}(2016)}]{2016Eldridge}
{Eldridge}, J.~J. \& {Stanway}, E.~R. 2016, \mnras, 462, 3302

\bibitem[{{Fahed} \& {Moffat}(2012)}]{2012Fahed}
{Fahed}, R. \& {Moffat}, A.~F.~J. 2012, \mnras, 424, 1601

\bibitem[{{Faherty} {et~al.}(2014){Faherty}, {Shara}, {Zurek}, {Kanarek}, \& {Moffat}}]{2014Faherty}
{Faherty}, J.~K., {Shara}, M.~M., {Zurek}, D., {Kanarek}, G., \& {Moffat}, A. F.~J. 2014, \aj, 147, 115

\bibitem[{{Figer} {et~al.}(1997){Figer}, {McLean}, \& {Najarro}}]{1997Figer}
{Figer}, D.~F., {McLean}, I.~S., \& {Najarro}, F. 1997, \apj, 486, 420

\bibitem[{{Fitzpatrick}(1999)}]{1999Fitzpatrick}
{Fitzpatrick}, E.~L. 1999, \pasp, 111, 63

\bibitem[{{Frost} {et~al.}(2022){Frost}, {Bodensteiner}, {Rivinius}, {Baade}, {Merand}, {Selman}, {Abdul-Masih}, {Banyard}, {Bordier}, {Dsilva}, {Hawcroft}, {Mahy}, {Reggiani}, {Shenar}, {Cabezas}, {Hadrava}, {Heida}, {Klement}, \& {Sana}}]{2022Frost}
{Frost}, A.~J., {Bodensteiner}, J., {Rivinius}, T., {et~al.} 2022, \aap, 659, L3

\bibitem[{{Frost} {et~al.}(2024){Frost}, {Sana}, {Mahy}, {Wade}, {Barron}, {Le Bouquin}, {M{\'e}rand}, {Schneider}, {Shenar}, {Barb{\'a}}, {Bowman}, {Fabry}, {Farhang}, {Marchant}, {Morrell}, \& {Smoker}}]{2024Frost}
{Frost}, A.~J., {Sana}, H., {Mahy}, L., {et~al.} 2024, Science, 384, 214

\bibitem[{{Gallenne} {et~al.}(2015){Gallenne}, {M{\'e}rand}, {Kervella}, {Monnier}, {Schaefer}, {Baron}, {Breitfelder}, {Le Bouquin}, {Roettenbacher}, {Gieren}, {Pietrzy{\'n}ski}, {McAlister}, {ten Brummelaar}, {Sturmann}, {Sturmann}, {Turner}, {Ridgway}, \& {Kraus}}]{2015Gallenne}
{Gallenne}, A., {M{\'e}rand}, A., {Kervella}, P., {et~al.} 2015, \aap, 579, A68

\bibitem[{{Gamen} \& {Niemela}(2003)}]{2003Gamen}
{Gamen}, R.~C. \& {Niemela}, V.~S. 2003, in IAU Symposium, Vol. 212, A Massive Star Odyssey: From Main Sequence to Supernova, ed. K.~{van der Hucht}, A.~{Herrero}, \& C.~{Esteban}, 184

\bibitem[{{Gamow}(1943)}]{1943Gamow}
{Gamow}, G. 1943, \apj, 98, 500

\bibitem[{{Gaposchkin}(1949)}]{1949Gapo}
{Gaposchkin}, S. 1949, Peremennye Zvezdy, 7, 36

\bibitem[{{Georgy} {et~al.}(2015){Georgy}, {Ekstr{\"o}m}, {Hirschi}, {Meynet}, {Groh}, \& {Eggenberger}}]{2015Georgy}
{Georgy}, C., {Ekstr{\"o}m}, S., {Hirschi}, R., {et~al.} 2015, in Wolf-Rayet Stars, ed. W.-R. {Hamann}, A.~{Sander}, \& H.~{Todt}, 229--232

\bibitem[{{Gosset} {et~al.}(2009){Gosset}, {Naz{\'e}}, {Sana}, {Rauw}, \& {Vreux}}]{2009Gosset}
{Gosset}, E., {Naz{\'e}}, Y., {Sana}, H., {Rauw}, G., \& {Vreux}, J.~M. 2009, \aap, 508, 805

\bibitem[{{Gosset} {et~al.}(1991){Gosset}, {Remy}, {Manfroid}, {Vreux}, {Balona}, {Sterken}, \& {Franco}}]{1991Gosset}
{Gosset}, E., {Remy}, M., {Manfroid}, J., {et~al.} 1991, Information Bulletin on Variable Stars, 3571, 1

\bibitem[{{Haiman} \& {Loeb}(1997)}]{1997Haiman}
{Haiman}, Z. \& {Loeb}, A. 1997, \apj, 483, 21

\bibitem[{{Hamann} {et~al.}(2019){Hamann}, {Gr{\"a}fener}, {Liermann}, {Hainich}, {Sander}, {Shenar}, {Ramachandran}, {Todt}, \& {Oskinova}}]{2019Hamann}
{Hamann}, W.~R., {Gr{\"a}fener}, G., {Liermann}, A., {et~al.} 2019, \aap, 625, A57

\bibitem[{{Hartkopf} {et~al.}(1999){Hartkopf}, {Mason}, {Gies}, {ten Brummelaar}, {McAlister}, {Moffat}, {Shara}, \& {Wallace}}]{1999Hartkopf}
{Hartkopf}, W.~I., {Mason}, B.~D., {Gies}, D.~R., {et~al.} 1999, \aj, 118, 509

\bibitem[{{Hartkopf} {et~al.}(2012){Hartkopf}, {Tokovinin}, \& {Mason}}]{2012Hartkopf}
{Hartkopf}, W.~I., {Tokovinin}, A., \& {Mason}, B.~D. 2012, \aj, 143, 42

\bibitem[{{Hendrix} {et~al.}(2016){Hendrix}, {Keppens}, {van Marle}, {Camps}, {Baes}, \& {Meliani}}]{2016Hendrix}
{Hendrix}, T., {Keppens}, R., {van Marle}, A.~J., {et~al.} 2016, \mnras, 460, 3975

\bibitem[{{Higgins} {et~al.}(2021){Higgins}, {Sander}, {Vink}, \& {Hirschi}}]{2021Higgins}
{Higgins}, E.~R., {Sander}, A.~A.~C., {Vink}, J.~S., \& {Hirschi}, R. 2021, \mnras, 505, 4874

\bibitem[{{Hill} {et~al.}(2002){Hill}, {Moffat}, \& {St-Louis}}]{2002Hill}
{Hill}, G.~M., {Moffat}, A.~F.~J., \& {St-Louis}, N. 2002, \mnras, 335, 1069

\bibitem[{{Hill} {et~al.}(2018){Hill}, {Moffat}, \& {St-Louis}}]{2018Hill}
{Hill}, G.~M., {Moffat}, A.~F.~J., \& {St-Louis}, N. 2018, \mnras, 474, 2987

\bibitem[{{Hillier}(1987)}]{1987Hillier}
{Hillier}, D.~J. 1987, \apjs, 63, 965

\bibitem[{{Hillier}(2015)}]{2015Hillier}
{Hillier}, D.~J. 2015, in Wolf-Rayet Stars, ed. W.-R. {Hamann}, A.~{Sander}, \& H.~{Todt}, 65--70

\bibitem[{{Hopkins} {et~al.}(2014){Hopkins}, {Kere{\v{s}}}, {O{\~n}orbe}, {Faucher-Gigu{\`e}re}, {Quataert}, {Murray}, \& {Bullock}}]{2014Hopkins}
{Hopkins}, P.~F., {Kere{\v{s}}}, D., {O{\~n}orbe}, J., {et~al.} 2014, \mnras, 445, 581

\bibitem[{{Janssens} {et~al.}(2022){Janssens}, {Shenar}, {Sana}, {Faigler}, {Langer}, {Marchant}, {Mazeh}, {Sch{\"u}rmann}, \& {Shahaf}}]{2022Janssens}
{Janssens}, S., {Shenar}, T., {Sana}, H., {et~al.} 2022, \aap, 658, A129

\bibitem[{{Justham} {et~al.}(2014){Justham}, {Podsiadlowski}, \& {Vink}}]{2014Justham}
{Justham}, S., {Podsiadlowski}, P., \& {Vink}, J.~S. 2014, \apj, 796, 121

\bibitem[{{Koenigsberger} {et~al.}(2010){Koenigsberger}, {Georgiev}, {Hillier}, {Morrell}, {Barb{\'a}}, \& {Gamen}}]{2010Koenigsberger}
{Koenigsberger}, G., {Georgiev}, L., {Hillier}, D.~J., {et~al.} 2010, \aj, 139, 2600

\bibitem[{{Lachaume} {et~al.}(2019){Lachaume}, {Rabus}, {Jord{\'a}n}, {Brahm}, {Boyajian}, {von Braun}, \& {Berger}}]{2019Lachaume}
{Lachaume}, R., {Rabus}, M., {Jord{\'a}n}, A., {et~al.} 2019, \mnras, 484, 2656

\bibitem[{{Lamberts} {et~al.}(2017){Lamberts}, {Millour}, {Liermann}, {Dessart}, {Driebe}, {Duvert}, {Finsterle}, {Girault}, {Massi}, {Petrov}, {Schmutz}, {Weigelt}, \& {Chesneau}}]{2017Lamberts}
{Lamberts}, A., {Millour}, F., {Liermann}, A., {et~al.} 2017, \mnras, 468, 2655

\bibitem[{{Lamers} {et~al.}(1991){Lamers}, {Maeder}, {Schmutz}, \& {Cassinelli}}]{1991Lamers}
{Lamers}, H.~J.~G.~L.~M., {Maeder}, A., {Schmutz}, W., \& {Cassinelli}, J.~P. 1991, \apj, 368, 538

\bibitem[{{Langer}(2012)}]{2012Langer}
{Langer}, N. 2012, \araa, 50, 107

\bibitem[{{Langer} {et~al.}(1994){Langer}, {Hamann}, {Lennon}, {Najarro}, {Pauldrach}, \& {Puls}}]{1994Langer}
{Langer}, N., {Hamann}, W.~R., {Lennon}, M., {et~al.} 1994, \aap, 290, 819

\bibitem[{{Langer} {et~al.}(2020){Langer}, {Sch{\"u}rmann}, {Stoll}, {Marchant}, {Lennon}, {Mahy}, {de Mink}, {Quast}, {Riedel}, {Sana}, {Schneider}, {Schootemeijer}, {Wang}, {Almeida}, {Bestenlehner}, {Bodensteiner}, {Castro}, {Clark}, {Crowther}, {Dufton}, {Evans}, {Fossati}, {Gr{\"a}fener}, {Grassitelli}, {Grin}, {Hastings}, {Herrero}, {de Koter}, {Menon}, {Patrick}, {Puls}, {Renzo}, {Sander}, {Schneider}, {Sen}, {Shenar}, {Sim{\'o}n-D{\'\i}as}, {Tauris}, {Tramper}, {Vink}, \& {Xu}}]{2020Langer}
{Langer}, N., {Sch{\"u}rmann}, C., {Stoll}, K., {et~al.} 2020, \aap, 638, A39

\bibitem[{{Lapeyrere} {et~al.}(2014){Lapeyrere}, {Kervella}, {Lacour}, {Azouaoui}, {Garcia-Dabo}, {Perrin}, {Eisenhauer}, {Perraut}, {Straubmeier}, {Amorim}, \& {Brandner}}]{2014Lapeyrere}
{Lapeyrere}, V., {Kervella}, P., {Lacour}, S., {et~al.} 2014, in Society of Photo-Optical Instrumentation Engineers (SPIE) Conference Series, Vol. 9146, Optical and Infrared Interferometry IV, ed. J.~K. {Rajagopal}, M.~J. {Creech-Eakman}, \& F.~{Malbet}, 91462D

\bibitem[{{Lau} {et~al.}(2020{\natexlab{a}}){Lau}, {Eldridge}, {Hankins}, {Lamberts}, {Sakon}, \& {Williams}}]{2020Lau}
{Lau}, R.~M., {Eldridge}, J.~J., {Hankins}, M.~J., {et~al.} 2020{\natexlab{a}}, \apj, 898, 74

\bibitem[{{Lau} {et~al.}(2022){Lau}, {Hankins}, {Han}, {Argyriou}, {Corcoran}, {Eldridge}, {Endo}, {Fox}, {Garcia Marin}, {Gull}, {Jones}, {Hamaguchi}, {Lamberts}, {Law}, {Madura}, {Marchenko}, {Matsuhara}, {Moffat}, {Morris}, {Morris}, {Onaka}, {Ressler}, {Richardson}, {Russell}, {Sanchez-Bermudez}, {Smith}, {Soulain}, {Stevens}, {Tuthill}, {Weigelt}, {Williams}, \& {Yamaguchi}}]{2022Lau}
{Lau}, R.~M., {Hankins}, M.~J., {Han}, Y., {et~al.} 2022, Nature Astronomy, 6, 1308

\bibitem[{{Lau} {et~al.}(2020{\natexlab{b}}){Lau}, {Hankins}, {Han}, {Endo}, {Moffat}, {Ressler}, {Sakon}, {Sanchez-Bermudez}, {Soulain}, {Stevens}, {Tuthill}, \& {Williams}}]{2020aLau}
{Lau}, R.~M., {Hankins}, M.~J., {Han}, Y., {et~al.} 2020{\natexlab{b}}, \apj, 900, 190

\bibitem[{{Lefever} {et~al.}(2023){Lefever}, {Sander}, {Shenar}, {Poniatowski}, {Dsilva}, \& {Todt}}]{2023Lefever}
{Lefever}, R.~R., {Sander}, A.~A.~C., {Shenar}, T., {et~al.} 2023, \mnras, 521, 1374

\bibitem[{{Li} {et~al.}(2024){Li}, {Zhu}, {L{\"u}}, {Li}, {Liu}, {Guo}, {Yu}, \& {Lu}}]{2024Li}
{Li}, Z., {Zhu}, C., {L{\"u}}, G., {et~al.} 2024, \apj, 969, 160

\bibitem[{{Lima} {et~al.}(2014){Lima}, {Bica}, {Bonatto}, \& {Saito}}]{2014Lima}
{Lima}, E.~F., {Bica}, E., {Bonatto}, C., \& {Saito}, R.~K. 2014, \aap, 568, A16

\bibitem[{{Limongi} \& {Chieffi}(2018)}]{2018Limongi}
{Limongi}, M. \& {Chieffi}, A. 2018, \apjs, 237, 13

\bibitem[{{Luehrs}(1997)}]{1997Luehrs}
{Luehrs}, S. 1997, \pasp, 109, 504

\bibitem[{{Mahy} {et~al.}(2022){Mahy}, {Lanthermann}, {Hutsem{\'e}kers}, {Kluska}, {Lobel}, {Manick}, {Miszalski}, {Reggiani}, {Sana}, \& {Gosset}}]{2022Mahy}
{Mahy}, L., {Lanthermann}, C., {Hutsem{\'e}kers}, D., {et~al.} 2022, \aap, 657, A4

\bibitem[{Marchant \& Bodensteiner(2024)}]{2023Marchant}
Marchant, P. \& Bodensteiner, J. 2024, Annual Review of Astronomy and Astrophysics

\bibitem[{{Martins} \& {Plez}(2006)}]{2006Martins}
{Martins}, F. \& {Plez}, B. 2006, \aap, 457, 637

\bibitem[{{Menon} {et~al.}(2024){Menon}, {Ercolino}, {Urbaneja}, {Lennon}, {Herrero}, {Hirai}, {Langer}, {Schootemeijer}, {Chatzopoulos}, {Frank}, \& {Shiber}}]{2024Menon}
{Menon}, A., {Ercolino}, A., {Urbaneja}, M.~A., {et~al.} 2024, \apjl, 963, L42

\bibitem[{{M{\'e}rand}(2022)}]{2022SPIE12183E..1NM}
{M{\'e}rand}, A. 2022, in Society of Photo-Optical Instrumentation Engineers (SPIE) Conference Series, Vol. 12183, Optical and Infrared Interferometry and Imaging VIII, ed. A.~{M{\'e}rand}, S.~{Sallum}, \& J.~{Sanchez-Bermudez}, 121831N

\bibitem[{{Millour}(2014)}]{2014Millour}
{Millour}, F. 2014, in EAS Publications Series, Vol. 69-70, EAS Publications Series, 17--52

\bibitem[{{Moe} \& {Di Stefano}(2017)}]{2017Moe}
{Moe}, M. \& {Di Stefano}, R. 2017, \apjs, 230, 15

\bibitem[{{Moffat} \& {Seggewiss}(1977)}]{1977Moffat}
{Moffat}, A.~F.~J. \& {Seggewiss}, W. 1977, \aap, 54, 607

\bibitem[{{Monnier} {et~al.}(2007){Monnier}, {Tuthill}, {Danchi}, {Murphy}, \& {Harries}}]{2007Monnier}
{Monnier}, J.~D., {Tuthill}, P.~G., {Danchi}, W.~C., {Murphy}, N., \& {Harries}, T.~J. 2007, \apj, 655, 1033

\bibitem[{{Monnier} {et~al.}(2011){Monnier}, {Zhao}, {Pedretti}, {Millan-Gabet}, {Berger}, {Traub}, {Schloerb}, {ten Brummelaar}, {McAlister}, {Ridgway}, {Sturmann}, {Sturmann}, {Turner}, {Baron}, {Kraus}, {Tannirkulam}, \& {Williams}}]{2011Monnier}
{Monnier}, J.~D., {Zhao}, M., {Pedretti}, E., {et~al.} 2011, \apjl, 742, L1

\bibitem[{{Montes} {et~al.}(2011){Montes}, {Gonz{\'a}lez}, {Cant{\'o}}, {P{\'e}rez-Torres}, \& {Alberdi}}]{2011Montes}
{Montes}, G., {Gonz{\'a}lez}, R.~F., {Cant{\'o}}, J., {P{\'e}rez-Torres}, M.~A., \& {Alberdi}, A. 2011, \aap, 531, A52

\bibitem[{{Montes} {et~al.}(2009){Montes}, {P{\'e}rez-Torres}, {Alberdi}, \& {Gonz{\'a}lez}}]{2009Montes}
{Montes}, G., {P{\'e}rez-Torres}, M.~A., {Alberdi}, A., \& {Gonz{\'a}lez}, R.~F. 2009, \apj, 705, 899

\bibitem[{{Naz{\'e}} {et~al.}(2021{\natexlab{a}}){Naz{\'e}}, {Gosset}, \& {Marechal}}]{2021Naze}
{Naz{\'e}}, Y., {Gosset}, E., \& {Marechal}, Q. 2021{\natexlab{a}}, \mnras, 501, 4214

\bibitem[{{Naz{\'e}} {et~al.}(2021{\natexlab{b}}){Naz{\'e}}, {Rauw}, \& {Gosset}}]{2021bNaze}
{Naz{\'e}}, Y., {Rauw}, G., \& {Gosset}, E. 2021{\natexlab{b}}, \mnras, 502, 5038

\bibitem[{{Naz{\'e}} {et~al.}(2023){Naz{\'e}}, {Rauw}, {Johnson}, {Gosset}, \& {Hoffman}}]{2023Naze}
{Naz{\'e}}, Y., {Rauw}, G., {Johnson}, R., {Gosset}, E., \& {Hoffman}, J.~L. 2023, \mnras, 526, 2167

\bibitem[{{Naz{\'e}} {et~al.}(2013){Naz{\'e}}, {Rauw}, {Sana}, \& {Corcoran}}]{2013Naze}
{Naz{\'e}}, Y., {Rauw}, G., {Sana}, H., \& {Corcoran}, M.~F. 2013, \aap, 555, A83

\bibitem[{{Neugent} \& {Massey}(2014)}]{2014Neugent}
{Neugent}, K.~F. \& {Massey}, P. 2014, \apj, 789, 10

\bibitem[{{Niemela} {et~al.}(1985){Niemela}, {Mandrini}, \& {Mendez}}]{1985Niemela}
{Niemela}, V.~S., {Mandrini}, C.~H., \& {Mendez}, R.~H. 1985, \rmxaa, 11, 143

\bibitem[{{Nomoto} {et~al.}(2013){Nomoto}, {Kobayashi}, \& {Tominaga}}]{2013Nomoto}
{Nomoto}, K., {Kobayashi}, C., \& {Tominaga}, N. 2013, \araa, 51, 457

\bibitem[{{Paczy{\'n}ski}(1967)}]{1967Paczynski}
{Paczy{\'n}ski}, B. 1967, \actaa, 17, 355

\bibitem[{{Parkin} \& {Gosset}(2011)}]{2011Parkin}
{Parkin}, E.~R. \& {Gosset}, E. 2011, \aap, 530, A119

\bibitem[{{Pauli} {et~al.}(2022){Pauli}, {Langer}, {Aguilera-Dena}, {Wang}, \& {Marchant}}]{2022Pauli}
{Pauli}, D., {Langer}, N., {Aguilera-Dena}, D.~R., {Wang}, C., \& {Marchant}, P. 2022, \aap, 667, A58

\bibitem[{{Pecaut} \& {Mamajek}(2013)}]{2013Pecaut}
{Pecaut}, M.~J. \& {Mamajek}, E.~E. 2013, \apjs, 208, 9

\bibitem[{{Rajagopal}(2010)}]{2010Rajagopal}
{Rajagopal}, J. 2010, in Revista Mexicana de Astronomia y Astrofisica Conference Series, Vol.~38, Revista Mexicana de Astronomia y Astrofisica Conference Series, 54--58

\bibitem[{{Rauw} {et~al.}(1996){Rauw}, {Gosset}, {Manfroid}, {Vreux}, \& {Claeskens}}]{1996Rauw}
{Rauw}, G., {Gosset}, E., {Manfroid}, J., {Vreux}, J.~M., \& {Claeskens}, J.~F. 1996, \aap, 306, 783

\bibitem[{{Richardson} {et~al.}(2024){Richardson}, {Daly}, {Williams}, {Hill}, {Shenavrin}, {Endo}, {Chen{\'e}}, {Karnath}, {Lau}, {Moffat}, \& {Weigelt}}]{2024Richardson}
{Richardson}, N.~D., {Daly}, A.~R., {Williams}, P.~M., {et~al.} 2024, \apj, 969, 140

\bibitem[{{Richardson} {et~al.}(2021){Richardson}, {Lee}, {Schaefer}, {Shenar}, {Sander}, {Hill}, {Fullard}, {Monnier}, {Anugu}, {Davies}, {Gardner}, {Lanthermann}, {Kraus}, \& {Setterholm}}]{2021Richardson}
{Richardson}, N.~D., {Lee}, L., {Schaefer}, G., {et~al.} 2021, \apjl, 908, L3

\bibitem[{{Richardson} {et~al.}(2016){Richardson}, {Shenar}, {Roy-Loubier}, {Schaefer}, {Moffat}, {St-Louis}, {Gies}, {Farrington}, {Hill}, {Williams}, {Gordon}, {Pablo}, \& {Ramiaramanantsoa}}]{2016Richardson}
{Richardson}, N.~D., {Shenar}, T., {Roy-Loubier}, O., {et~al.} 2016, \mnras, 461, 4115

\bibitem[{{Rosslowe} \& {Crowther}(2015)}]{2015RossloweCrowther}
{Rosslowe}, C.~K. \& {Crowther}, P.~A. 2015, \mnras, 447, 2322

\bibitem[{{Rosslowe} \& {Crowther}(2018)}]{2018Rosslowe}
{Rosslowe}, C.~K. \& {Crowther}, P.~A. 2018, \mnras, 473, 2853

\bibitem[{{Sana} {et~al.}(2012){Sana}, {de Mink}, {de Koter}, {Langer}, {Evans}, {Gieles}, {Gosset}, {Izzard}, {Le Bouquin}, \& {Schneider}}]{2012Sana}
{Sana}, H., {de Mink}, S.~E., {de Koter}, A., {et~al.} 2012, Science, 337, 444

\bibitem[{{Sana} {et~al.}(2014){Sana}, {Le Bouquin}, {Lacour}, {Berger}, {Duvert}, {Gauchet}, {Norris}, {Olofsson}, {Pickel}, {Zins}, {Absil}, {de Koter}, {Kratter}, {Schnurr}, \& {Zinnecker}}]{2014Sana}
{Sana}, H., {Le Bouquin}, J.~B., {Lacour}, S., {et~al.} 2014, \apjs, 215, 15

\bibitem[{{Sander} {et~al.}(2019){Sander}, {Hamann}, {Todt}, {Hainich}, {Shenar}, {Ramachandran}, \& {Oskinova}}]{2019Sander}
{Sander}, A.~A.~C., {Hamann}, W.~R., {Todt}, H., {et~al.} 2019, \aap, 621, A92

\bibitem[{{Sander} {et~al.}(2023){Sander}, {Lefever}, {Poniatowski}, {Ramachandran}, {Sabhahit}, \& {Vink}}]{2023Sander}
{Sander}, A.~A.~C., {Lefever}, R.~R., {Poniatowski}, L.~G., {et~al.} 2023, \aap, 670, A83

\bibitem[{{Sander} {et~al.}(2020){Sander}, {Vink}, \& {Hamann}}]{2020Sander}
{Sander}, A. A.~C., {Vink}, J.~S., \& {Hamann}, W.~R. 2020, \mnras, 491, 4406

\bibitem[{{Schootemeijer} \& {Langer}(2018)}]{2018Schootemeijer}
{Schootemeijer}, A. \& {Langer}, N. 2018, \aap, 611, A75

\bibitem[{{Schootemeijer} {et~al.}(2024){Schootemeijer}, {Shenar}, {Langer}, {Grin}, {Sana}, {Gr{\"a}fener}, {Sch{\"u}rmann}, {Wang}, \& {Xu}}]{2024Schootemeijer}
{Schootemeijer}, A., {Shenar}, T., {Langer}, N., {et~al.} 2024, \aap, 689, A157

\bibitem[{{Shara} {et~al.}(2022){Shara}, {Howell}, {Furlan}, {Gnilka}, {Moffat}, {Scott}, \& {Zurek}}]{2022Shara}
{Shara}, M.~M., {Howell}, S.~B., {Furlan}, E., {et~al.} 2022, \mnras, 509, 2897

\bibitem[{{Shenar} {et~al.}(2020){Shenar}, {Gilkis}, {Vink}, {Sana}, \& {Sand er}}]{2020Shenar}
{Shenar}, T., {Gilkis}, A., {Vink}, J.~S., {Sana}, H., \& {Sand er}, A.~A.~C. 2020, \aap, 634, A79

\bibitem[{{Shenar} {et~al.}(2019){Shenar}, {Sablowski}, {Hainich}, {Todt}, {Moffat}, {Oskinova}, {Ramachandran}, {Sana}, {Sander}, {Schnurr}, {St-Louis}, {Vanbeveren}, {G{\"o}tberg}, \& {Hamann}}]{2019Shenar}
{Shenar}, T., {Sablowski}, D.~P., {Hainich}, R., {et~al.} 2019, \aap, 627, A151

\bibitem[{{Shenar} {et~al.}(2023){Shenar}, {Wade}, {Marchant}, {Bagnulo}, {Bodensteiner}, {Bowman}, {Gilkis}, {Langer}, {Nicolas-Chen{\'e}}, {Oskinova}, {Van Reeth}, {Sana}, {St-Louis}, {de Oliveira}, {Todt}, \& {Toonen}}]{2023Shenar}
{Shenar}, T., {Wade}, G.~A., {Marchant}, P., {et~al.} 2023, Science, 381, 761

\bibitem[{{Shylaja}(1990)}]{1990Shylaja}
{Shylaja}, B.~S. 1990, \apss, 164, 63

\bibitem[{{Siess} {et~al.}(2000){Siess}, {Dufour}, \& {Forestini}}]{2000Siess}
{Siess}, L., {Dufour}, E., \& {Forestini}, M. 2000, \aap, 358, 593

\bibitem[{{Skinner} {et~al.}(2021){Skinner}, {Schmutz}, {G{\"u}del}, \& {Zhekov}}]{2021Skinner}
{Skinner}, S.~L., {Schmutz}, W., {G{\"u}del}, M., \& {Zhekov}, S.~A. 2021, Research Notes of the American Astronomical Society, 5, 125

\bibitem[{{Skinner} {et~al.}(2010){Skinner}, {Zhekov}, {G{\"u}del}, {Schmutz}, \& {Sokal}}]{2010Skinner}
{Skinner}, S.~L., {Zhekov}, S.~A., {G{\"u}del}, M., {Schmutz}, W., \& {Sokal}, K.~R. 2010, \aj, 139, 825

\bibitem[{{Skinner} {et~al.}(2012){Skinner}, {Zhekov}, {G{\"u}del}, {Schmutz}, \& {Sokal}}]{2012Skinner}
{Skinner}, S.~L., {Zhekov}, S.~A., {G{\"u}del}, M., {Schmutz}, W., \& {Sokal}, K.~R. 2012, \aj, 143, 116

\bibitem[{{Smette} {et~al.}(2015){Smette}, {Sana}, {Noll}, {Horst}, {Kausch}, {Kimeswenger}, {Barden}, {Szyszka}, {Jones}, {Gallenne}, {Vinther}, {Ballester}, \& {Taylor}}]{2015A&A...576A..77S}
{Smette}, A., {Sana}, H., {Noll}, S., {et~al.} 2015, \aap, 576, A77

\bibitem[{{Smith}(1968)}]{1968aSmith}
{Smith}, L.~F. 1968, \mnras, 138, 109

\bibitem[{{Smith} {et~al.}(1996){Smith}, {Shara}, \& {Moffat}}]{1996Smith}
{Smith}, L.~F., {Shara}, M.~M., \& {Moffat}, A. F.~J. 1996, \mnras, 281, 163

\bibitem[{{Smith} {et~al.}(1994){Smith}, {Crowther}, \& {Prinja}}]{1994Smith}
{Smith}, L.~J., {Crowther}, P.~A., \& {Prinja}, R.~K. 1994, \aap, 281, 833

\bibitem[{{Smith}(2014)}]{2014Smith}
{Smith}, N. 2014, \araa, 52, 487

\bibitem[{{Smith} {et~al.}(2020){Smith}, {E Andrews}, {Moe}, {Milne}, {Bilinski}, {Kilpatrick}, {Fong}, {Badenes}, {Filippenko}, {Kasliwal}, \& {Silverman}}]{2020Smit}
{Smith}, N., {E Andrews}, J., {Moe}, M., {et~al.} 2020, \mnras, 492, 5897

\bibitem[{{Sota} {et~al.}(2014){Sota}, {Ma{\'\i}z Apell{\'a}niz}, {Morrell}, {Barb{\'a}}, {Walborn}, {Gamen}, {Arias}, \& {Alfaro}}]{2014Sota}
{Sota}, A., {Ma{\'\i}z Apell{\'a}niz}, J., {Morrell}, N.~I., {et~al.} 2014, \apjs, 211, 10

\bibitem[{{Soulain} {et~al.}(2023){Soulain}, {Lamberts}, {Millour}, {Tuthill}, \& {Lau}}]{2023Soulain}
{Soulain}, A., {Lamberts}, A., {Millour}, F., {Tuthill}, P., \& {Lau}, R.~M. 2023, \mnras, 518, 3211

\bibitem[{{St-Louis} {et~al.}(2009){St-Louis}, {Chen{\'e}}, {Schnurr}, \& {Nicol}}]{2009StLouis}
{St-Louis}, N., {Chen{\'e}}, A.~N., {Schnurr}, O., \& {Nicol}, M.~H. 2009, \apj, 698, 1951

\bibitem[{{Turner}(1982)}]{1982Turner}
{Turner}, D.~G. 1982, in IAU Symposium, Vol.~99, Wolf-Rayet Stars: Observations, Physics, Evolution, ed. C.~W.~H. {De Loore} \& A.~J. {Willis}, 57--60

\bibitem[{{van der Hucht}(2001)}]{2001vanderHucht}
{van der Hucht}, K.~A. 2001, \nar, 45, 135

\bibitem[{{van der Hucht} {et~al.}(1988){van der Hucht}, {Hidayat}, {Admiranto}, {Supelli}, \& {Doom}}]{1988Hucht}
{van der Hucht}, K.~A., {Hidayat}, B., {Admiranto}, A.~G., {Supelli}, K.~R., \& {Doom}, C. 1988, \aap, 199, 217

\bibitem[{{Vanbeveren} \& {Conti}(1980)}]{1980VanbeverenConti}
{Vanbeveren}, D. \& {Conti}, P.~S. 1980, \aap, 88, 230

\bibitem[{{Vanbeveren} {et~al.}(1998){Vanbeveren}, {De Donder}, {Van Bever}, {Van Rensbergen}, \& {De Loore}}]{1998Vanbeveren}
{Vanbeveren}, D., {De Donder}, E., {Van Bever}, J., {Van Rensbergen}, W., \& {De Loore}, C. 1998, \na, 3, 443

\bibitem[{{Woosley} {et~al.}(2002){Woosley}, {Heger}, \& {Weaver}}]{2002Woosley}
{Woosley}, S.~E., {Heger}, A., \& {Weaver}, T.~A. 2002, Reviews of Modern Physics, 74, 1015

\end{thebibliography}

\begin{appendix}

\counterwithin{figure}{section}

\section{Summary of Observations}

\begin{table}[ht]
    \caption[]{Summary of GRAVITY Observations of WRs.}
    \renewcommand{\arraystretch}{1.15}
\begin{tabular*}{\columnwidth}{@{\extracolsep{\fill}} cccc }
\hline
\begin{tabular}{@{}c@{}} WR \\ \#  \end{tabular} 
& \begin{tabular}{@{}c@{}} Calibration \\ Star ID \end{tabular}
& \begin{tabular}{@{}c@{}} MJD of \\ Observation \end{tabular}
& \begin{tabular}{@{}c@{}} Spectral \\ Resolution \\ \end{tabular} \\

\hline
\hline

P111.24JN \\

\hline

8 & HD 64545 & 60058.062 & Medium \\

9 & HD 61103 & 60295.214 & Medium \\

12 & HD 75027 & 60088.029 & Medium \\

14 & HD 75027 & 60058.013 & Medium \\ 

15 & HD 80347 & 60264.312 & Medium \\

16 & HD 85779 & 60064.099 & Medium \\

18 & HD 88687 & 60066.125 & Medium \\

21 & HD 88687 & 60326.230 & Medium \\

22 & HD 90274 & 60089.074 & High \\

23 & HD 93808 & 60060.064 & Medium \\

24 & HD 91652 & 60351.197 & High \\

31 & HD 96010 & 60332.294 & Medium \\

31a & HD 95096 & 60351.397 & Medium \\

42 & HD 98677 & 60089.029 & Medium \\

47 & HD 107837 & 60089.140 & Medium \\

48 & HD 116703 & 60065.182 & High \\

52 & HD 116140 & 60088.192 & Medium \\

55 & HD 116471 & 60332.327 & Medium \\

57 & HD 116007 & 60332.375 & Medium \\

66 & HD 135378 & 60146.056 & Medium \\

75 & HD 147155 & 60180.008 & Medium \\

79 & HD 151790 & 60109.202 & High \\

79b & HD 152370 & 60109.227 & Medium \\

81 & HD 153951 & 60147.186 & Medium \\

85 & HD 156641 & 60176.170 & Medium \\

87 & HD 155808 & 60176.116 & Medium \\

89 & HD 155808 & 60175.149 & Medium \\

92 & HD 156291 & 60146.097 & Medium \\

93 & HD 158021 & 60116.186 & High \\

97 & HD 159488 & 60146.181 & Medium \\

98 & HD 159488 & 60179.001 & Medium \\

108 & HD 166031 & 60182.147 & Medium \\

110 & HD 166240 & 60148.158 & Medium \\

111 & HD 166031 & 60177.158 & Medium \\

113 & HD 169783 & 60181.999 & High \\

114 & HD 167811 & 60179.153 & Medium \\

115 & HD 169783 & 60180.176 & Medium \\

\hline
\hline

P109.23CN \\

\hline

78 & HD 96568 & 59715.111 & High \\

79a & HD 160263 & 59794.043 & High \\

\hline

\end{tabular*}

\medskip
\textbf{Notes:} The columns list the WR number, corresponding calibration star, time of observation and GRAVITY spectral resolution used (Medium R$\sim$500 or High R$\sim$4000). Also mentioned are the ESO Programs 111.24JN and 109.23CN, under which all WRs were observed.

    \label{tab:allobs}

    \end{table}

\newpage

\section{Qualitative Analysis of GRAVITY Data}
\label{sec:qual}

Three interferometric observables of interest captured by GRAVITY are |V|, T3PHI and DPHI, each capable of storing information regarding the size and structure of the target to varying extents. In addition, GRAVITY also provides a $K$-band spectrum, which is normalized and referred to as NFLUX. For a qualitative view, just inspecting the general features in these observables can motivate our modeling approach.

The visibility amplitude (|V|) is indicative of the extent to which the target is resolved. At its extremes, we have |V| = 1 at all baselines for an unresolved source, whereas |V| = 0 at all baselines for a fully resolved source. Partially resolved sources typically show intermediate behavior between the two extremes. In case the target is comprised of two components, the |V| data is sensitive to their flux ratio. In a simplified case of two components, the combined complex visibility follows the equation:

\begin{equation}
    \label{eq:1}
    V = \frac{F_1 V_1 + F_2 V_2}{F_1 + F_2}.
\end{equation}

\begin{figure*}
  \includegraphics[width=\textwidth]{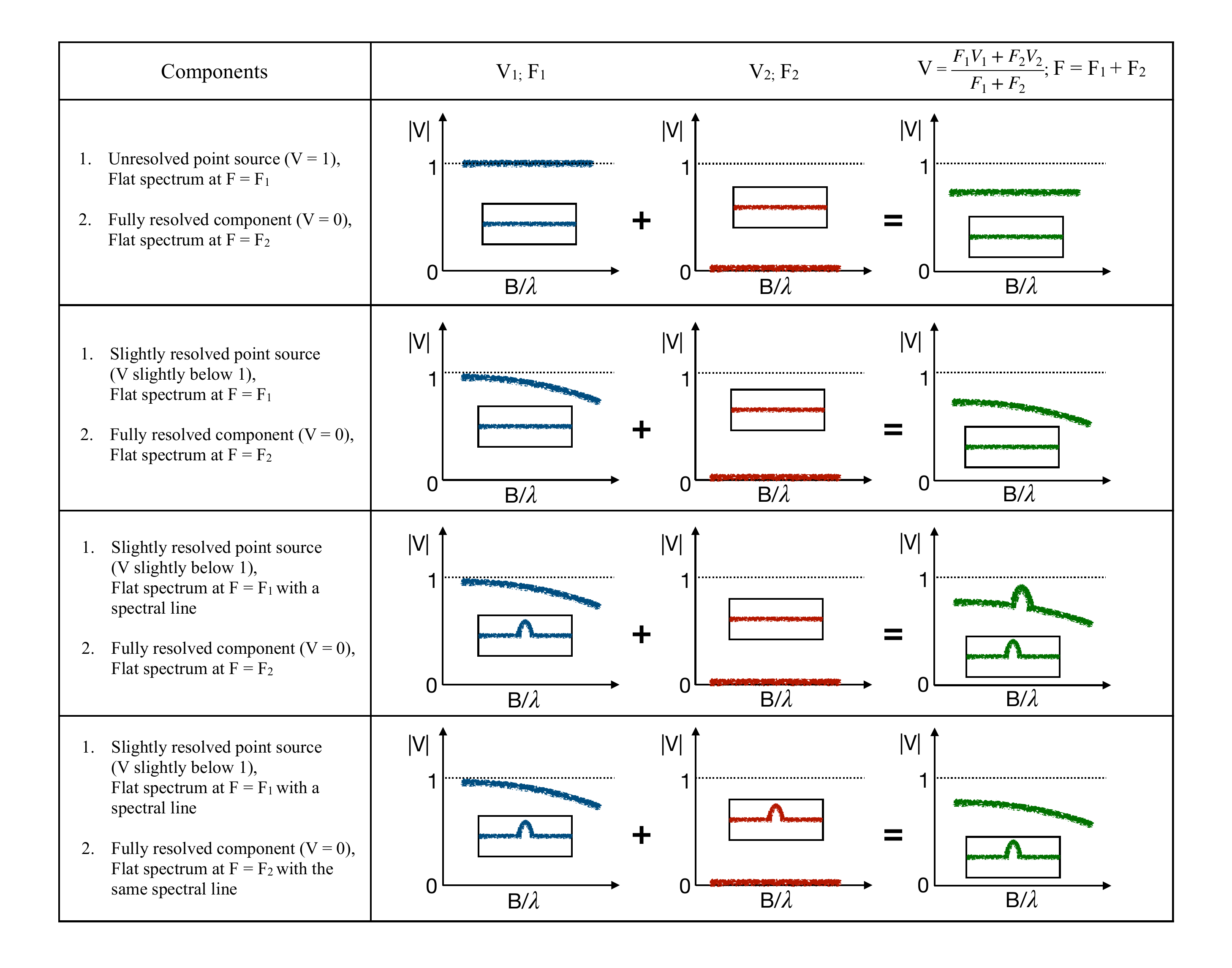}
  \caption{Four simple examples of |V| plots for two-component targets where one component is unresolved or slightly resolved and the other component is fully resolved. The first column describes the two components along with their flux properties, while the second column shows how the components add up together resulting in the combined |V|. The boxes in each plot show the spectrum. In interferometric data, we only have access to the final (green) plots and this table describes a qualitative approach to guessing the components present in the target based on |V|. Although the four examples serve as a good representative set, actual data and corresponding models can be much more complex, warranting the addition of more parameters.}
  \label{fig:int-basics}
\end{figure*}

Figure\,\ref{fig:int-basics} shows four simple examples of two-component systems to get a qualitative understanding of how |V| manifests for such targets if they are composed of a combination of unresolved or partially resolved (i.e. V$\sim$1) and a fully resolved (V=0) component. Because the latter has zero visibility, we can ignore the phase term and the linearity of the complex visibility applies to the amplitude directly. 

It is important to note here that the examples shown are for simple cases, while it is possible for |V| to be much more complex owing to different spatial and/or spectral components in the system. When it comes to searching for appropriate models, a more comprehensive inspection is necessary. For example, in case of a binary system consisting of two distinct point sources, |V| depends on their flux ratio, separation as well as orientation. Such a system can only be modeled by keeping the unknown parameters as free variables and searching a grid in the parameter space to find the best-fit solution.

The closure phase (T3PHI) data is sensitive to departure from point asymmetry within the target being observed. A point-symmetric source typically displays zero closure phase. A non-zero closure phase will be observed for an object with a resolved non-point like symmetry, such as a non-equal binary. T3PHI is a strong indicator of a target harboring a binary system, and is fit along with |V| data using the grid search described above.

The differential phase (DPHI) data can help detect asymmetry at a certain wavelength relative to a different wavelength at scales much smaller than those accessible with |V|-T3PHI. A non-zero differential phase (DPHI) in a line indicates a wavelength-dependent shift in the photocenter. For example, in a very close binary system in which one star has a certain spectral line which the other does not, DPHI data might detect a signal at the spectral line wavelength, even though |V|-T3PHI are consistent with a single unresolved point source. Although DPHI is a powerful tool to detect such asymmetry, detailed modeling requires one to know the properties of the system beforehand, especially when it is unresolved in |V|-T3PHI data. Consequently, DPHI is mostly used for qualitative evaluation of targets, and only invoked in modeling when a system is resolved in |V|-T3PHI as well.

When modeling interferometric data, it is ideal to have a consistent fit across all observables - |V|, T3PHI, DPHI and NFLUX. In case there is no spectral feature in NFLUX and hence in |V|, and DPHI = 0, it is likely that |V| and T3PHI alone can be sufficient for modeling - the first two examples in Fig.\,\ref{fig:int-basics} fall in this category. If we do see a spectral feature in NFLUX as well as |V|, it is important to include NFLUX and DPHI for modeling the system - the third example in Fig.\,\ref{fig:int-basics} falls in this category. Lastly, if we see a spectral feature in NFLUX but not in |V|, and DPHI = 0, it is likely that the spectral feature is present in multiple components and is somehow suppressed in |V| due to the flux ratio following Equation\,\ref{eq:1}. Such cases can be approached in two ways - 1. assuming flat spectra for all components without taking NFLUX into account, and simply modeling |V|-T3PHI to obtain general properties of the system; 2. performing detailed modeling including spectral lines and finding a solution to replicate NFLUX as well as |V| consistently, along with T3PHI and DPHI. The first approach is often sufficient for a companion search since |V|-T3PHI data alone can clearly indicate the presence (or lack) of a resolved binary or multiple system. It does not require introducing unknown/uncertain parameters to force a better fit to the data. The second approach is more informative, but also more challenging due to a higher number of parameters and potential degeneracies introduced with them. Consequently, the second approach is favoured only for specific cases where we have physically motivated parameters to add to the model.

\clearpage

\section{Grid Search for Resolved Companions}
\label{sec:grid}

The prominent signatures of a binary in interferometric data include a non-zero closure phase indicating asymmetry, and an approximately sinusoidally modulating visibility amplitude. The latter can be expressed analytically as 

\begin{equation}
    |V| = \sqrt{\frac{1+f^{2}+2f\,\cos\left(\frac{\overrightarrow{\rho}.\overrightarrow{x_{0}}}{\lambda}\right)}{1+f^{2}}},
\end{equation}

where $f$ is the flux ratio of the secondary to the primary, $\overrightarrow{\rho}$ is the spatial frequency ($\rho = B/\lambda$) and $\overrightarrow{x_{0}}$ is the vector pointing from the primary to the secondary \citep{2014Millour}. While it is not possible to find an accurate binary model for interferometric data using a purely analytical approach, it can still approximately inform us on the orientation and separation of the binary. Consequently, a qualitative look at GRAVITY data can motivate the approach we take to find the best binary model.

Only four WRs in our sample (WR 89, WR 115, WR 48 and WR 93) displayed potential signatures of a resolved binary. Nevertheless, we performed a grid search for companions on all 39 WRs in our GRAVITY sample. The flux ratio of the secondary to the primary was left as a free parameter. We started with a coarse initial grid spanning the range $\Delta{\rm E}$ = --100 to 100 mas and $\Delta{\rm N}$ = --100 to 100 mas, with steps of 20 mas each. According to Equation C.1, the wider a binary (higher the $x_{0}$), the more sinusoidally modulated its |V| data. In case the |V| data for a WR were mostly flat and the coarse grid search did not result in a good model, we performed a second, finer grid search much closer to the primary. This second search spanned the range $\Delta{\rm E}$ = --15 to 15 mas and $\Delta{\rm N}$ = --15 to 15 mas, with steps of 3 mas each. As an extension to the discussion about $\chi^{2}_{\rm red}$ in Section\,\ref{sec:data}, we note here that including NFLUX and DPHI data for modeling while not taking every single spectral feature into account can lead to large values of $\chi^{2}_{\rm red}$. While we ensured that our solutions were consistent even when using |V|-T3PHI data alone, we report the results including NFLUX and DPHI for a more complete picture of the systems.

Based on the two searches described above, 35 WRs in our sample were found to have no wide companion detections that were significant. Here, we discuss the remaining four WRs.

\subsection{WR 89}

\begin{figure}
  \includegraphics[width=\columnwidth]{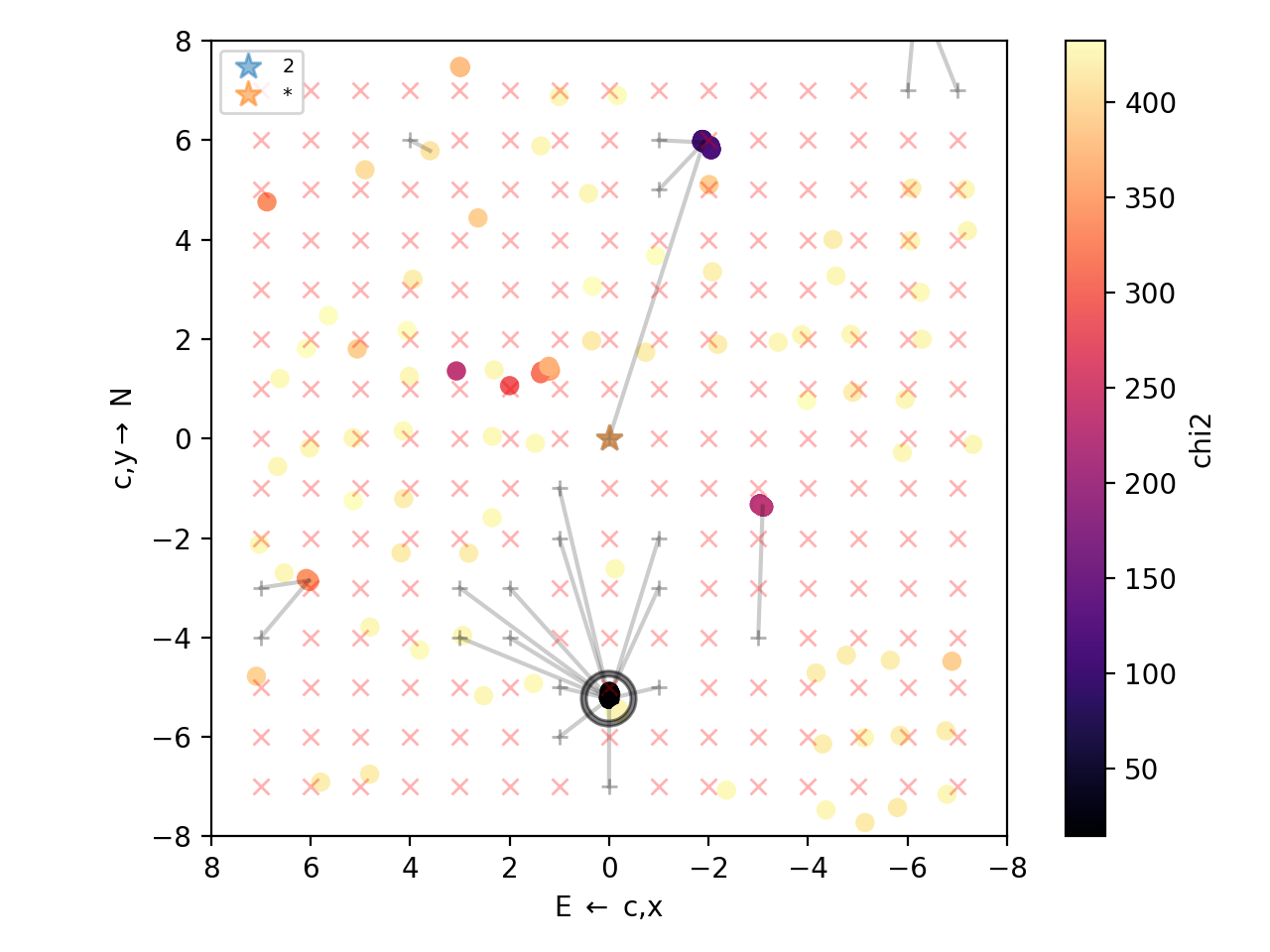}
  \caption{The fine grid search for a companion for WR 89. We found the global minimum for $\chi^{2}_{\rm red}$ at $\Delta{\rm E}$ = $0.02\pm0.08$ mas and $\Delta{\rm N}$ = $-5.18\pm0.10$ mas with the primary star set at the origin. The color scale denotes $\chi^{2}_{\rm red}$ across the grid.}
  \label{fig:wr89_grid}
\end{figure}

For WR 89, we found the best-fit model in the fine grid search. The companion was detected at $\Delta{\rm E}$ = $0.02\pm0.08$ mas and $\Delta{\rm N}$ = $-5.18\pm0.10$ mas relative to the central primary, as shown in Figure\,\ref{fig:wr89_grid}.

\subsection{WR 115}

The GRAVITY data for WR 115 displayed a high amount of sinusoidal modulation in |V| (see Fig.\,\ref{wr115_2}). The coarse as well as fine grid searches failed to obtain an accurate binary model for WR 115. We turned to Equation C.1 for a qualitative estimate of a suitable model. The sinusoidal modulation in |V| is sensitive to the binary separation ($x_{0}$) as well as its orientation with respect to the baseline ($\overrightarrow{\rho}.\overrightarrow{x_{0}}$), with the highest modulation seen for baselines aligned with the binary. From Figure\,\ref{wr115_2}, it is evident that the highest modulation is present in the J2A0 and G1A0 baselines. 

\begin{figure}
  \includegraphics[width=\columnwidth]{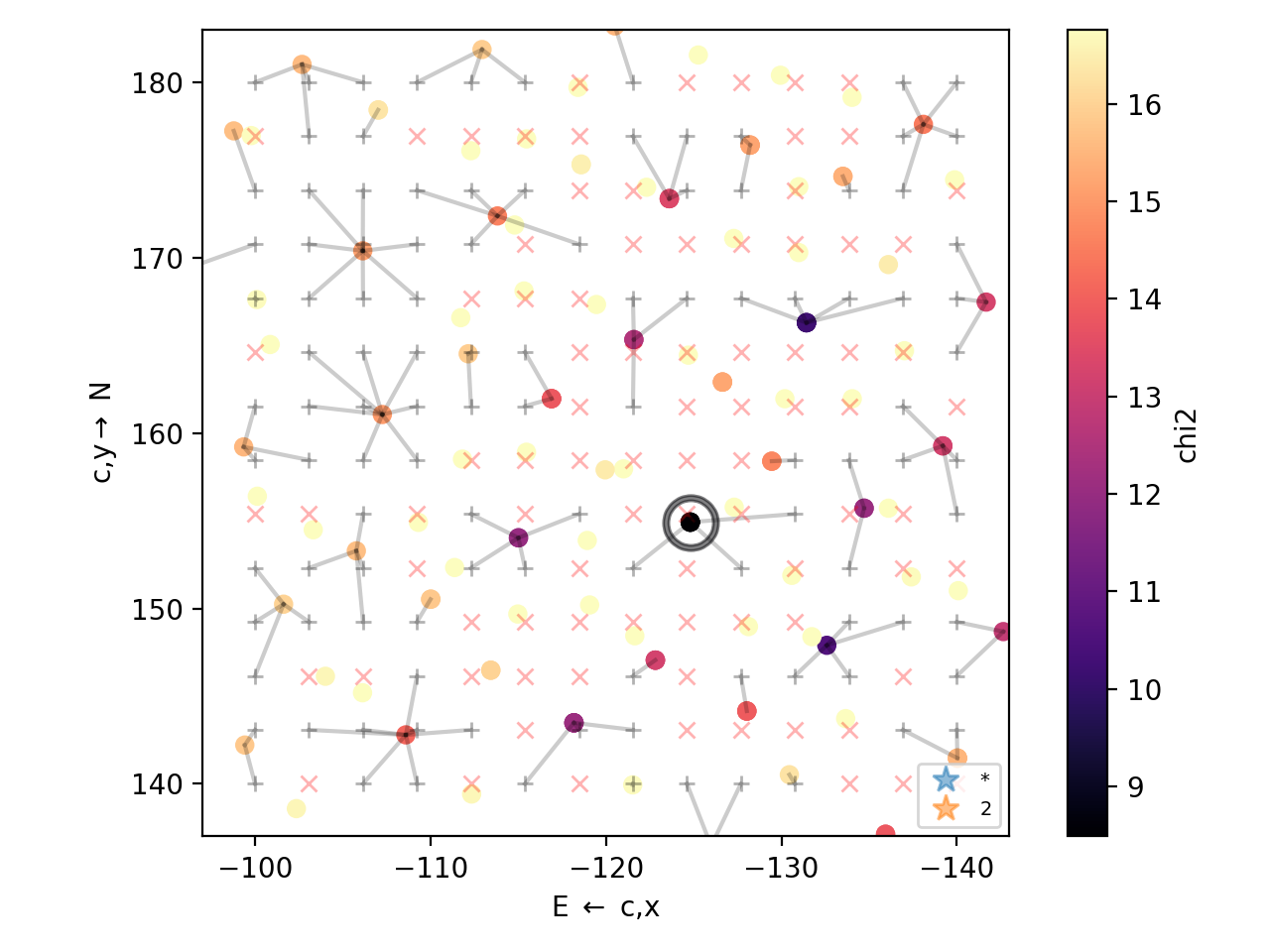}
  \caption{The fine grid search for a companion for WR 115. We found the global minimum for $\chi^{2}_{\rm red}$ at $\Delta{\rm E}$ = $-124.80\pm0.04$ mas and $\Delta{\rm N}$ = $154.93\pm0.02$ mas with the primary star set at the origin. The color scale denotes $\chi^{2}_{\rm red}$ across the grid.}
  \label{fig:wr115_1}
\end{figure}

Based on these inputs, we performed another coarse grid search in the ranges $\Delta{\rm E}$ = --200 to --100 mas and $\Delta{\rm N}$ = 100 to 200 mas; and $\Delta{\rm E}$ = 100 to 200 mas and $\Delta{\rm N}$ = --200 to --100 mas, with steps of 20 mas each. We found an approximate detection around $\Delta{\rm E}$ = --120 mas and $\Delta{\rm N}$ = 150 mas. Consequently, we performed a fine grid search closer to the approximate solution, and detected the companion at $\Delta{\rm E}$ = $-124.80\pm0.04$ mas and $\Delta{\rm N}$ = $154.93\pm0.02$ (see Fig.\,\ref{fig:wr115_1}).

\subsection{WR 48}

\begin{figure}
  \includegraphics[width=\columnwidth]{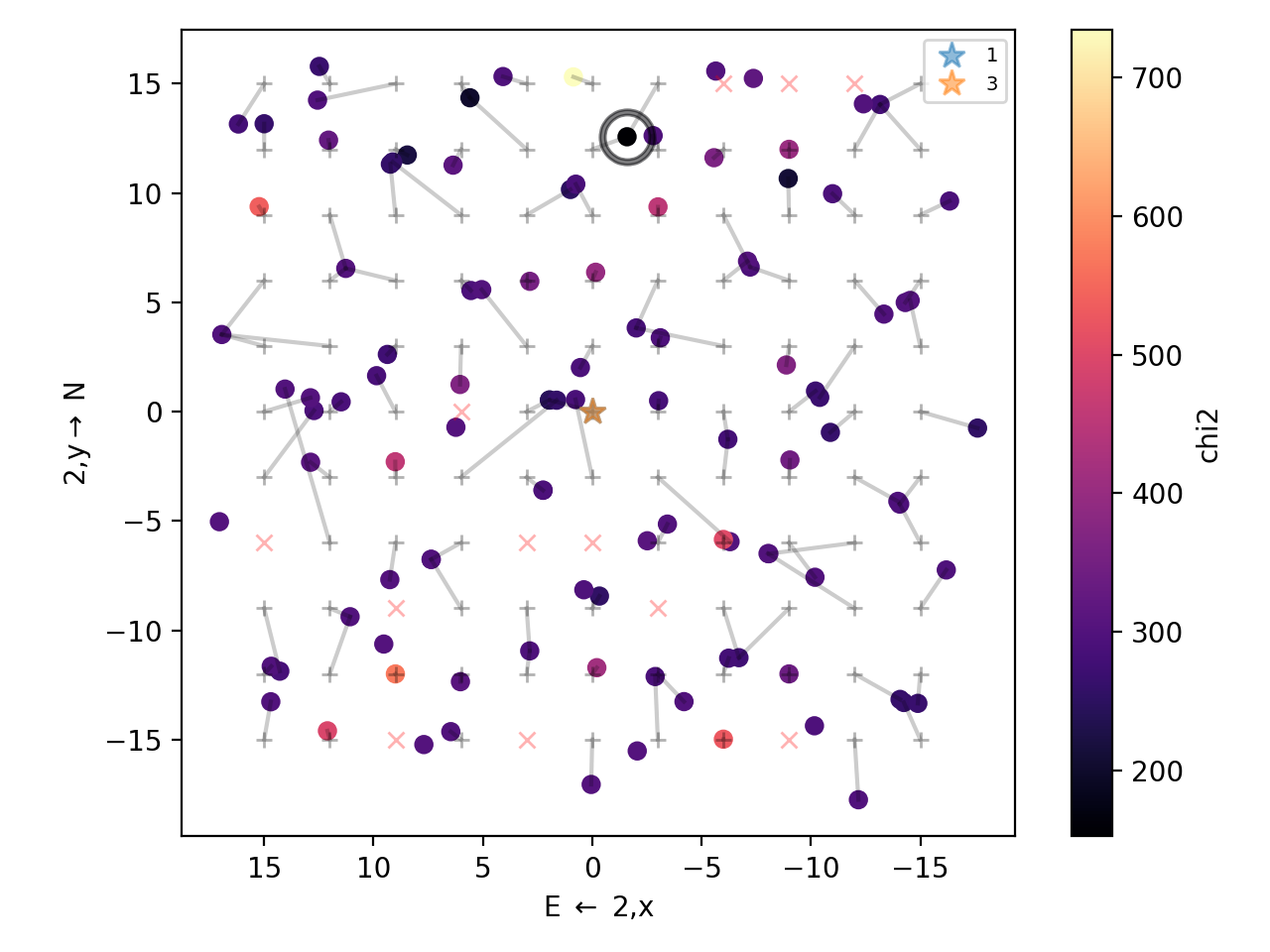}
  \caption{The fine grid search for a companion for WR 48. Here, the wide companion is brighter than the WR, so we fixed the companion at the origin and searched the WR position across the grid. We found the global minimum for $\chi^{2}_{\rm red}$ at $\Delta{\rm E}$ = $1.59\pm0.03$ mas and $\Delta{\rm N}$ = $-12.56\pm0.06$ mas with the wide companion set at the origin. The color scale denotes the $\chi^{2}_{\rm red}$ across the grid.}
  \label{fig:wr48_1}
\end{figure}

For WR 48, as mentioned in the main text of the paper, a known outer O supergiant companion is reported in \citet{1999Hartkopf}, which is brighter than the central object. Although our GRAVITY data for WR 48 is incomplete, we see clear indications of this companion (see Figures\,\ref{wr48_1},\,\ref{wr48_2}). In this case, we fixed the O supergiant at the origin and performed a grid search for the WR. We found the best-fit model in a fine grid search, where the WR was detected at $\Delta{\rm E}$ = $-1.59\pm0.03$ mas and $\Delta{\rm N}$ = $12.56\pm0.06$ mas relative to the central O supergiant (see Fig.\,\ref{fig:wr48_1}). The main text of the paper reports the position of the O supergiant companion relative to the WR.

\subsection{WR 93}

\begin{figure}
  \includegraphics[width=\columnwidth]{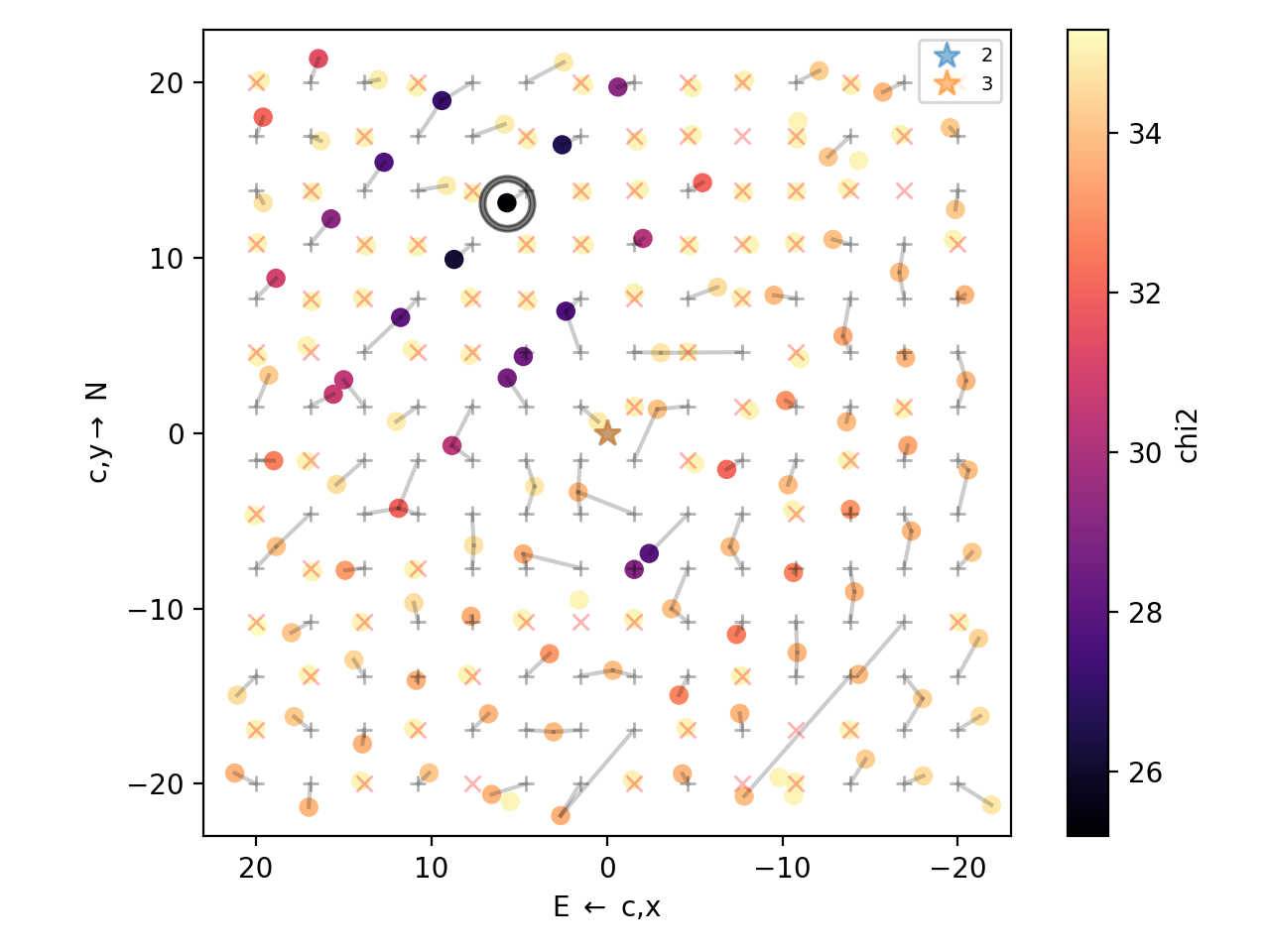}
  \caption{The fine grid search for a companion for WR 93. We found the global minimum for $\chi^{2}_{\rm red}$ at $\Delta{\rm E}$ = $5.60\pm0.03$ mas and $\Delta{\rm N}$ = $13.19\pm0.03$ mas with the primary star set at the origin. The color scale denotes $\chi^{2}_{\rm red}$ across the grid.}
  \label{fig:wr93_1}
\end{figure}

The GRAVITY data for WR 93 showed strong indications of a wide companion (see Figures\,\ref{wr93_1},\,\ref{wr93_2}), and were analysed similarly to WR 89. We found the best-fit model in the fine grid search. The companion was detected at $\Delta{\rm E}$ = $5.60\pm0.03$ mas and $\Delta{\rm N}$ = $13.19\pm0.03$ mas relative to the central primary, as shown in Figure\,\ref{fig:wr93_1}.

\vspace{5mm}
\noindent More details about the companions such as flux ratios, physical separations and potential spectral types are discussed in the main text of the paper.

\clearpage

\section{Comments on Individual Objects}
\label{sec:all}

\begin{figure}[H]
\centering
  \includegraphics[height=60mm]{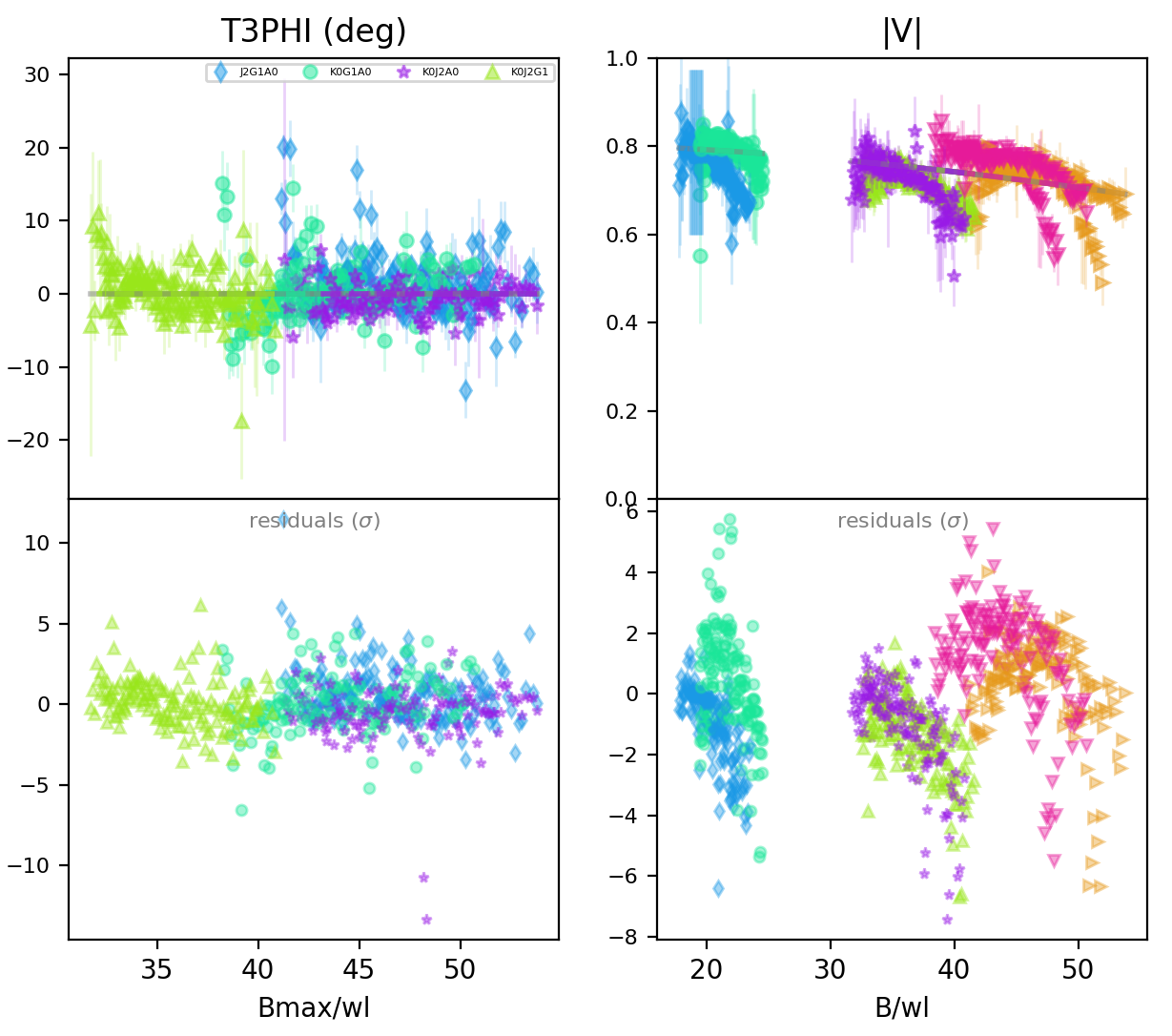}
  \caption{|V|-T3PHI data (top panels) for WR 8 fit with a central point source + fully resolved component, along with corresponding residuals (bottom panels).}
  \label{wr8}
\end{figure}

\begin{figure}[H]
\centering
  \includegraphics[height=60mm]{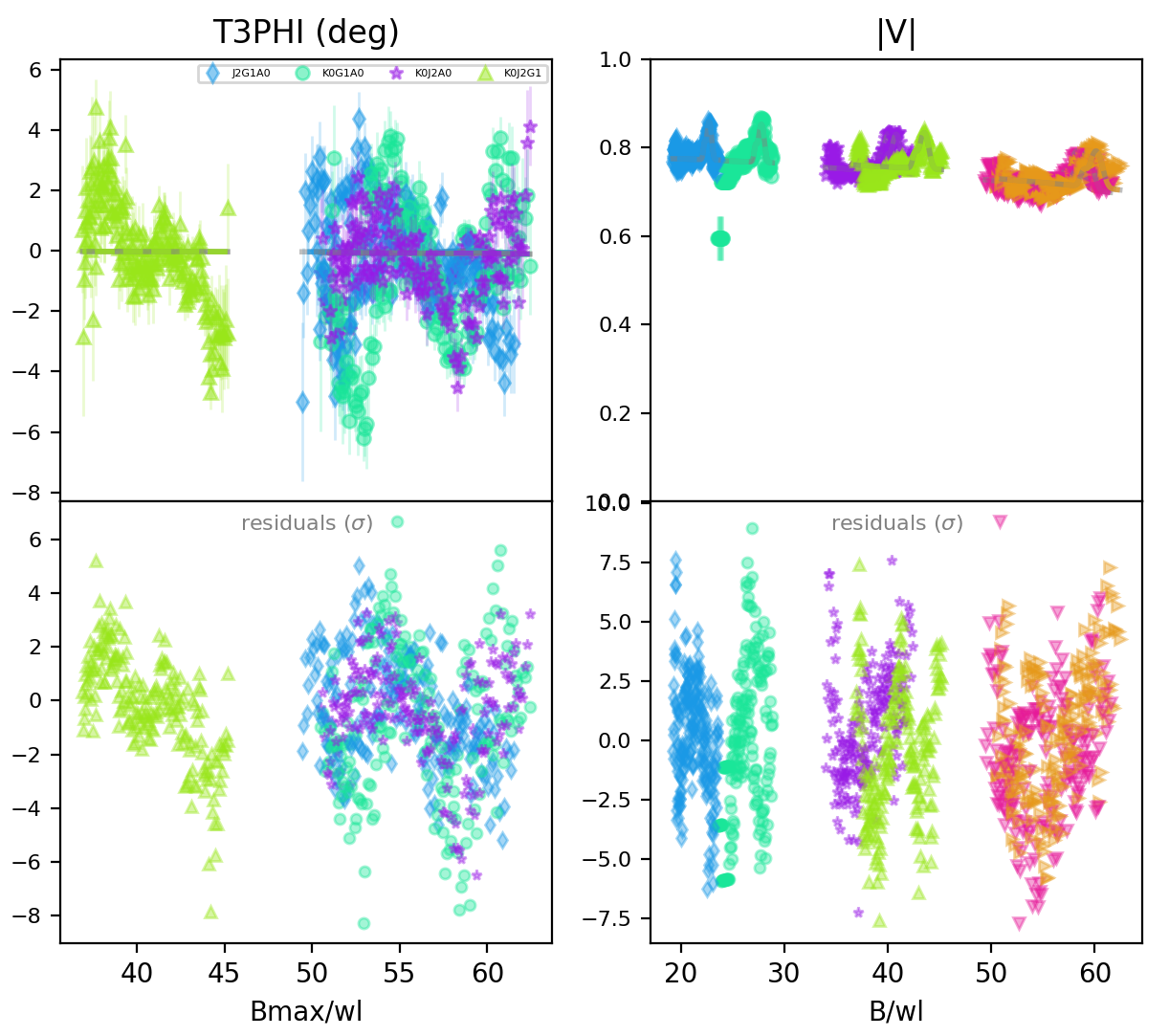}
  \caption{|V|-T3PHI data (top panels) for WR 9 fit with a central point source + fully resolved component, along with corresponding residuals (bottom panels).}
  \label{wr9_1}
\end{figure}

\begin{figure}[H]
\centering
  \includegraphics[height=60mm]{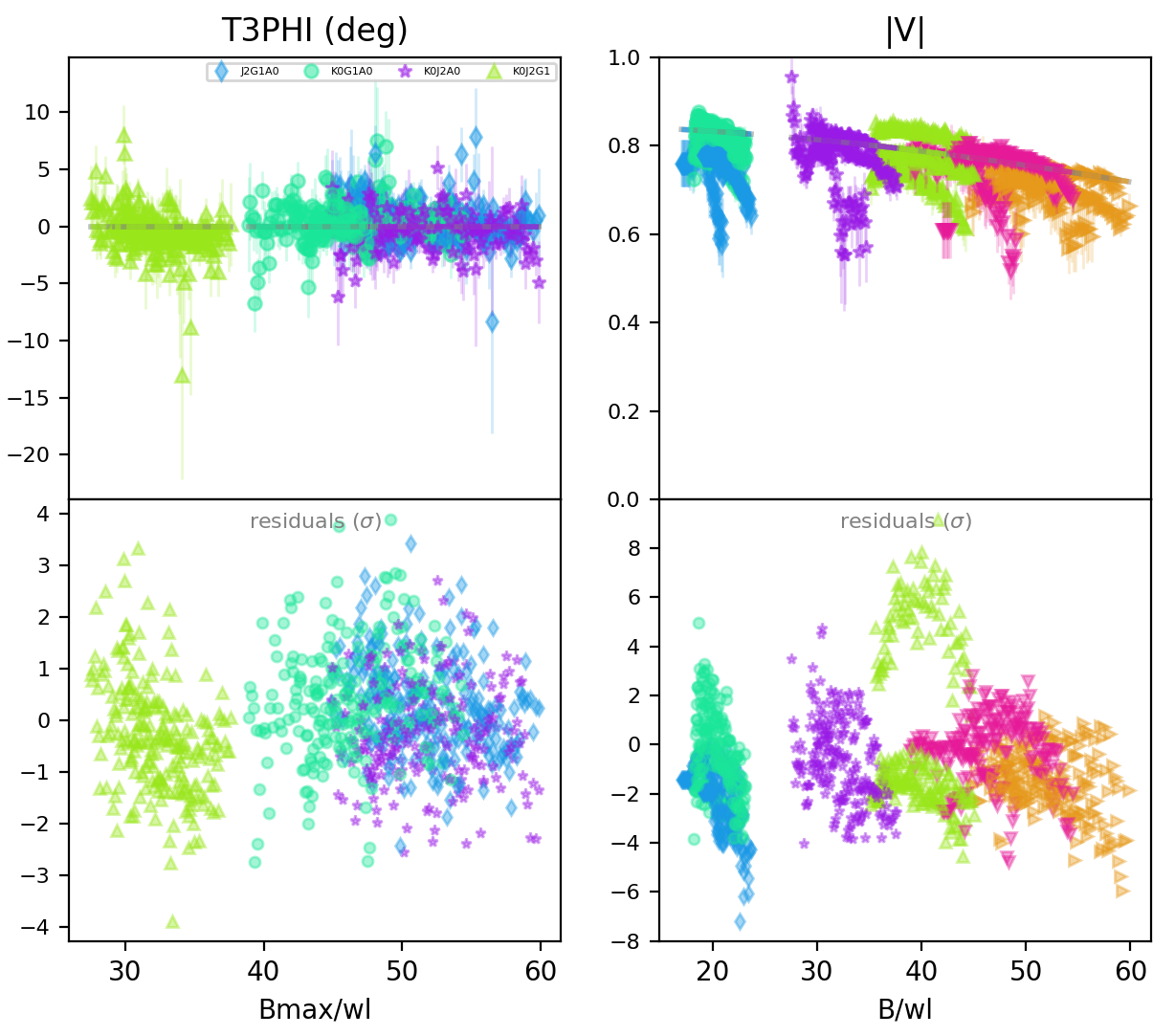}
  \caption{|V|-T3PHI data (top panels) for WR 12 fit with a central point source + fully resolved component, along with corresponding residuals (bottom panels).}
  \label{wr12}
\end{figure}

\begin{figure}[H]
\centering
  \includegraphics[height=60mm]{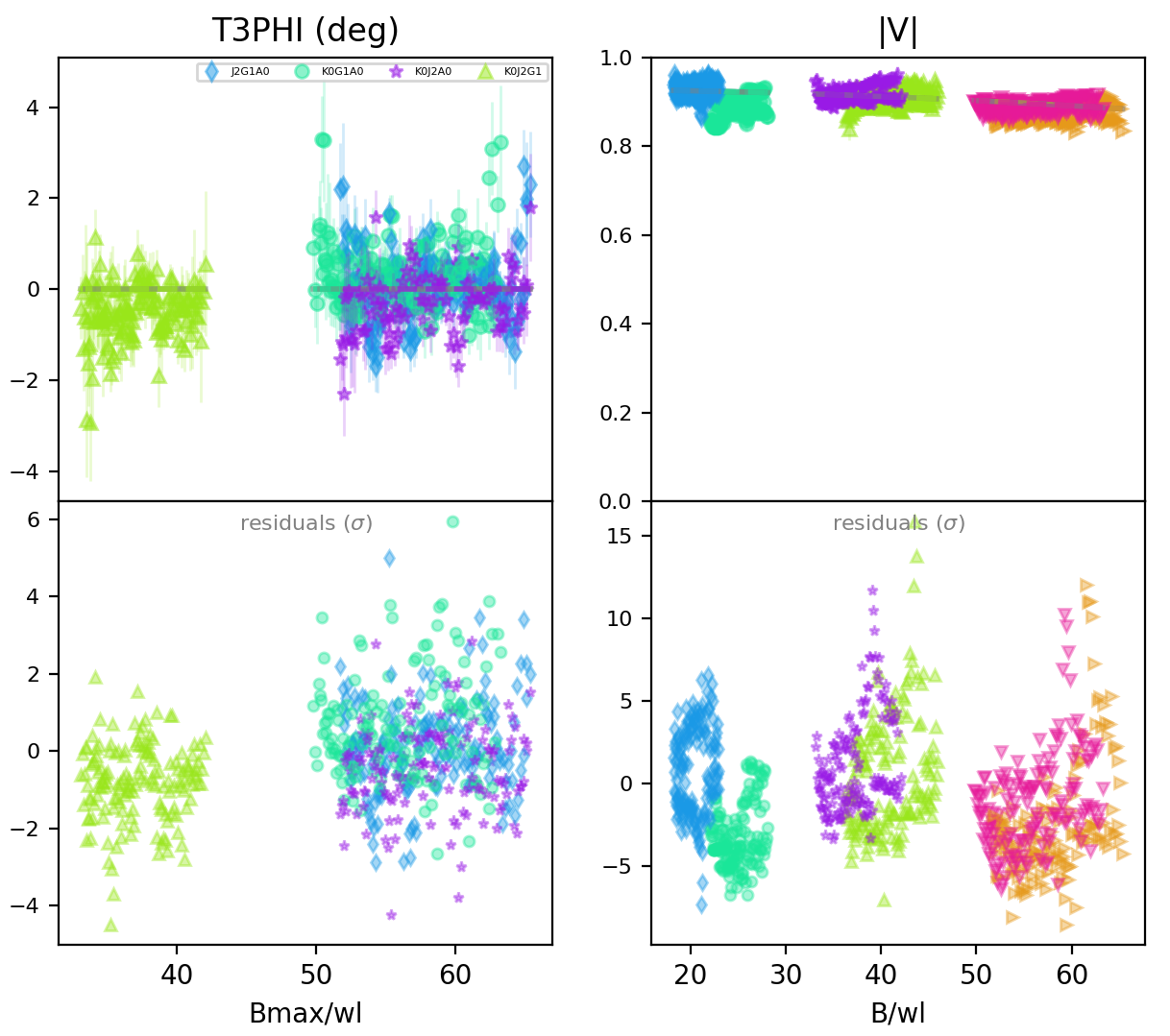}
  \caption{|V|-T3PHI data (top panels) for WR 14 fit with a central point source + fully resolved component, along with corresponding residuals (bottom panels).}
  \label{wr14}
\end{figure}

\begin{figure}[H]
\centering
  \includegraphics[height=60mm]{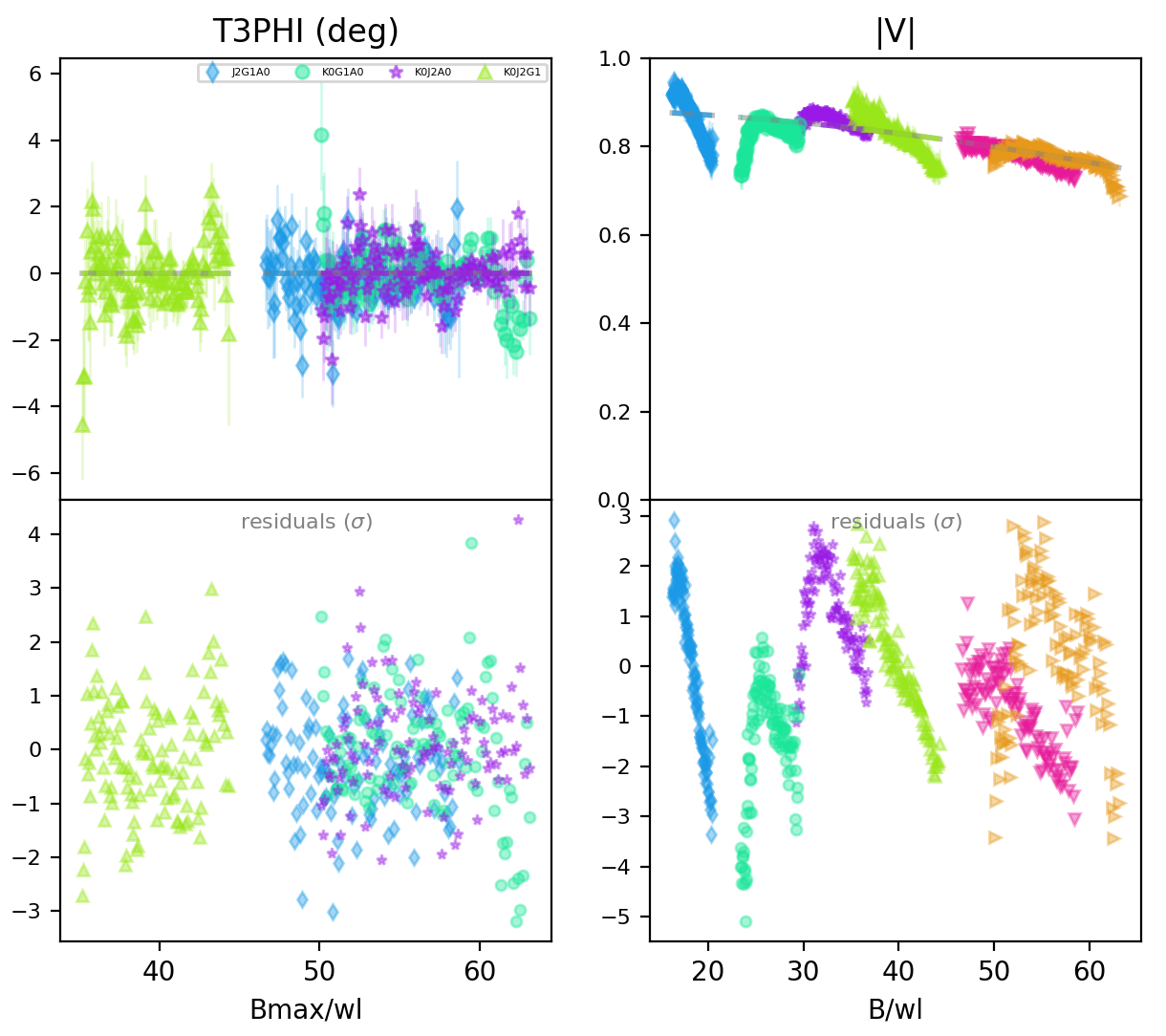}
  \caption{|V|-T3PHI data (top panels) for WR 15 fit with a central point source + fully resolved component, along with corresponding residuals (bottom panels).}
  \label{wr15}
\end{figure}

\begin{figure}[H]
\centering
  \includegraphics[height=60mm]{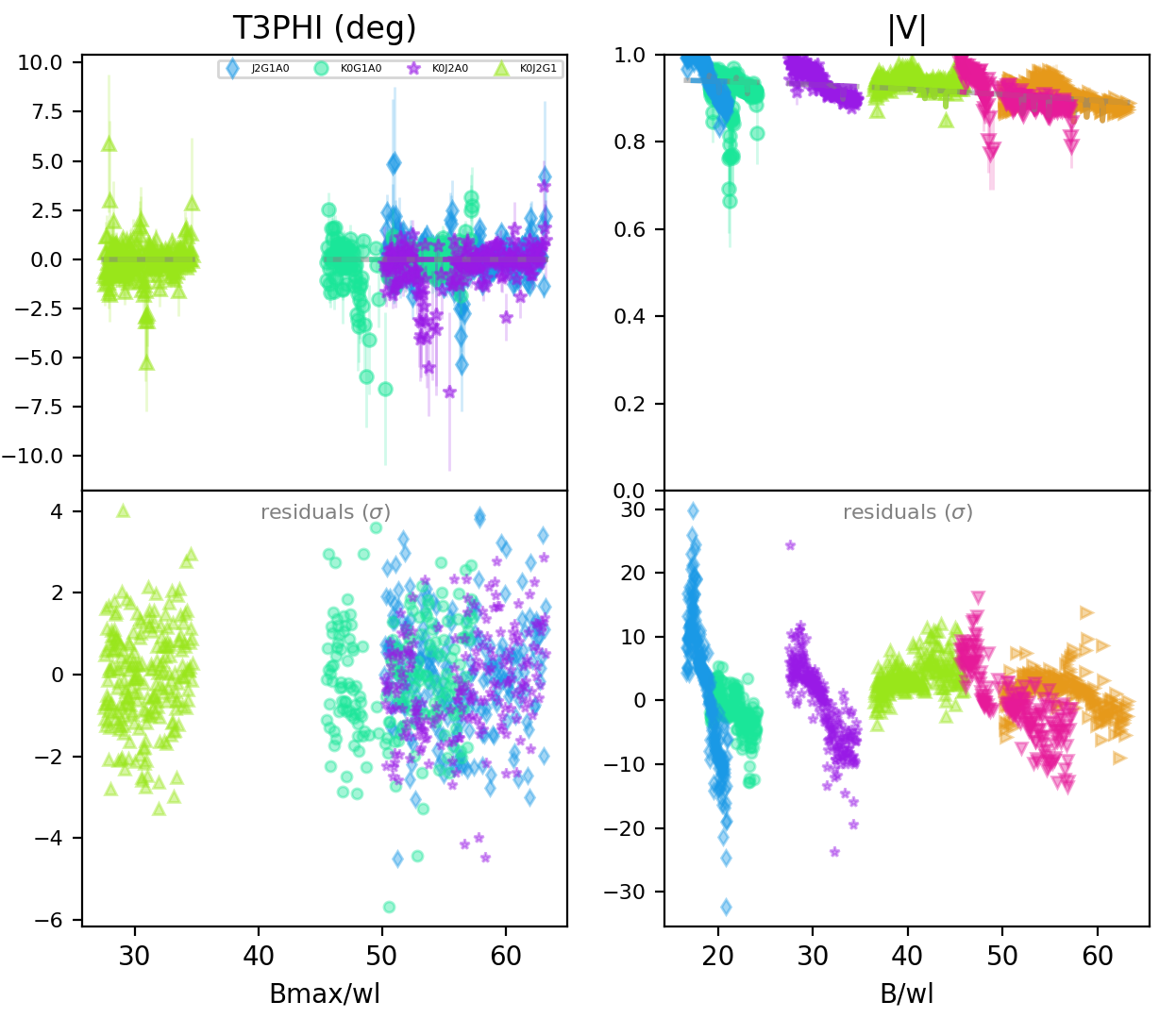}
  \caption{|V|-T3PHI data (top panels) for WR 16 fit with a central point source + fully resolved component, along with corresponding residuals (bottom panels).}
  \label{wr16_1}
\end{figure}

\begin{figure}[H]
\centering
  \includegraphics[height=60mm]{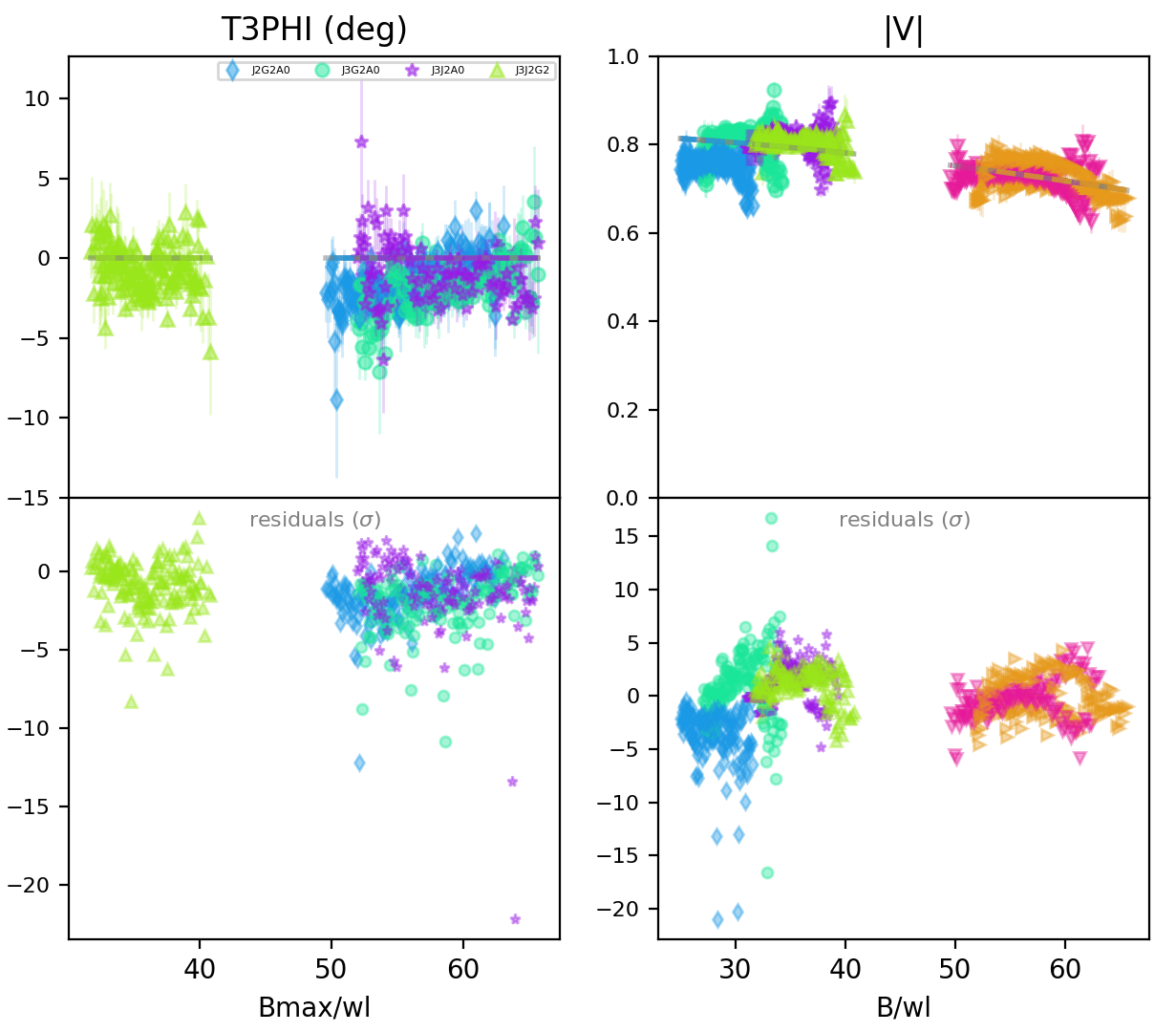}
  \caption{|V|-T3PHI data (top panels) for WR 21 fit with a central point source + fully resolved component, along with corresponding residuals (bottom panels).}
  \label{wr21}
\end{figure}

\begin{figure}[H]
\centering
  \includegraphics[height=60mm]{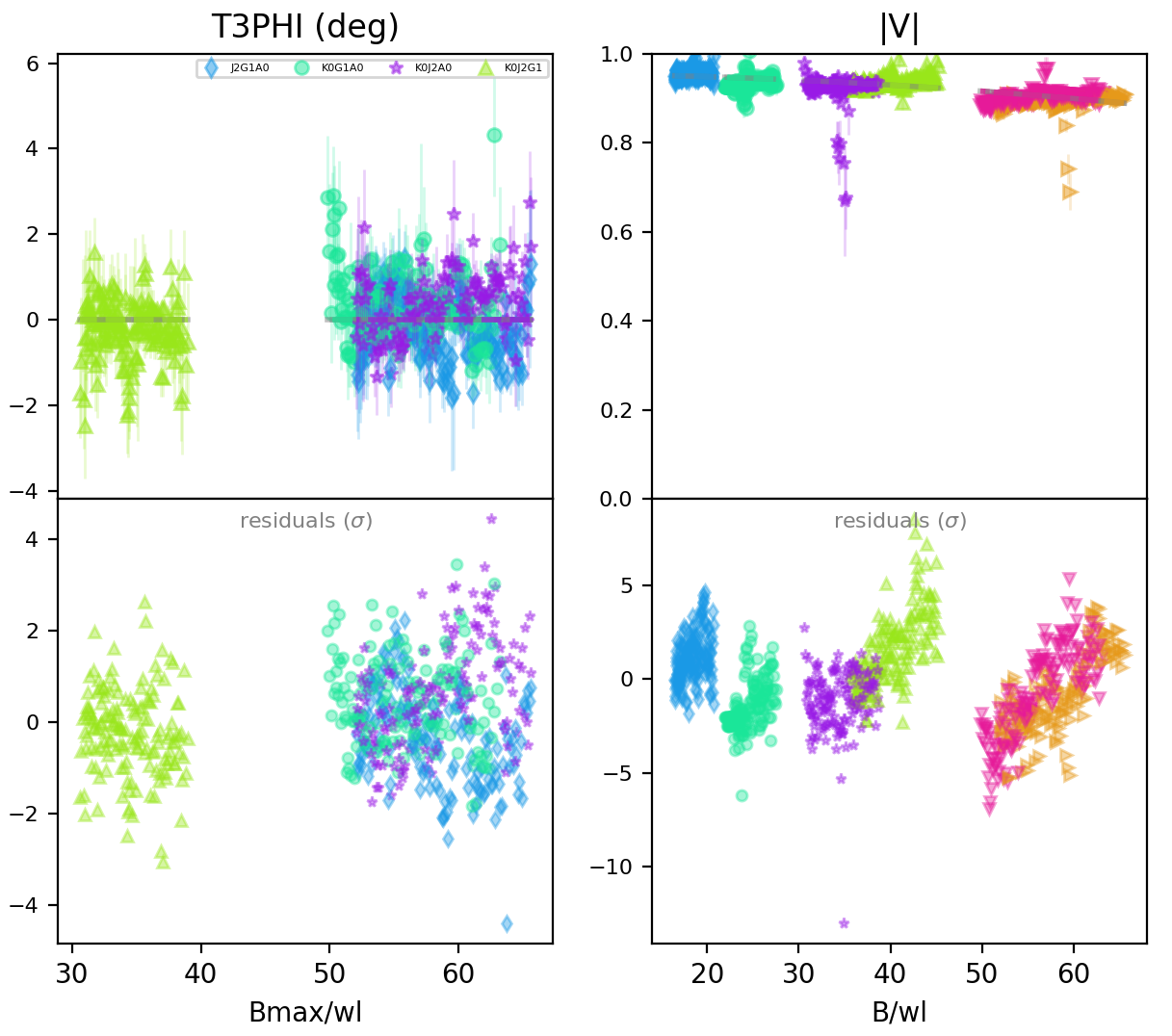}
  \caption{|V|-T3PHI data (top panels) for WR 23 fit with a central point source + fully resolved component, along with corresponding residuals (bottom panels).}
  \label{wr23}
\end{figure}

\begin{figure}[H]
\centering
  \includegraphics[height=60mm]{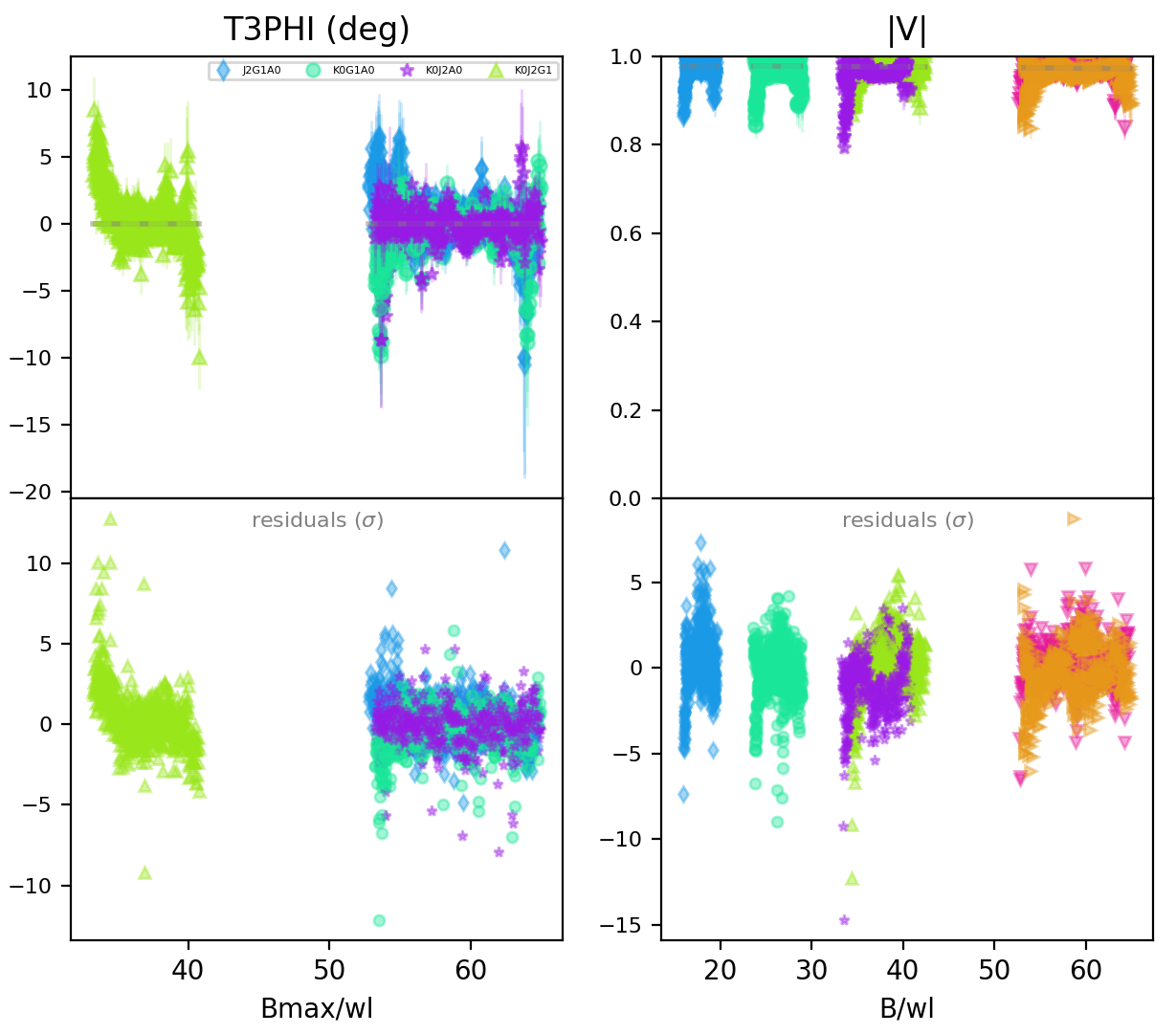}
  \caption{|V|-T3PHI data (top panels) for WR 24 fit with a central point source + fully resolved component, along with corresponding residuals (bottom panels).}
  \label{wr24}
\end{figure}

\begin{figure}[H]
\centering
  \includegraphics[height=60mm]{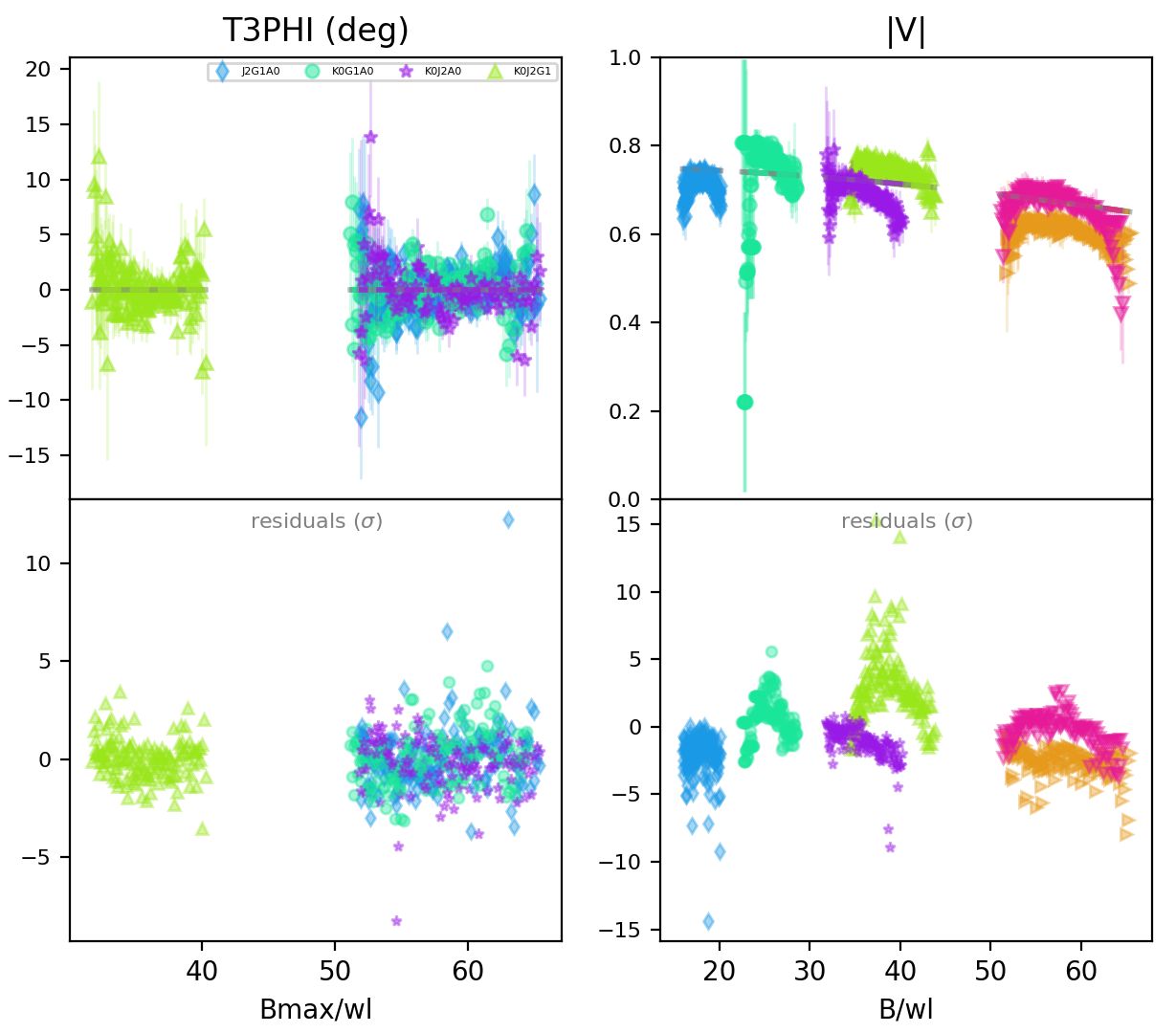}
  \caption{|V|-T3PHI data (top panels) for WR 31 fit with a central point source + fully resolved component, along with corresponding residuals (bottom panels).}
  \label{wr31}
\end{figure}

\begin{figure}[H]
\centering
  \includegraphics[height=60mm]{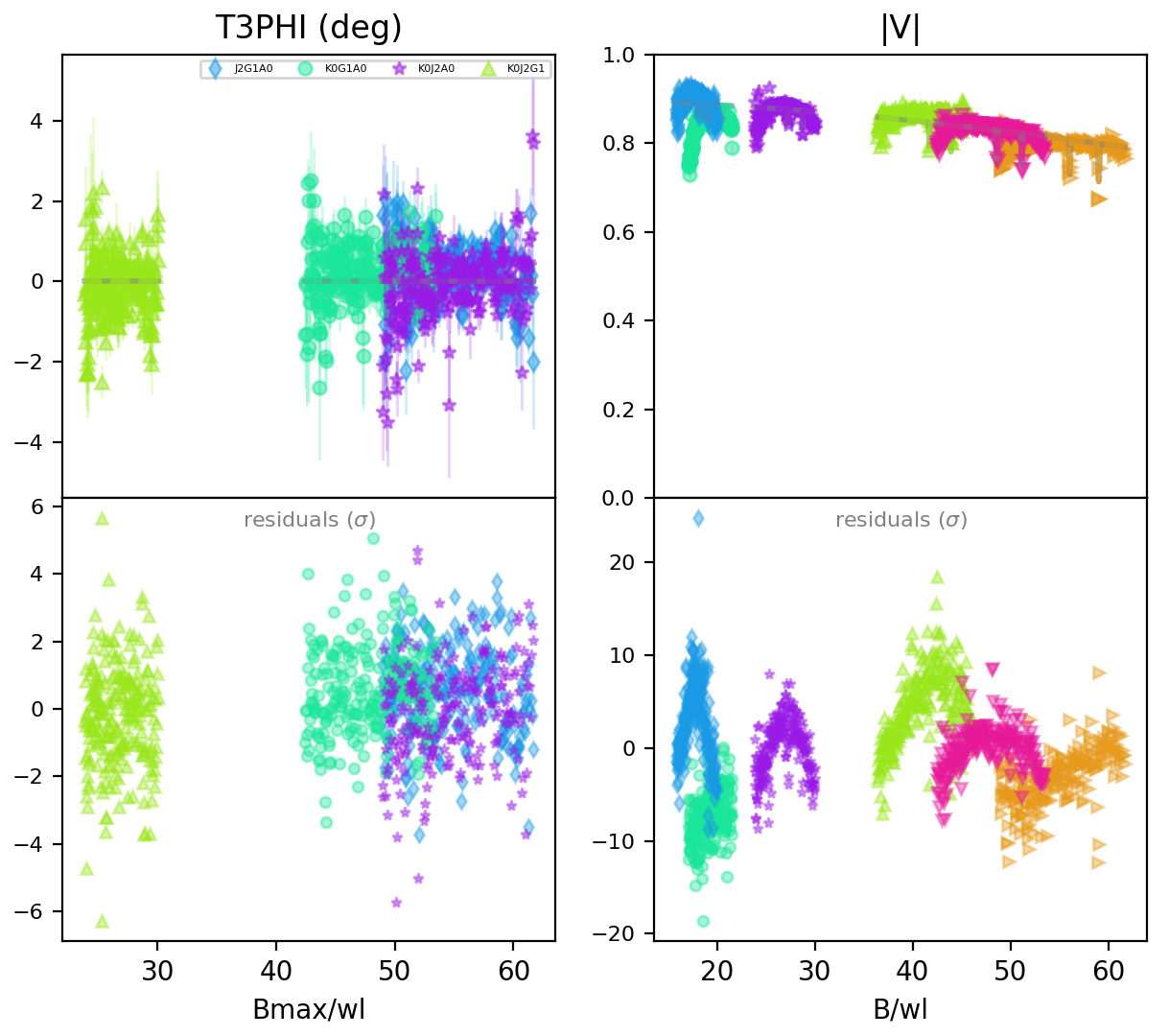}
  \caption{|V|-T3PHI data (top panels) for WR 31a fit with a central point source + fully resolved component, along with corresponding residuals (bottom panels).}
  \label{wr31a_1}
\end{figure}

\begin{figure}[H]
\centering
  \includegraphics[height=60mm]{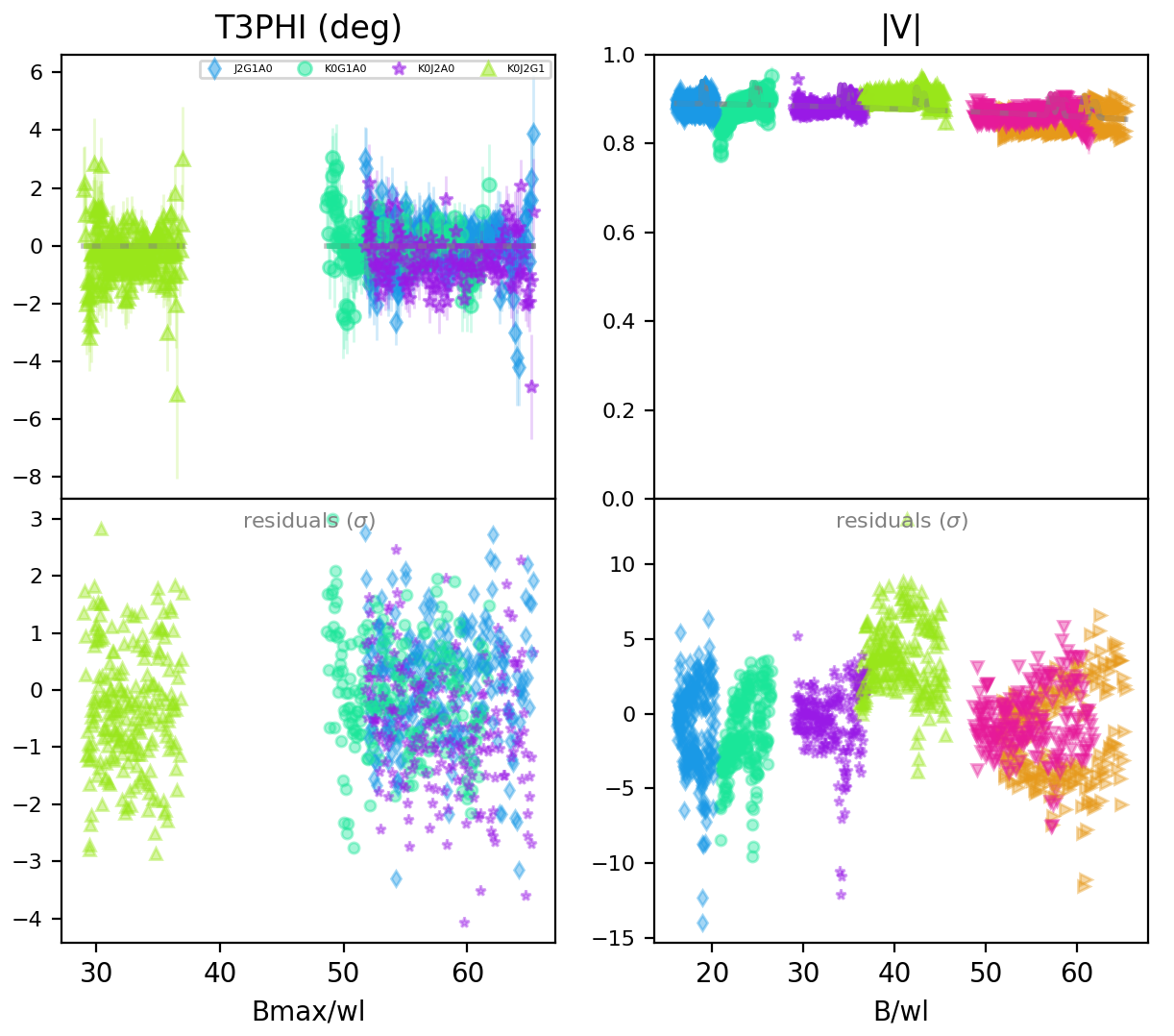}
  \caption{|V|-T3PHI data (top panels) for WR 42 fit with a central point source + fully resolved component, along with corresponding residuals (bottom panels).}
  \label{wr42_1}
\end{figure}

\begin{figure}[H]
\centering
  \includegraphics[height=60mm]{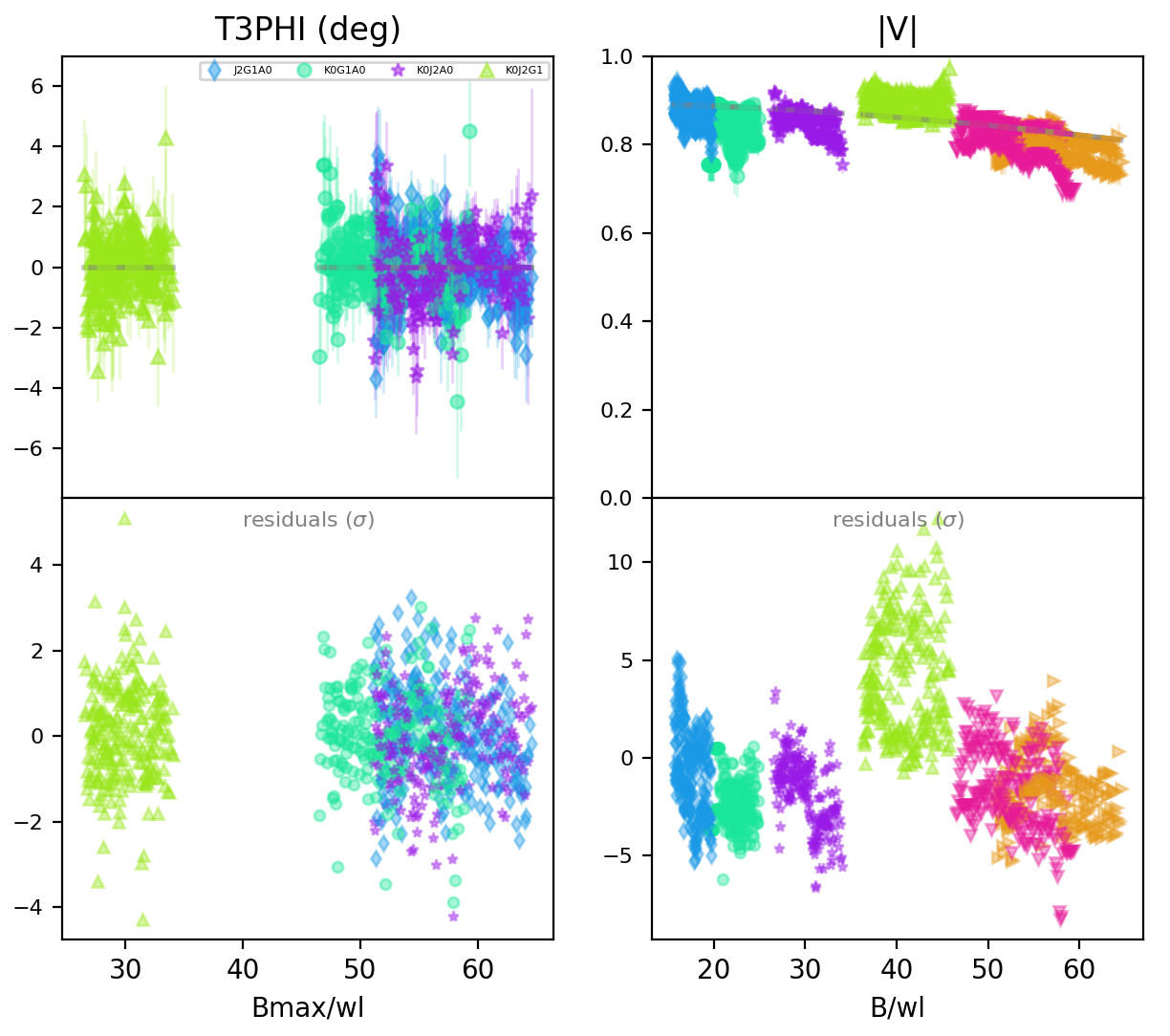}
  \caption{|V|-T3PHI data (top panels) for WR 47 fit with a central point source + fully resolved component, along with corresponding residuals (bottom panels).}
  \label{wr47_1}
\end{figure}

\begin{figure}[H]
\centering
  \includegraphics[height=60mm]{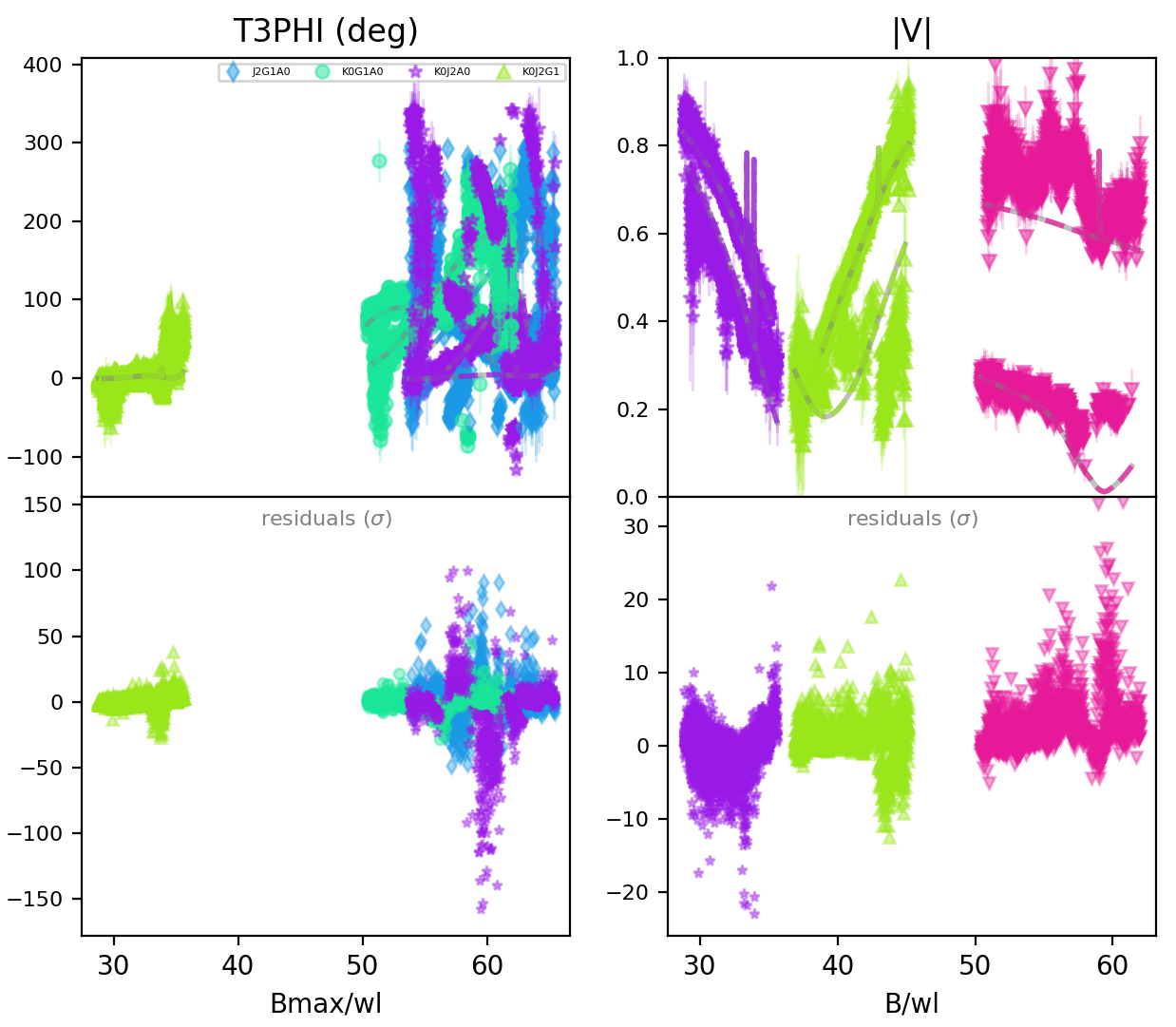}
  \caption{|V|-T3PHI data (top panels) for WR 48 fit with two point sources to model the binary, along with a fully resolved component, and the corresponding residuals (bottom panels).}
  \label{wr48_1}
\end{figure}

\begin{figure}[H]
\centering
  \includegraphics[height=60mm]{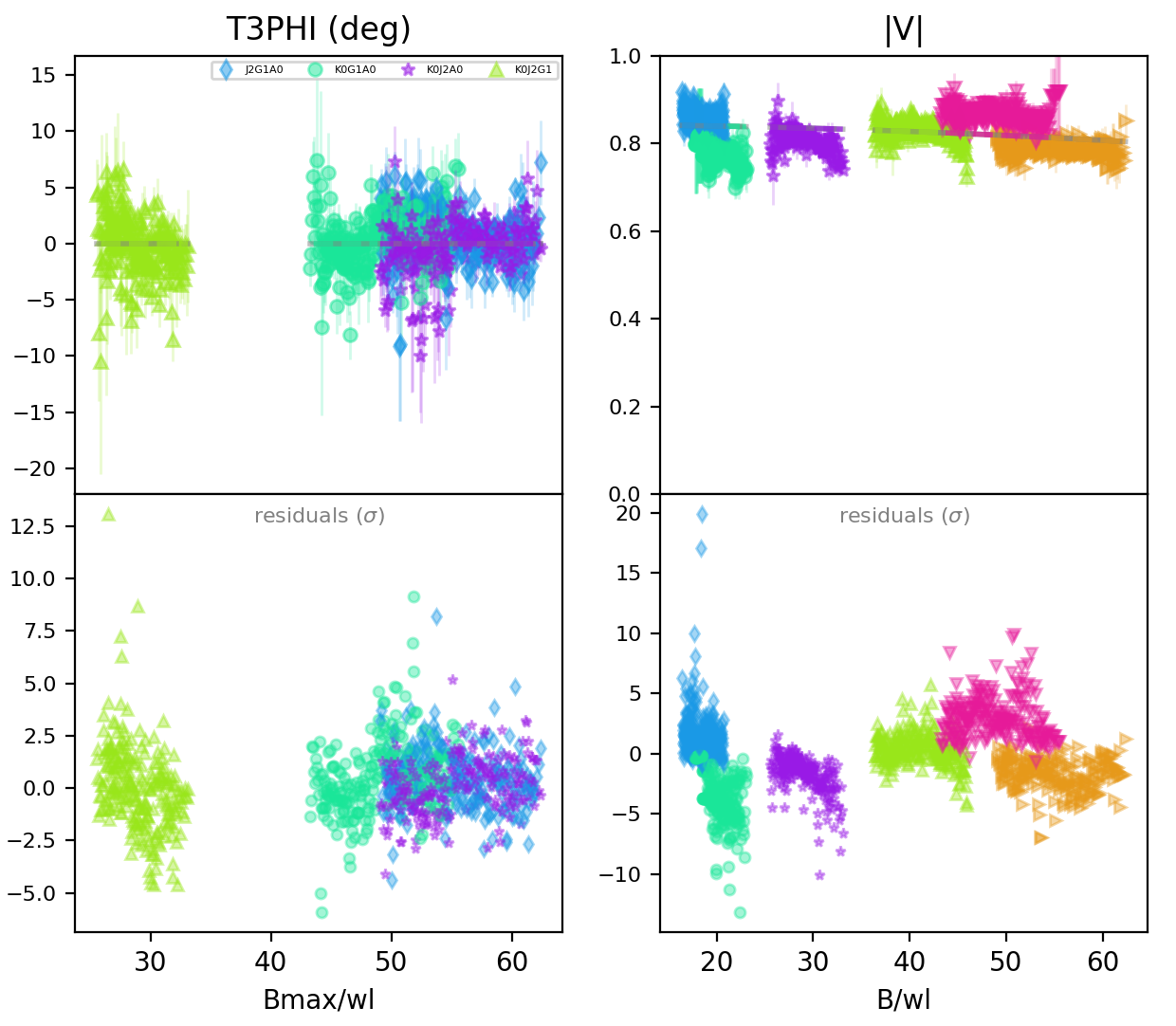}
  \caption{|V|-T3PHI data (top panels) for WR 52 fit with a central point source + fully resolved component, along with corresponding residuals (bottom panels).}
  \label{wr52}
\end{figure}

\begin{figure}[H]
\centering
  \includegraphics[height=60mm]{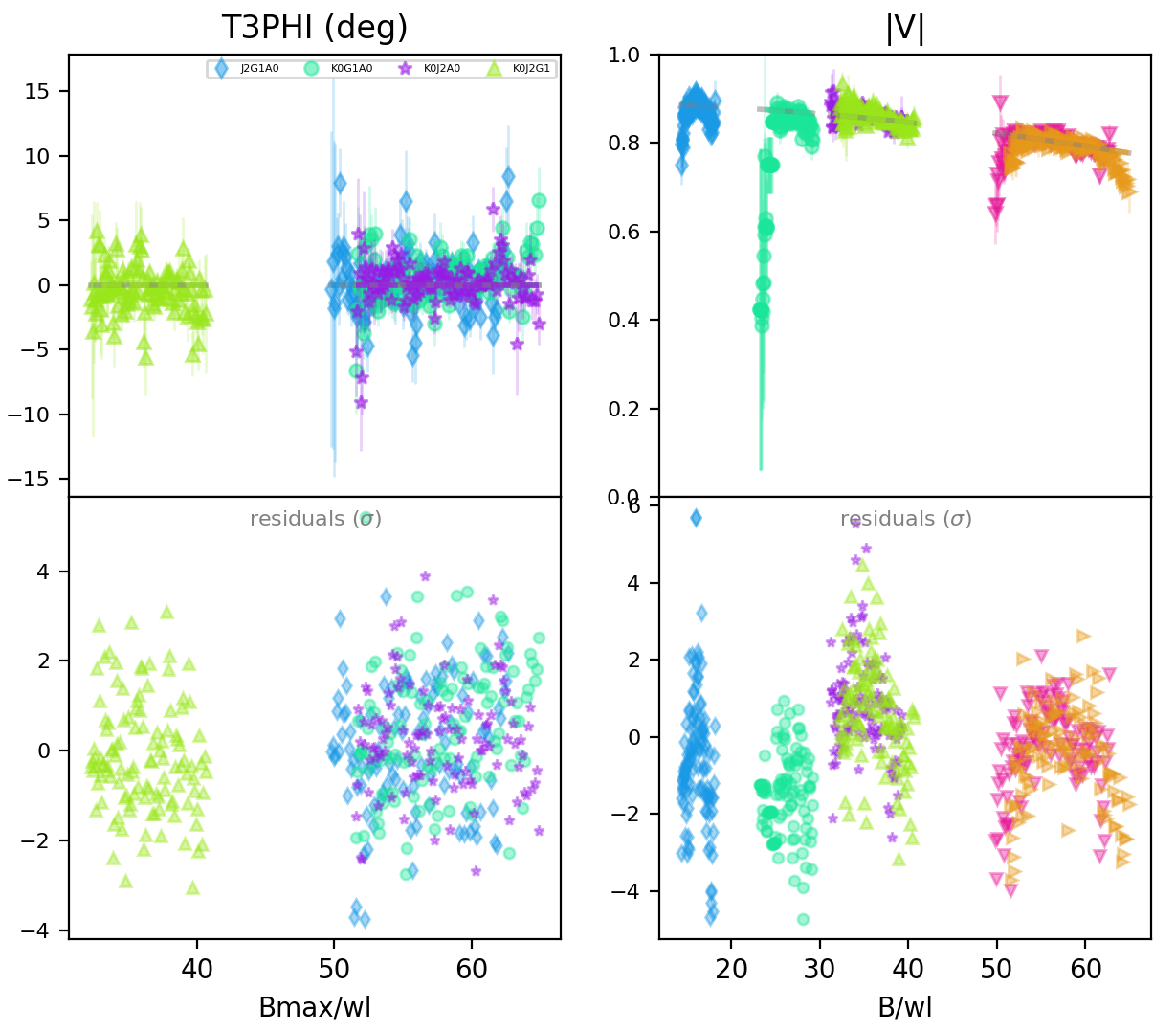}
  \caption{|V|-T3PHI data (top panels) for WR 55 fit with a central point source + fully resolved component, along with corresponding residuals (bottom panels).}
  \label{wr55}
\end{figure}

\begin{figure}[H]
\centering
  \includegraphics[height=60mm]{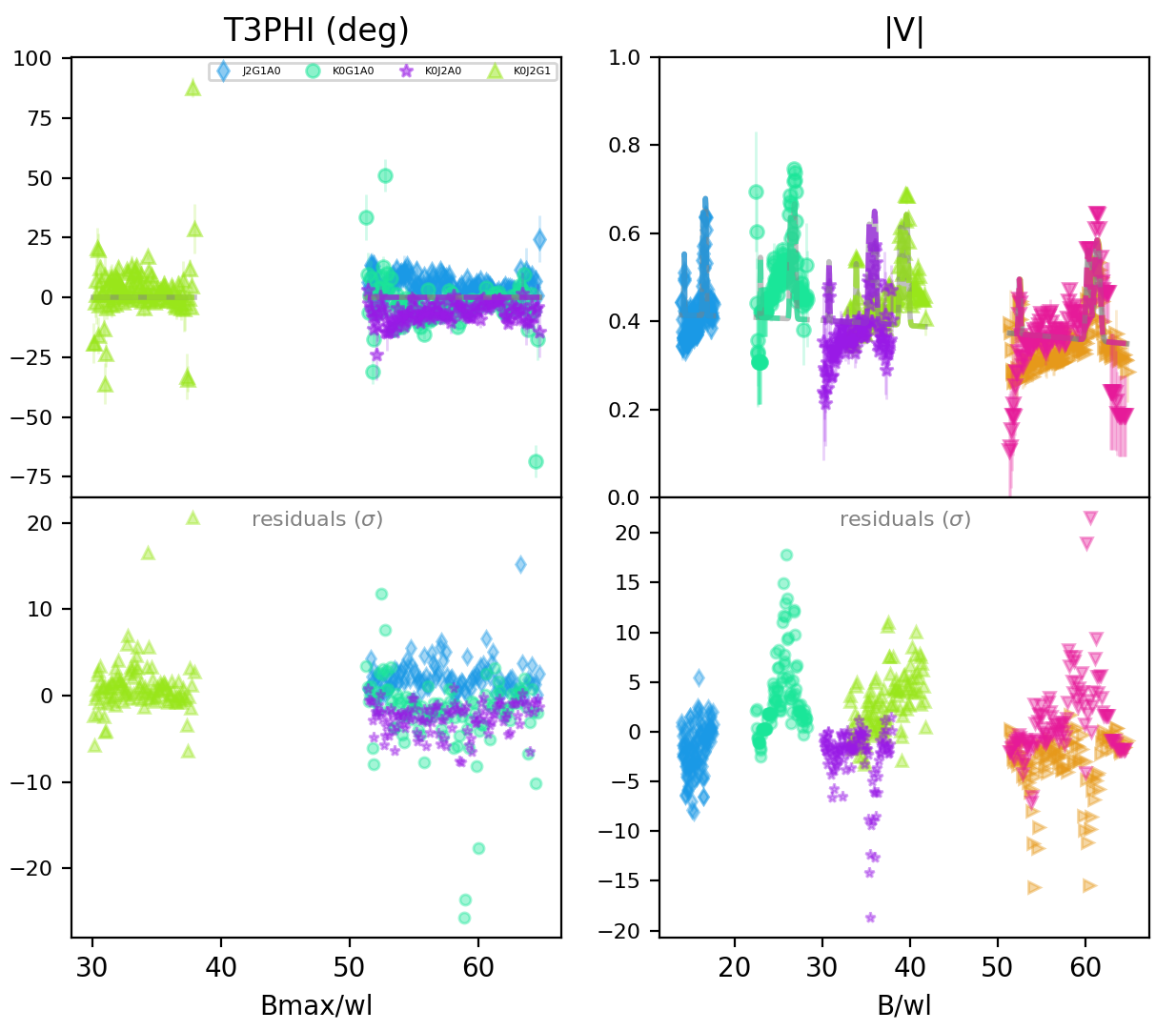}
  \caption{|V|-T3PHI data (top panels) for WR 57 fit with a central point source + fully resolved component, along with corresponding residuals (bottom panels).}
  \label{wr57_1}
\end{figure}

\begin{figure}[H]
\centering
  \includegraphics[height=60mm]{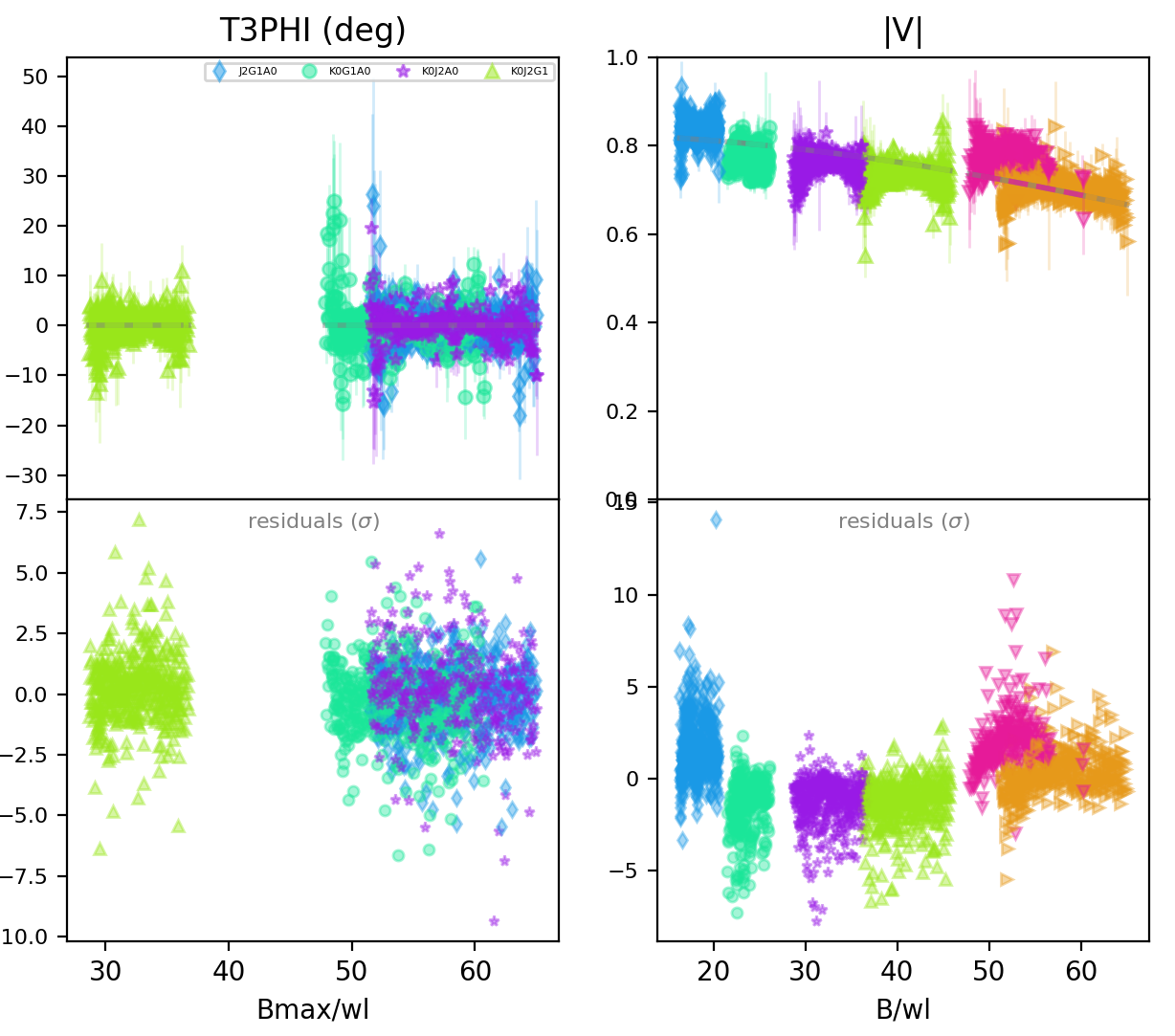}
  \caption{|V|-T3PHI data (top panels) for WR 66 fit with a central point source + fully resolved component, along with corresponding residuals (bottom panels).}
  \label{wr66}
\end{figure}

\begin{figure}[H]
\centering
  \includegraphics[height=60mm]{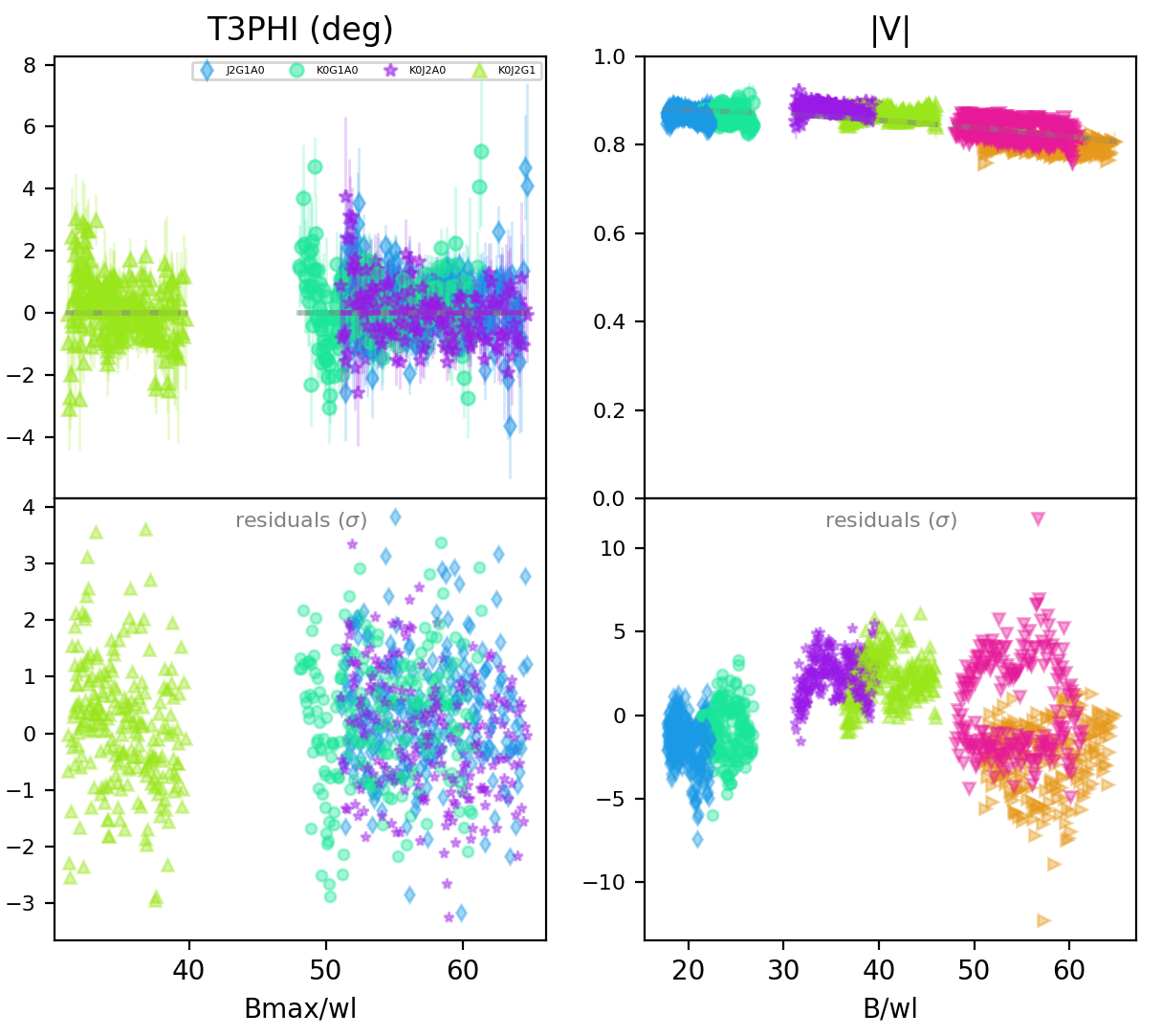}
  \caption{|V|-T3PHI data (top panels) for WR 75 fit with a central point source + fully resolved component, along with corresponding residuals (bottom panels).}
  \label{wr75}
\end{figure}

\begin{figure}[H]
\centering
  \includegraphics[height=60mm]{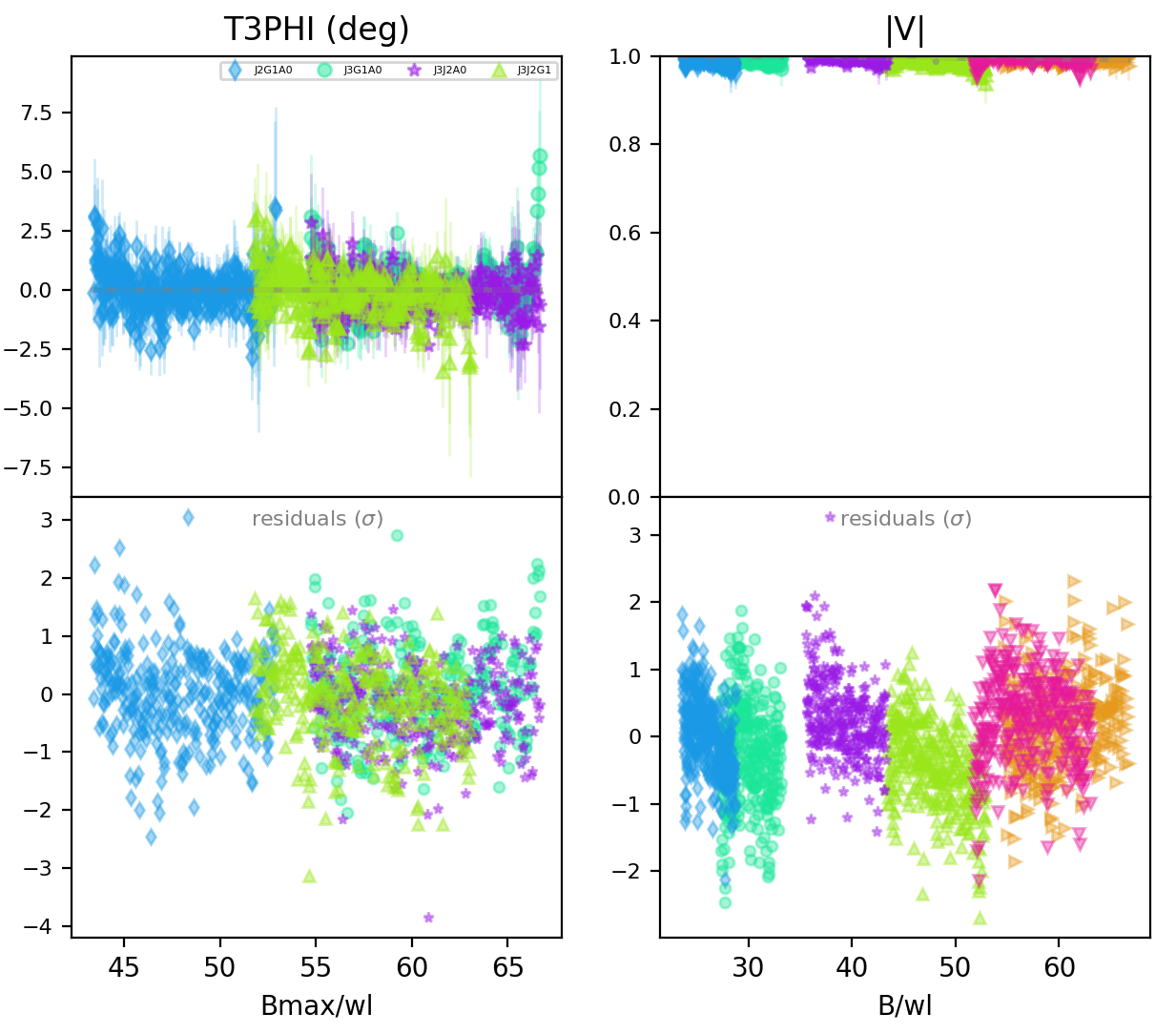}
  \caption{|V|-T3PHI data (top panels) for WR 78 fit with only a point source, along with corresponding residuals (bottom panels).}
  \label{wr78}
\end{figure}

\begin{figure}[H]
\centering
  \includegraphics[height=60mm]{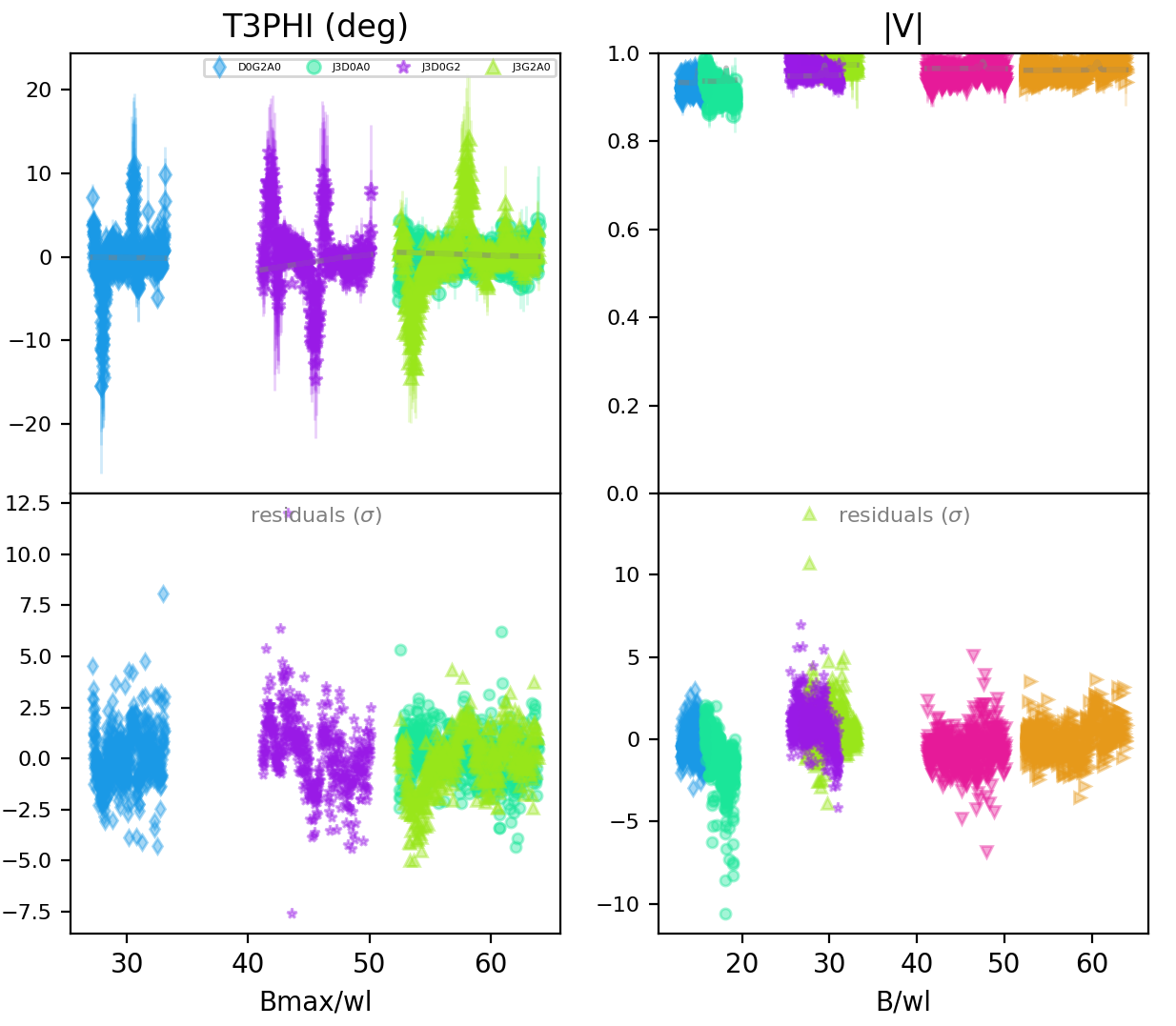}
  \caption{|V|-T3PHI data (top panels) for WR 79 fit with a point source + fully resolved component, along with corresponding residuals (bottom panels).}
  \label{wr79_1}
\end{figure}

\begin{figure}[H]
\centering
  \includegraphics[height=60mm]{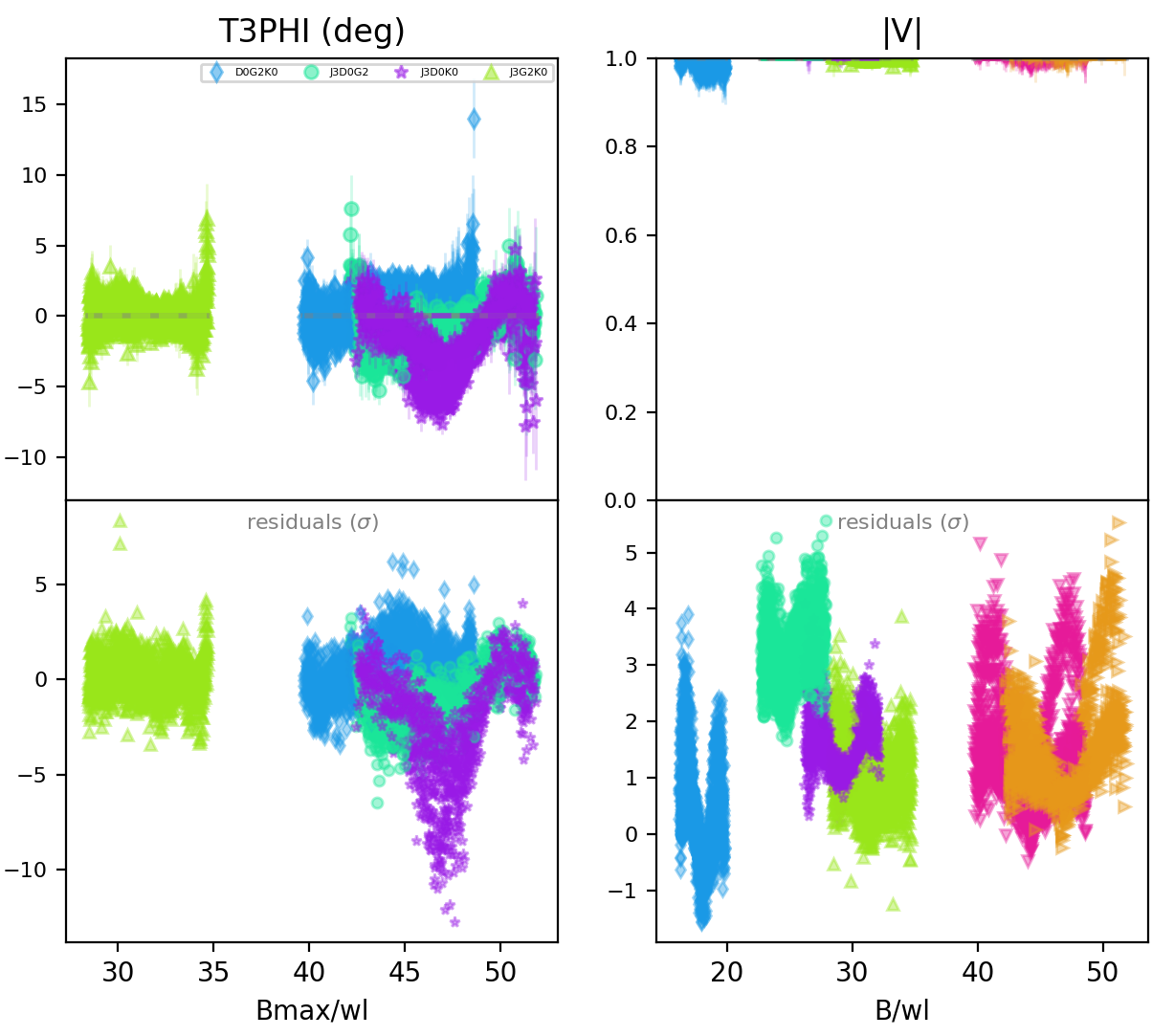}
  \caption{|V|-T3PHI data (top panels) for WR 79a fit with an unresolved point source, along with corresponding residuals (bottom panels).}
  \label{wr79a}
\end{figure}

\begin{figure}[H]
\centering
  \includegraphics[height=60mm]{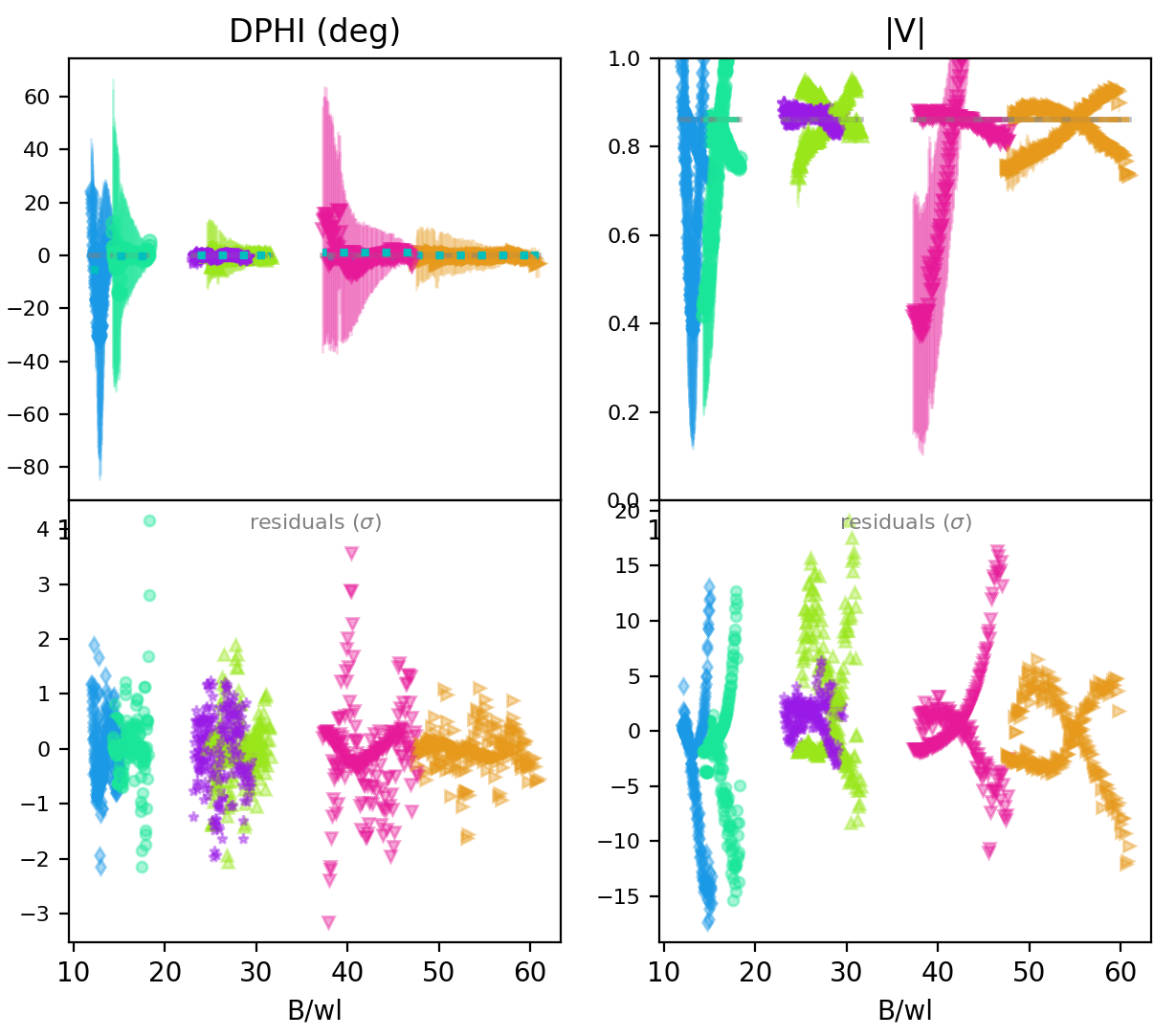}
  \caption{|V|-T3PHI data (top panels) for WR 79b fit with an unresolved point source + fully resolved component, along with corresponding residuals (bottom panels).}
  \label{wr79b}
\end{figure}

\begin{figure}[H]
\centering
  \includegraphics[height=60mm]{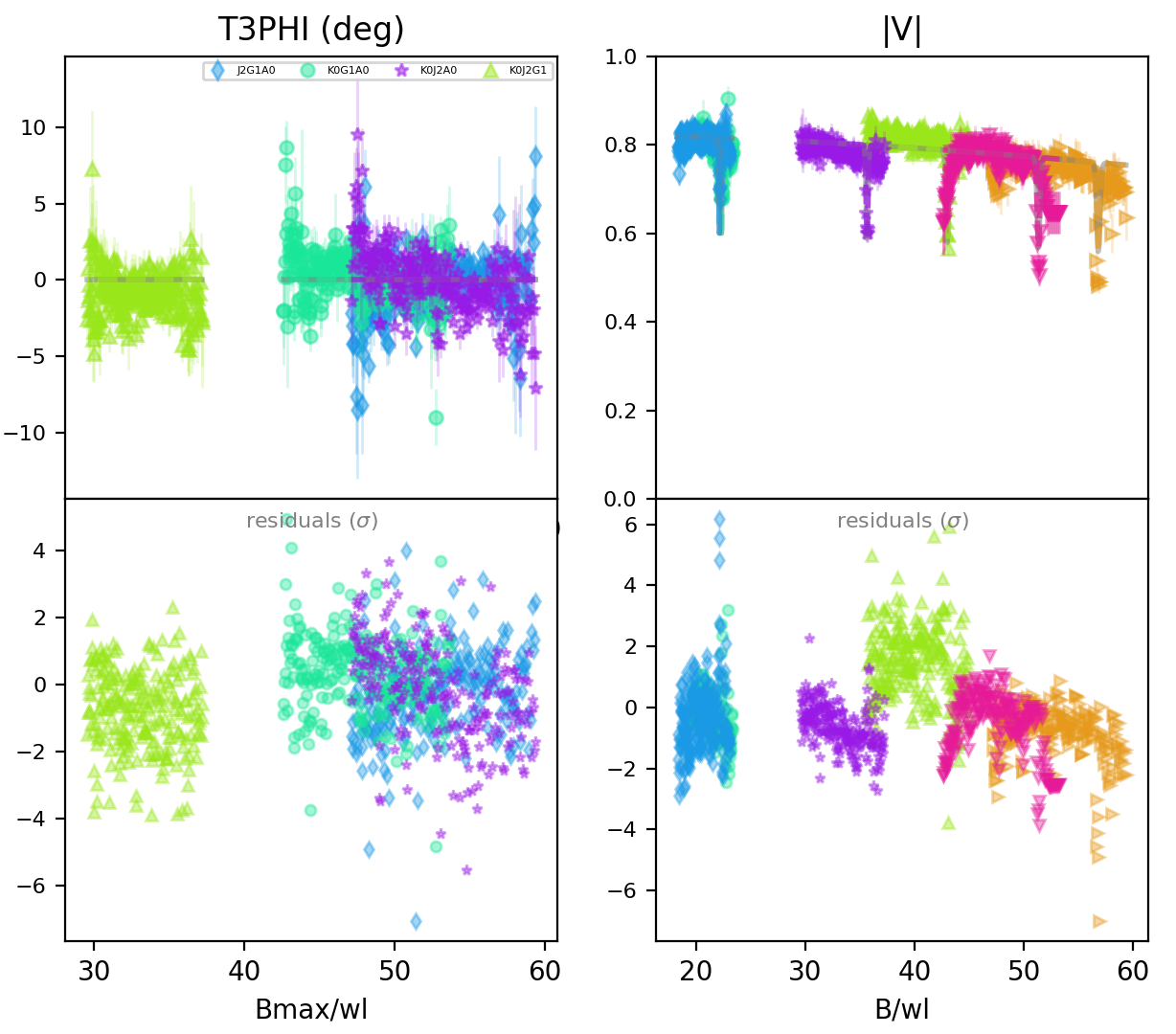}
  \caption{|V|-T3PHI data (top panels) for WR 81 fit with a central point source + fully resolved component, along with corresponding residuals (bottom panels).}
  \label{wr81_1}
\end{figure}

\begin{figure}[H]
\centering
  \includegraphics[height=60mm]{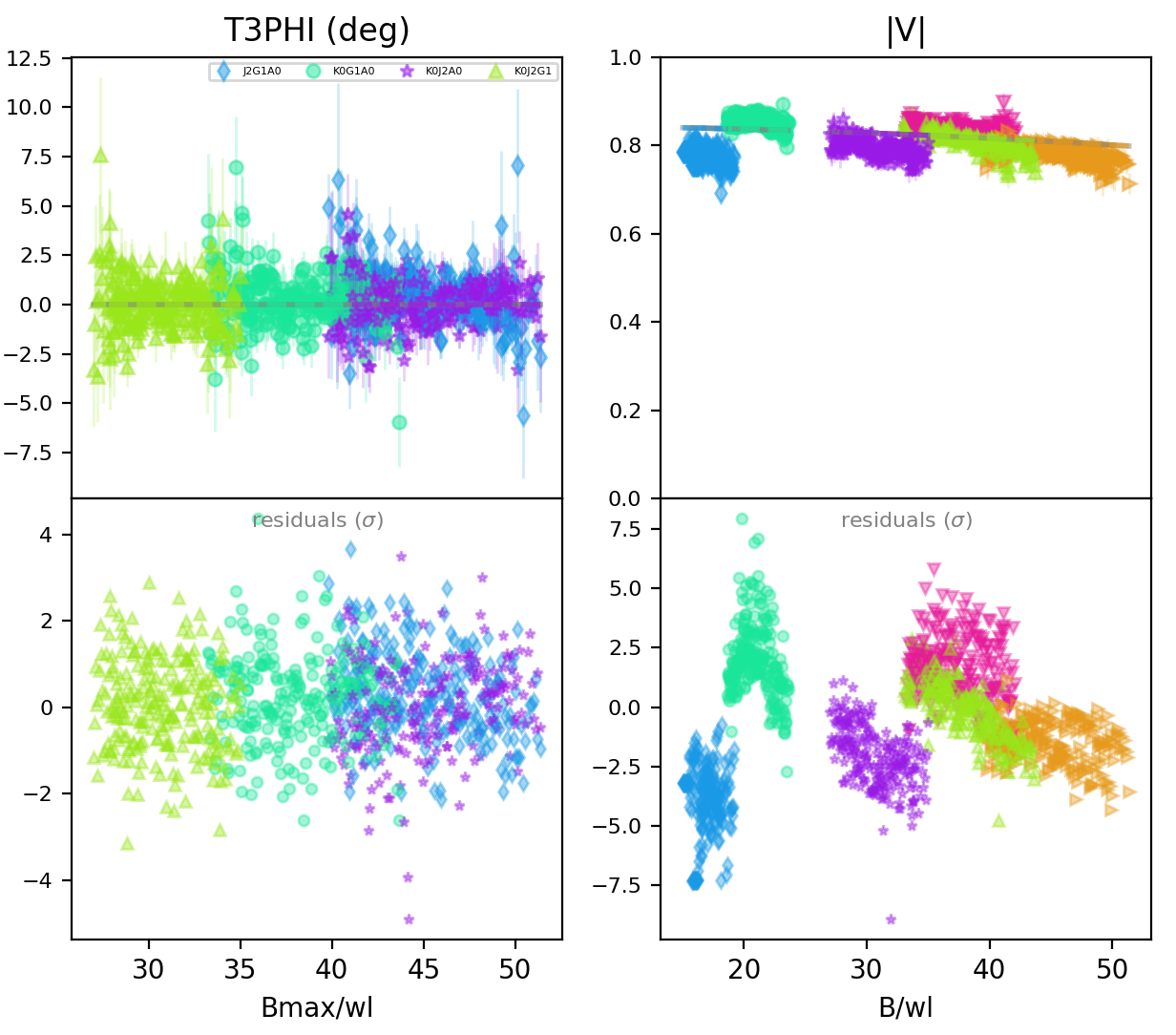}
  \caption{|V|-T3PHI data (top panels) for WR 85 fit with a central point source + fully resolved component, along with corresponding residuals (bottom panels).}
  \label{wr85}
\end{figure}

\begin{figure}[H]
\centering
  \includegraphics[height=60mm]{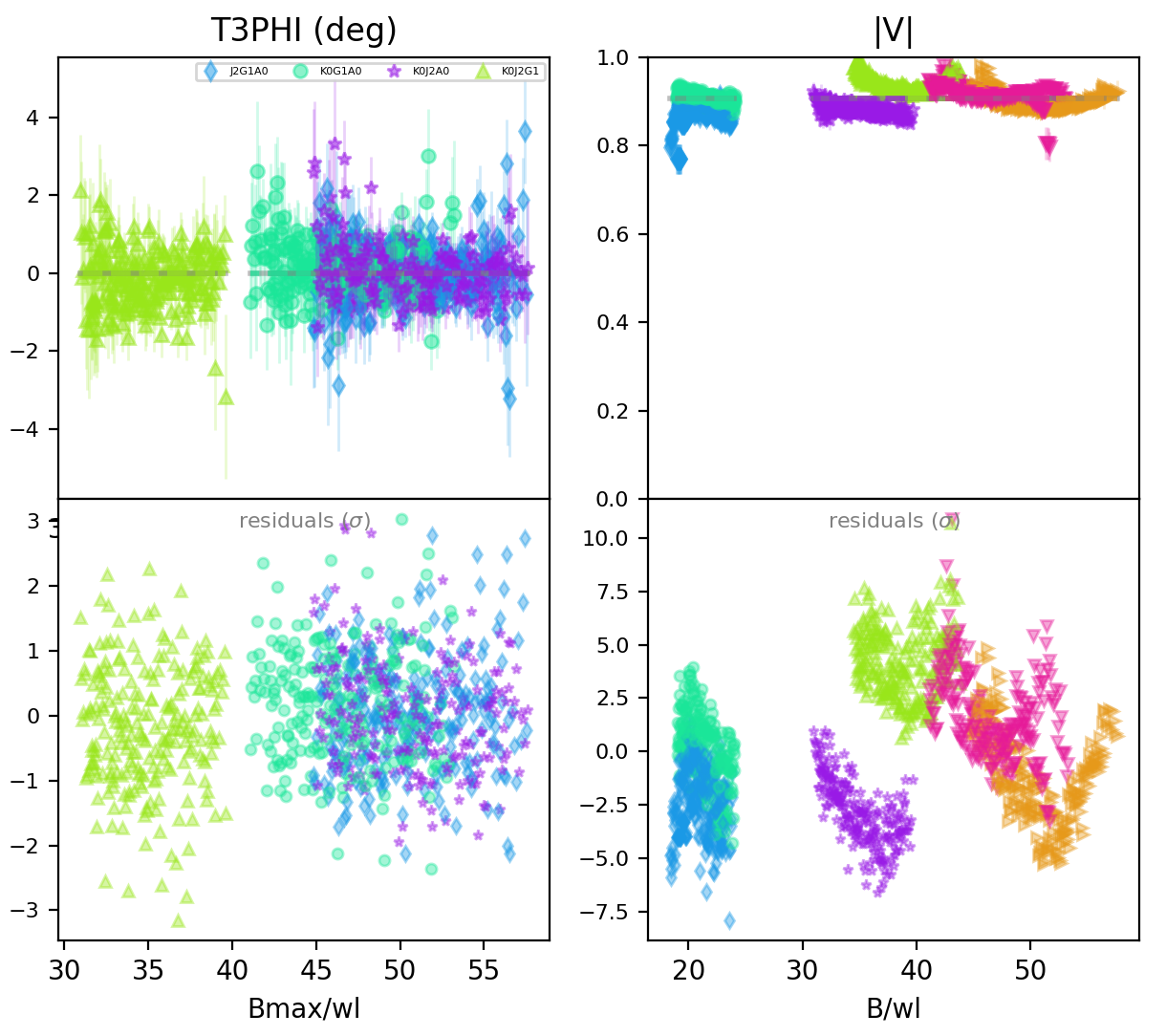}
  \caption{|V|-T3PHI data (top panels) for WR 87 fit with a central point source + fully resolved component, along with corresponding residuals (bottom panels).}
  \label{wr87}
\end{figure}

\begin{figure}[H]
\centering
  \includegraphics[height=60mm]{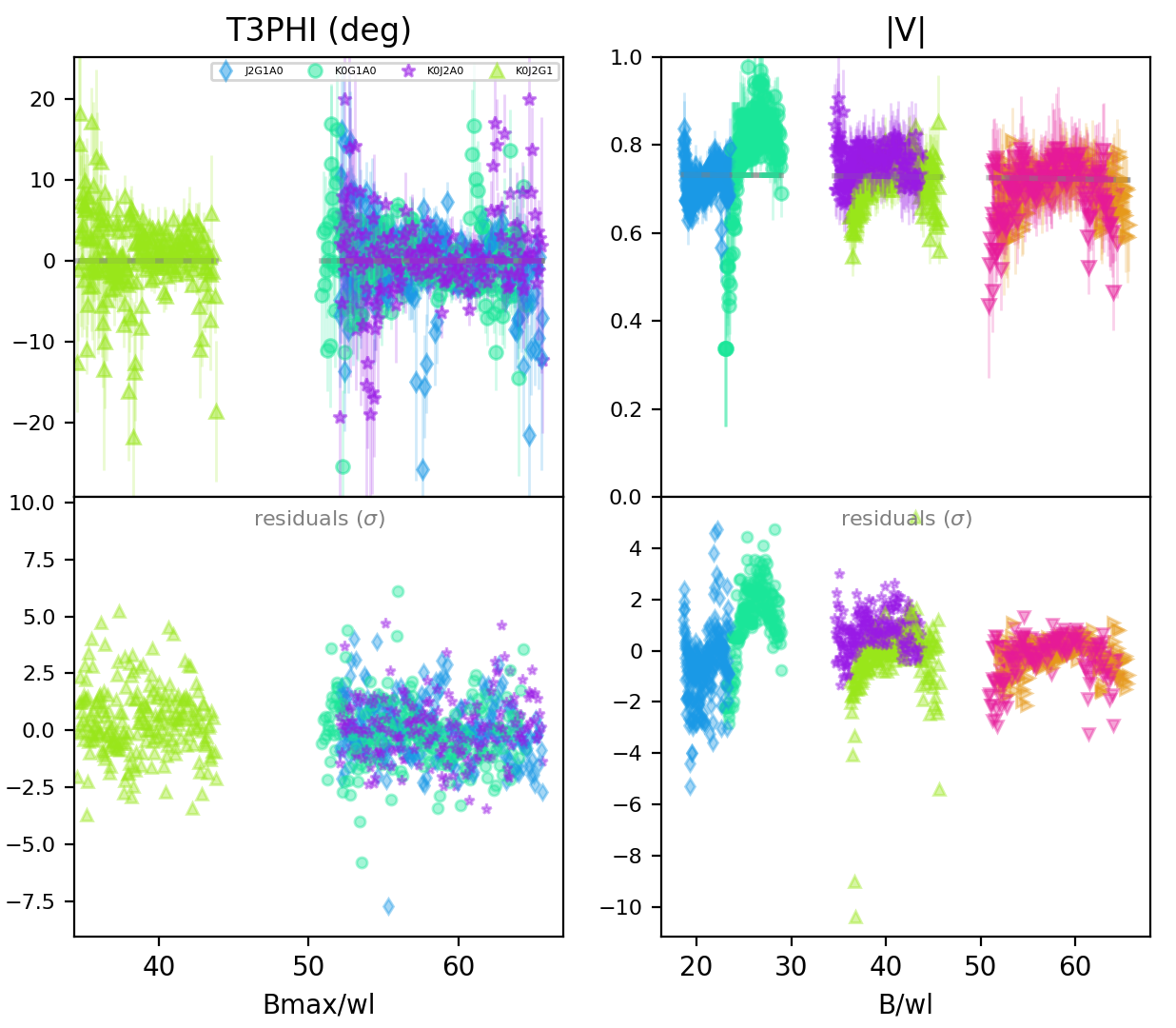}
  \caption{|V|-T3PHI data (top panels) for WR 92 fit with a central point source + fully resolved component, along with corresponding residuals (bottom panels).}
  \label{wr92_1}
\end{figure}

\begin{figure}[H]
\centering
  \includegraphics[height=60mm]{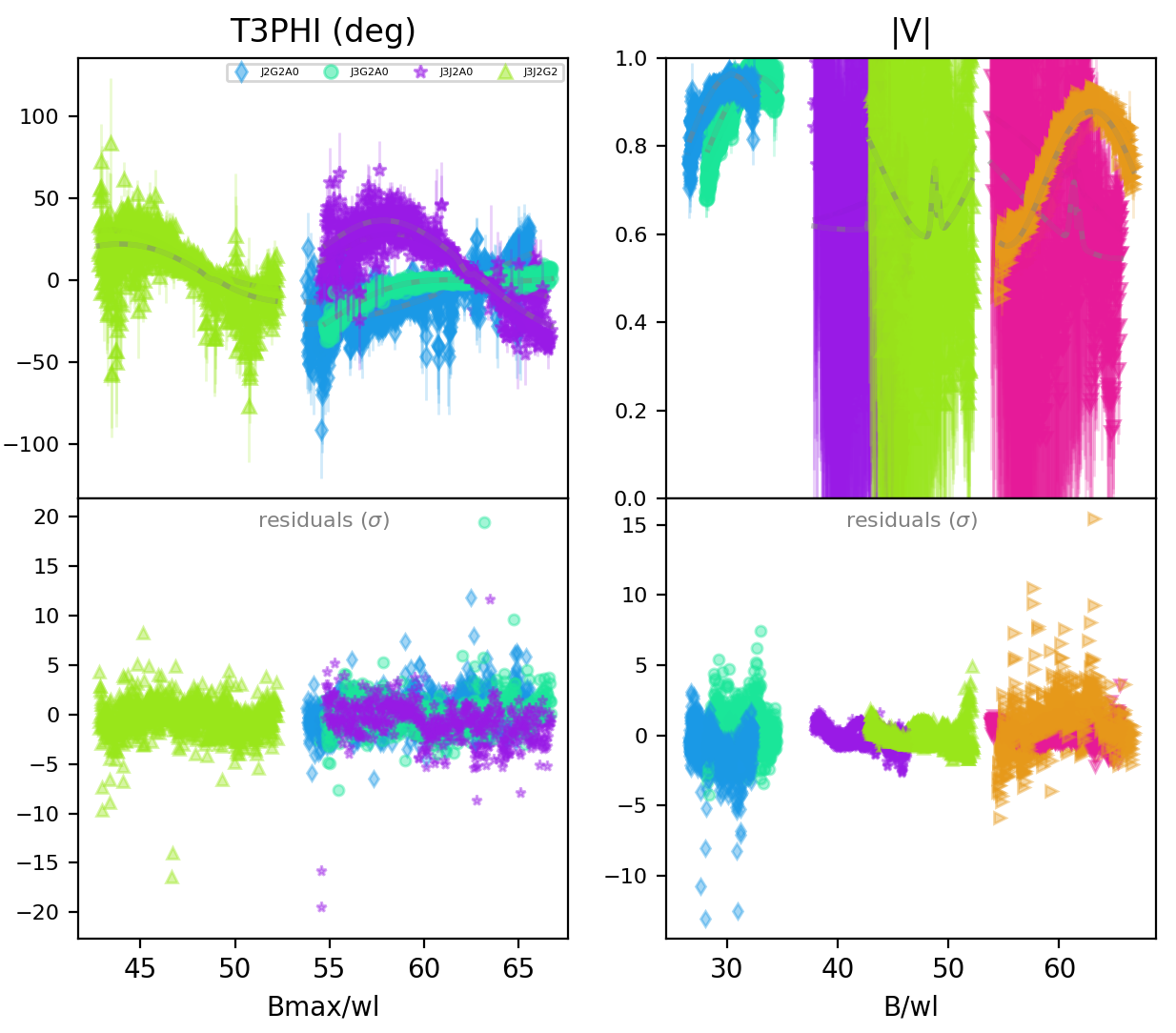}
  \caption{|V|-T3PHI data (top panels) for WR 93 fit with two point sources to model the binary, along with a fully resolved component, and the corresponding residuals (bottom panels).}
  \label{wr93_1}
\end{figure}

\begin{figure}[H]
\centering
  \includegraphics[height=60mm]{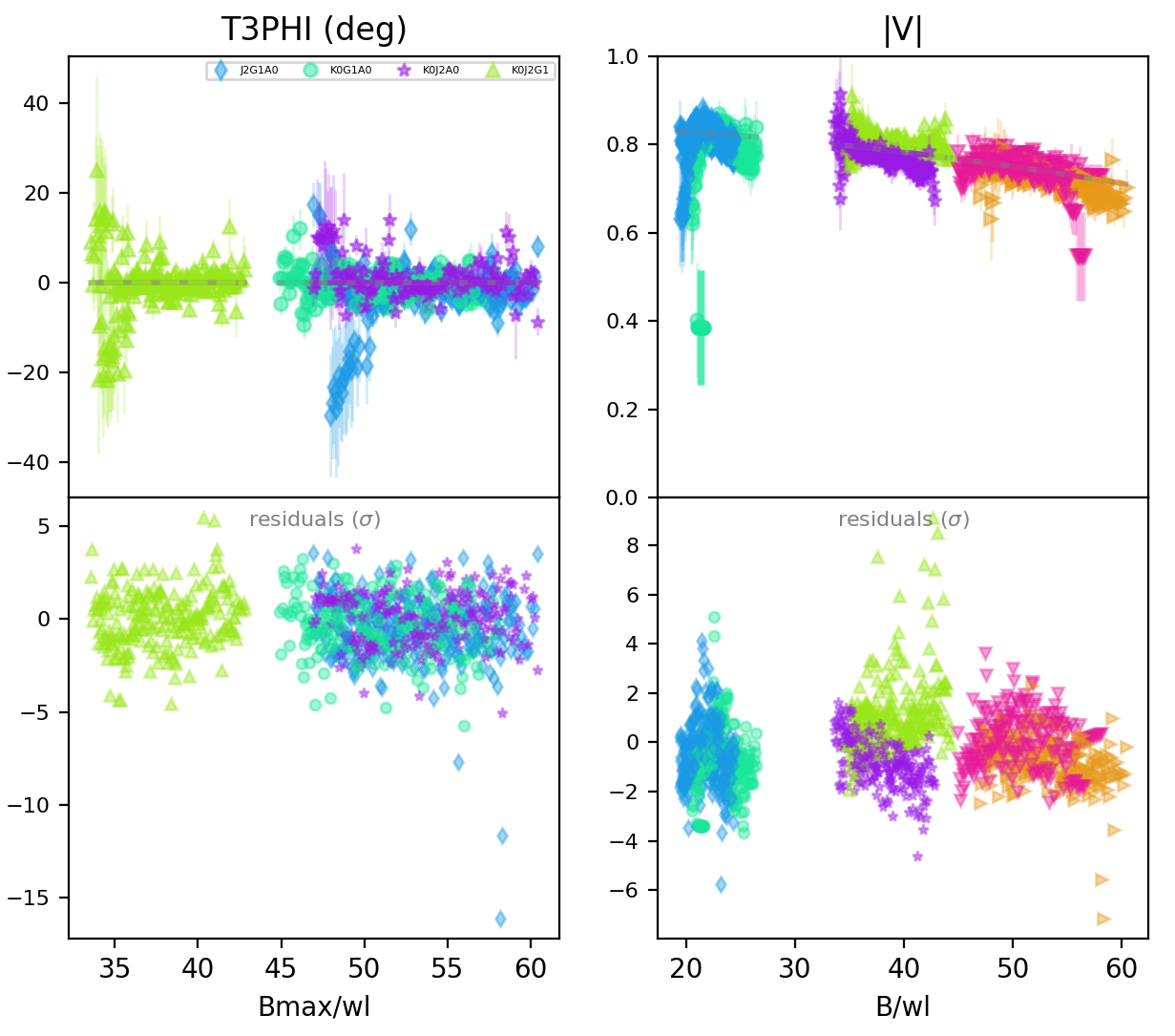}
  \caption{|V|-T3PHI data (top panels) for WR 97 fit with a central point source + fully resolved component, along with corresponding residuals (bottom panels).}
  \label{wr97}
\end{figure}

\begin{figure}[H]
\centering
  \includegraphics[height=60mm]{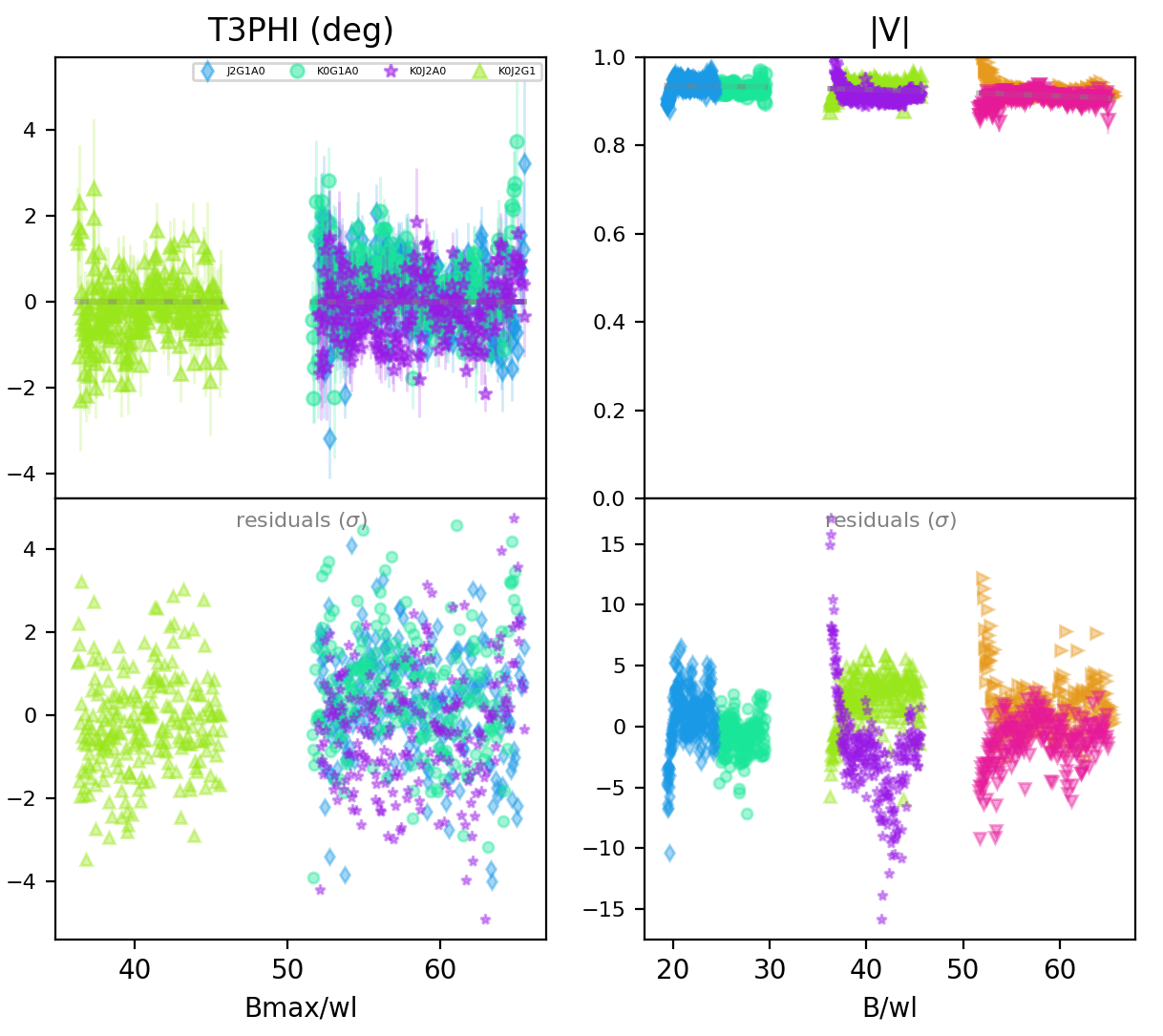}
  \caption{|V|-T3PHI data (top panels) for WR 98 fit with a central point source + fully resolved component, along with corresponding residuals (bottom panels).}
  \label{wr98_2}
\end{figure}

\begin{figure}[H]
\centering
  \includegraphics[height=60mm]{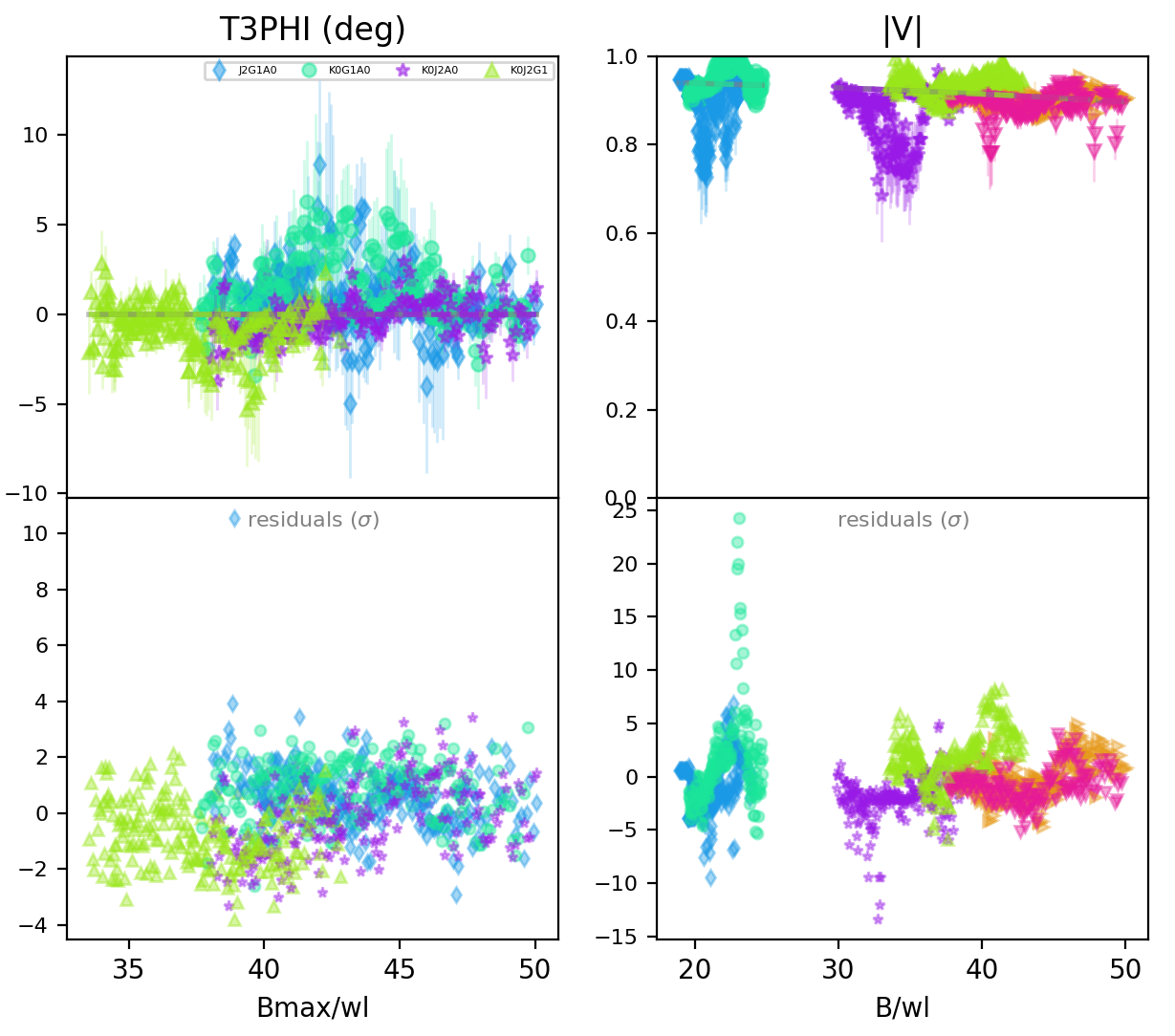}
  \caption{|V|-T3PHI data (top panels) for WR 108 fit with a central point source + fully resolved component, along with corresponding residuals (bottom panels).}
  \label{wr108}
\end{figure}

\begin{figure}[H]
\centering
  \includegraphics[height=60mm]{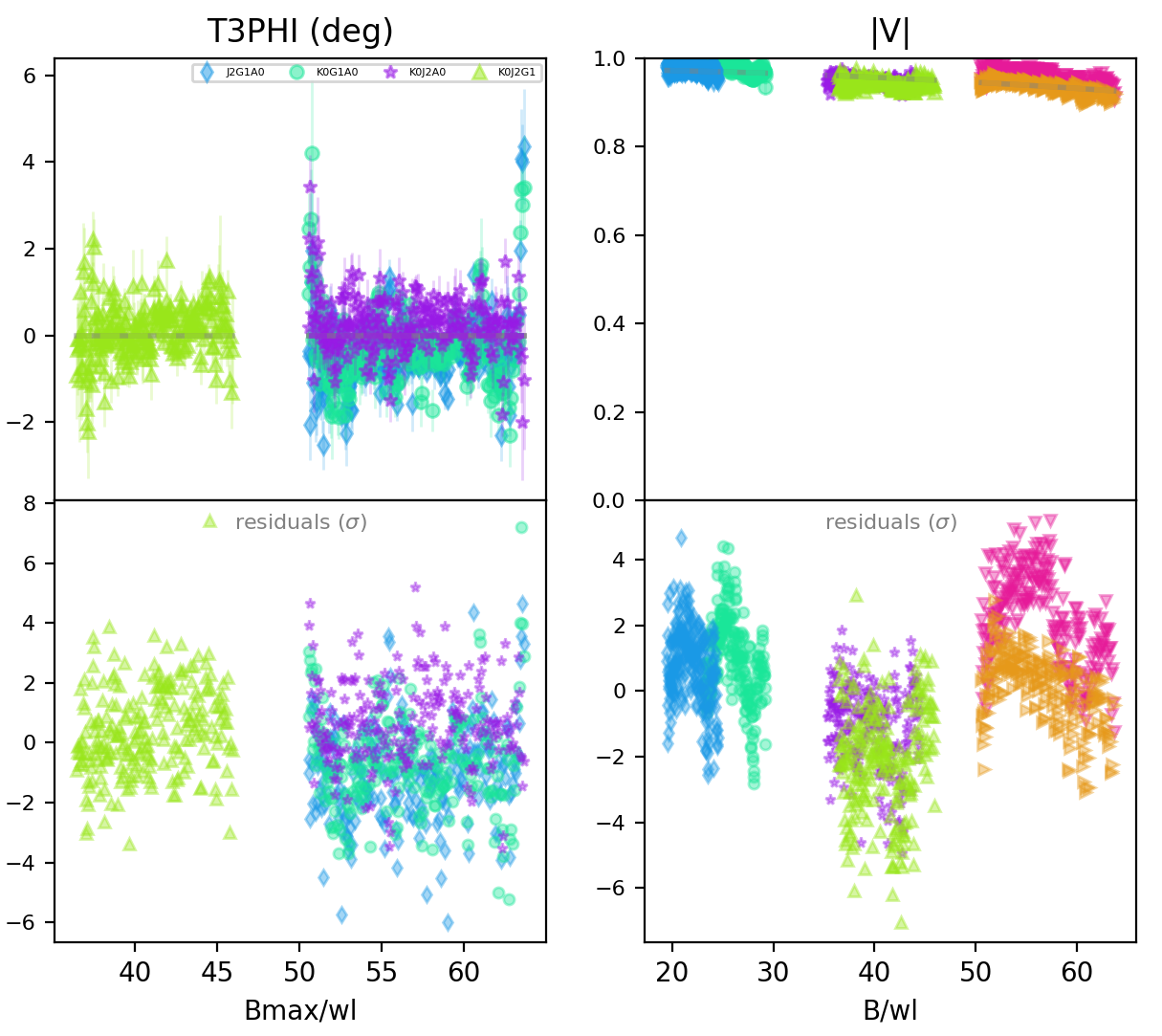}
  \caption{|V|-T3PHI data (top panels) for WR 110 fit with a central point source + fully resolved component, along with corresponding residuals (bottom panels).}
  \label{wr110_1}
\end{figure}

\begin{figure}[H]
\centering
  \includegraphics[height=60mm]{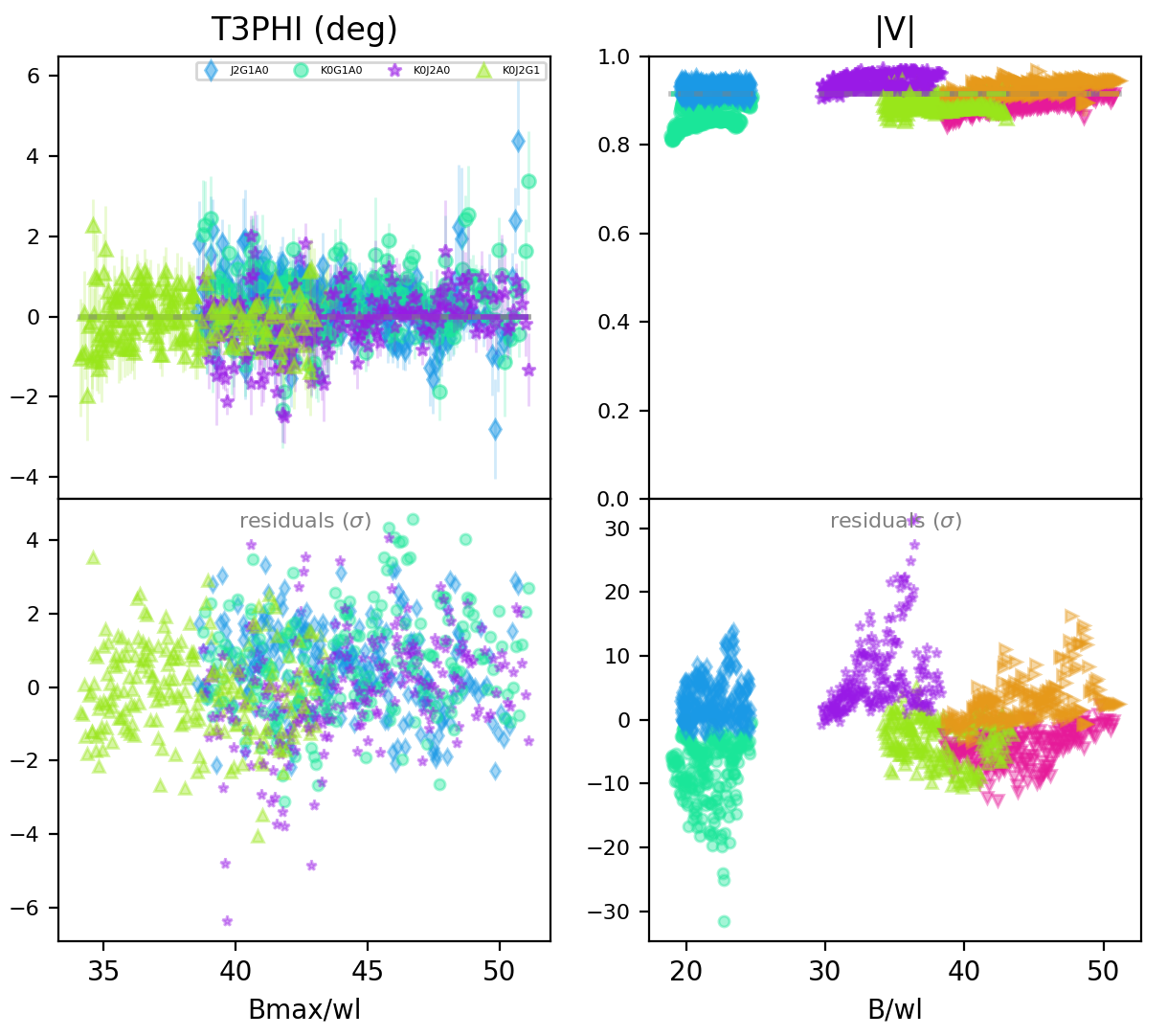}
  \caption{|V|-T3PHI data (top panels) for WR 111 fit with a central point source + fully resolved component, along with corresponding residuals (bottom panels).}
  \label{wr111}
\end{figure}

\begin{figure}[H]
\centering
  \includegraphics[height=60mm]{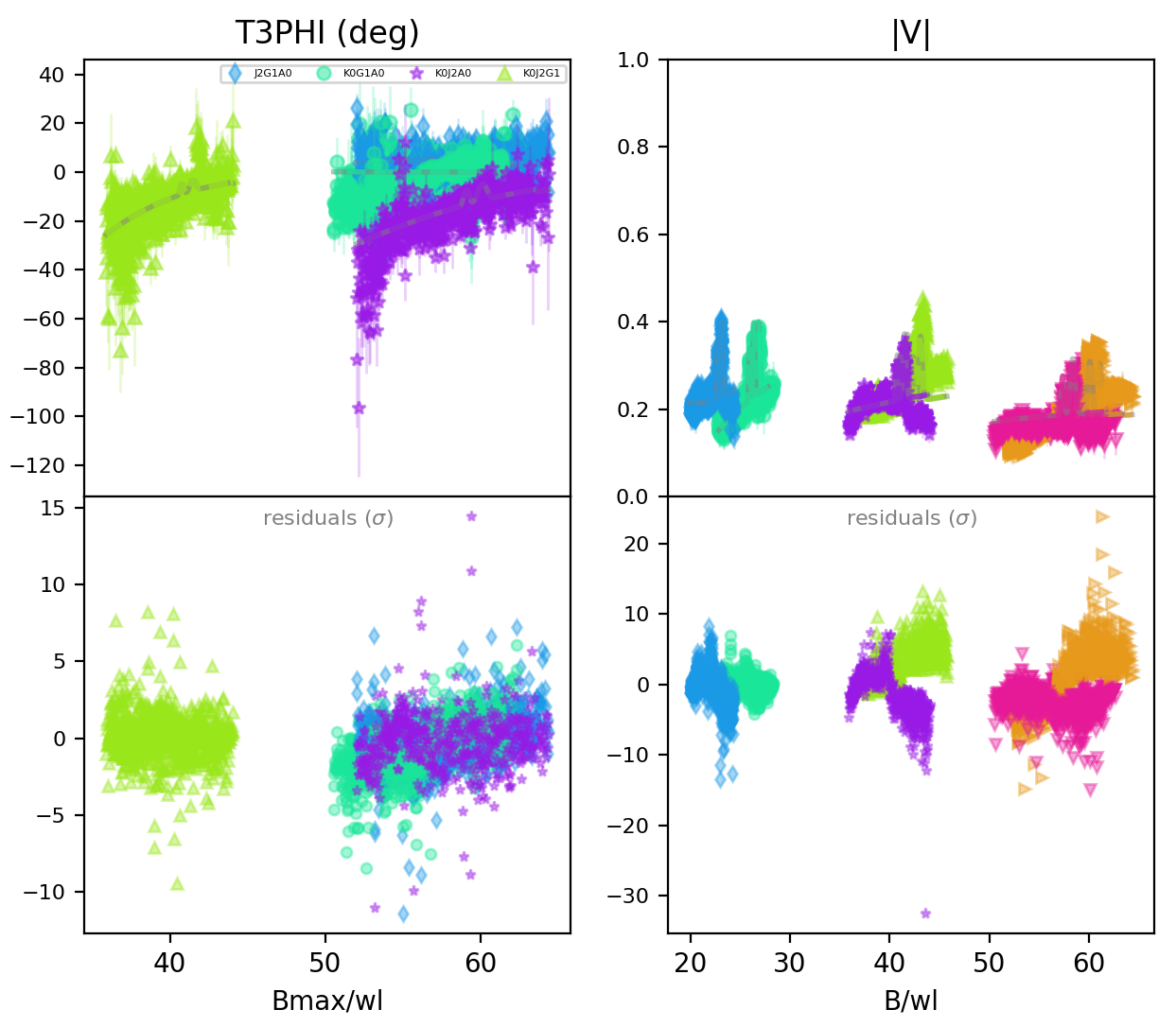}
  \caption{|V|-T3PHI data (top panels) for WR 113 fit with a central point source + a very significant fully resolved component, along with corresponding residuals (bottom panels).}
  \label{wr113}
\end{figure}

\begin{figure}[H]
\centering
  \includegraphics[height=60mm]{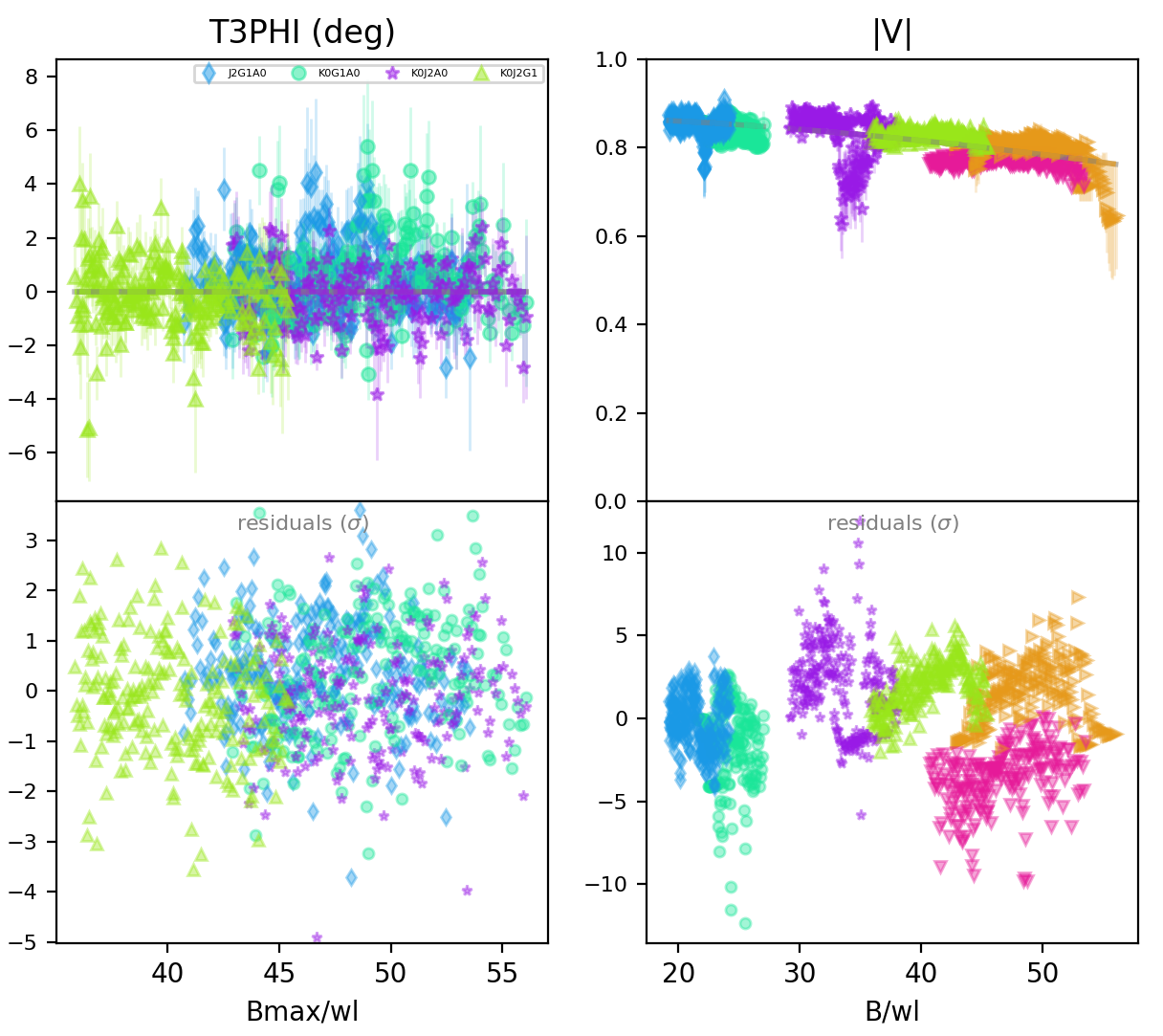}
  \caption{|V|-T3PHI data (top panels) for WR 114 fit with a central point source + fully resolved component, along with corresponding residuals (bottom panels).}
  \label{wr114}
\end{figure}

\begin{figure}[H]
\centering
  \includegraphics[height=60mm]{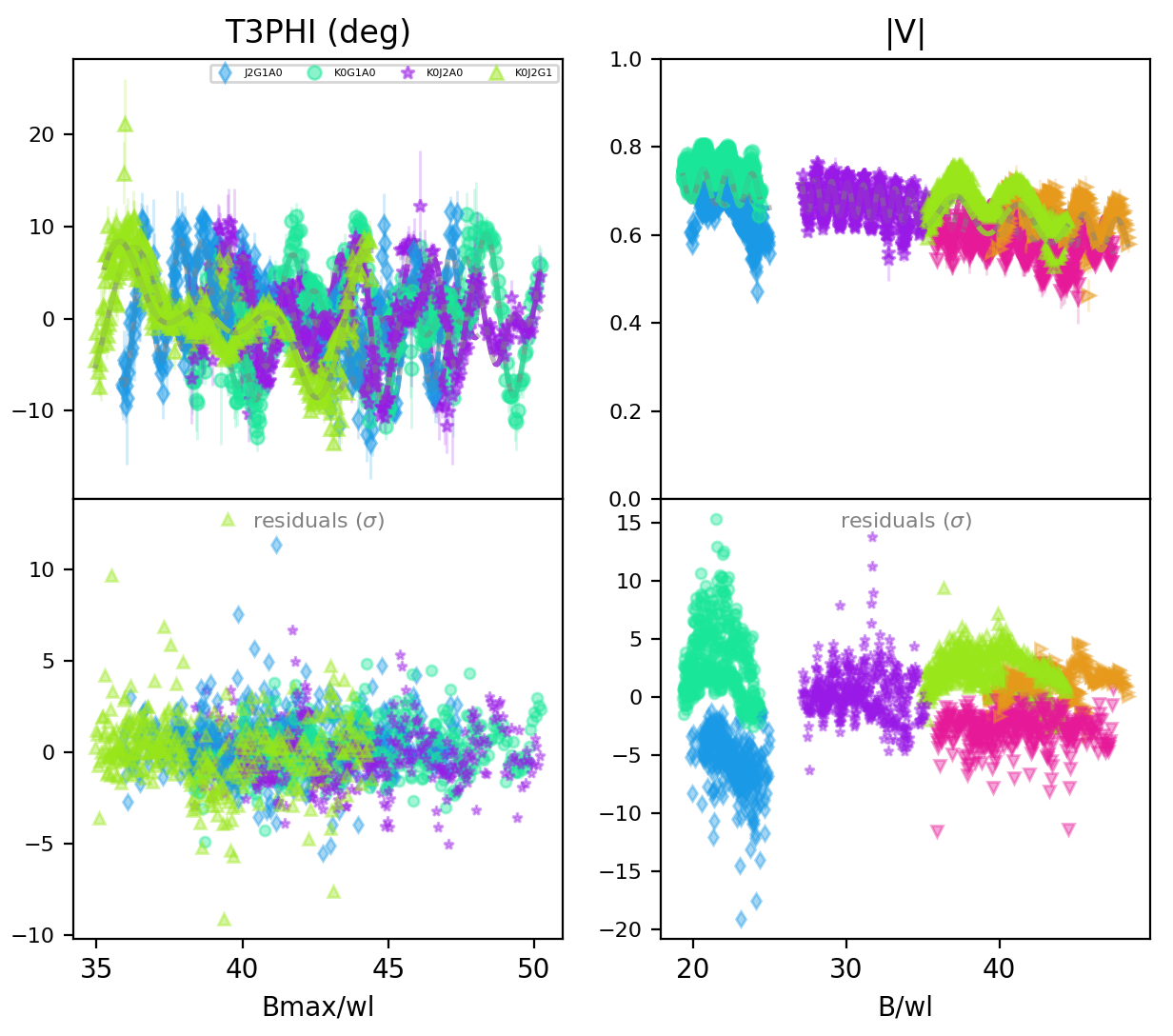}
  \caption{|V|-T3PHI data (top panels) for WR 115 fit with two point sources to model the binary, along with a fully resolved component, and the corresponding residuals (bottom panels).}
  \label{wr115_1}
\end{figure}

\begin{figure*}
\centering
  \includegraphics[trim={0.5cm 1.5cm 0.5cm 1.5cm}, clip, height=100mm]{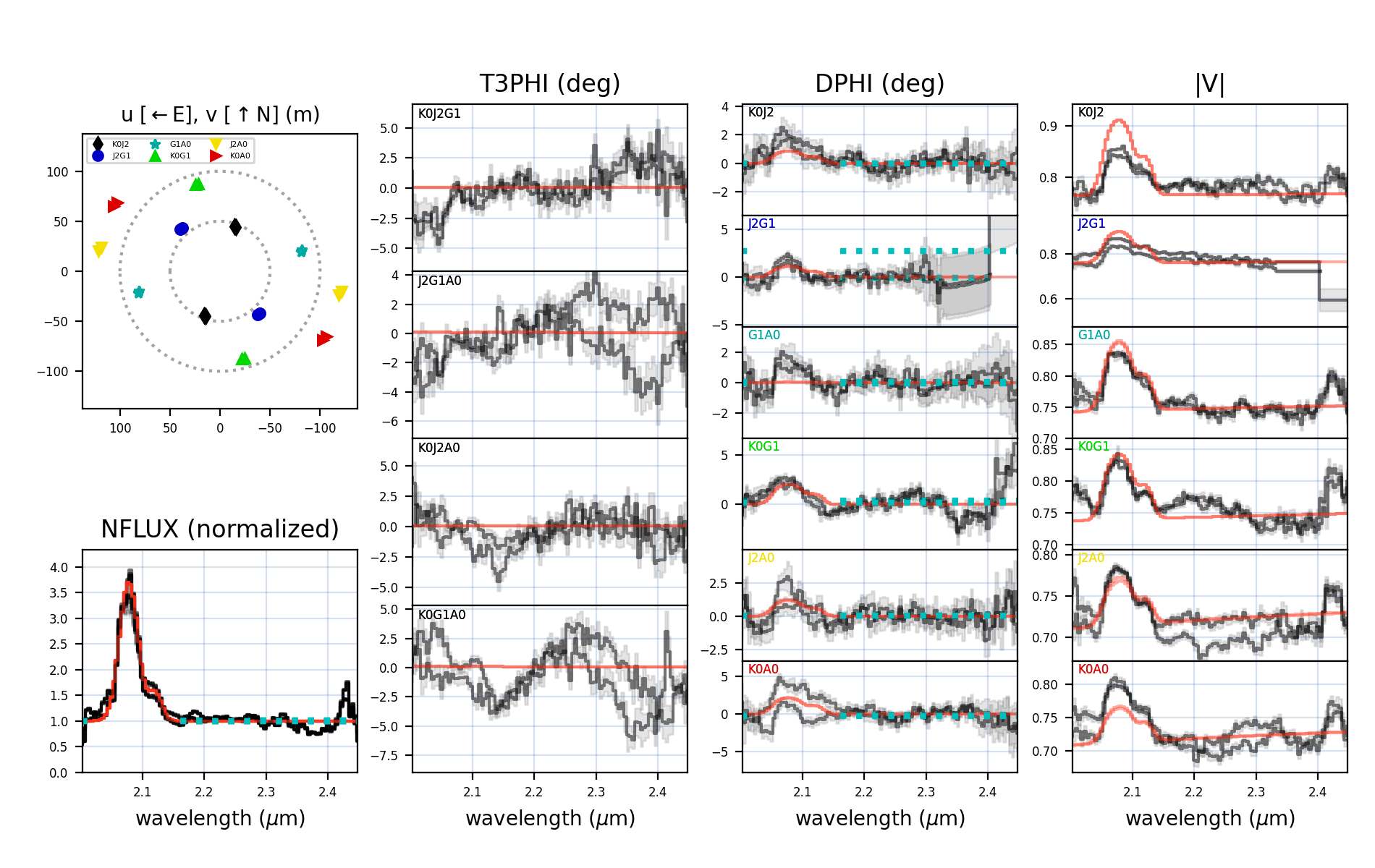}
  \caption{Complete spectro-interferometric data for WR 9 with the best-fit model, similar to Figure\,\ref{fig:wr98_1}.}
  \label{wr9_2}
\end{figure*}

\begin{figure*}
\centering
  \includegraphics[trim={0.5cm 1.5cm 0.5cm 1.5cm}, clip, height=100mm]{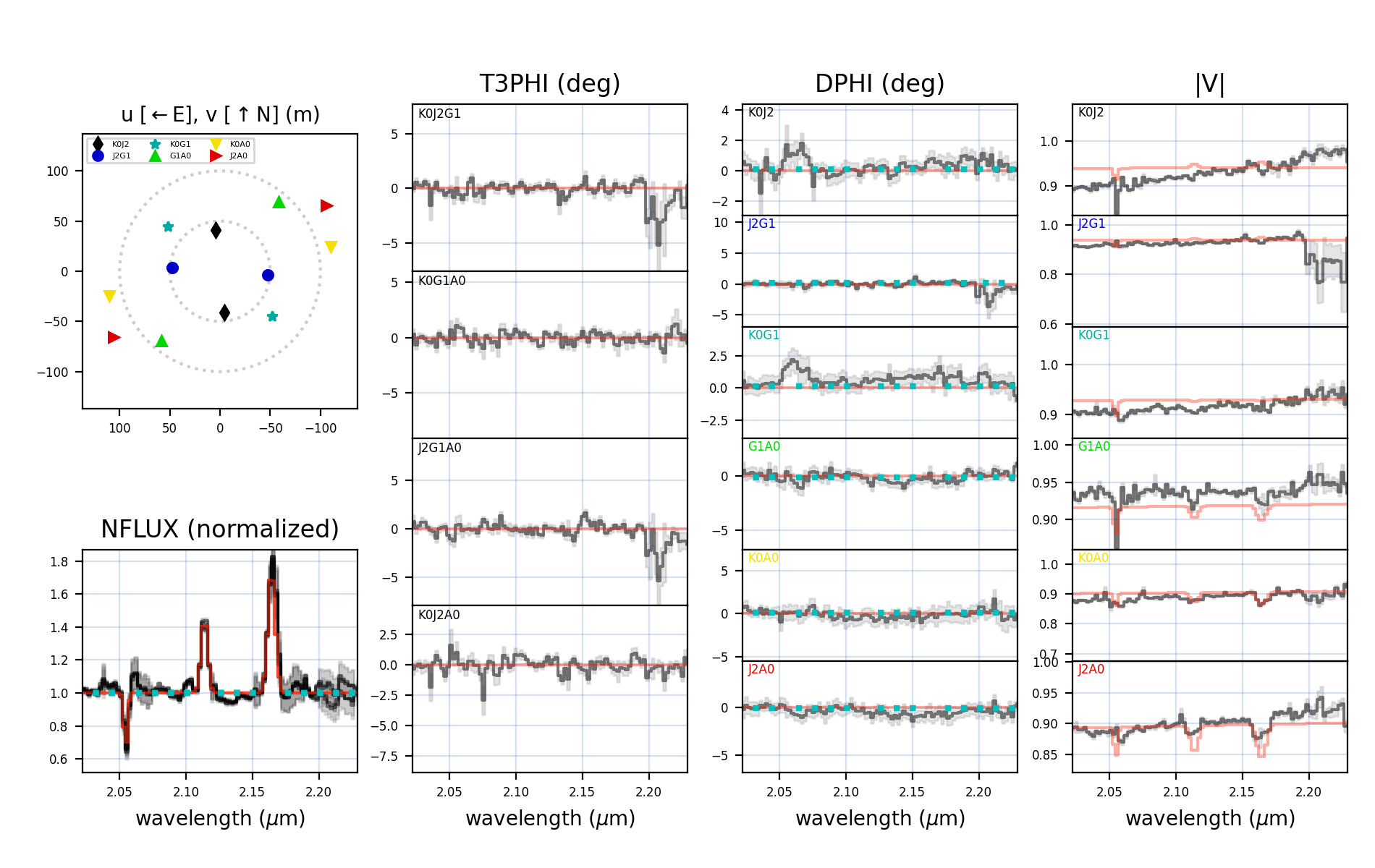}
  \caption{A zoomed-in snippet of complete spectro-interferometric data for WR 16 with the best-fit model, similar to Figure\,\ref{wr78_1}.}
  \label{wr16_2}
\end{figure*}

\begin{figure*}
\centering
  \includegraphics[trim={0.5cm 1.5cm 0.5cm 1.5cm}, clip, height=100mm]{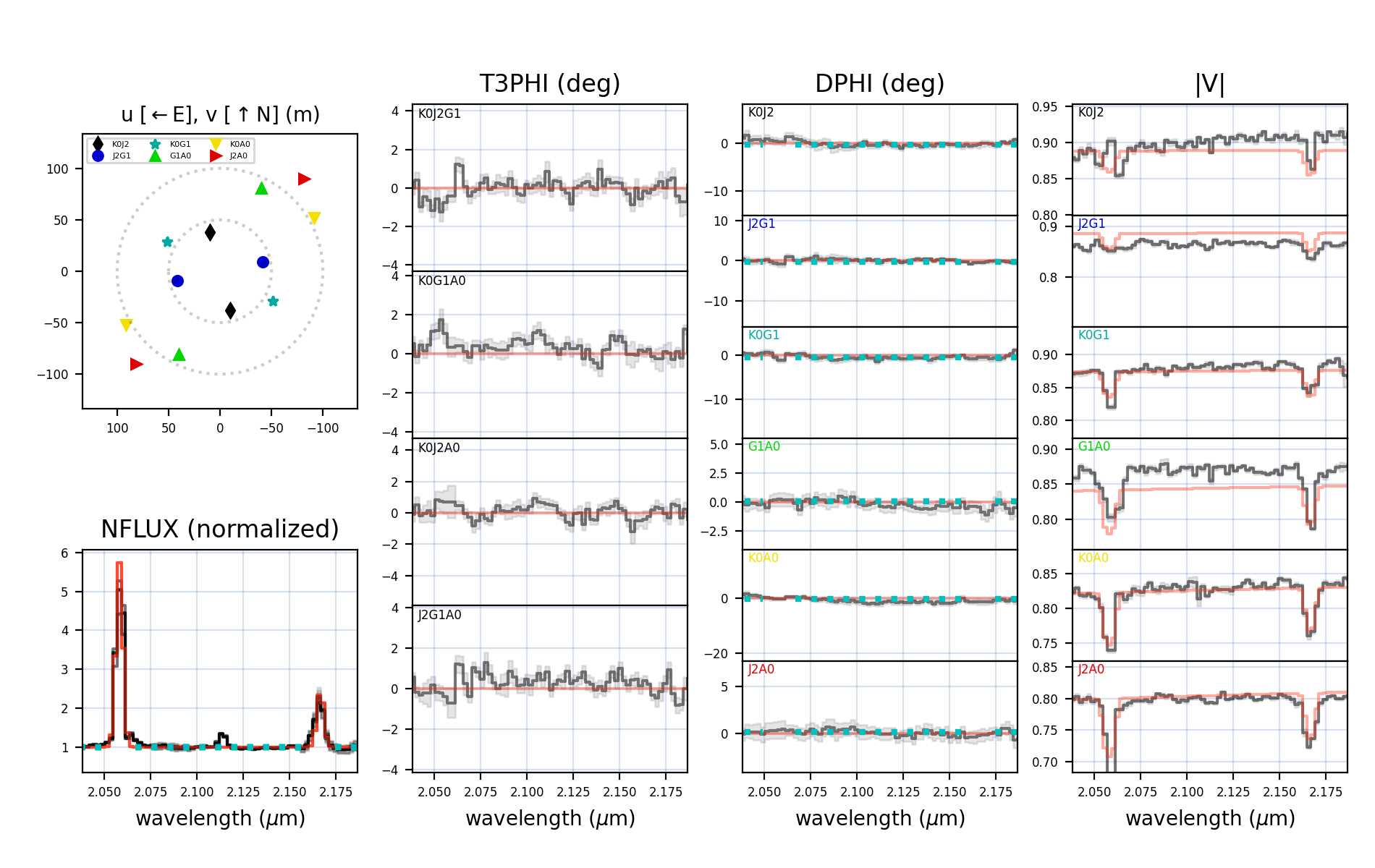}
  \caption{A zoomed-in snippet of complete spectro-interferometric data for WR 31a with the best-fit model, similar to Figure\,\ref{wr78_1}.}
  \label{wr31a_2}
\end{figure*}

\begin{figure*}
\centering
  \includegraphics[trim={0.2cm 0.5cm 0.2cm 0.5cm}, clip, height=100mm]{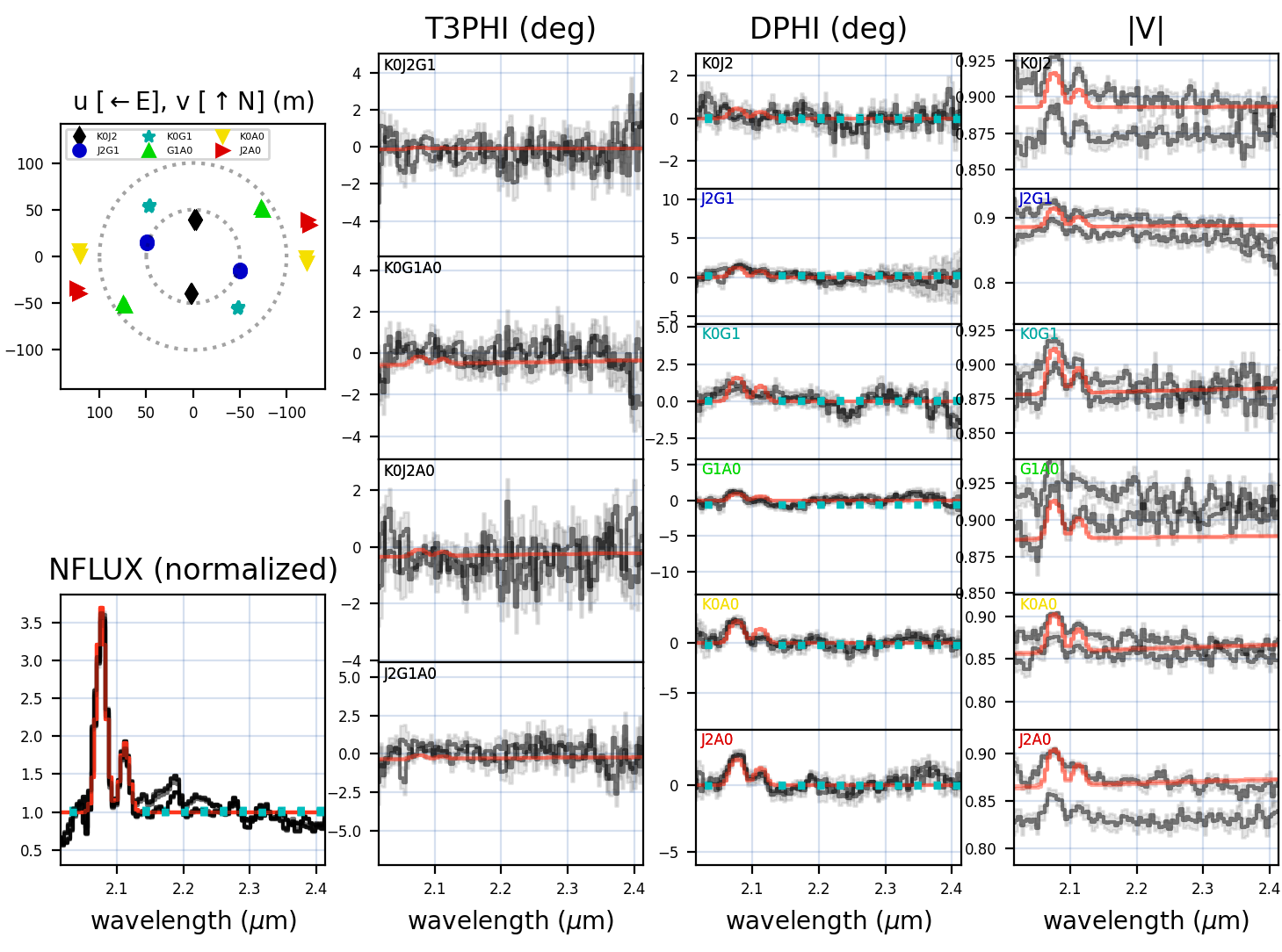}
  \caption{Complete spectro-interferometric data for WR 42 with the best-fit model, similar to Figure\,\ref{fig:wr98_1}.}
  \label{wr42_2}
\end{figure*}

\begin{figure*}
\centering
  \includegraphics[trim={0 0.5cm 0 0.5cm}, clip, height=100mm]{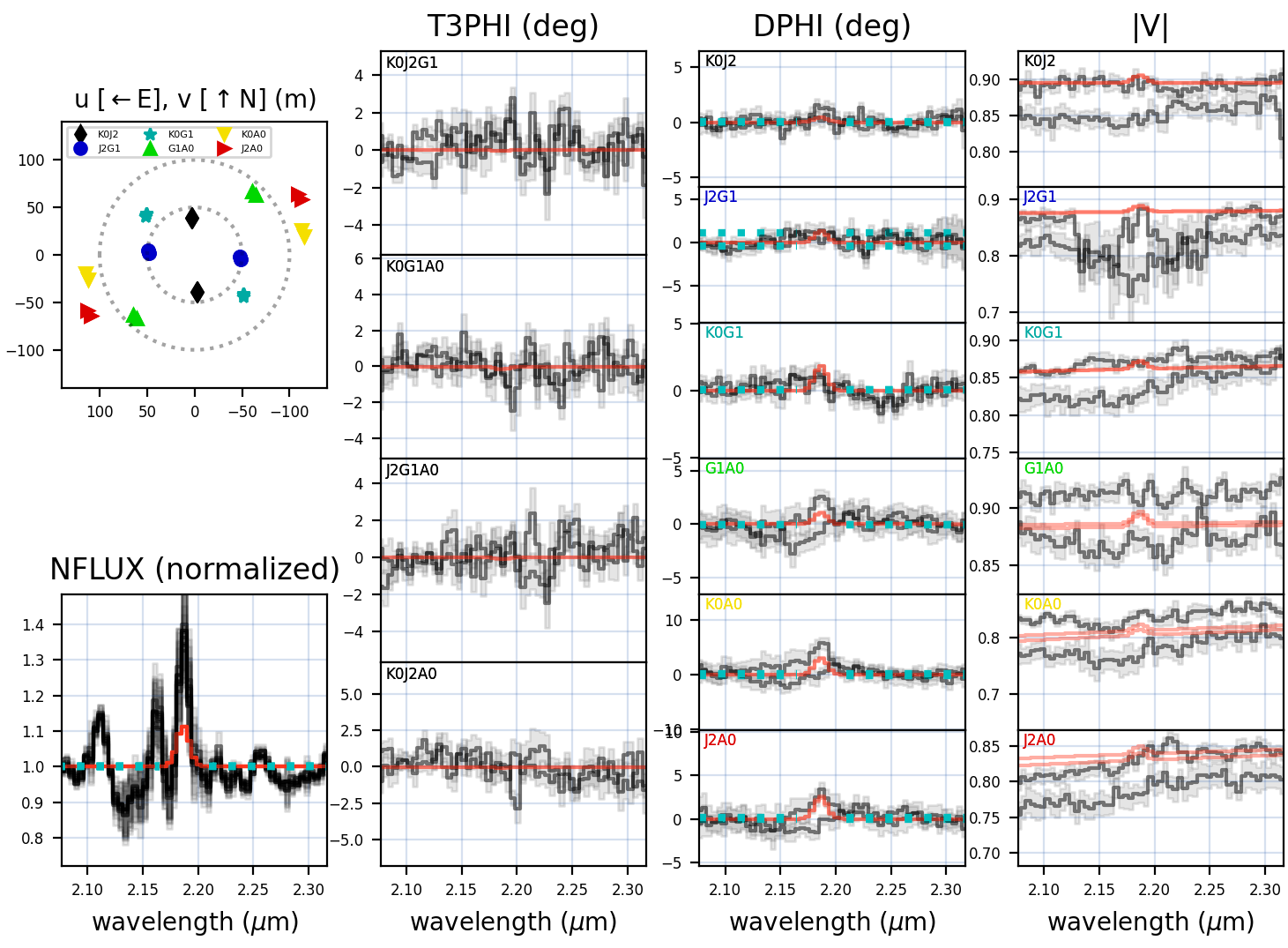}
  \caption{Complete spectro-interferometric data for WR 47 with an approximate model (similar to Fig.\,\ref{fig:wr98_1}) over-plotted.}
  \label{wr47_2}
\end{figure*}

\begin{figure*}
\centering
  \includegraphics[trim={0.2cm 0.5cm 0.2cm 0.5cm}, clip, height=100mm]{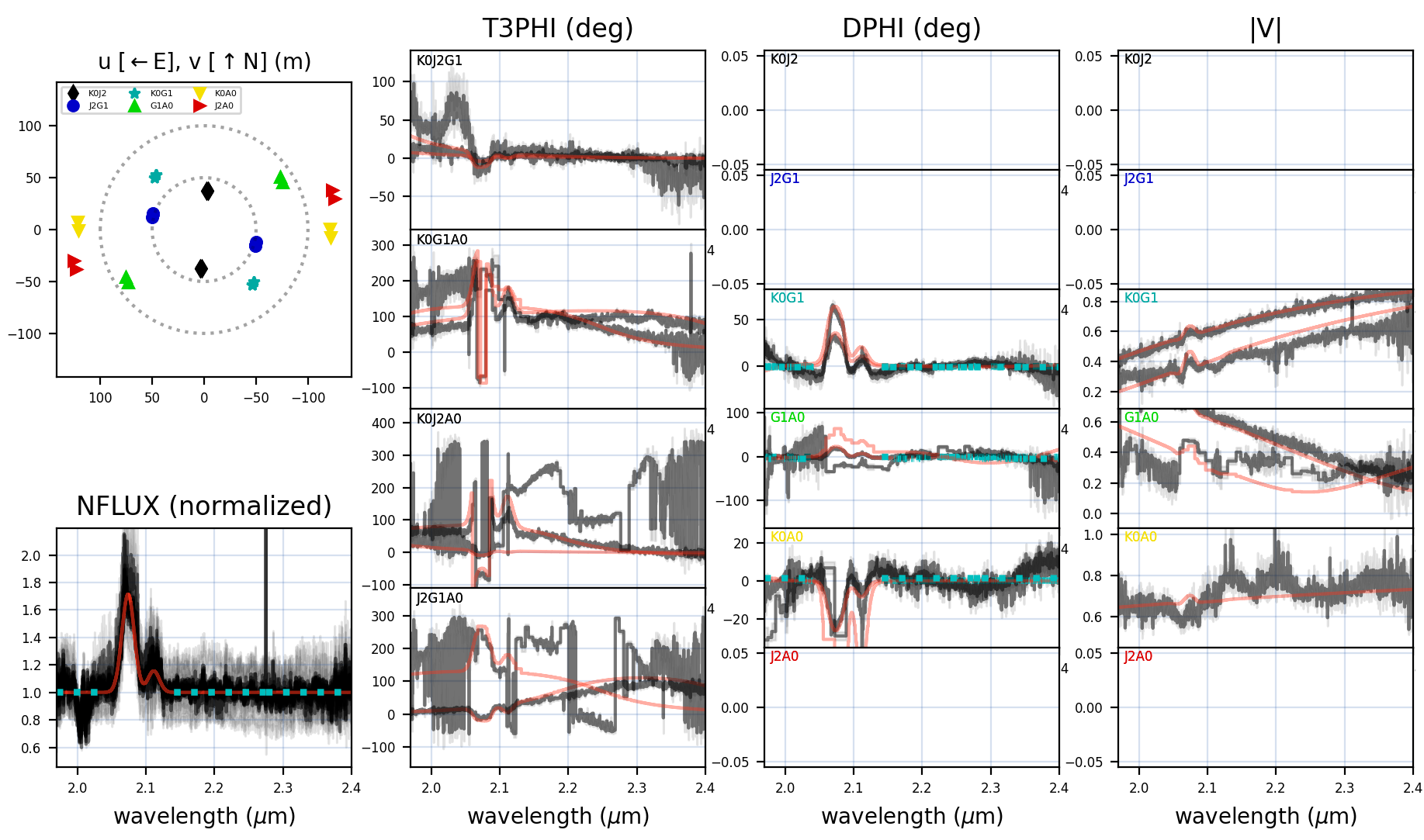}
  \caption{Complete spectro-interferometric data for WR 48 with the best-fit binary model.}
  \label{wr48_2}
\end{figure*}

\begin{figure*}
\centering
  \includegraphics[trim={0.2cm 0.5cm 0.2cm 0.5cm}, clip, height=100mm]{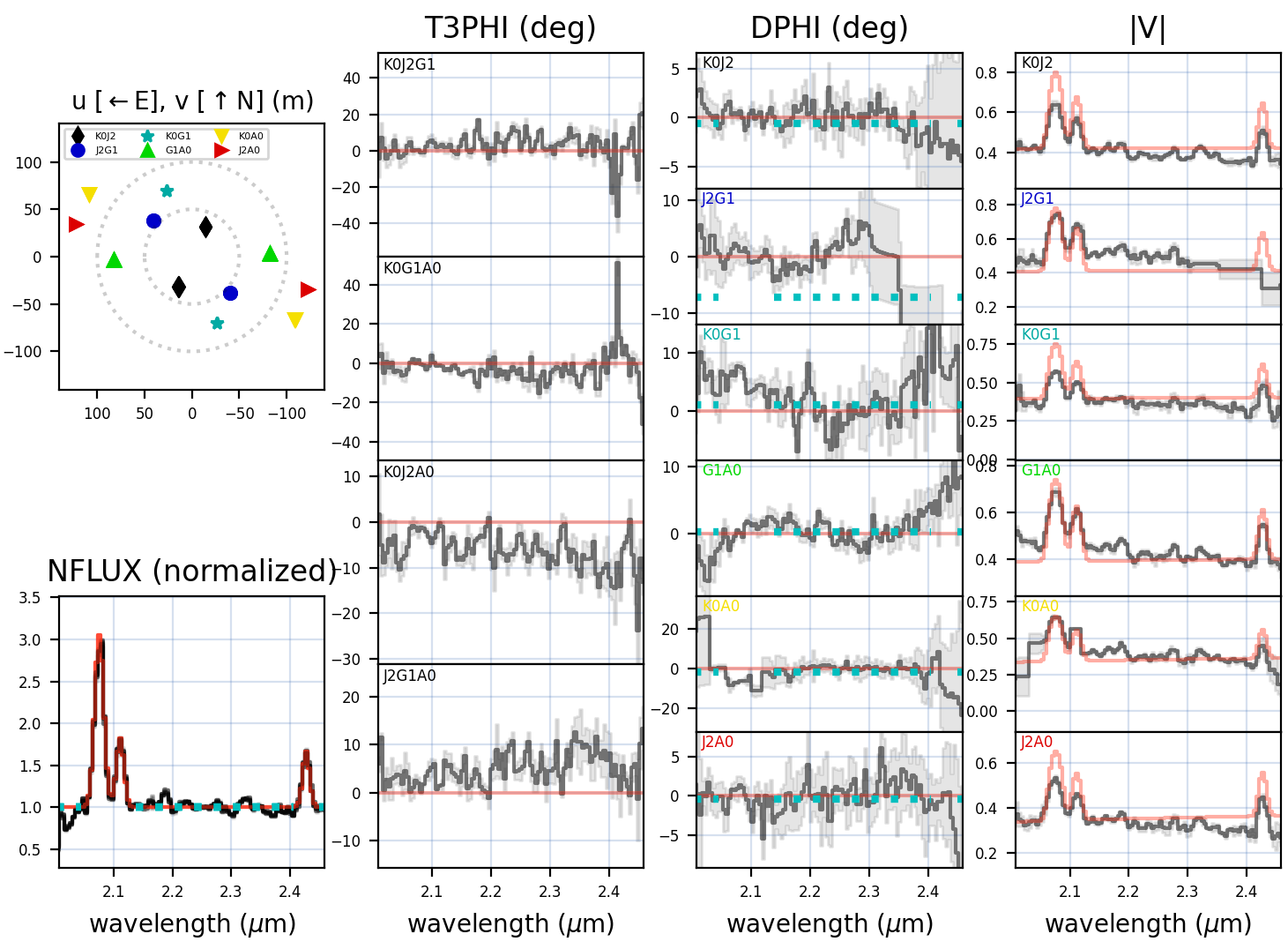}
  \caption{Complete spectro-interferometric data for WR 57 with the best-fit model.}
  \label{wr57_2}
\end{figure*}

\begin{figure*}
\centering
  \includegraphics[trim={0.5cm 1.5cm 0.5cm 1.5cm}, clip, height=100mm]{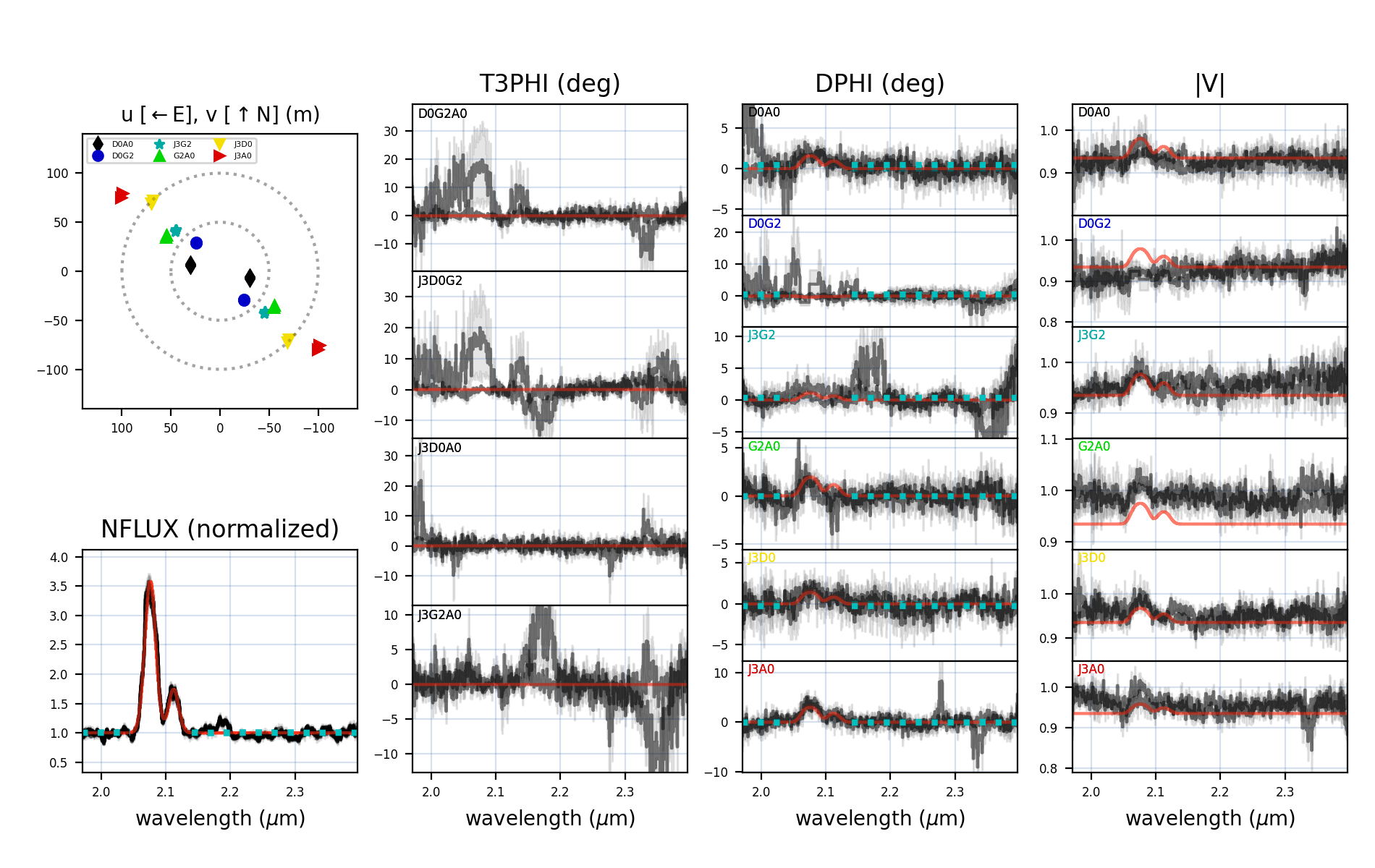}
  \caption{Complete spectro-interferometric data for WR 79 with the best-fit model, similar to Figure\,\ref{fig:wr98_1}.}
  \label{wr79_2}
\end{figure*}

\begin{figure*}
\centering
  \includegraphics[trim={0.5cm 1.5cm 0.5cm 1.5cm}, clip, height=100mm]{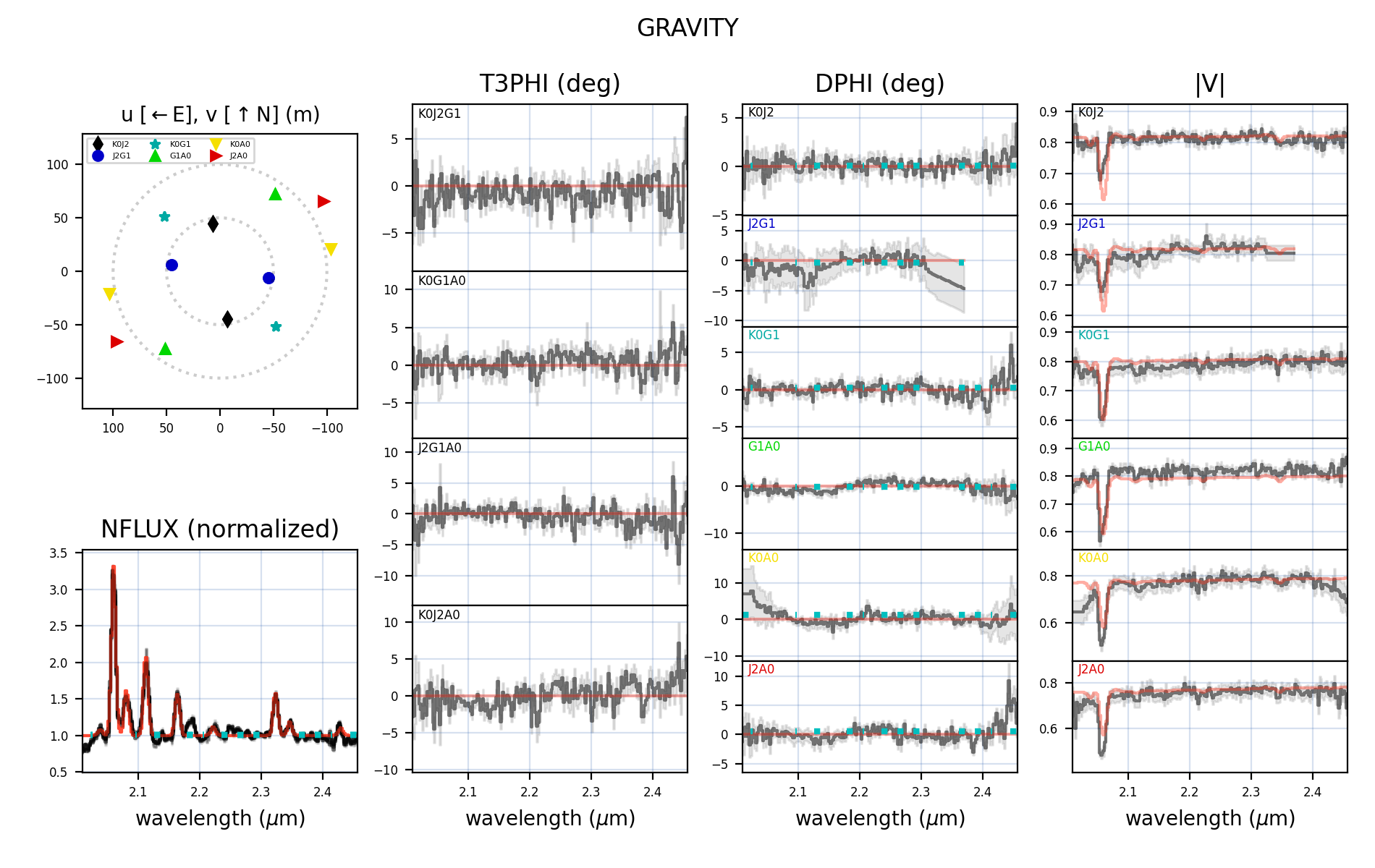}
  \caption{Complete spectro-interferometric data for WR 81 with the best-fit model.}
  \label{wr81_2}
\end{figure*}

\begin{figure*}
\centering
  \includegraphics[trim={0.5cm 1.5cm 0.5cm 1.5cm}, clip, height=100mm]{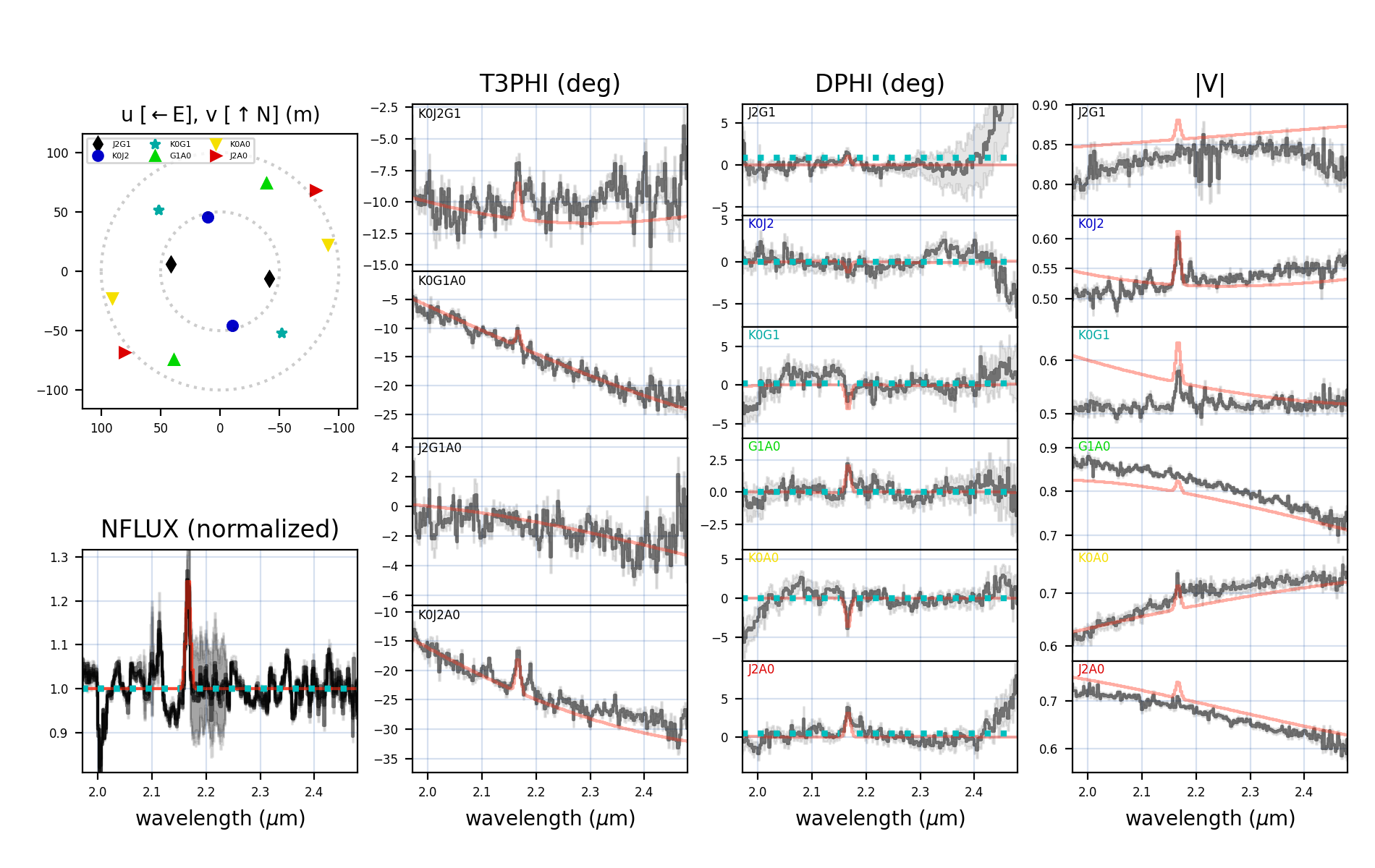}
  \caption{Complete spectro-interferometric data for WR 89 with the best-fit binary model.}
  \label{wr89_2}
\end{figure*}

\begin{figure*}
\centering
  \includegraphics[trim={0.5cm 1.5cm 0.5cm 1.5cm}, clip, height=100mm]{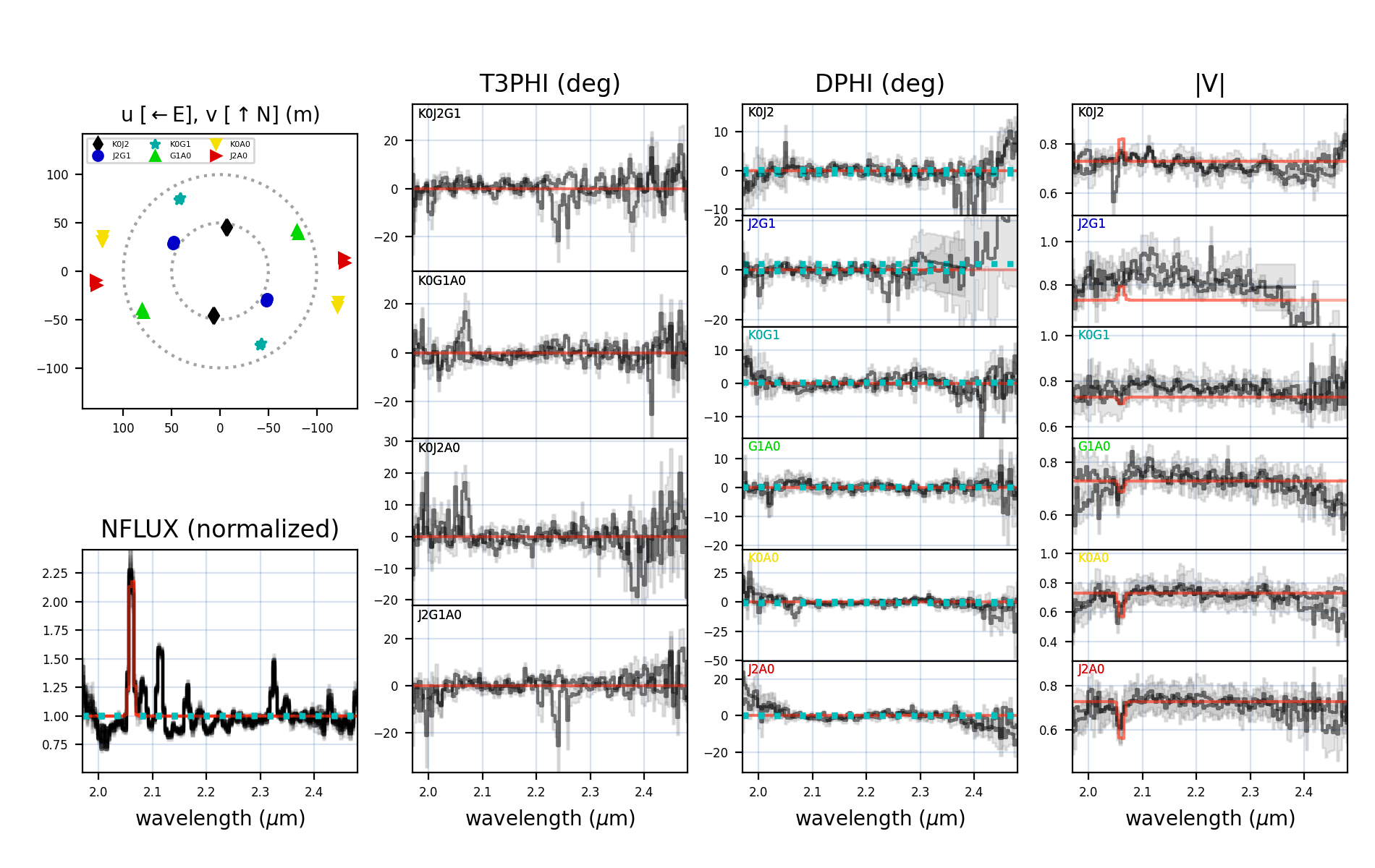}
  \caption{Complete spectro-interferometric data for WR 92 with the best-fit model.}
  \label{wr92_2}
\end{figure*}

\begin{figure*}
\centering
  \includegraphics[trim={0.5cm 1.5cm 0.5cm 1.5cm}, clip, height=100mm]{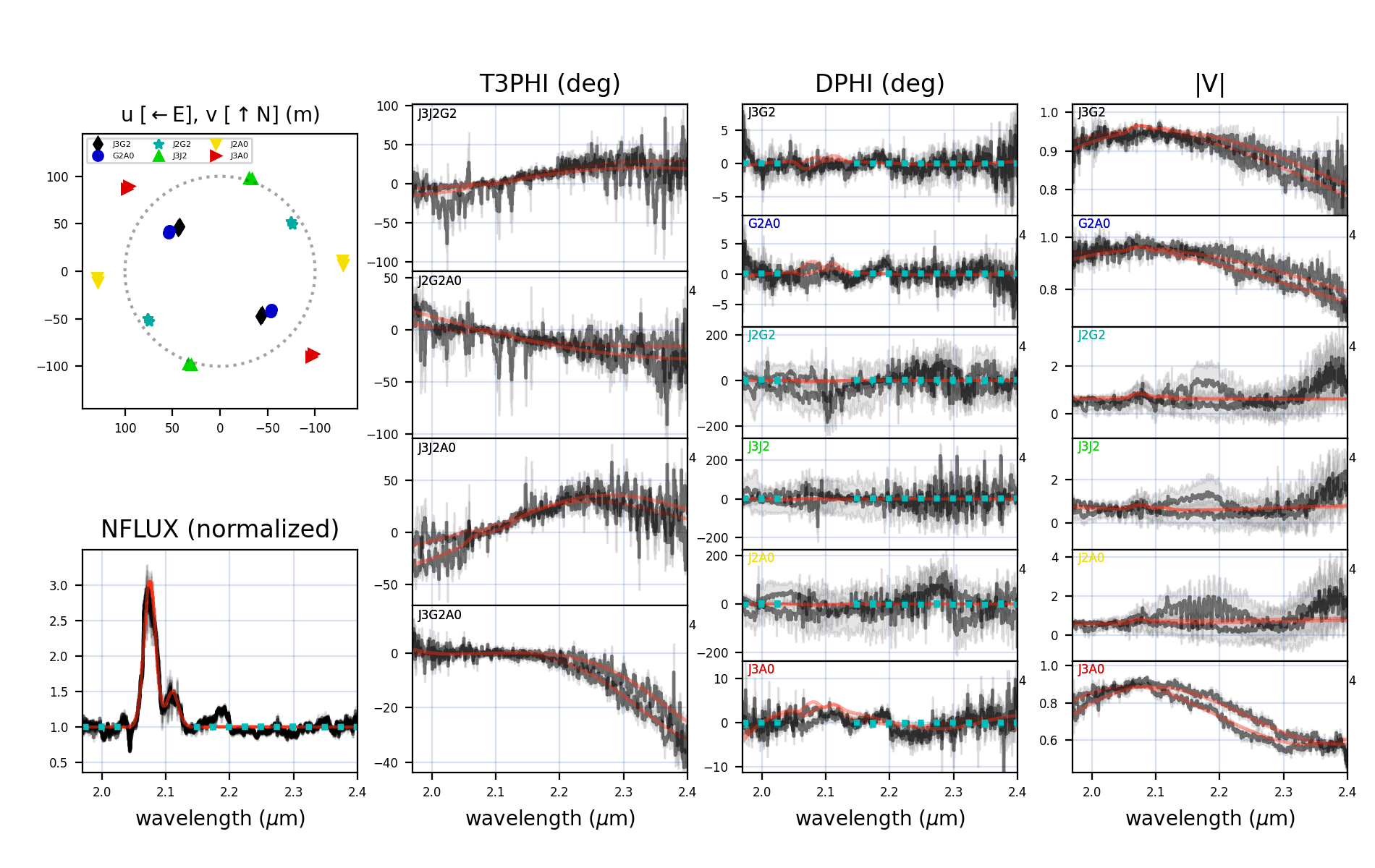}
  \caption{Complete spectro-interferometric data for WR 93 with the best-fit binary model.}
  \label{wr93_2}
\end{figure*}

\begin{figure*}
\centering
  \includegraphics[trim={0.5cm 1.5cm 0.5cm 1.5cm}, clip, height=100mm]{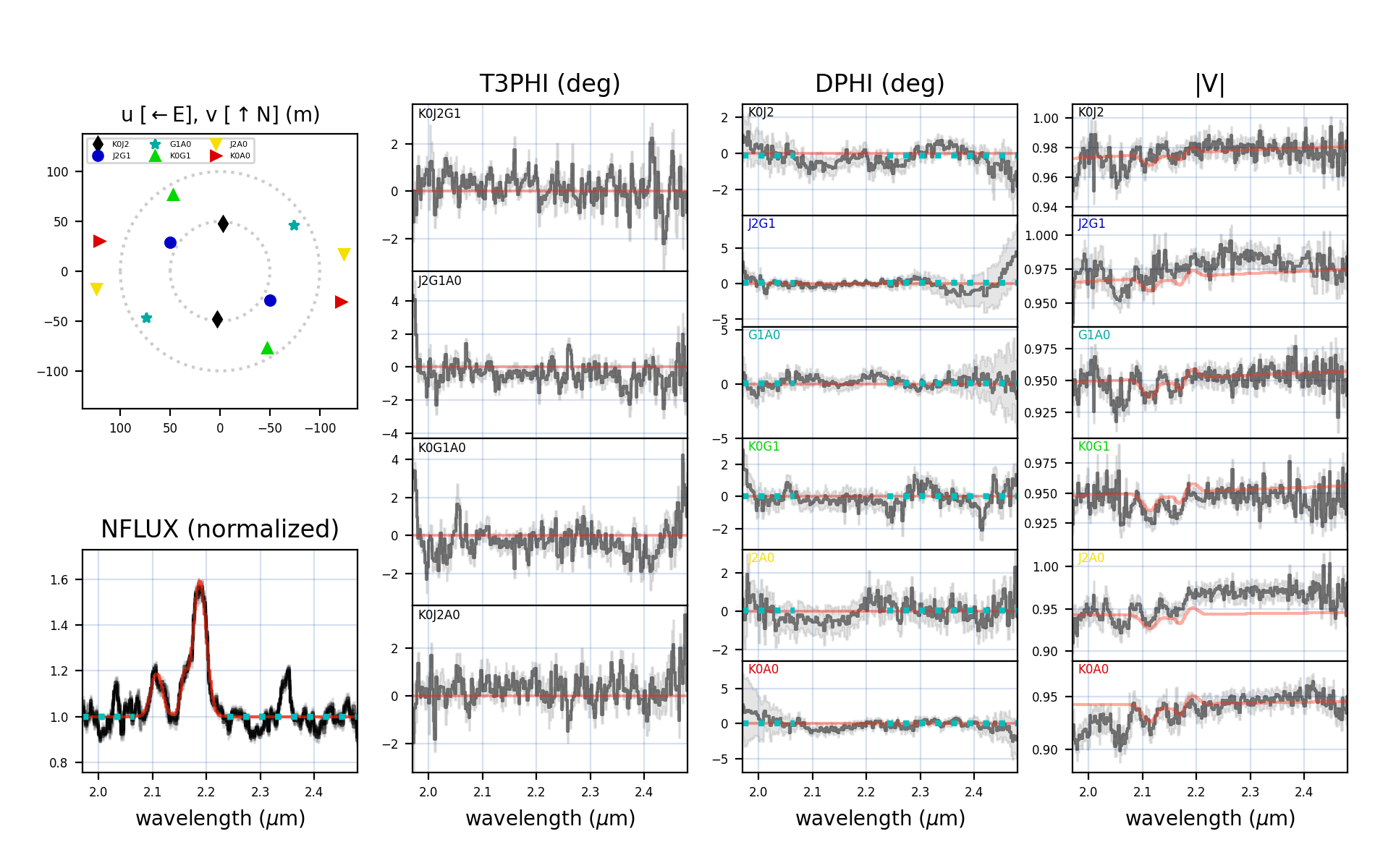}
  \caption{Complete spectro-interferometric data for WR 110 with an approximate model (similar to WR 78) over-plotted.}
  \label{wr110_2}
\end{figure*}

\begin{figure*}
\centering
  \includegraphics[trim={0.5cm 1.5cm 0.5cm 1.5cm}, clip, height=100mm]{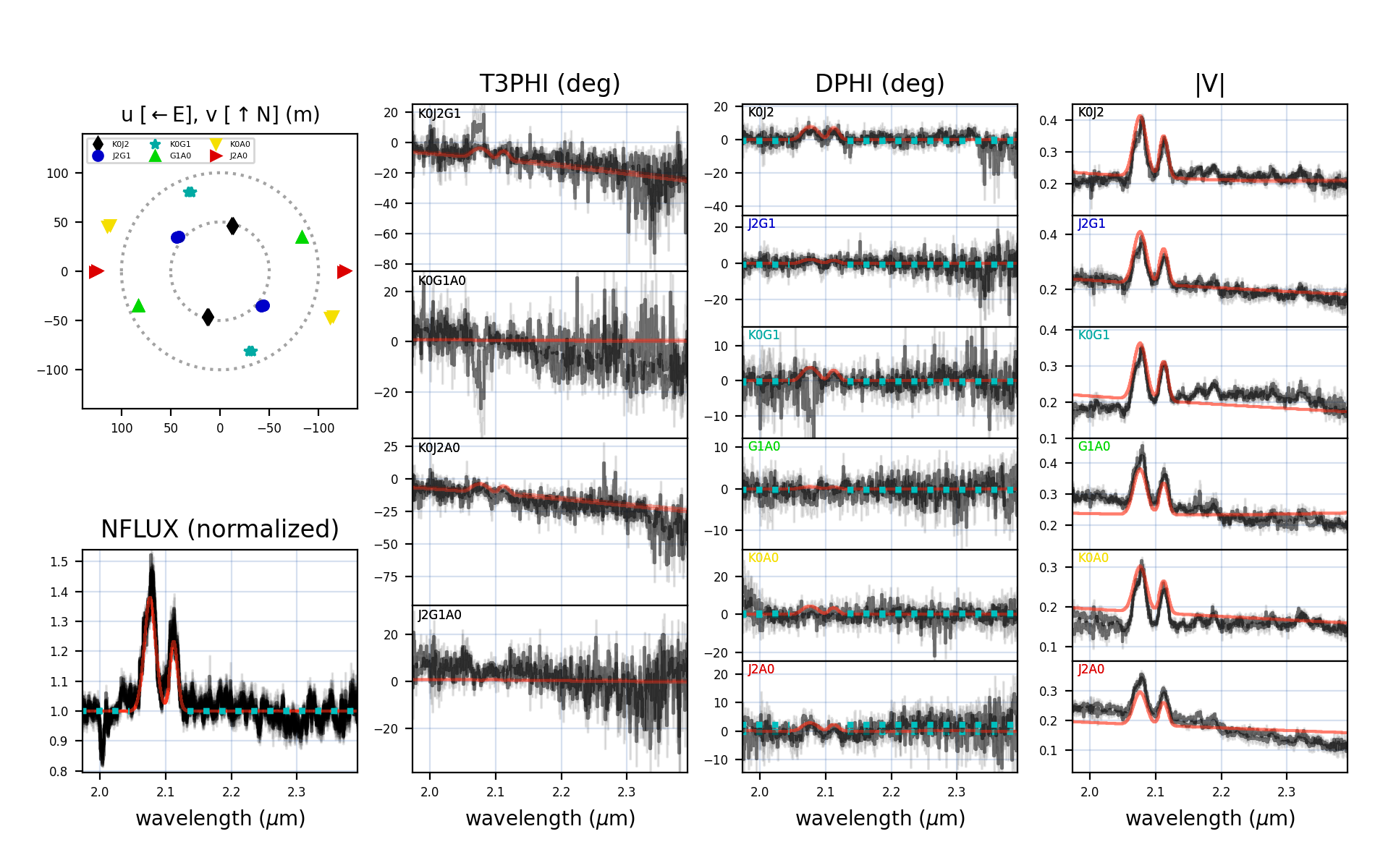}
  \caption{Complete spectro-interferometric data for WR 113 with the best-fit model.}
  \label{wr113_2}
\end{figure*}

\begin{figure*}
\centering
  \includegraphics[trim={0.5cm 1.5cm 0.5cm 1.5cm}, clip, height=100mm]{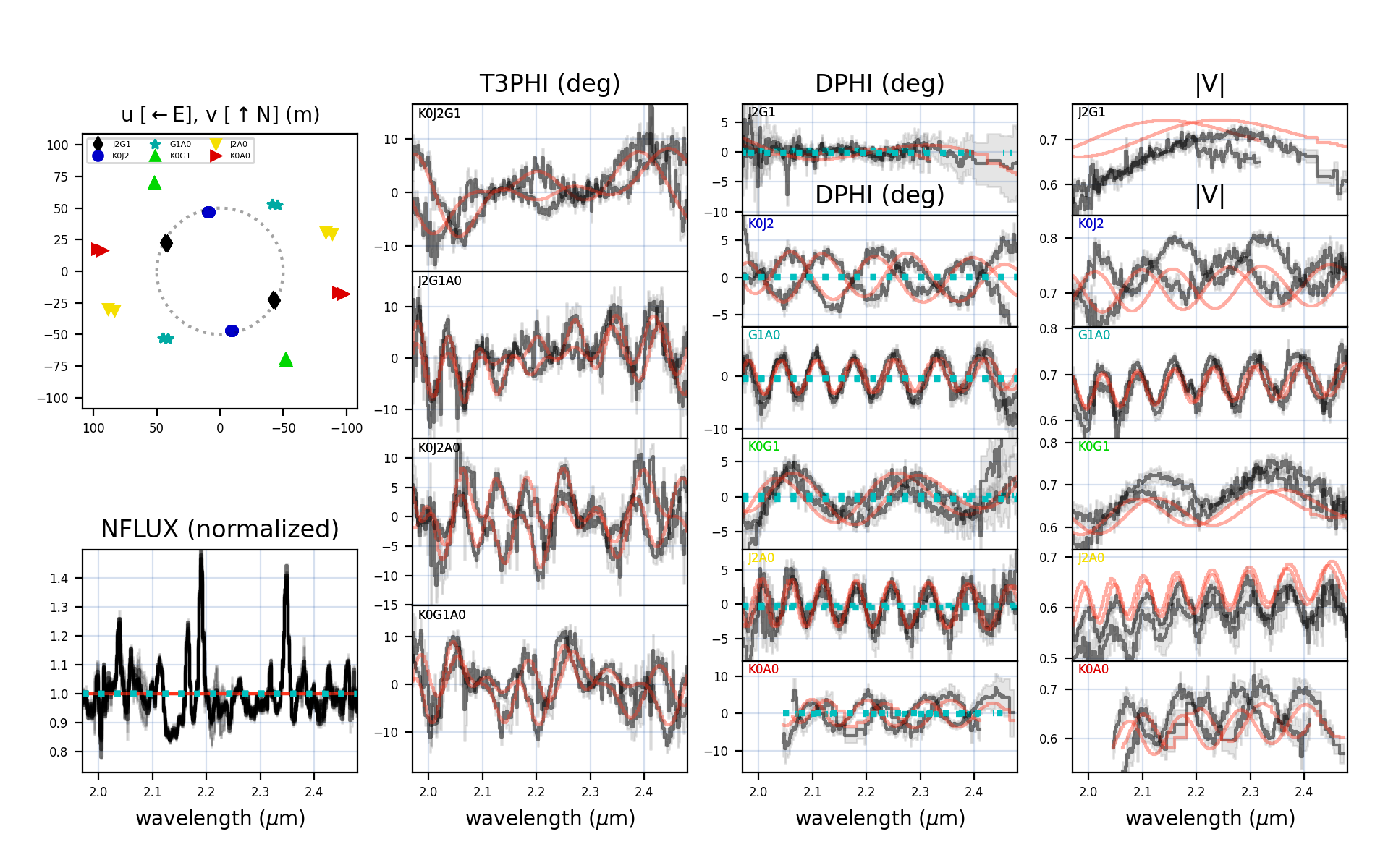}
  \caption{Complete spectro-interferometric data for WR 115 with the best-fit binary model.}
  \label{wr115_2}
\end{figure*}

Here we provide some details for each star in our sample, including the known properties and new results from interferometry. Following the descriptions for all WRs are the corresponding figures (\ref{wr8} to \ref{wr115_2}), and Table\,\ref{tab:summ} summarising the best-fit geometric models for all stars.

\vspace{5mm}

\textbf{WR 8} is classified as a single WN6/WC4 star by \citet{1995crowther}. It displays corotating-interaction-region-type (CIR-type) variability in its spectral lines \citep{2011CheneStLouis}. GRAVITY data for WR 8 did not show signs of binarity in any of the interferometric observables. The data were fit reasonably well with a simple point source and a fully resolved component, as shown in Figure\,\ref{wr8}.

\textbf{WR 9} is classified as a WC5 + O7 spectroscopic binary with a 14.3-day orbital period \citep{1988Hucht}. The GRAVITY data were best modeled using a point source and a fully resolved component (see Fig.\,\ref{wr9_1}). Although the T3PHI residuals seemed to have a signal, we did not a find a binary or even a triple system model that could explain it. However, features corresponding to spectral lines could be seen in the |V| and DPHI data, likely indicating a close binary as discussed in Section\,\ref{sec:DPHI}. Figure\,\ref{wr9_2} shows the complete spectro-interferometric data for WR 9, along with the best-fit model consisting of two close point sources and a fully resolved component. The position of the companion is essentially unconstrained.

\textbf{WR 12} is classified as a WN8h + OB spectroscopic binary with a 23.9-day orbital period, also known to harbor colliding winds \citep{2012Fahed}. The |V| and T3PHI data were modeled sufficiently well with a point source and a fully resolved component as shown in Figure\,\ref{wr12}.

\textbf{WR 14} is a WC7 initially classified as a potential SB1 spectroscopic binary \citep{1990Shylaja}, which was later classified as single \citep{1992Drissen}. The GRAVITY data were best modeled using a central point source and fully resolved component, the latter not being significant (Fig.\,\ref{wr14}).

\textbf{WR 15} is a WC6 classified as a single star \citep{2019Sander}. The GRAVITY data were best modeled using a point source and a fully resolved component, although leading to significant residuals (Fig.\,\ref{wr15}). A companion search for a binary companion or even an additional tertiary companion was not able to fit the data. We therefore speculate the origin of the residuals to be either some structure within the fully resolved component or poor visibility calibration.

\textbf{WR 16} is a WN8h classified as a single star \citep{2019Hamann}, also faintly detected in X-rays \citep{2012Skinner}. The GRAVITY data for WR 16 largely resembled a point source with a slight contribution from a fully resolved flux (see Fig.\,\ref{wr16_1}). Although the data were noisy, on close inspection, we found features in |V| data corresponding to the spectral lines in NFLUX data. 

As discussed in Section\,\ref{sec:LER}, the case of WR 16 can be similarly explained by a two component model for the central star - an unresolved component for the continuum and 2.06 {\textmu}m spectral line; and a partially resolved component for the 2.11 and 2.16 {\textmu}m spectral lines. The partially resolved component was modeled as a Gaussian with an fwhm of $0.98\pm0.11$ mas. Figure\,\ref{wr16_2} shows the best-fit model for the complete spectro-interferometric GRAVITY data for WR 16.

\textbf{WR 18} is a WN4 classified as a single star \citep{2019Hamann}, also known to emit X-rays \citep{2010Skinner}. The interferometric data were best modeled using a point source primary and a fully resolved diffuse component, as shown in Figure\,\ref{fig:wr18_1} in the main text.

\textbf{WR 21} is a WN5 classified as a spectroscopic binary with an O4-6 companion and a 8.25-day orbital period \citep{2012Fahed}, showing the colliding wind phenomenon \citep{2023Naze}. GRAVITY data (particularly |V|) for WR 21 show the presence of a significant fully resolved component. The |V|-T3PHI were modeled adequately with a point source and a fully resolved component (see Fig.\,\ref{wr21}). The data show slight deviation from point source behaviour, although not significant enough to consider binarity. Higher SNR data might resolve this ambiguity.

\textbf{WR 22} is a WN7h classified as an eclipsing spectroscopic binary with an O9III-V companion and an orbital period of about 80 days \citep{1991Gosset}. It has been studied extensively as an X-ray source as well \citep{2009Gosset,2011Parkin}. As discussed in Section\,\ref{aps} and shown in Figure\,\ref{fig:wr22},  WR 22 was consistent with a simple point source model, with a fully resolved component essentially absent.

\textbf{WR 23} is a WC6 classified as a single star \citep{2019Sander}. The GRAVITY data were best modeled by a point source and a fully resolved emission component (see Fig.\,\ref{wr23}).

\textbf{WR 24} is a WN6ha-w classified as a single star \citep{2019Hamann}, also showing X-ray emission \citep{2010Skinner}. The GRAVITY data were consistent with a point source as shown in Figure\,\ref{wr24}, with negligible contribution from a fully resolved component.

\textbf{WR 31} is a WN4 classified as a spectroscopic binary with an O8V companion and a 4.8-day orbital period \citep{1985Niemela}, also showing the colliding wind phenomenon \citep{2012Fahed,2023Naze}. The GRAVITY data for WR 31 were best modeled by a point source and a fully resolved flux contributing significantly (see Fig.\,\ref{wr31}).

\textbf{WR 31a} is a WN11h/candidate LBV classified as a single star \citep{1994Smith}. The GRAVITY data for WR 31a were largely consistent with a point source and fully resolved component model as seen in Figure\,\ref{wr31a_1}. On close inspection however, similar to WR 16, it shows spectral features in |V|, indicating resolved line-emitting regions (see Sec.\,\ref{sec:LER}). In particular, the emission lines around 2.06 and 2.16 {\textmu}m showed significant features in |V| data. 

We modeled WR 31a using two components: an unresolved component for the continuum; and a partially resolved component for the 2.06 and 2.16 {\textmu}m spectral lines. The partially resolved component was modeled as a Gaussian with an fwhm of $0.91\pm0.07$ mas. Figure\,\ref{wr31a_2} shows the best-fit model for WR 31a.

\textbf{WR 42} is a WC7 classified as a spectroscopic binary with an O7V companion and a 7.89-day orbital period \citep{1981Davis}. The GRAVITY data for WR 42 resembled a point source with a slight contribution from a fully resolved component (see Fig.\,\ref{wr42_1}. The |V| and DPHI data, however, did show some spectral features corresponding to the emission lines in the NFLUX data (see Fig.\,\ref{wr42_2}). 

Similar to WR 9, this behaviour can be explained by the presence of a very close companion that is unresolved in a |V|-T3PHI binary search, but slightly resolved in DPHI. This is consistent with the already confirmed binary nature of WR 42, but cannot be reliably used to determine the position of the companion.

\textbf{WR 47} is a WN6 classified as a spectroscopic binary with an O5V companion and a 6.2-day orbital period, also displaying the colliding wind phenomenon \citep{2012Fahed}. The GRAVITY data for WR 47 were modeled sufficiently well with a point source and a fully resolved component (see Fig.\,\ref{wr47_1}).

With a more detailed approach, we found that the DPHI data show a spectral feature. The data are considerably noisy, and we tried an approximate model similar to WR 42, which could explain the DPHI feature. Figure\,\ref{wr47_2} shows the spectro-interferometric data for WR 47, along with an approximate model over-plotted on the data.

\textbf{WR 48} is discussed in detail in the main text of the paper. Figures\,\ref{wr48_1} and \ref{wr48_2} show the |V|-T3PHI and complete GRAVITY data respectively for WR 48, along with the best-fit model. Despite the data being noisy (and, for some baselines, absent), we found a good fit with a model consisting of two point sources and a negligible fully resolved component. A more detailed analysis of WR 48 will be covered in Gosset et al. (in prep.).

\textbf{WR 52} is a WC4 classified as a single star \citep{2019Sander}. The GRAVITY data were consistent with a point source as shown in Figure\,\ref{wr52}, with some contribution from a fully resolved component. The |V| data do show some features, although no simple model was found to perfectly describe them. Nevertheless, a wide binary was effectively ruled out after a thorough grid search.

\textbf{WR 55} is a WN7 classified as a single star \citep{2019Hamann}. The interferometric data were best modeled using a point source primary and a fully resolved diffuse component, as shown in Figure\,\ref{wr55}.

\textbf{WR57} is a WC8 classified as a single star \citep{2019Sander}. GRAVITY data for WR 57, |V| data in particular, reveal peculiar features. Figure\,\ref{wr57_1} shows the |V|-T3PHI data modeled using a point source and a fully resolved component, the latter contributing to a much larger extent than other WRs. The closure and differential phases were both, however, flat at zero. Figure\,\ref{wr57_2} shows the complete interferometric data with the model including spectral lines of the WR. 

WR 113 (later in this Appendix) was the only other WR in our sample that approximately resembles the |V| data for WR 57. The former is known to produce dust \citep{1975Cohen}, while the latter is not. More observations to improve the SNR, or even perform image reconstruction, were deemed necessary to understand the true origin of the significant fully resolved component in WR 57.

\textbf{WR 66} is a WN8h classified as a single star \citep{2019Hamann}. It is known to show a 3.5-4 hr periodic photometric variability \citep{1995Antokhin,1996Rauw,2021bNaze} and faint X-ray emission \citep{2021Skinner}. The source of variability is not well understood, with the presence of a compact companion being one possibility among few others. The GRAVITY data for WR 66 were best modeled by a point source and a fully resolved component (see Fig.\,\ref{wr66}).

\textbf{WR 75} is a WN6 classified as a single star \citep{2019Hamann}. The GRAVITY data were consistent with a point source as shown in Figure\,\ref{wr75}, with some contribution from a fully resolved component.

\textbf{WR 78} is a WN7h classified as a single star \citep{2019Sander}, also known to emit X-rays \citep{2012Skinner}. The GRAVITY data for WR 78 show features of spatially resolved line-emitting region for certain spectral lines, as described in detail in the main text of the paper. In context of companion search using |V|-T3PHI data, WR 78 was best modeled as a point source with no contribution from a fully resolved component (see Fig.\,\ref{wr78}).

\textbf{WR79} is a WC7 classified as a spectroscopic binary with an O5-8 companion and a 8.9-day orbital period, also showing the colliding wind phenomenon \citep{1997Luehrs}. The |V|-T3PHI data were largely consistent with a point source and a slightly contributing fully resolved component, as shown in Figure\,\ref{wr79_1}. On closer inspection of the complete spectro-interferometric data, we found small spectral features in DPHI as well as |V| corresponding to the WR spectral lines in NFLUX. This can be explained by a close binary as described in Section\,\ref{sec:DPHI}. Figure\,\ref{wr79_2} shows the best-fit model for WR 79 consisting of a very close binary and a slight fully resolved component. Similar to other WRs of this kind, the position of the companion is not constrained.

\textbf{WR 79a} is a WN9ha/O8:Iafpe classified as a single star \citep{1997CrowtherBohannanOf}, also known to emit X-rays \citep{2010Skinner}. The visibilities are more than unity due to poor visibility calibration. The best-fit model for GRAVITY data of WR 79a was an unresolved point source by itself (Fig.\,\ref{wr79a}). A binary search with or without the bad |V| data did not fit the data well, and we classified it as a single star.

\textbf{WR 79b} is a WN9ha/O6:Iafpe classified as a single star \citep{2014Sota}. Although the GRAVITY data are noisy, they were best explained with an unresolved point source and a fully resolved component (Fig.\,\ref{wr79b}). The visibility residuals seem to contain some signal, but similar to WR 15, it can be attributed to the resolved component or poor visibility calibration.

\textbf{WR 81} is a WC9 classified as a single star \citep{2019Sander}. The GRAVITY data for WR 81, particularly |V|-T3PHI, were well explained by a point source and a fully resolved component (see Fig.\,\ref{wr81_1}). On a closer look, the |V| data have a prominent spectral feature corresponding to the 2.06 {\textmu}m line of the WR. Such a feature was only seen in WR 81 and WR 92 (see below). We modeled it by including the same spectral features in the fully resolved component (see Figure\,\ref{wr81_2}), something that explains the data but is currently non-trivial to reconcile with any physical interpretation. Higher S/N data in the future can help constrain the properties of the fully resolved component in greater detail.

\textbf{WR 85} is a WN6h classified as a single star \citep{2019Hamann}. The interferometric data were best fit by a point source and a fully resolved component (see Fig.\,\ref{wr85}). The |V| data show some residual signal which can be attributed to some structure in the fully resolved component.

\textbf{WR 87} is a WN7h+abs classified as a single star, where abs stands for an absorption component \citep{1996Smith}. It is also a known X-ray source, although the origin of X-rays is uncertain \citep{2013Naze}. The GRAVITY data for WR 87 were best fit by a point source and a slight fully resolved component, as shown in Figure\,\ref{wr87}.

\textbf{WR 89} is discussed in detail in the main text of the paper. Figure\,\ref{wr89_2} shows the complete spectro-interferometric data for WR 89 along with the best-fit model.

\textbf{WR 92} is a WC9 classified as a single star \citep{2019Sander}. The |V|-T3PHI data for WR 92 were best modeled using a point source and a fully resolved component (see Fig.\,\ref{wr92_1}). The |V| data, however, have a prominent spectral feature corresponding to the 2.06 {\textmu}m line of the WR similar to WR 81, and were also modeled similarly (see Figure\,\ref{wr92_2}).

\textbf{WR 93} is discussed in detail in the main text of the paper. Figures\,\ref{wr93_1} and\,\ref{wr93_2} show the |V|-T3PHI and complete GRAVITY data respectively for WR 93, along with the best-fit model.

\textbf{WR 97} is a WN5b classified as a spectroscopic binary with an O7 companion and a 12.6-day orbital period \citep{1996Smith}, also known to exhibit the colliding wind phenomenon \citep{2021Naze}. The GRAVITY data were best modeled by a central point source and a fully resolved component, as shown in Figure\,\ref{wr97}.

\textbf{WR 98} is a WN8/WC7 classified as a spectroscopic binary with an O8-9 companion and a 47.8-day period \citep{2003Gamen,2014Faherty}. It is also known to emit in the radio band, potentially due to colliding winds \citep{2011Montes}. The |V|-T3PHI data for WR 98 could be explained well by a point source model, with only a slight contribution from a fully resolved component (see Fig.\,\ref{wr98_2}). Including DPHI and NFLUX does, however, reveal spectral features indicating the presence of a close binary, as described in Section\,\ref{sec:DPHI}. Figure\,\ref{fig:wr98_1} shows the complete spectro-interferometric data along with the best-fit model.

\textbf{WR 108} is a WN9ha classified as a single star \cite{1996Smith}. The GRAVITY data for WR 108 were best fit by a point source and a fully resolved component, as shown in Figure\,\ref{wr108}.

\textbf{WR 110} is a WN5-6b classified as a single star, potentially hosting corotating interaction regions in its wind \citep{2011Chene}. The GRAVITY data for WR 110 were consistent with a point source model and a negligible fully resolved component (see Fig.\,\ref{wr110_1}). {The |V| data, however, show features of a spatially resolved line-emitting region for multiple spectral lines, similar to WR 78 described in the main text of the paper. In this case however, the spectral features could not be constrained as simply as other similar cases.} Figure\,\ref{wr110_2} shows the complete spectro-interferometric data for WR 110, along with an approximate model similar to WR 78. More observations with higher SNR were deemed necessary to better constrain its properties.

\textbf{WR 111} is a WC5 classified as a single star \citep{2019Sander}. The GRAVITY data were best fit by a point source and a slight fully resolved component, as shown in Figure\,\ref{wr111}.

\textbf{WR 113} is a WC8d classified as a spectroscopic binary with an O8-9IV companion and a 29.7-day orbital period \citep{1949Gapo}, also known to produce dust \citep{2018Hill}. \citet{2022Shara} resolved a wide companion candidate at a separation of 1.16 arcseconds, although whether or not it is gravitationally bound the central binary is not known.

\citet{2007Monnier} studied the near-infrared sizes of dusty WRs including WR 113, finding its $K$-band ``size'' to be $27\pm3$ mas. In our GRAVITY data, we see strong evidence of this dusty disk reflecting in the strong |V| offset at all baselines (Figure\,\ref{wr113}). The DPHI data shows spectral features, although in this case they could potentially arise due to the presence of a dusty disk. Constraining the size of the dusty disk was not possible based on the current data. Nevertheless, our best-fit model WR 113 was a central point source and a slightly off-center gaussian disk with an fwhm of around 15 mas (Figure\,\ref{wr113_2}). To constrain the latter at a higher precision, more GRAVITY data would be necessary.

\textbf{WR 114} is a WC5 classified as a single star \citep{2019Sander}. The GRAVITY data for WR 114 were best modeled with a point source and a fully resolved component (see Fig.\,\ref{wr114}).

\textbf{WR 115} is discussed in detail in the main text of the paper. Figures\,\ref{wr115_1} and\,\ref{wr115_2} show the |V|-T3PHI and complete GRAVITY data respectively for WR 115, along with the best-fit model.

\clearpage

\begin{table*}
    \caption[]{Summary of parametric modeling of all 39 WRs in our GRAVITY sample using the PMOIRED tool.}
    \label{tab:summ}
    \renewcommand{\arraystretch}{1.15}
\begin{tabular*}{\textwidth}{@{\extracolsep{\fill}} ccccccc }
\hline
WR \#  & Spectral Type  
& \begin{tabular}{@{}c@{}}Point Source \\ fwhm \\ (mas)\end{tabular}
& \begin{tabular}{@{}c@{}}Fully Resolved \\ Component \\ (flux \% w.r.t WR)\end{tabular}
& \begin{tabular}{@{}c@{}}Resolved \\ Line-emitting \\ Region (mas)\end{tabular}
& \begin{tabular}{@{}c@{}}Wide Binary \\ Separation \\ (mas)\end{tabular} 
& \begin{tabular}{@{}c@{}}Data \\ Quality \end{tabular} \\

\hline

24 & WN6ha-w & $0.12\pm0.06$ & $2\pm1$ & - & - & Average \\

85 & WN6h & $0.46\pm0.19$ & $18\pm2$ & - & - & Average \\

22 & WN7h + O9III-V & $0.12\pm0.02$ & $0$ & - & - & Average \\

78 & WN7h & $0.002$ (fixed) & $0$ & $0.51\pm0.02$ & - & Good \\

87 & WN7h+abs & $0.00$ & $10\pm1$ & - & - & Average \\

12 & WN8h + OB & $0.74\pm0.05$ & $18\pm1$ & - & - & Poor \\

16 & WN8h & $0.002$ (fixed) & $26\pm9$ & $0.98\pm0.11$ & - & Poor \\

66 & WN8h & $0.78\pm0.08$ & $21\pm2$ & - & - & Average \\

89 & WN8h+abs & $0.65\pm0.32$ & $15\pm7$ & - & $5.18\pm0.10$ & Good \\

79a & WN9ha/O8:Iafpe & 0.00 & 0 & - & - & Poor \\

79b & WN9ha/O6:Iafpe & 0.00 & $16\pm1$ & - & - & Poor \\

108 & WN9ha & $0.49\pm0.10$ & $6\pm1$ & - & - & Poor \\

31a & WN11h & $0.002$ (fixed) & $23\pm3$ & $0.91\pm0.07$ & - & Average \\

\hline

47 & WN6 + O5V & $0.53\pm0.05$ & $12\pm1$ & - & - & Average \\

75 & WN6 & $0.51\pm0.05$ & $13\pm1$ & - & - & Average \\

115 & WN6-w & $1.02\pm0.16$ & $31\pm4$ & - & $198.94\pm0.03$ & Good \\

8 & WN6 & $0.81\pm0.07$ & $24\pm2$ & - & - & Average \\

55 & WN7 & $0.63\pm0.02$ & $12\pm1$ & - & - & Good \\

98 & WN8 + O8-9 & $0.30\pm0.06$ & $7\pm1$ & - & - & Good \\

\hline

18 & WN4 & $0.70\pm0.02$ & $16\pm1$ & - & - & Good \\

31 & WN4 + O8V & $0.64\pm0.15$ & $32\pm4$ & - & - & Average \\

21 & WN5 + O4-6 & $0.70\pm0.06$ & $20\pm2$ & - & - & Average \\

97 & WN5b + O7 & $0.75\pm0.04$ & $18\pm1$ & - & - & Average \\

110 & WN5-6b & $0.002$ (fixed) & $0$ & Not Constrained & - & Good \\

\hline

52 & WC4 & $0.31\pm0.15$ & $19\pm2$ & - & - & Average \\

9 & WC5 + O7 & $0.51\pm0.07$ & $35\pm8$ & - & - & Average \\

111 & WC5 & $0.12\pm0.10$ & $9\pm1$ & - & - & Average \\

114 & WC5 & $0.72\pm0.06$ & $14\pm1$ & - & - & Average \\

15 & WC6 & $0.70\pm0.06$ & $13\pm1$ & - & - & Good \\

23 & WC6 & $0.45\pm0.01$ & $5\pm1$ & - & - & Average \\

48 & WC6 + O6-7V & $0.57\pm0.31$ & $11\pm7$ & - & $12.66\pm0.06$ & Poor \\

14 & WC7 & $0.38\pm0.04$ & $8\pm1$ & - & - & Average \\

42 & WC7 + O7V & $0.30\pm0.08$ & $12\pm1$ & - & - & Average \\

79 & WC7 + O5-8 & 0.00 & 0 & - & - & Average \\

93 & WC7 + O7-9 & $0.62\pm0.09$ & $3\pm1$ & - & $14.33\pm0.03$ & Good \\

57 & WC8 & $0.40\pm0.38$ & $123\pm13$ & - & - & Good \\

113 & WC8d + O8-9IV & $1.32\pm0.03$ & $335\pm7^\ast$ & - & - & Good \\

81 & WC9 & $0.62\pm0.14$ & $21\pm2$ & - & - & Good \\

92 & WC9 & $0.20\pm0.17$ & $36\pm3$ & - & - & Average \\

\hline
\end{tabular*}

\medskip
\textbf{Notes:} Shown are the WR number (column 1), spectral type (column 2), fwhm of the point source used to model the WR (column 3) and contribution of the fully resolved component (column 4). For WRs with resolved line-emitting regions, column 5 provides the fwhm of the resolved component, while column 6 provides projected separations for resolved wide binaries. Lastly, column 7 gives a purely qualitative evaluation of GRAVITY data for each WR determined on case-by-case inspection. ($\ast$ In case of WR 113, a Gaussian disk was used instead of the fully resolved component)

   \end{table*}

\end{appendix}

\end{document}